\documentclass[a4paper,12pt]{book}

\usepackage[utf8]{inputenc}
\usepackage[greek,english]{babel}
 
\usepackage{amsmath}
\usepackage{array}
\usepackage{amssymb}
\usepackage{latexsym}
\usepackage{graphicx}
\usepackage{epsfig}
\usepackage{color}
\usepackage{alphabeta}

\usepackage{cite}

\graphicspath{{images/}}

\usepackage[a4paper]{geometry}
\allowdisplaybreaks[1] 

\usepackage{afterpage}  

\newcommand\blankpage{%
    \null
    \thispagestyle{empty}%
    \addtocounter{page}{-1}%
    \newpage}

\usepackage{fancyhdr}

\usepackage[colorlinks=true,
            linkcolor=blue,
            urlcolor=cyan,
            citecolor=red,
            bookmarksnumbered=true]{hyperref}

\newcommand{\nocontentsline}[3]{}
\newcommand{\tocless}[2]{\bgroup\let\addcontentsline=\nocontentsline#1{#2}\egroup}

\def\beq{\begin{equation}}
\def\eeq{\end{equation}}
\def\bea{\begin{eqnarray}}
\def\eea{\end{eqnarray}}

\def\a{\alpha}
\def\b{\beta}
\def\r{\rho}
\def\d{\delta}
\def\m{\mu}
\def\n{\nu}
\def\h{\eta}
\def\s{\sigma}

\def\l{\lambda}
\def\L{\Lambda}
\def\g{\gamma}
\def\G{\Gamma}
\def\F{\Phi}
\def\f{\phi}
\def\p{\pi}
\def\saa{\sqrt{\alpha}}
\def\sign{\text{sign}}

\def\mE{\mathcal{E}}
\def\mJ{\mathcal{J}}
\def\mF{\mathcal{F}}
\def\mK{\mathcal{K}}

\def\hh{\tilde{h}}

\def\da{\partial^2}
\def\aa{\partial}

\def\2{\;\;}
\def\4{\;\;\;\;}

\def\tela{\textlatin}

\def\nn{\nonumber}

\begin{document}

\begin{titlepage}
\centering

 \includegraphics[width=1cm,height=2cm]{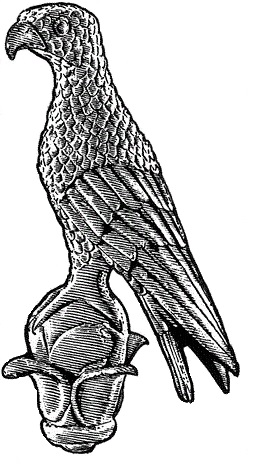}\\ [0.5cm]

 {\bf UNIVERSITY OF IOANNINA}\\
\bf SCHOOL OF NATURAL SCIENCES\\
  PHYSICS DEPARTMENT \\[3.5cm]

{\LARGE \bf  Black holes and wormholes in the\\[2mm] Einstein-scalar-Gauss-Bonnet generalized theories of gravity } \\[3.5cm]

\bf {Athanasios Bakopoulos } \\[2.5cm]

PH.D. THESIS\\[2.5cm]

\vfill

{ IOANNINA 2020}

  \afterpage{\blankpage}
  \clearpage
  \thispagestyle{empty}

\includegraphics[width=1cm,height=2cm]{pouli.jpg}\\ [0.5cm]

 {\bf UNIVERSITY OF IOANNINA}\\
\bf SCHOOL OF NATURAL SCIENCES\\
  PHYSICS DEPARTMENT \\[3.5cm]

{\LARGE \bf  Black holes and wormholes in the\\[2mm] Einstein-scalar-Gauss-Bonnet generalized theories of gravity } \\[3.5cm]

\bf {Athanasios Bakopoulos } \\[2.5cm]

PH.D. THESIS\\[2.5cm]

\vfill

{ IOANNINA 2020}

 \afterpage{\blankpage} 
\end{titlepage}


\clearpage
\thispagestyle{empty}


\clearpage
\thispagestyle{empty}

\noindent\textbf{This work is an updated version of my Ph.D. dissertation. The version I defend in the university of Ioannina my be found here:}\\  \href{http://www.physics.uoi.gr/el/node/1079}{http://www.physics.uoi.gr/el/node/1079}
  
  \vspace{2cm}
\noindent This research is co-financed by Greece and the European Union (European Social Fund- ESF) through the Operational Programme «Human Resources Development, Education and Lifelong Learning» in the context of the project “Strengthening Human Resources Research Potential via Doctorate Research – 2nd Cycle” (MIS-5000432), implemented by the State Scholarships Foundation (ΙΚΥ).

\noindent  Part of this work is implemented through the Operational Program “Human Resources Development, Education and Lifelong Learning” and is co-financed by the European Union (European Social Fund) and Greek national funds (MIS code: 5006022)

\vspace{5cm}

\includegraphics[height=.1\textheight, angle =0]{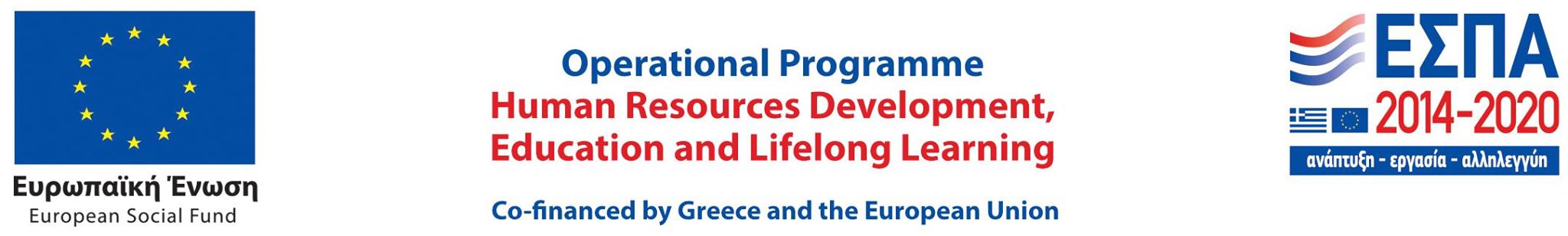}

\afterpage{\blankpage}

\clearpage
\thispagestyle{empty}


\clearpage
\thispagestyle{empty}
\noindent{\textbf{Three-member advisory committee}:
\begin{itemize}
\item P.Kanti,  Professor, Physics Department, University of Ioannina (Supervisor).
\item K. Tamvakis, Professor, Physics Department, University of Ioannina.
\item L. Perivolaropoulos, Professor, Physics Department, University of Ioannina. 
\end{itemize}
\vspace*{0.5cm}
 \textbf{Seven-member PhD thesis examination committee}:
\begin{itemize}
\item P. Kanti,  Professor, Physics Department, University of Ioannina (Supervisor).
\item K. Tamvakis, Professor, Physics Department, University of Ioannina.
\item L. Perivolaropoulos, Professor, Physics Department, University of Ioannina. 
\item G. Leontaris, Professor, Physics Department, University of Ioannina.  
\item I. Rizos, Professor, Physics Department, University of Ioannina.  
\item A. Dedes, Professor, Physics Department, University of Ioannina.  
\item C. Charmousis, Professor \tela{Laboratoire de Physique Theorique, CNRS, Universite Paris-Sud, Universite Paris-Saclay, Orsay}, France.
\end{itemize} 
} 
\afterpage{\blankpage}

\clearpage
\thispagestyle{empty}
\vspace*{4cm}
\hfill{ To my Family}
\afterpage{\blankpage}

\clearpage
\thispagestyle{empty}
\vspace*{4cm}
   
\begin{center}
{\bf Acknowledgements}\\[1cm]
\end{center}

First of all, I thank my supervisor Professor Panagiota Kanti for her continuous guidance and support all  these  years.  I  would  also  like  to  thank  the  rest  of  my  collaborators  Georgios Antoniou, Jutta Kunz, Burkhard Kleihaus and Nikolaos Pappas for the fruitful collaboration. I thank the members of  the three member advisory comittee  (P. Kanti, K. Tamvakis and L. Perivolaropoulos) for their guidance.   

I would also thank the Greek State Scholarships Foundation (ΙΚΥ) for financial support the last two years of my Ph.D. studies. 

Last but not least, many thanks to my family and friends for their love and support

 \clearpage
\thispagestyle{empty}


\pagestyle{fancy}
\fancyhead{}
\fancyhead[LO]{Contents}
\fancyhead[LE]{Contents}

\tableofcontents

\clearpage
\thispagestyle{empty}


\latintext
\chapter*{Publications List}
\label{puu1}
\addcontentsline{toc}{chapter}{\nameref{puu1}}
\fancyhead{}
\fancyhead[LO]{}
\fancyhead[LE]{Publications List}

Within  the  context  of  my  Ph.D.  dissertation,  four  peer-reviewed  articles  have  been  produced.   They  are listed here for reference in descending chronological order:

\begin{enumerate}

\item \textbf{Evasion of No-Hair Theorems and Novel Black-Hole Solutions in Gauss-Bonnet Theories}, Georgios Antoniou, Athanasios Bakopoulos and Panagiota Kanti \cite{ABK1}

\item \textbf{Black-Hole Solutions with Scalar Hair in Einstein-Scalar-Gauss-Bonnet Theories}, Georgios Antoniou, Athanasios Bakopoulos and Panagiota Kanti \cite{ABK2} 

\item \textbf{Novel Black-Hole Solutions in Einstein-Scalar-Gauss-Bonnet Theories with a Cosmological Constant}, Athanasios Bakopoulos, Georgios Antoniou, and Panagiota Kanti \cite{ABK3}

\item \textbf{Novel Einstein-scalar-Gauss-Bonnet wormholes without exotic matter}, Georgios Antoniou, Athanasios Bakopoulos, Panagiota Kanti, Burkhard Kleihaus and Jutta Kunz \cite{ABK4}

\end{enumerate}
\noindent
During the early and final stages of my Ph.D. studies, I also worked in parallel on the following three articles which do not consist part of my Ph.D. dissertation:

\begin{enumerate}
  \setcounter{enumi}{4}

\item \textbf{Novel Ansatzes and Scalar Quantities in Gravito-Electromagnetism},\\ Athanasios Bakopoulos and Panagiota Kanti \cite{bakopgem2}

\item \textbf{Existence of solutions with a horizon in pure scalar-Gauss-Bonnet theories}, Athanasios Bakopoulos, Panagiota Kanti and Nikolaos Pappas \cite{bakoppgb}

\item \textbf{Large and ultracompact Gauss-Bonnet black holes with a self-interacting scalar field}, Athanasios Bakopoulos, Panagiota Kanti and Nikolaos Pappas \cite{bakoppk}

\end{enumerate}

\pagebreak
\noindent
Finally, during the period of my Ph.D. studies we published three articles in conference proceedings:

\begin{enumerate}
 \setcounter{enumi}{7}

\item \textbf{Novel black hole solutions with scalar hair in Einstein-scalar-Gauss-Bonnet theories}, Athanasios Bakopoulos, Georgios Antoniou and Panagiota Kanti \cite{bak1}

\item \textbf{Scalar-Gauss-Bonnet Theories: Evasion of No-Hair Theorems and novel black-hole solutions}, Panagiota Kanti, Athanasios Bakopoulos and Nikolaos Pappas \cite{bak2}

\item \textbf{Scalar-Gauss-Bonnet theories: evasion of no-hair theorems and novel  black-hole solutions}, Panagiota Kanti, Athanasios Bakopoulos and Nikolaos Pappas \cite{bak3}

\end{enumerate}

\clearpage
\thispagestyle{empty}

\chapter*{ }
\addcontentsline{toc}{chapter}{Abstract}
\begin{center}
{\bf Abstract}\\[1cm]
\end{center}

\noindent
In this Ph.D. dissertation we study the emergence of black-hole and wormhole solutions in the framework of the Einstein-scalar-Gauss-Bonnet (EsGB) theory.
Particularly we study a family of theories where the coupling function $f(\f)$ between the scalar field of the theory and the quadratic Gauss-Bonnet gravitational term has an arbitrary form. At first, we analytically derive that the aforementioned family of theories may evade the constraints imposed by Bekenstein's No-Scalar Hair theorems and new solutions for black holes may be found. Then, using numerical integration methods we find solutions for black holes for many different forms of the coupling function. Also we derive their physical characteristics namely their mass,   scalar charge, horizon area and entropy as well. Subsequently, by introducing a cosmological constant in the theory we investigate the existence of novel black-hole solutions. Specifically, assuming that the cosmological constant may be positive or negative we find numerical solutions which are asymptotically de Sitter or anti-de Sitter. In addition, as in the case of the asymptotically flat black holes, for each case we derive their physical characteristics. Finally, in the framework of the EsGB theory we derive novel wormhole solutions. The Gauss-Bonnet wormholes are traversable, may have a single or a double throat and do not demand the existence of exotic matter. 


\clearpage
\thispagestyle{empty}

\latintext
\chapter*{Preface}
\label{1}
\addcontentsline{toc}{chapter}{\nameref{1}}
\fancyhead{}
\fancyhead[LO]{}
\fancyhead[LE]{Preface}

As the ultimate theory of Quantum Gravity, that would robustly describe gravitational interactions at high energies and facilitate their unification with the other forces, is still eluding us, the interest in generalized gravitational theories remains unabated in the scientific literature.  The construction of generalised gravitational theories, with the inclusion of extra fields or higher-curvature terms in the action, has attracted an enormous interest during the last decades \cite{Stelle,General1,General2}. The main reason is that these theories may provide the framework  in the  context of which several problems of the traditional General Relativity may be resolved. Therefore, in the context of these modified gravitational theories, different aspects of gravity, from black-hole solutions to  cosmological solutions, have been re-addressed and, in several occasions, shown to lead to novel, interesting solutions. 

In this spirit, generalised gravitational theories containing scalar fields were among the
first to be studied. However, the quest for novel black-hole solutions -- beyond the three
well-known families of GR -- was abruptly stopped when the no-hair theorem was
formulated \cite{NH-scalar1,NH-scalar2}, that forbade the existence of a static solution of this form
with a non-trivial scalar field associated with it. Nevertheless, counter-examples appeared
in the years that followed and included black holes with Yang-Mills \cite{YM1,YM2,YM3,YM4}, Skyrme fields
\cite{Skyrmions1,Skyrmions2} or with a conformal coupling to gravity \cite{Conformal1,Conformal2}. A novel
formulation of the no-hair theorem was proposed in 1995 \cite{Bekenstein} but
this was, too, evaded within a year with the discovery of the dilatonic black holes
found in the context of the Einstein-Dilaton-Gauss-Bonnet theory \cite{DBH1,DBH2}
(for some earlier studies that paved the way, see \cite{Gibbons, Callan, Campbell1,Campbell2,
Mignemi, Kanti1995}). The coloured black holes were found next in the context of the
same theory completed by the presence of a Yang-Mills field \cite{Torii, KT}, and
higher-dimensional \cite{Guo1,Guo2,Guo3,Guo4} or rotating versions \cite{Kleihaus1, Kleihaus2, Pani1, Pani2, Herdeiro, Ayzenberg,Maselli:2015tta,Kleihaus:2014lba,Kunz,Ayzen,Pani:2009wy}
were also constructed (for a number of interesting reviews on the topic, see
\cite{Win-review, Charmousis, Herdeiro-review, Blazquez}). This second wave of black-hole solutions were derived in the context of theories inspired
by superstring theory \cite{Metsaev}. During the last decade, though, the construction
of generalised gravitational theories was significantly enlarged via the revival of the 
Horndeski \cite{Horndeski} and Galileon \cite{Galileon} theories. Accordingly, novel
formulations of the no-hair theorems were proposed that covered the case of standard
scalar-tensor theories \cite{SF} and Galileon fields \cite{HN}. However, these recent forms
were also evaded \cite{SZ} and concrete black-hole solutions were constructed
\cite{Babichev, Benkel1,Benkel2,Benkel3, Yunes2011}. 

 Motivated by string theory with the dilaton as the scalar field,
the Einstein-dilaton-Gauss-Bonnet (EdGB) theory features an exponential
coupling between the scalar field and the GB term
\cite{Zwiebach:1985uq,Gross:1986mw,Metsaev}.
Black-hole solutions arising in the context of the EdGB theory differ
from the Schwarzschild or Kerr black holes since they possess a non-trivial
dilaton field and thus carry dilaton hair 
\cite{DBH1, DBH2,Torii,KT,Guo1,Guo2,Guo3,Guo4,Pani:2009wy,Pani1,Pani2,Kleihaus1,Ayzen,Ayzenberg, Herdeiro,Maselli:2015tta,Kleihaus:2014lba,Kleihaus2,Kunz}.
The extended Einstein-scalar-Gauss-Bonnet (EsGB) theories, where the coupling function $f(\f)$ may have any other form has attracted recently considerably attention \cite{SZ,Benkel1,ABK1,Doneva,Silva,ABK2,Witek2018,Minamitsuji:2018xde,Don-Kunz1,Silva:2018qhn,ABK3,Brihaye:2018grv,Macedo:2019sem,Doneva:2019vuh,Myung:2019wvb,Cunha:2019dwb,Brihaye:2019dck}. The Einstein-scalar-Gauss-Bonnet theory represent interesting alternative theories of gravity. They belong to the class of {\sl quadratic gravitational theories}  that contain higher-curvature gravitational terms.  These terms are treated as small deformations that nevertheless complete Einstein's General Relativity and may modify its predictions at regimes of strong gravity.  The resulting field equations are of second order, avoiding Ostrogradski instability and ghosts  \cite{Horndeski,Charmousis:2011bf,Kobayashi:2011nu}. In addition, this quadratic theory has so far survived the constraints set by the detection of gravitational waves emitted during the binary mergers, when the coupling function allows to set the scalar field to zero in the cosmological context, and thus lead to the same solutions as the standard cosmological $\Lambda$CDM model \cite{Sakstein:2017xjx}. The study of the types of solutions that this theory admits  is therefore of paramount importance.

In this thesis we will focus on the derivation of new local solutions in the framework of the Einstein-scalar-Gauss-Bonnet theory. A few years ago, we demonstrated \cite{ABK1, ABK2} that for any form of the coupling function, the Einstein-scalar-Gauss-Bonnet theory admits novel black-hole solutions
with a non-trivial scalar hair. Almost simultaneously, two independent groups \cite{Doneva, Silva} --examining specific forms of the coupling function-- proposed that in the framework of the EsGB theory the black holes may be spontaneously scalarized. These spontaneously scalarized solutions  are included in the analysis of \cite{ABK1} as special cases since it includes both the spontaneously and the induced (naturally) scalarized solutions.  In a general theoretical argument, that we presented in
\cite{ABK1, ABK2}, it was shown that the presence of the Gauss-Bonnet term was of paramount importance
for the evasion of the novel no-hair theorem \cite{Bekenstein}. In addition, the exact form
of the coupling function $f(\phi)$ between the scalar field and the GB term played no
significant role for the emergence of the solutions: as long as the first derivative of the
scalar field $\phi_h'$ at the horizon obeyed a specific constraint, an asymptotic solution
describing a regular black-hole horizon with a non-trivial scalar field could always be
constructed. Employing, then, several different forms of the coupling function $f(\phi)$,
a large number of asymptotically-flat black-hole solutions with scalar hair were determined
\cite{ABK1, ABK2}. Additional studies presenting novel black holes or compact objects in 
generalised gravitational theories have appeared \cite{Charmousis1a, Correa, Doneva-NS,
Motohashi1, Motohashi2, Radu1, Radu2, Radu3, Doneva-Papa, Butler, Danila, Stetsko} as well as further studies of the properties
of these novel solutions \cite{Ayzen, Dolan, Kunz, Chakra1,Chakra2, Tatter, Mukherjee, Chakra, Berti,
Brihaye, Prabhu, Myung, Don-Kunz1,Don-Kunz2, Benkel2018, Iorio, Ovalle1,Ovalle2, Barack, Gao, Lee, Witek2018, Moto,bbh1,bbh2,bbh3,bbh4,bbh5,bbh6,bbh7,bbh8,bbh9,bbh10,bbh11,bbh12,
bbh13,bbh14,bbh15,bbh16,bbh17,bbh18,bbh19,bbh20,bbh21,bbh22,bbh23,bbh24,
bbh25,bbh26,bbh27,bbh28,bbh29,bbh30,bbh31,bbh32,bbh33,bbh34,
bbh35,bbh36,bbh37,bbh38,bbh39,bbh40,bbh41,bbh42,bbh43,bbh44,bbh45,
bbh46,bbh47,bbh48,bbh49,bbh50,bbh51,bbh52,bbh53,bbh54,bbh55,bbh56,
bbh57,bbh58,bbh59,bbh60,bbh61,bbh62,bbh63,bbh64,bbh65,bbh66,bbh67,
bbh68,bbh69,bbh70,bhh71,bhh72,bhh73,bhh74,bhh75,bhh76,bhh77,bhh78,
bhh79,bhh80,bhh81,bhh82,bhh83,bhh84,bhh85,bhh86,bhh87,bhh88,bhh89,
bhh90,bhh91,bhh92,bhh93,bhh94,bhh95,bhh96,bhh97,bhh98,bhh99,bhh100,
bhh101,bhh102,bhh103,bhh104}. 
Finally the stability of scalarized black holes of EsGB theories 
has been addressed in detail by
analyzing their radial perturbations and revealing a distinct dependence on
the coupling function \cite{Don-Kunz1,Silva:2018qhn}. 

The last decades, the existence of black-hole solutions in the context of   scalar-tensor theories has attracted the interest of many scientists and many new solutions have been produced \cite{Narita, Winstanley-Nohair1,Winstanley-Nohair2, Bhatta2007, Bardoux, Kazanas}. While in the presence of a negative cosmological constant a substantial number of solutions --with an asymptotically Anti-de-Sitter behaviour-- have been found
 \cite{Martinez1, Martinez2, Martinez3, Radu-Win, Anabalon, Hosler,
Kolyvaris1, Kolyvaris2, Ohta1, Ohta2, Ohta3, Ohta4, Saenz, Caldarelli, Gonzalez, Gaete1, Gaete2, Giribet, BenAchour}, in the case of a  positive cosmological constant, the existing studies predominantly excluded the presence of a regular, black-hole solution \cite{Martinez-deSitter,Harper}. 
In this direction, we extended our previous analyses \cite{ABK1,ABK2}, by introducing in our theory a cosmological
constant $\Lambda$, either positive or negative \cite{ABK3}. Our aim was to investigate whether, and under which conditions, new solutions for black holes may produced in the presence of $\L$. 
  In this work we performed a comprehensive study of the existence of black-hole solutions  in 
the context of the Einstein-scalar-Gauss-Bonnet theory. As in our previous works, we keep arbitrary the form of the coupling function $f(\f)$ in order to keep our analysis as general as possible and we 
%
look for regular black-hole solutions with a non-trivial scalar hair and an (Anti-)de-Sitter expansion at infinity. 
 We  thus repeated our analytical calculations both in
the small and large-$r$ regimes to examine how the presence of $\Lambda$ affects
the asymptotic solutions both near and far away from the black-hole horizon. 
For any form of the coupling function and for both signs of the cosmological constant, our theoretical analysis indicates the existence of regular approximate solutions in both asymptotic regions. Using numerical methods we constructed both asymptotically  de-Sitter and  anti-de-Sitter solutions. 
We presented a large number of novel black-hole
solutions with a regular black-hole horizon and a non-trivial scalar field. In the production of these solutions we used a variety of forms of the coupling function $f(\phi)$: exponential, polynomial
(even or odd), inverse polynomial (even or odd) and logarithmic. Then, we also examined
their physical properties such as the temperature, entropy, and horizon area. We also
investigated features of the asymptotic profile of the scalar field, namely its effective potential
and rate of change at large distances since this greatly differs from the asymptotically-flat case.

A particularly interesting property of all the black holes solutions found in the framework of the Einstein-scalar-Gauss-Bonnet theory is the presence of regions with negative {\sl effective} energy density near the horizon.  This
may attributed  to the presence of the higher-curvature Gauss-Bonnet term and is therefore
of purely gravitational nature \cite{DBH1,DBH2,Kanti:2011jz}. Actually, this violation of the energy conditions near the horizon leads to the evasion of the No-scalar-Hair theorems. On the other hand, is known that the traversable wormholes  in General Relativity also violate the energy conditions \cite{Visser:1995cc, Morris:1988cz}.
Consequently, the Einstein-scalar-Gauss-Bonnet theory may allow the existence for Lorentzian, traversable wormhole
solutions. 
But whereas in General Relativity the violation is typically 
achieved by a phantom field 
\cite{Ellis:1973yv,Ellis:1979bh,Bronnikov:1973fh,Kodama:1978dw,Kleihaus:2014dla,Chew:2016epf},
in Einstein-scalar-Gauss-Bonnet theories it is the effective stress-energy tensor 
that allows for this violation \cite{Kanti:2011jz,Kanti:2011yv}.
Thus, since the violation of the energy condition is due to the Gauss-Bonnet term, it should not  associated with the presence of exotic matter. The first traversable wormhole solutions in the framework of the Einstein-scalar-Gauss-Bonnet theory was found about ten years ago in the special case with an exponential coupling function \cite{Kanti:2011jz,Kanti:2011yv}. 
Since, for any form of the coupling function we may construct black hole solutions,
it is tempting to conjecture that the more general Einstein-scalar-Gauss-Bonnet theories should also
allow for traversable wormhole solutions. 
Therefore, in our last work \cite{ABK4} we  considered
a general class of Einstein-scalar-Gauss-Bonnet theories with an arbitrary coupling function for the
scalar field. We first readdressed the case of the exponential coupling function,
and show that the theory is even richer than previously thought, since
it features also wormhole solutions with the new property of  a double throat and an equator in
between. Then, we considered alternative forms of the scalar coupling function,
and demonstrate that the Einstein-scalar-Gauss-Bonnet theories 
always allow for traversable wormhole solutions,
featuring both single and double throats. The scalar field may vanish or be
finite at infinity, and it may have nodes. 
We also mapped the domain of existence 
of these wormholes in various exemplifications, evaluated their global charges
and throat areas and demonstrated that the throat remains open without the need for any exotic matter.

The outline of this thesis is as follows:   chapter \ref{2} will serve as an introductory chapter. In this we will present an introduction to General Relativity and to static black hole solutions. We then discuss  the existence of traversable solutions for wormholes in General Relativity as well. Also, we mention the need for generalized gravitational theories and focus on the scalar-tensor theories and especially on the Einstein-scalar-Gauss-Bonnet theory. Finally, we present a brief introduction to the No-scalar-Hair theorem. In chapter \ref{3} we examine the evasion of the No-scalar-Hair theorems in the framework of the Einstein-scalar-Gauss-Bonnet theory and we derive the constraints for the existence of black hole solutions. In chapter \ref{4} we derive numerically asymptotically flat black hole solutions in the framework of the EsGB theory. In chapter \ref{5} we investigate the existence of black hole solutions in the EsGB theory and in the presence of a cosmological constant. Also, we derive numerical solutions in the case of a negative cosmological constant. In chapter \ref{6} we produce numerical solutions for wormhole solutions in EsGB theory. These wormholes are traversable, do not need exotic mater and may be characterized by a single or even a double throat.
Finally, in chapter \ref{7} we present our conclusions.

\clearpage
\thispagestyle{empty}

%

\chapter{Introduction}\label{2}

\fancyhead{}
\fancyhead[LO]{\slshape\nouppercase{\rightmark}}
\fancyhead[LE]{\slshape\nouppercase{\leftmark}}
\fancyfoot{}
\fancyfoot[CE,CO]{\thepage}

\section{Geometry of General Relativity}\label{A0}

Einstein’s General theory of Relativity (GR) \cite{einstgr} is the theory that describes gravity. It is a geometrical theory in which the gravitational interactions are related with the geometry of spacetime. Einstein’s theory not only includes (as a limit) the traditional Newtonian gravitational theory, which describes weak gravitational fields, but also extends it providing a framework for the description of strong gravitational interactions. Over the 100 years of its existence, it counts many experimental successes such as the explanation of the excess advance in the perihelion of Mercury, the prediction and observation of compact objects like the black holes \cite{schbh, reibh, norbh, kerrbh, newbh1, newbh2} and the neutron stars, the gravitational redshift and more recently the observations of gravitational waves \cite{gw1} and black hole shadows \cite{shad1}.

Before we proceed with the formulation of the General Relativity, we will first    discuss its mathematical framework \cite{carroll, inverno, mtw, hartle}; that of differential geometry which constitutes the generalization of the Euclidean geometry on curved spaces. The geometry in General Relativity is described by the metric tensor $g_{\m\n}$ which is a second rank\footnote{The rank of a tensor is the total number of its indices.} symmetric tensor. The metric tensor is appearing in the well-known ``first fundamental form" of the differential geometry which is also known as the line element
\begin{equation}
ds^2=g_{\m\n}dx^\m dx^\n,
\end{equation}  
where $x^\mu=(ct, \vec{x})$. Here we introduce the Einstein's summation convention in which an upper and a lower index (in the same side of the equation) denoted by the same letter are summed over.  The above equation is a generalized form of the well-known line element of the Special Theory of Relativity  
\begin{equation}
ds^2=\h_{\m\n}dx^\m dx^\n,
\end{equation} 
where $\h_{\m\n}$ is the metric tensor of the flat  Minkowski space-time\footnote{Throughout this work, we will use the $(-1,+1,+1,+1)$ signature 
for the Minkowski tensor $\eta_{\mu\nu}$.}. 

The most useful mathematical objects in General Relativity are the tensors. Tensors are objects with indices, which are the generalization of vectors, that obey a specific transformation rule under a coordinate change $x^\m\rightarrow x'^\m=x'^\m(x^\n)$, namely 
\begin{equation}
T'^{\a_1 \a_2 ... \a_n}_{\4\4\4\4\,\b_1\b_2...\b_m}=\frac{\partial x'^{a_1}}{\partial x^{\l_1}}...\frac{\partial x^{\rho_m}}{\partial x'^{\b_m}}T^{\l_1 \l_2 ... \l_n}_{\4\4\4\4\,\r_1\r_2...\r_m}.\label{transf}
\end{equation}
Not all objects with indices are tensors, two known examples are the Kronecker delta symbol $\d^\m_\n$ and the Levi-Civita symbol $\varepsilon^{\m\n\r\s}$. In this framework, it is very interesting to see how the partial derivative  $\left(\partial_\m=\frac{\partial}{\partial x^\m}\right)$  of a tensor transforms in a curved spacetime:
\begin{align}
\left(\partial_\m A_\n\right)'=\frac{\partial x^{\a}}{\partial x'^{\m}}\partial_\a\frac{\partial x^{\b}}{\partial x'^{\n}}A_\b=\frac{\partial x^{\a}}{\partial x'^{\m}}\frac{\partial x^{\b}}{\partial x'^{\n}}\partial_\a A_\b =\frac{\partial x^{\a}}{\partial x'^{\m}}\frac{\partial x^{\b}}{\partial x'^{\n}}\partial_\a A_\b+\frac{\partial x^{\a}}{\partial x'^{\m}}A_\b\frac{\partial}{\partial x^\a}\frac{\partial x^{\b}}{\partial x'^{\n}}. 
\end{align}
From the above equation it is clear that we need to generalize the definition of the partial derivative in order to preserve its tensorial behavior. The generalized derivative is called covariant\footnote{In the language of Differential Geometry the upper indices of a tensor are called contravariant indices while the lower ones are called covariant. Here, for simplicity we will use only the terms upper and lower with only exception the   covariant derivative.} derivative. For a vector  $A^\n$  is defined as: 
\begin{align}
\nabla_\m A^\n&=\partial_\m A^\n+\Gamma^\n_{\m\l}A^\l,\\[2mm]
\nabla_\m A_\n&=\partial_\m A_\n-\Gamma^\l_{\m\n}A_\l,
\end{align}
and it is easy to generalized it for tensors with more indices. In the above equation the objects $\Gamma^\r_{\m\n}$ are known as metric connections and it is clear that they are not tensors since their transformation law is: 
\begin{equation}
 \Gamma'^\r_{\m\n}=\frac{\partial x'^\r}{\partial x^\l}\frac{\partial x^\a}{\partial x'^\m}\frac{\partial x^\b}{\partial x'^\n}\Gamma^\l_{\a\b}-\frac{\partial x^\a}{\partial x'^\m}\frac{\partial x^\b}{\partial x'^\n}\frac{\partial}{\partial x^\a}\frac{\partial x'^\r}{\partial x^\b}.
\end{equation}
The metric connections are not unique. In the usual formulation of General Relativity they take the form of the Christoffel's symbols which  may be expressed  by the following combination of  the metric tensor and its derivatives
\begin{equation}
\G^{\r}_{\m\n}=\frac{1}{2}g^{\l\r} \left( \aa_\m g_{\l\n} + \aa_\n g_{\l\m} - \aa_\l g_{\m\n} \right).\label{chris}
\end{equation}
%
 Returning to the metric tensor, we may define its inverse through the relation $g^{\m\l}g_{\n\l}=\d^\m_\n$. Using the definitions of the covariant derivative and the Cristoffel's symbols it is easy to show that the covariant derivative of the metric tensor vanishes: $\nabla_\l g_{\m\n}=0$ and $\nabla^\l g_{\m\n}=0$, a property called metric compatibility. Also, we can use the metric tensor in order to change the position of an index by following the rules: $A_\m=g_{\m\n}A^\n$ or $A^\m=g^{\m\n} A_\n$. Finally, zeroth order tensors, also known as scalars, are of particular interest in physics since they are invariant under a coordinate transformation ($A'=A$). All the measurable quantities in physics are related with scalars. 

We have already mentioned that the geometry of the general relativity is a generalization of the flat Minkowski space in curved spacetimes. Although all the information about the curvature (and the geometry) of the spacetime is encoded inside the metric tensor, it is useful to define a curvature tensor.  The curvature tensor is known as Riemann's  curvature  tensor and is defined as:
\begin{equation}
R^\r_{\2\,\s\m\n}= \aa_\m  \G^{\r}_{\2\,\n\s} - \aa_\n  \G^{\r}_{\2\,\m\s} + \G^{\r}_{\2\,\m\l}\G^{\l}_{\2\,\n\s} -  \G^{\r}_{\2\,\n\l}\G^{\l}_{\2\,\m\s}\,,\label{rimm}
\end{equation}
in terms of the metric through the Christoffel's symbols. In the framework of the Differential Geometry the Riemman tensor is related with the so called Gaussian curvature.  
Two useful geometrical quantities for the General Relativity, which can be extracted from the Riemann  tensor, are the Ricci  tensor which is defined as 
\begin{equation}\label{ricc}
R_{\m\n}=g^{\l\r}R_{\l\m\r\n}=R^\r_{\2\,\m\r\n}= \aa_\r  \G^{\r}_{\2\,\m\n} - \aa_\m  \G^{\r}_{\2\,\r\n} + \G^{\r}_{\2\,\r\l}\G^{\l}_{\2\,\m\n} -  \G^{\r}_{\2\,\m\l}\G^{\l}_{\2\,\r\n}\,,
\end{equation}
and its trace, also known as Ricci scalar,  defined as
\begin{equation}\label{riccsca}
R=g^{\m\n}R_{\m\n}.
\end{equation}
From Eq. (\ref{chris}) and from the metric properties, it is easy to conclude that the Christoffel's symbols $\G^\r_{\m\n}$ are symmetric under the change of the two lower indices  $\G^\r_{\m\n}=\G^\r_{\n\m}$. Then,  from Eq. (\ref{rimm}) we may find the symmetries of the Riemann tensor. The Riemann tensor is antisymmetric under the change of its first two indices or under the change of its last two indices while it is symmetric under the change  of the first pair of indices with the  second: 
\begin{equation}
R_{\m\n\r\s}=-R_{\n\m\r\s}, \4\4 R_{\m\n\r\s}=-R_{\m\n\s\r},\4\4\text{and}\4\4 R_{\m\n\r\s}=R_{\r\s\m\n}.
\end{equation}
Using the above equations we find that the Ricci tensor is symmetric $R_{\m\n}=R_{\n\m}$. Finally, a very useful property of the Riemann tensor is the Bianchi identity: 
\begin{equation}
\nabla_\l R_{\r\s\m\n}+\nabla_\r R_{\s\l\m\n}+\nabla_\s R_{\l\r\m\n}=0.\label{bianc}
\end{equation}
Contracting twice the  Bianchi identity we find
\begin{align}
g^{\n\s}g^{\m\l}\left(\nabla_\l R_{\r\s\m\n}+\nabla_\r R_{\s\l\m\n}+\nabla_\s R_{\l\r\m\n}\right)&=0,\nn\\[2mm]
\nabla^\m R_{\r\m}-\frac{1}{2}\nabla_\r R &=0,\nn\\[2mm]
\nabla^\m\left(R_{\m\n}-\frac{1}{2}g_{\m\n}R\right)&=0.\label{estt}
\end{align}
The quantity inside the brackets in the above equation $G_{\m\n}\equiv R_{\m\n}-\frac{1}{2}g_{\m\n}R$ is known as Einstein's tensor and as we will see in the next section it is very important in General Relativity.

\section{The General Theory of Relativity}\label{Aa1}

Newton's gravitational theory consists of two elements: An equation which relates the gravitational field with the matter distribution and an equation for the gravitational force that a test particle feels when it moves inside the gravitational field. In order to formulate these two elements it is convenient to use the language of the gravitational potential   $\F$. The potential of the gravitational field is related with the density $\r$ of the mass distribution by Poisson's equation: 
\begin{equation}
\nabla^2\F=4\p G \rho, \label{pois}
\end{equation}
while the equation for the gravitational force is: 
\begin{equation}
\vec{F}_g=- m_g\vec{\nabla}\F. \label{gravf}
\end{equation}
In the above equation, with $G$ we denote the Newton's constant and the coefficient $m_g$  is called the gravitational mass and can be thought of as the gravitational charge of the test particle. Inserting Eq. (\ref{gravf}) into the Second Law of Newtonian mechanics we get $m_I \vec{a}=- m_g\vec{\nabla}\F$, where $m_I$ is the inertial mass.  These two masses in general could be very different. For example, if this was true the statement of Aristotelian physics that between two falling objects the heavier object falls faster would be also true. However, experiments   performed by Galileo showed that  for free falling objects the gravitational mass is equal to the inertial mass $m_g=m_I$. The equality of the two masses is known as \textit{Equivalence Principle} and for the gravitational field we may write
\begin{equation}
\vec a = -\vec\nabla \F.\label{eqm}
\end{equation}
The motion and the orbit of freely falling test particles is universal and do not depend on their mass. The validity of the Equivalence Principle was experimentally tested first by E\"{o}tv\"{o}s \cite{eo} and by other modern experiments \cite{eo1, eo2, eo3, eo4, eo5, eo6, eo7, eo8}. 

The universality of the movement of massive particles in a gravitational field as implied by the Equivalence Principle denotes that: \textit{In small enough regions of spacetime there is no way for a freely falling observer to distinguish the effect of a gravitational field from that of a uniform acceleration.} This universality also states that there is a special class of trajectories where unaccelerated  particles move. These trajectories are known as inertial trajectories and by unaccelerated we mean that a particle which travels under the effect of a gravitational field (i.e. freely falling) does not ``feel" its acceleration. This led to the deduction that gravity should not be considered as a force since a force is an effect that produces an acceleration. However, after the formulation of special relativity and the reconsideration of the concept of the mass, which now is related with the energy and momentum, Einstein thought that the Equivalence Principle should be also generalized. This generalization is  known as Einstein's Equivalence Principle and states that: \textit{The outcome of any local experiment in a freely falling laboratory does not depend on its velocity or on its location in spacetime}. We cannot detect the existence of a gravitational field using local experiments. The laws of physics  do not depend on the choice of the coordinate system and in small enough regions of spacetime reduce to those of special relativity.  
 
This strange universality of the gravitational field  led Einstein to the idea that the gravitational interaction is not a true force but instead  is related with the curvature of spacetime. In many cases, but not always, we can generalize the laws of physics in a curved background following a simple recipe. First, in order to formulate our equations in a coordinate independent form we express them using tensors and then we simply change all partial derivatives with covariant derivatives. Using this method, called the minimal coupling principle, we can easily find the equation that governs the motion of a test particle in general relativity. In Newtonian physics a free falling particle moves in a straight line. The parametric equation $x^\m(\xi)$ of a straight line obeys the differential equation: 
\begin{equation}
\frac{d^2 x^\m}{d\xi^2}=0.
\end{equation}
In a curved background, while $d x^\m/d\xi$ is invariant under a coordinate change  the  $d^2 x^\m/d\xi^2$  is not. We could easily see that by using the chain rule: 
\begin{equation}
\frac{d^2 x^\m}{d\xi^2}=\frac{dx^\n}{d\xi}\partial_\n\frac{dx^\m}{d\xi}.
\end{equation}
Now, in order to generalize the above equation in a curved background, we replace the partial derivative with a covariant one: 
\begin{equation}
 \frac{dx^\n}{d\xi}\partial_\n\frac{dx^\m}{d\xi}=0\4\2\rightarrow\4\2  \frac{dx^\n}{d\xi}\nabla_\n\frac{dx^\m}{d\xi}=\frac{d^2 x^\m}{d\xi^2}+\G^\m_{\r\s}\frac{dx^\r}{d\xi}\frac{dx^\s}{d\xi}=0.
\end{equation}
The above equation is known as geodesics equation and in General Relativity describes the motion of a particle which moves only under the gravitational interaction. In a curved background the geodesic curves are the generalization of the straight lines; they are the curves of minimum length that connect two distinct points. In General Relativity it is useful to express the geodesics equation using the proper time parameter $\tau$ (with $d\tau^2=-ds^2$) instead of the random parameter $\xi$: 
\begin{equation} \label{geodesics-GEM}
\frac{d^2 x^\rho}{d\tau^2} + \Gamma^\rho_{\mu\nu}\,\frac{dx^\mu}{d\tau}
\frac{dx^\nu}{d\tau}=0. 
\end{equation}
As in the Newton's theory of gravity, the motion is independent of the particle's mass.

While the geodesic equation  is the generalization of the straight line equation, in curved spacetimes we cannot just claim that it describes gravity. If the geodesic equation describes the gravitational field, it should reproduce Newtonian gravity Eq. (\ref{eqm}) at the weak field limit \cite{carroll, inverno, mtw, hartle}. In order to do this we use the weak field approximation. In this, the metric tensor may be written as 
\begin{equation}
g_{\mu\nu} (x^\mu) =\eta_{\mu\nu} + h_{\mu\nu}(x^\mu)\,,
\label{metricc}
\end{equation}
where $h_{\mu\nu}$ are small perturbations over the Minkowski space-time. In the context of General Relativity, the metric perturbations
$h_{\mu\nu}$  are sourced by gravitating bodies and  obey the
inequality $|h_{\mu\nu}|\ll 1$. As a result a linear-approximation
analysis may be followed. Here we will briefly review the corresponding formalism and we will present the equations of General Relativity 
in the linear-order approximation (for a more detailed analysis, see for
example  \cite{inverno, carroll, mtw, hartle, bakopgem1, bakopmaster, bakopgem2}). If we use Eq. (\ref{metricc}) and keep only terms 
linear in the perturbation $h_{\mu\nu}$, we easily find that the Christoffel
symbols take the following form
\begin{equation}\label{Christoffel}
\Gamma^{\alpha}_{\mu\nu}=\frac{1}{2}\,\eta^{\alpha\rho}\left(
h_{\mu\rho,\nu}+h_{\nu\rho,\mu}-h_{\mu\nu,\rho}\right),
\end{equation}
where we introduce the comma notation $(A,_\m\equiv\partial_\m A)$.  In the linear approximation, the tensor indices are raised and lowered by
the Minkowski metric $\eta_{\mu\nu}$.  Also, the proper time 
\begin{align}
d\tau^2=-ds^2=c^2dt^2\left(1-\frac{d\vec{r}^2}{c^2dt^2}\right) \4\2\Rightarrow\4\2d\tau^2=c^2\g^2dt^2,
\end{align}
 at the Newtonian limit --where the velocities are small enough compared with the speed of light $\left( u\ll c\right)$  so $\g\simeq 1$-- takes the simple approximate form $d\tau=cdt$. Using this, the spatial part of the geodesics equation Eq. (\ref{geodesics-GEM}) takes the explicit form\footnote{Here we use the usual convention that Greek indices take spacetime values $(0,1,2,3)$ while Latin indices take only spatial values $(1,2,3)$.} 
\begin{equation}
\frac{d^2x^i}{dt^2}+c^2\Gamma^i_{00}+2c\Gamma^i_{0j}\frac{dx^j}{dt}+\G^i_{jk}\frac{dx^j}{dt}\frac{dx^k}{dt}=0,  \4\2\Rightarrow\4 a^i\simeq-c^2\Gamma^i_{00}. \label{prenn}
\end{equation}
From Eq. (\ref{Christoffel}) and using the fact that the Newtonian gravitational field is static we find that
\begin{equation}
\Gamma^{i}_{00}=\frac{1}{2}\,\eta^{ij}\left(
h_{0j,0}+h_{0j,0}-h_{00,j}\right)=-\frac{1}{2}\,\eta^{ij}\partial_j h_{00}.
\end{equation}
Replacing the above equation in Eq. (\ref{prenn}) we find $a^i=\frac{1}{2}c^2\,\eta^{ij}\partial_j h_{00}$ and comparing with the Newtonian equation (\ref{eqm}) we find
\begin{equation}
h_{00}=-\frac{2 \F}{c^2}, 
\end{equation}
where $\F$ is the Newtonian gravitational potential. In terms of the metric tensor we have $g_{00}=-\left(1+\frac{2\F}{c^2}\right).$ Thus, we have shown that the curvature of the spacetime may describe the gravitational force at the Newtonian limit. In the strong field regime the theory must be experimentally tested\footnote{Actually, over the last century, the General Relativity counts many experimental successes; from the explanation of the excess advance in the perihelion of Mercury up to the recent observations of the gravitational waves or the black hole shadows.  }.

Unfortunately, in order to find the second gravitation equation  (the field equation)  we cannot start from Eq. (\ref{pois}) and use our simple recipe. However, we know that the generalization of the gravitational potential is the metric tensor $g_{\m\n}$ which is a second rank symmetric tensor. Thus, the density $\r$ of the matter distribution should be also generalized to a second rank symmetric tensor $T_{\m\n}$ which is known as energy-momentum tensor. We also know that the majority of the equations in physics contain up to second order derivatives. Therefore, we search for an equation of the form $A_{\m\n}\left(g_{\m\n},\partial g_{\m\n},\partial^2g_{\m\n}\right)=kT_{\m\n}$. Finally, we know that the energy is a conserved quantity. In terms of the energy-momentum tensor this conservation is expressed by the following equation: 
\begin{equation}
\nabla^\m T_{\m\n}=0. \label{encon}
\end{equation} 
Thus, the $A_{\m\n}$ tensor should also obey the same equation $\nabla^\m A_{\m\n}=0$. Instead of $A_{\m\n}$ we could use $\nabla_\r\nabla^\r g_{\m\n}$ but we  already know that the covariant derivative of the metric tensor vanishes $\nabla_{\r}g_{\m\n}=0$. However, we note that the Riemann tensor $R_{\m\n\r\s}$   is constructed by first and second derivatives of the metric tensor and also the Ricci tensor $R_{\m\n}$, which is produced from the Riemann tensor, is a second rank symmetric tensor. Furthermore, we recall from differential geometry that any tensor which contains up to second order derivatives of the metric tensor may be expressed using the Riemann tensor and its contractions. Finally, in order to find the explicit form of the tensor $A_{\m\n}$ we observe that the Einstein's tensor $G_{\m\n}=R_{\m\n}-\frac{1}{2}g_{\m\n}R$ justifies all the above criteria. Concluding, the field equations of general relativity are given by the following equation: 
\begin{equation}\label{fieldeq}
G_{\m\n}=k\, T_{\m\n},
\end{equation}
where $k$ is a constant. 
 
In order to determine the proportionality constant $k$, we again use the weak field approximation and demand that Eq. (\ref{fieldeq}) reproduces the Newtonian field equation Eq. (\ref{pois}). 
Using Eqs. (\ref{metricc}-\ref{Christoffel}), we can easily show that the Ricci tensor  Eq. (\ref{ricc})  in the linear approximation assumes the form
\begin{equation}\label{RicciT}
R_{\mu\nu}=\frac{1}{2}\left({h^\rho}_{\mu,\nu\rho}+
{h^\rho}_{\nu,\mu\rho}-\da h_{\mu\nu} - h_{,\mu\nu}\right),
\end{equation}
where $\da=\h^{\mu\nu}\aa_\m\aa_\n$. In the above equation, we have also defined
the trace of the metric perturbations $h=\h^{\m\n}h_{\m\n}$. Correspondingly, the trace of the Ricci
tensor (Eq. (\ref{riccsca})) is found to be
\begin{equation}\label{RicciS}
R={h^{\m\n}}_{,\m\n} - \da h\,.
\end{equation}
If we combine the Ricci tensor and its trace, the Einstein tensor   takes the following form
\begin{equation}\label{Einstein}
G_{\m\n}=\frac{1}{2}\left({h^\a}_{\m,\n\a}+ {h^\a}_{\n,\m\a}-
\da h_{\m\n}-h_{,\m\n}-\h_{\m\n}\,{h^{\a\b}}_{,\a\b}+\h_{\m\n}\,\da h \right).
\end{equation}
Usually in gravity,  in the linear approximation, one may define the so-called trace-inverse perturbations $\hh_{\m\n}$ in terms of the original perturbations $h_{\m\n}$:
\begin{equation}\label{newh}
\hh_{\m\n}=h_{\m\n} - \frac{1}{2}\,\h_{\m\n}\,h\,.
\end{equation}
In terms of the new perturbations,  the Einstein tensor simplifies to:
\begin{equation}\label{Einstein_full}
G_{\m\n}=\frac{1}{2}\left({\hh^{\a}}_{\2\m,\n\a}+{\hh^{\a}}_{\2\n,\m\a}-
\da \hh_{\m\n}-\h_{\m\n}\,{\hh^{\a\b}}_{\4,\a\b}\right).
\end{equation}
The above tensor satisfies Einstein's field equations Eq. (\ref{fieldeq}) which now are written as
\begin{equation}\label{field_eqs_full}
{\hh^{\a}}_{\2\m,\n\a}+{\hh^{\a}}_{\2\n,\m\a}-
\da \hh_{\m\n}-\h_{\m\n}\,{\hh^{\a\b}}_{\4,\a\b}=2k\,T_{\mu\nu}\,.
\end{equation}

 If we make a coordinate transformation of the form $x^\m\rightarrow x'^\m = x^\m - \g^\m$, we can easily see that the metric perturbations transform as $h_{\m\n} \rightarrow h'_{\m\n} = h_{\m\n} + \g_{\m,\n} + \g_{\n,\m} $. This transformation leaves Einstein tensor unchanged and hence the field equation. Transformations of this form, which do not alter  observable quantities, are called {\it gauge transformations}.  When we solve a problem, it is often convenient  to reduce the gauge freedom by imposing constraints to the perturbations $h_{\m\n}$. These constraints are called  {\it gauge conditions}. Usually, in the linear approximation, we use the so-called {\it transverse} gauge condition ${\hh^{\m\n}}_{\4,\n}=0$. Under the imposition of the transverse gauge condition the Einstein  tensor takes the simple form
\begin{equation}\label{Einstein_new}
G_{\m\n}=-\frac{1}{2} \da \hh_{\m\n},
\end{equation}
while the field equations are
\begin{equation}\label{field_eqs}
\da \hh_{\m\n}=-2k\,T_{\mu\nu}\,.
\end{equation}
If we multiply Eqs. (\ref{field_eqs}) with $\h^{\m\n}$, we get $ \da \hh=-2k\,T$ or $\da h=2k\,T$ and we can rewrite the field equations as:
\beq\label{field_alt}
\da h_{\m\n}=-2k\,\left(T_{\mu\nu}-\frac{1}{2}\h_{\m\n}T\right)\,.
\eeq
 
For a perfect fluid  matter distribution, the energy-momentum tensor  takes the form: 
\begin{equation}
T_{\m\n}=(\r+p)u_\m u_\n +p g_{\m\n},
\end{equation}
where $u_\m$ is the four velocity of the fluid, while $\r$ and $p$ its energy density and momentum density respectively. Since the pressure of a fluid is important only if its velocity is relativistic, in the Newtonian limit we may neglect it. In what follows we will assume that the distribution of energy in the system is described by the expression $T_{\mu\nu}=\rho\,u_\mu u_\nu$ which corresponds to  dust. We will also work in the rest frame of the fluid so the 4-velocity is $u^\mu=(u^0,u^i)=(c,0,0,0)$. Using this, we find that the trace of the energy-momentum tensor is $T=-c^2\r$ and the expression inside the brackets in Eq. (\ref{field_alt}) is $\left(T_{00}-\frac{1}{2}\h_{00}T\right)=\frac{c^2}{2}\r$. Using $h_{00}=-2\F/c^2$ and the fact that the Newtonian potential is static, Eq. (\ref{field_alt}) takes the form 
\begin{equation}
\partial_i\partial^i\F=\frac{k}{2}c^4\r. 
\end{equation}
Comparing the above equation with the Poisson equation (\ref{pois}), we identify the proportionality constant as $k=\frac{8 \pi G}{c^4}$ and the field equations of General Relativity therefore are: 
\begin{equation}
R_{\m\n}-\frac{1}{2}g_{\m\n}R=\frac{8\p G}{c^4}T_{\m\n}.
\end{equation}
For simplicity, from now on we will use the so called geometric unit system  in which we set $c=1$ and $G=1$. This unit system will also help us to avoid the confusion by using the same symbol for Einstein's tensor $G_{\m\n}$ and for Newton's constant $G$. We could easily confuse the Newton's constant $G$ with the trace of the Einstein's tensor $G=g^{\m\n}G_{\m\n}$. Finally, we note that in vacuum (i.e. $T_{\m\n}=0$) Eq. (\ref{field_alt}) takes the form $\partial^2h_{\m\n}=0$ which is a wave equation with propagation speed equal to  the speed of  light: 
\begin{equation}
\left(- \frac{1}{c^2}\frac{\partial^2}{\partial t^2} + \h^{ij}\frac{\partial}{\partial x^i}\frac{\partial}{\partial x^j}\right)h_{\m\n}=0. 
\end{equation} 
The General theory of Relativity predicts the existence of   gravitational waves which propagate with the speed of light \cite{carroll, inverno, mtw, hartle}. These waves were first detected a few years ago \cite{gw1, gw2, gw3, gw4}, exactly 100 years after the formulation of General relativity.  

\section{Lagrangian Formulation of General Relativity}

The General theory of Relativity, as every field theory, may be expressed using the Lagrangian Formulation. Usually, when we construct a Lagrangian for a field we start from a kinetic term and a potential and we add possible interactions. However, in General Relativity our field is the metric tensor. We mentioned in the previous section that any tensor which contains up to second order derivatives of the metric tensor may be expressed in terms of the Riemann tensor and its contractions. Therefore the only non trivial scalar we could use as Lagrangian density is the Ricci scalar: 
\begin{equation}
S=\int d^4x\,\sqrt{-g}\,R,
\end{equation} 
where $g$ is the determinant of the metric tensor. The above action is known as Hilbert action. 

In order to find the field equations we have to vary the above action with respect to the metric tensor $g^{\m\n}$. Using the relation $g_{\m\a}g^{\n\a}=\d_\m^\n$ and the fact that the Kronecker delta remains unchanged under a variation, we find $\d g_{\m\n}=-g_{\m\r}g_{\n\s}\d g^{\r\s}$. Using $R=g^{\m\n}R_{\m\n}$ we may write the variation of the action as: 
\begin{equation}
\d S= \d S_1 + \d S_2 + \d S_3, \label{varr1}
\end{equation}
where 
\begin{align}
\d S_1&=\int d^4 x \sqrt{-g}g^{\m\n} \d R_{\m\n},\\[2mm]
\d S_2&=\int d^4 x \sqrt{-g}R_{\m\n} \d g^{\m\n},\\[2mm]
\d S_3&=\int d^4 x R\, \d \sqrt{-g}.
\end{align}
The second term $\d S_2$ has the required form so we have to work out only the other two terms. Starting from the first term we first note from Eqs. (\ref{rimm})-(\ref{ricc}) that the Riemann (and the Ricci) tensor is expressed in terms of the Christoffel symbols. In order to find the variation of the Riemann tensor with respect to the metric tensor it is easier to find first its variation with respect to the Christoffel symbols. We define the variation  of the Christoffel symbols as: 
\begin{equation}
\G'^\r_{\m\n}=\G^\r_{\m\n}+\d \G^\r_{\m\n}.
\end{equation} 
The variation of the Christoffel symbols $(\d \G^\r_{\m\n})$ as the difference of two Christoffel symbols is a tensor and as a result we may define its covariant derivative: 
\begin{equation}
\nabla_\l\left(\d \G^\r_{\m\n}\right)=\partial_\l\left(\d \G^\r_{\m\n}\right)+\G^\r_{\l\s}\d\G^\s_{\m\n}-\G^\s_{\l\n}\d\G^\r_{\s\m}-\G^\s_{\l\m}\d\G^\r_{\n\s}.
\end{equation}
All the covariant derivatives in the above equation are taken with   respect to the metric tensor $g_{\m\n}$. After some algebraic manipulation we find that the variation of the Riemann tensor assumes the form: 
\begin{equation}
\d R^\r_{\2\m\l\n}=\nabla_\l(\d \G^\r_{\m\n})-\nabla_\n(\d \G^\r_{\l\m}).
\end{equation}
The substitution of the above equation in the first term of the action's variation $\d S_1$ yields 
\begin{equation}
\d S_1=\int d^4 x \sqrt{-g}\,\nabla_\s \left[  g^{\m\n}(\d \G^\s_{\m\n})-g^{\m\s}(\d \G^\l_{\l\m})  \right].
\end{equation}
The variation in the above equation has the form of a total derivative i.e. $\nabla_\m A^\m$. By using the Stokes's theorem the contribution of a total derivative, in the field equations, is equal to a boundary contribution at infinity. By making the variation vanish at infinity we may eliminate the boundary contribution. Thus, a total derivative does not contribute to the field equations and we may set $\d S_1 =0$. 
 
For the third term $\d S_3$ we need the following relation which holds for any square matrix $M$ with a non-trivial determinant: $\ln(\det M)=\text{Tr} (\ln M)$. The variation of this equation is
\begin{equation}
\frac{1}{\det M}\d (\det M)=\text{Tr}\left(M^{-1}\d M\right). 
\end{equation}
Using the above equation with $\det M = g$ we find that $\d g = -g \,g_{\m\n}\d g^{\m\n}$. Now, we can easily find the variation of the $\sqrt{-g}$ term appearing in the $\d S_3$ 
\begin{equation}
\d \sqrt{-g} = -\frac{1}{2\sqrt{-g}}\d g=-\frac{1}{2}\sqrt{-g}g_{\m\n}\d g^{\m\n}. 
\end{equation}
By replacing in Eq. (\ref{varr1}) we find: 
\begin{equation}
\d S= \frac{\d S}{\d g^{\m\n}}\d g^{\m\n}=\int d^4x \,\sqrt{-g}\left(R_{\m\n}-\frac{1}{2}g_{\m\n}R\right)\d g^{\m\n}.\label{varr2} 
\end{equation}
From the above equation we derive the vacuum field equations of General Relativity $G_{\m\n}=0$. As we will see in the following sections the advantage of the Lagrangian formulation is that we can easily modify General Relativity using it. In order to find the complete set of the field equations we have to add an extra term in the action which is related with the matter:
\begin{equation}
S=\frac{1}{16\p} S_H+S_M, 
\end{equation}
where with $S_H$ we denote the Hilbert action and with $S_M$ the matter term. Using the middle term from Eq. (\ref{varr2})  we can extract the definition of the energy-momentum tensor: 
\begin{equation}
T_{\m\n}=-\frac{2}{\sqrt{-g}}\frac{\d S_M}{\d g^{\m\n}}. 
\end{equation}
Therefore, the complete field equations of General Relativity are
\begin{equation}
R_{\m\n}-\frac{1}{2}g_{\m\n}R=8\p \,T_{\m\n}. \label{feqq1}
\end{equation}

\section{The Schwarzschild Solution}

The first exact solution of General Relativity was derived by Karl Schwarzschild a few months after the presentation of General Relativity \cite{schbh}. The Schwarzschild Solution describes the gravitational field around a spherical, static and non rotating  mass distribution. The solution is very important since it describes, with high accuracy, slowly rotating astrophysical objects such as black holes, stars and planets. The most general form for the metric of a spherically symmetric and time dependent spacetime is the following: 
\begin{equation}
ds^2=-e^{A(r,t)} dt^2+e^{B(r,t)}dr^2 + 2 e^{\frac{\L(r,t)}{2}}dt\,dr+r^2 e^{F(r,t)}\left(d\theta^2 +\sin^2\theta d\varphi^2 \right).
\end{equation}
We may eliminate the unknown function  in front of the angular part of the metric by redefining the radial coordinate as $r'=re^{F/2}$. The metric then yields: 
\begin{equation}
ds^2=-e^{A(r,t)} dt^2+e^{B(r,t)}dr^2 + 2 e^{\frac{\L(r,t)}{2}}dt\,dr+r^2\left(d\theta^2 +\sin^2\theta\, d\varphi^2 \right),
\end{equation}
where, for simplicity, we have ignored the primes. By redefining the time coordinate as: 
\begin{equation}
dt' = e^{\frac{H(r,t)}{2}}\left( e^{A(r,t)}dt - e^{\frac{\L(r,t)}{2}}dr \right),
\end{equation}
we may eliminate the cross term: 
\begin{equation}
ds^2=-e^{-A(r,t')+H(r,t')} dt'^2+\left( e^{B(r,t')}+e^{-A(r,t')+\L(r,t')} \right)dr^2 +r^2\left(d\theta^2 +\sin^2\theta\, d\varphi^2 \right),
\end{equation}
Finally, by defining two new $A$ and $B$ functions in the place of the complicated coefficients in front of the terms $dt'^2$ and $dr^2$ and by ignoring again the primes, we find that the metric takes the simplified form:  
\begin{equation}
ds^2=-e^{A(r,t)} dt^2+e^{B(r,t)}dr^2+r^2\left(d\theta^2 +\sin^2\theta\, d\varphi^2 \right). \label{met}
\end{equation}

Substituting  the above metric into Eq. (\ref{chris}) we find that the non trivial components of the Christoffel symbols are the following:
\begin{equation}
\begin{array}{*3{>{\displaystyle}l}}
\G^t_{tt}= \frac{1}{2}\dot A,   & \G^t_{tr}= \frac{1}{2}A', &\G^t_{rr}=\frac{1}{2}e^{B-A}\dot B,\\[4mm]
\G^r_{tt}= \frac{1}{2} e^{A-B}A'  & \G^r_{tr}=\frac{1}{2}\dot B, &\G^r_{rr}=\frac{1}{2}B',\\[4mm]
\G^\theta_{r\theta}=\frac{1}{r}, &\G^r_{\theta\theta}=-re^{-B}, &\G^\varphi_{r\varphi}=\frac{1}{r},\\[4mm]
\G^r_{\varphi\varphi}=-re^{-B}\sin^2\theta,\2\, &\G^\theta_{\varphi\varphi}=-\sin\theta\,\cos\theta,\2\, &\G^\varphi_{\varphi\theta}=\frac{\cos\theta}{\sin\theta}.
\end{array}
\end{equation}
In the above equations, the prime denotes differentiation with respect to the radial coordinate $r$, while the dot denotes differentiation with respect to the time coordinate $t$. Using the components of the Christoffel symbols we may find the non-vanishing components of the Ricci tensor: 
\begin{align}
R_{tt}&=\frac{1}{4}\left(-2 \ddot B -\dot{B}^2+\dot A \dot B \right)+\frac{1}{4}e^{A-B}\left( 2 A''+A'^2-A'B'+\frac{4}{r}A'\right)\label{tte}\\[3mm]
R_{rr}&=\frac{1}{4}e^{B-A}\left(2\ddot B + \dot{B}^2-\dot A \dot B \right) -\frac{1}{4}\left( 2A''+A'^2-A'B'-\frac{4}{r}B'\right)\label{rre}\\[3mm]
R_{tr}&=\frac{1}{r}\dot B\label{rte}\\[3mm]
R_{\theta\theta}&=\frac{1}{2}e^{-B}\left(rB'-rA'-2\right)+1,\label{the}\\[3mm]
R_{\varphi\varphi}&=R_{\theta\theta}\,\sin^2\theta.\label{phe}
\end{align}

The Schwarzschild solution is a vacuum solution, so, we set $T_{\m\n}=0$ in Einstein's field equations (\ref{feqq1}). However taking the trace of the field equations we find $R=-8\p T$ and since $T=0$ we also have $R=0$; thus the vacuum field equations are $R_{\m\n}=0$.  From the equation $R_{tr}=0$ we find $\dot B =0$. Substituting $\dot{B}=0$ in the components of the Ricci tensor and taking the time derivative of the equation $R_{\theta\theta}=0$ we find $(\dot{A})'=0$. These two equations may be easily integrated to yield: 
\begin{align}
B(r,t)&=B(r),\\[3mm]
A(r,t)&=A(r)+F(t).
\end{align} 
Thus, the first term of the metric is $-e^{A(r)}e^{F(t)}dt^2$. However, since we can always define a new time coordinate, we can absorb the time function inside a new time coordinate $dt'=e^{F(t)/2}$. As a result, the metric takes the following form:
\begin{equation}
ds^2=-e^{A(r)} dt^2+e^{B(r)}dr^2+r^2\left(d\theta^2 +\sin^2\theta \,d\varphi^2 \right), \label{met1}
\end{equation}
where again, for simplicity we are neglecting the primes. We observe that  a spherically symmetric spacetime is always static.

 We may easily calculate the components of the Ricci tensor for the new metric (\ref{met1}) by setting $\dot A=\dot B =0$ in Eqs. (\ref{tte}-\ref{phe}). Taking the combination $-e^{-A}R_{tt}-e^{-B}R_{rr}=0$ we find $A'+B'=0$ which is integrated to give: 
\begin{equation}
A(r)=-B(r)+c, \label{e2}
\end{equation}
where $c$ is an integration constant. In order to find the value of the constant $c$ we have to implement the boundary conditions for our solution. The spacetime far away from a local object is the Minkowski spacetime of Special Relativity. This property is known as asymptotic flatness.  Therefore  as the radial coordinate approaches infinity $g_{rr}\approx-g_{tt}\simeq 1$ or equivalently $A\approx-B\simeq 0$, thus, the integration constant is $c=0$. Substituting Eq. (\ref{e2}) (with $c=0$) into the equation $R_{\theta\theta}=0$ we find $\left(r e^A\right)'=1$ which may be easily integrated to
\begin{equation}
e^A=1-\frac{a}{r},
\end{equation}
where $a$ is an integration constant. In order to find the value of the integration constant we will use again the boundary conditions. In the previous section we showed that at the weak field approximation the $(tt)$ component of the metric tensor is $g_{tt}=-1-2\F$, where $\F$ is the Newtonian potential. For a spherical symmetric mass distribution with total mass $M$ the Newtonian potential assumes: 
\begin{equation}
\F(r)=-\frac{M}{r}. 
\end{equation}
Thus, we may identify the integration constant as $a=2M$. Finally, the metric for the Schwarzschild solution has the following form: 
\begin{equation}
ds^2=-\left(1-\frac{2M}{r}\right) dt^2+\left(1-\frac{2M}{r}\right)^{-1}dr^2+r^2\left(d\theta^2 +\sin^2\theta \,d\varphi^2 \right). \label{sch}
\end{equation}
Our analysis above 
agrees with the Birkhoff theorem. This theorem is a uniqueness theorem which states that in General Relativity the only spherically symmetric solution of the vacuum field equations is the Schwarzschild solution \cite{carroll, inverno, mtw, hartle, birk}.

An interesting property of the  Schwarzschild solution is that the $g_{rr}$ component of the metric vanishes at $r=0$ and diverges at $r=2M$. Singe $|g_{tt}|=1/g_{rr}$ at these two values of the radial coordinate, the $g_{tt}$ component also has an ill-defined behavior. This strange behavior may indicate pathological regions of the spacetime known as singularities or  may be attributed to a bad choice of the coordinate system. Since the components of the metric are coordinate dependent, in order to identify a singularity we need to examine the coordinate invariant quantities i.e. the scalar quantities. Using the Riemann tensor we may construct three fundamental scalars: the Ricci scalar $R$, the square of the Ricci tensor $R_{\m\n}R^{\m\n}$ and the square of the Riemann tensor $R_{\m\n\r\s}R^{\m\n\r\s}$ also known as Kretschmann scalar. The vacuum field equations are $R_{\m\n}=0$, as a result the Ricci scalar and the square of the Ricci tensor are vanishing identically for the Schwarzschild solution or any other vacuum solution. The calculation of the Kretschmann scalar leads to: 
\begin{equation}
R_{\m\n\r\s}R^{\m\n\r\s}=\frac{48 M^2}{r^6}. 
\end{equation}
Since the above quantity diverges only at $r=0$, the true singularity at the Schwarzschild metric is there. The divergence of the metric at $r=2M$ indicates that the coordinate system we use is pathological there and by using a different system we may bypass this apparent singularity.  However, as we will see in the following paragraph, the radius $r_s=2M$ --which is known as Schwarzschild radius-- has very interesting properties. We have mentioned that the Schwarzschild solution may describe  the geometry on the exterior of a non-rotating and spherically symmetric astrophysical object like   stars and   planets. However for these objects we do not need to worry about the singularity or the Schwarzschild radius because they are located in the interior of the object. For example for the earth the Schwarzschild radius is $r_s=9\,\, \rm{mm}$ while for the sun is $r_s=3\,\,\rm{Km}$. In the interior of the object we solve the field equations using, for any kind of object, the appropriate form for the energy momentum tensor. The system of the differential equations that describe the interior of an object is known as Tolman-Oppenheimer-Volkoff  (TOV) \cite{Tolman:1939jz, Oppenheimer:1939ne}. Then we find the inner solutions which are not described by the Schwarzschild solution and thus do not have horizons or singularities. The two solutions, the TOV inner solution and the Schwarzschild exterior solution, are joined at the surface of the object using the appropriate junction conditions. 

In order to investigate the properties of objects with radius $r<2M$ we will focus on the geodesics of the Schwarzschild solution and especially the null curves. Photons travel in null curves with $ds^2=0$ and we will consider the case of the radial null curves i.e. $\theta$ and $\varphi$ are constants thus $d\theta=d\varphi=0$. Setting $ds^2=d\theta^2=d\varphi^2=0$ in the metric, we find
\begin{equation}
\frac{dt}{dr}=\pm\left(1-\frac{2M}{r}\right)^{-1}. \label{slo1}
\end{equation}
The above equation defines the slope of the light cone in a $(t-r)$ spacetime diagram. At spatial infinity the slope is $\pm 1$, as in the Minkowski spacetime. This agrees with the fact that the Schwarzschild solution is an asymptotically flat solution. As we approach the Schwarzschild radius (from the infinity) the angular range of the light cone decreases, while at $r=2M$ vanishes. This means that a light ray moving from infinity to $r=2M$ never seems to reach the horizon and instead it is asymptotically approaching it. In other words a light ray, as observed from a lab at infinity, needs infinite time to arrive at the Schwarzschild radius. Of course this is happening due to the bad choice of the coordinate system and we may fix it by using the appropriate coordinate system. The surface $r=2M$ while it seems locally regular it has many interesting properties. In the region $r<2M$ the signs in Eq. (\ref{slo1}) are reversed and so the interpretation of the coordinates also changes, the time coordinate becoming space coordinate and vice versa. In this case any particle, even   photons, cannot escape to infinity, instead it is forced to move towards the singularity. This means that all the matter in the region $r<2M$ is located at one point, at the singularity at $r=0$. Due to the fact that from the region with   $r<2M$ particles cannot escape to infinity, this region is called a black hole. The surface which is defined from the Schwarzschild radius $r=2M$ and determines the boundary of the black hole is called event horizon or just horizon of the black hole since it ``covers'' the singularity. Once a particle crosses the horizon it is doomed to fall in the singularity. Thus, an observer at infinity may detect/observe the horizon but not the singularity.  

The Schwarzschild metric is well behaved only in the $r>2M$ region of the black hole. At the horizon the metric diverges while in the region $r<2M$ the time becomes spacelike while the radial coordinate becomes timelike  making the interior of the black hole dynamic. These problems may be easily bypassed by adopting the so called Kruskal-Szekeres coordinates: 
\begin{align}
\tilde t &=\left(\frac{r}{2M}-1\right)^{1/2}e^{r/4M}\sinh\left(\frac{t}{4M}\right),\label{sk1}\\[4mm]
\tilde r &=\left(\frac{r}{2M}-1\right)^{1/2}e^{r/4M}\cosh\left(\frac{t}{4M}\right),\label{sk2}\\[4mm]
\end{align}  
in the region $r\geq 2M$ and
\begin{align}
\tilde t &=\left(1-\frac{r}{2M} \right)^{1/2}e^{r/4M}\sinh\left(\frac{t}{4M}\right),\label{sk3}\\[4mm]
\tilde r &=\left(1-\frac{r}{2M} \right)^{1/2}e^{r/4M}\cosh\left(\frac{t}{4M}\right),\label{sk4} 
\end{align}  
in the region $r\leq 2M$. In terms of the new coordinates the Schwarzschild solution takes the following form: 
\begin{equation}
ds^2=\frac{32 M^3}{r}e^{-r/2M}\left( -d\tilde{t}^2+d\tilde{r}^2\right)+r^2\left(d\theta^2 +\sin^2\theta \,d\varphi^2 \right).\label{krum}
\end{equation}
From the above metric, we observe that in the new coordinate system the divergence at the horizon is removed while the divergence at $r=0$, as a true singularity, remains. Using the definition of the Kruskal-Szekeres coordinates Eqs. (\ref{sk1})-(\ref{sk4}) we find their relation with the Schwarzschild radial coordinate: 
\begin{equation}
\tilde{t}^2-\tilde{r}^2=\left(1-\frac{r}{2M}\right)e^{r/2M}.\label{krur}
\end{equation}
Setting $ds^2=0$ in Eq. (\ref{krum}) we find that the radial null curves obey the differential equation $d\tilde{t}/d\tilde{r}=\pm 1$. The slope of the light cone remains everywhere constant and as in the Minkowski spacetime the photons move along the curves $\tilde t =\pm \tilde r+\text{constant}$. In addition, setting $r=2M$ in Eq. (\ref{krur}) we find that the horizon of the black hole in  Kruskal-Szekeres coordinates is at the curve $\tilde{t}=\pm \tilde{r}$. From  Eq. (\ref{krur}) we also find that surfaces with constant $r$ are given by the equation $\tilde{t}^2-\tilde{r}^2=\text{constant}$ while from the definition of the coordinates we find that surfaces with constant $t$ are given by $\tilde{t}=\tilde{r}\tanh\left(t/4M\right)$. Likewise we may verify that the allowed region of the new coordinates is $-\infty\leq \tilde{r}\leq\infty$ and $\tilde{t}^2<\tilde{r}^2+1$.  Finally the singularity $r=0$, in Schwarzschild coordinates, corresponds to two singularities $\tilde{t}=\pm\sqrt{\tilde{r}^2+1}$ while the asymptotic region $r\gg 2M$ corresponds to two asymptotic regions in the new coordinates: $\tilde{r}\gg|\tilde{t}|$ and $\tilde{r}\ll-|\tilde{t}|$. The explanation for this is that the Schwarzschild coordinates cover only a part of the Schwarzschild spacetime. However, using the transformations (\ref{sk1})-(\ref{sk4}) we managed to maximally extend the Schwarzschild solution in order to cover all of the spacetime.
\begin{figure}[t!] 
\begin{center}
\hspace{0.0cm} \hspace{-0.6cm}
\includegraphics[height=.4\textheight, angle =0]{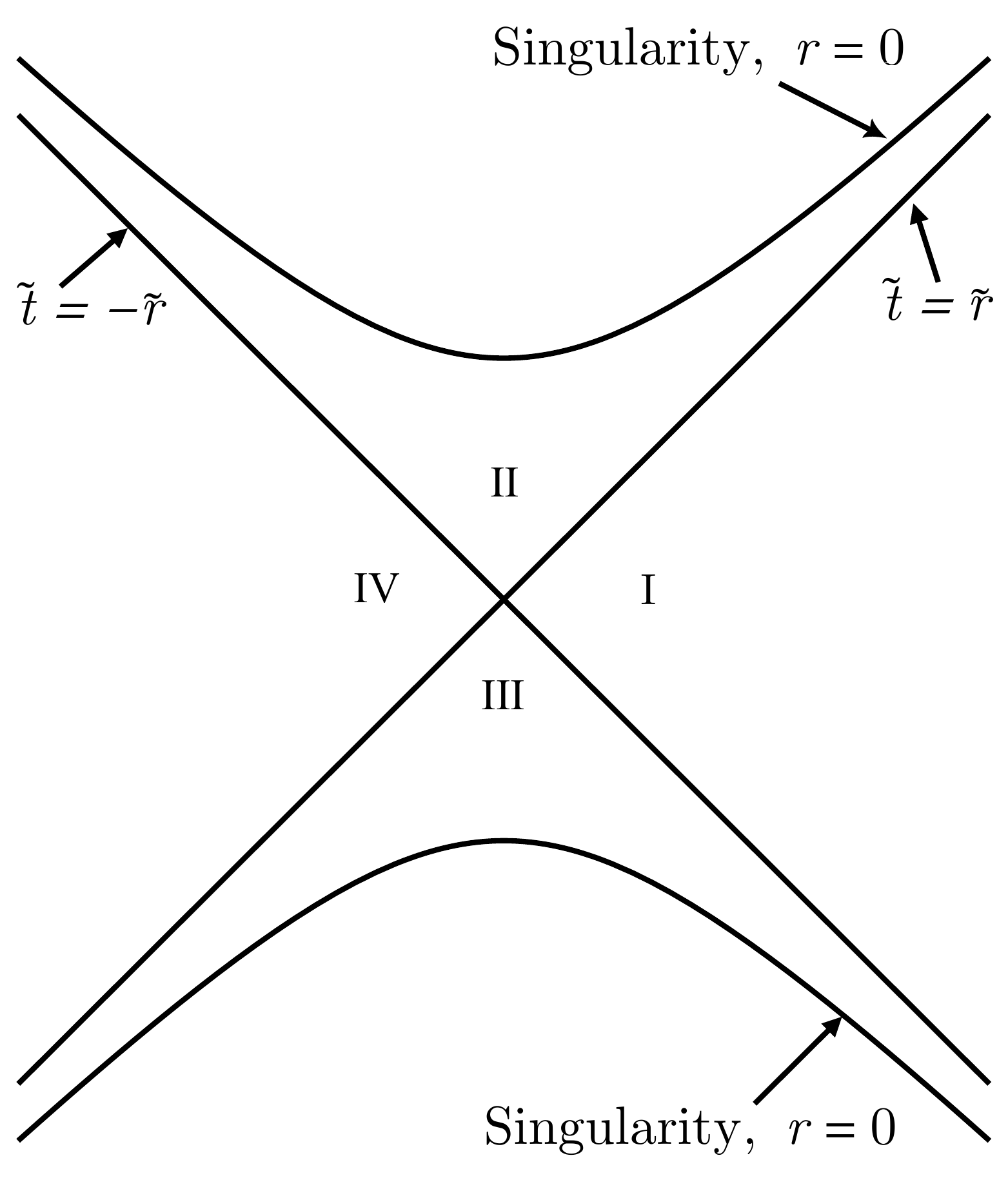}
\end{center}
\caption{The Schwarzschild solution in the Kruskal-Szekeres coordinates.
\label{fig0}}
\end{figure} 

We may draw the Schwarzschild spacetime in the Kruskal-Szekeres coordinates in a $\tilde{t}-\tilde{r}$ diagram as in   Fig. \ref{fig0} which is known as the Kruskal diagram. The region I in   Fig. \ref{fig0} corresponds to the exterior region $r>2M$ of the Schwarzschild solution. Thus, it is the asymptotically flat space time around the spherically symmetric black hole. An observer who falls freely and radially will eventually cross the horizon (the curve $\tilde t=\tilde r$ in the diagram) and   reach   region II of the Kruskal diagram. At this point he is doomed not only to stay in   region II, but to fall into the singularity $\tilde t = \sqrt{\tilde{r}^2+1}$.  Any information (emitted from the observer) heading to region I, even if it is constituted by photons, it is doomed to return and hit the singularity. For this reason, the region II corresponds  to the black hole. However, apart from the regions I and II, due to the maximall  extension, the Kruskal diagram contains two more regions. The region III is characterized by exactly the same properties with the region II but in a ``time reversed" way. This means that any observer in the region III emanates from the second singularity $\tilde t = -\sqrt{\tilde{r}^2+1}$ and is forced  to come out of it. The region III emits particles from its singularity and for this reason it is called a white hole. Finally, the region IV is an asymptotically flat region similar to the region I but   is located on the other side of the horizon $r=2M$. Particles from region IV cannot travel to our region I nor particles from our region may travel there.

\section{The Schwarzschild (Anti-)de Sitter Solution}

In classical physics, the  Euler–Lagrange equations are invariant under a constant shift in the Lagrangian $\mathcal{L}'=\mathcal{L}+\mathcal{C}$. In other words, in classical physics we do not care about the exact value of the energy of a body's state, but about the energy variation between the initial and the final state of the body. However in General Relativity the field equations are not invariant under a similar shift in the Lagrangian. The constant  $\mathcal{C}$ will couple with gravity through the term $\sqrt{-g}$ and will contribute to the field equations. This property indicates that we should include in the field equations the contribution of a vacuum energy which is the energy density of the empty space. Although the contribution of the vacuum energy is negligible in local experiments, in cosmological scales is significant. The last 20 years, cosmological measurements indicate that the expansion of the universe is accelerating. This accelerating expansion may be attributed to the presence of a constant and positive vacuum energy density. For historical reasons, the contribution of the vacuum energy in general relativity is introduced through the so called Cosmological Constant $\L$. The action of the General Relativity is modified as
\begin{equation}
S=\int d^4x\sqrt{-g}\left[  \frac{1}{16\p}\left(R-2\L\right)+\mathcal{L}_M  \right],
\end{equation}
which lead to the generalized field equations 
\begin{equation}
R_{\m\n}-\frac{1}{2}g_{\m\n}\,R+g_{\m\n}\,\L=8\p\, T_{\m\n}.\label{fel}
\end{equation}
Since the vacuum energy is related with the cosmological constant as $\r_\text{v}=\L/8\p$, a positive vacuum energy is equal to a positive cosmological constant. Therefore the accelerated expansion of the universe is ascribed to a positive cosmological constant. 

We already mentioned that the contribution of the cosmological constant is negligible for local experiments. However, we may construct black-hole solutions in the presence of a cosmological constant. These solutions are joined with a cosmological background at infinity which is created by the cosmological constant, instead of the non-realistic Minkowski spacetime. In order to find the black hole solutions we first need to find the matter-vacuum, i.e. $T_{\m\n}=0$, field equations. Taking the trace of   Eq. (\ref{fel}), we may express the Ricci scalar as  $R=4\L-8\p T$. Thus we may rewrite   Eq. (\ref{fel}) as 
\begin{equation}
R_{\m\n}-g_{\m\n}\L=8\p\left(T_{\m\n}-\frac{1}{2}g_{\m\n}T\right),
\end{equation}
and the matter-vacuum equations are: 
\begin{equation}
H_{\m\n}\equiv R_{\m\n}-g_{\m\n}\L=0.
\end{equation}
For simplicity, we have defined the vacuum equations as $H_{\m\n}=0$. As in the case of the Schwarzschild solution, we use a static and spherically symmetric form for the line element: 
\begin{equation}
ds^2=-e^{A(r)} dt^2+e^{B(r)}dr^2+r^2\left(d\theta^2 +\sin^2\theta \,d\varphi^2 \right).\label{met2}
\end{equation}
The components of the Ricci tensor for the metric (\ref{met2}) are given  in Eqs. (\ref{tte})-(\ref{phe}) with $\dot A=\dot B =0$ . Taking the combination $-e^{-A}H_{tt}-e^{-B}H_{rr}=0$ we find $A'+B'=0$ which is integrated to: 
\begin{equation}
A(r)=-B(r)+c, \label{e22}
\end{equation}
where $c$ is an integration constant. In order to find the value of the constant $c$ we have to implement the boundary conditions for our solution. The spacetime far away from the black hole  will be the (Anti-)de Sitter cosmological spacetime \cite{carroll, inverno, mtw, hartle} 
\begin{equation}
ds^2=-\left(1-\frac{\L}{3} r^2\right) dt^2+\left(1-\frac{\L}{3}  r^2\right)^{-1}dr^2+r^2\left(d\theta^2 +\sin^2\theta \,d\varphi^2 \right), \label{dss}
\end{equation}
which describes an 
evolving universe in static coordinates.
  Therefore,  since the spacetime is asymptotically (Anti-)de Sitter at infinity we have $1/g_{rr}=-g_{tt}$ or equivalently $A\approx-B$, thus the integration constant is $c=0$. Substituting Eq. (\ref{e22}) (with $c=0$) into the equation $H_{\theta\theta}=0$ we find $\left(r e^A+r^3\L/3\right)'=1$ which may be easily integrated to give
\begin{equation}
e^A=1-\frac{a}{r}-\frac{\L}{3}r^2,
\end{equation}
where $a$ is an integration constant.  
As in the case of the Schwarzschild solution, we may identify the integration constant as $a=2M$. Therefore, our  solution takes the following form: 
\begin{equation}
ds^2=-\left(1-\frac{2M}{r}-\frac{\L}{3}r^2\right) dt^2+\left(1-\frac{2M}{r}-\frac{\L}{3}r^2\right)^{-1}dr^2+r^2\left(d\theta^2 +\sin^2\theta \,d\varphi^2 \right). \label{schds}
\end{equation}
The above solution is known as Schwarzschild-(Anti-)de Sitter \cite{nari1, nari2} and describes a   static black hole in the cosmological backround of a 
 universe filled with a cosmological constant. The asymptotically de Sitter solution corresponds to a positive cosmological constant while the asymptotically Anti-de Sitter solution to a negative one.  

By calculating the square of the Riemann tensor  
\begin{equation}
R_{\m\n\r\s}R^{\m\n\r\s}=\frac{48M^2}{r^6}+\frac{8\L^2}{3}\label{rimcu}
\end{equation}
we easily verify that the Schwarzschild (Anti-)de Sitter spacetime has a black hole singularity at $r=0$ while it has a constant curvature at infinity $R_{\m\n\r\s}R^{\m\n\r\s} \simeq 8\L^2/3$. For an asymptotically de Sitter  black hole with $\L>0$,  the equation $1/g_{rr}=0$ may have two positive roots. In terms of the roots we may rewrite the components of the metric as 
\begin{equation}
|g_{tt}|=1/g_{rr}=\frac{\L}{3r}(r-r_h)(r_c-r)(r+r_c+r_h), 
\end{equation}
with 
\begin{equation}
\frac{3}{\L}=(r_c+r_h)^2-r_cr_h, \4\2\text{and}\4\2
\frac{6 M}{\L}=(r_c+r_h)r_cr_h.  
\end{equation}
The $r_h$ corresponds to the horizon of the black hole while the $r_c$ corresponds to the second root which is known as the cosmological horizon. Due to the accelerated expansion of a de Sitter universe, the speed at which a very distant object moves away from our reference point will eventually effectively exceed the speed of light.  Thus, after some time we will not be able to detect light signals emitted  from this object. The cosmological event horizon  
 determines the largest  distance from which light  emitted now from an object  can ever reach our reference point in the future. For an asymptotically Anti-de Sitter  black hole with $\L=-|\L|$,  the equation $1/g_{rr}=0$ may have only one  root. In terms of this root we may express the metric as 
\begin{equation}
|g_{tt}|=1/g_{rr}=\frac{|\L|}{3r}(r-r_h)(r^2+r_h \,r+\b), 
\end{equation}
with $\b=6M/(|\L|\,r_h)$ and
 
\begin{equation}
r_h=\frac{1}{\left(-3 \Lambda ^2
   M+\sqrt{9 \Lambda ^4 M^2+|\Lambda| ^3}\right)^{1/3}}-\frac{\left(-3 \Lambda ^2 M\sqrt{9 \Lambda ^4 M^2+|\Lambda|^3}\right)^{1/3}}{|\Lambda|
   }
\end{equation}
The Anti-de Sitter solution cannot describe an accelerated expanding universe thus is ruled out as an invalid cosmological model. However the last years, the Anti-de Sitter (AdS) spacetimes have been proved very important due to their application in the AdS/CFT (Conformal Field Theory) correspondence \cite{ads1,ads2,ads3}. In this, the physics in an Anti-de Sitter spacetime  is considered to  be holographically equivalent to the physics in a flat spacetime of one less  dimension \cite{ads4,ads5,ads6,ads7,ads8}. 

\begin{figure}[t!] 
\begin{center}
\minipage{0.55\textwidth}
  \includegraphics[width=\linewidth]{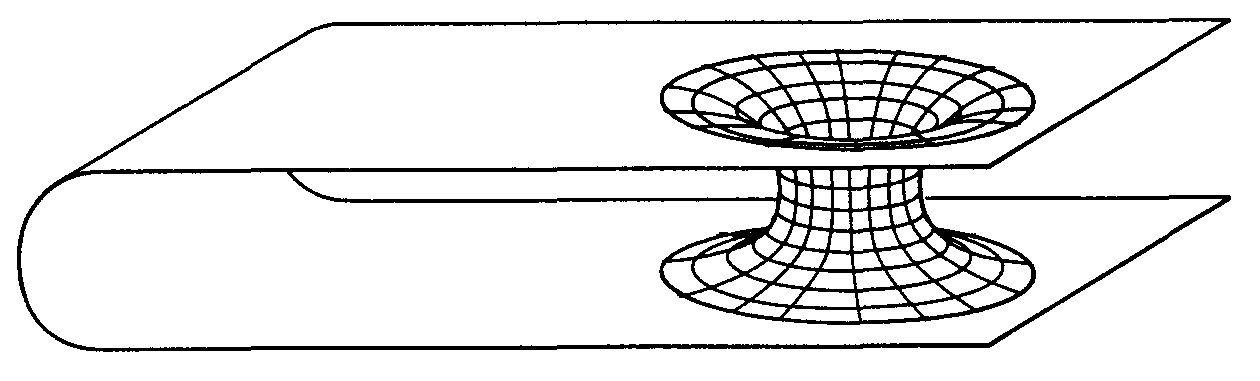}
\endminipage\hspace{3mm}
\minipage{0.39\textwidth}
  \includegraphics[width=\linewidth]{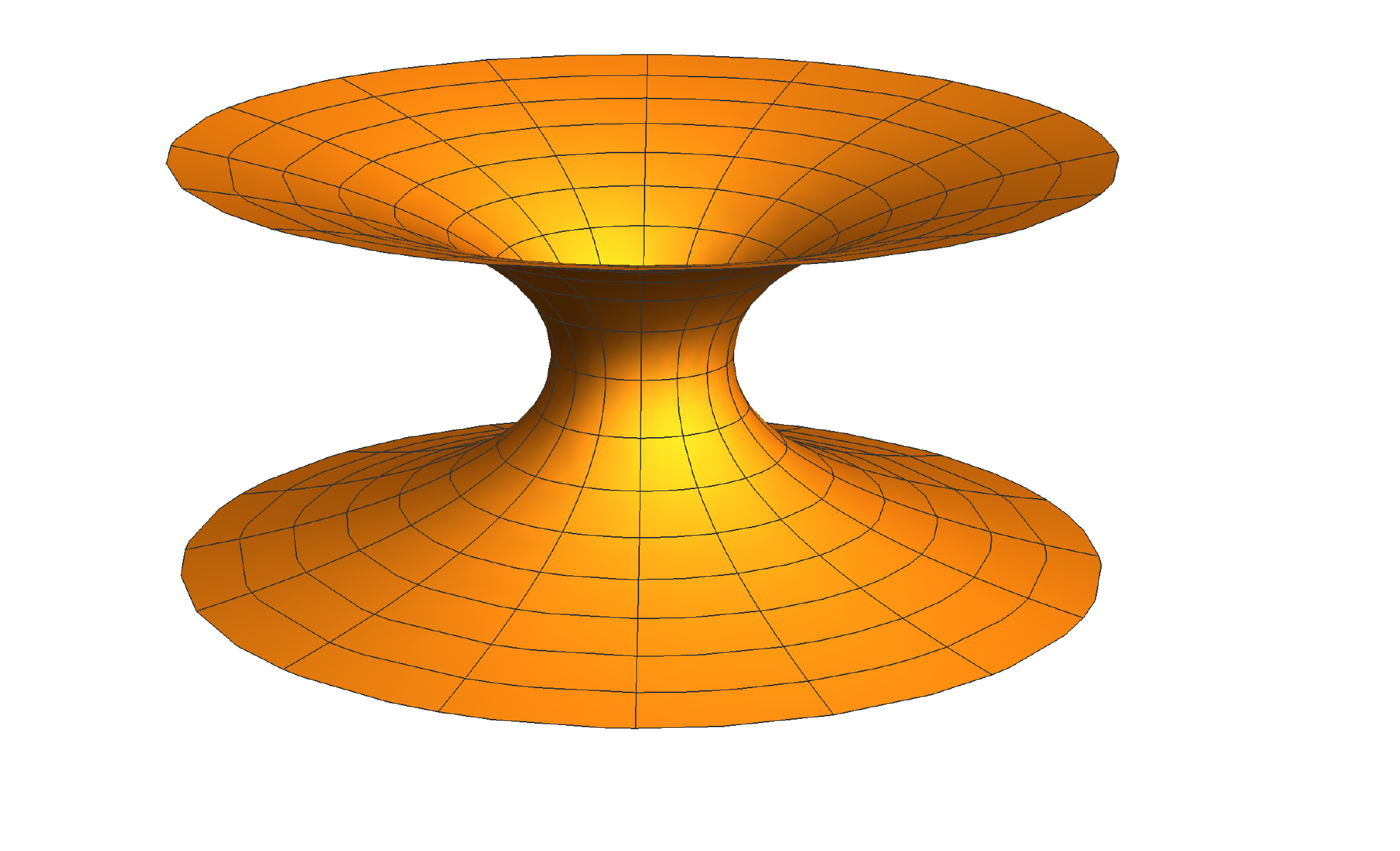}
\endminipage
\\
\hspace*{0.7cm} {(a)} \hspace*{7.5cm} {(b)}  \vspace*{-0.5cm}
\end{center}
\caption{\label{fig1b}
(a) An intra-universe wormhole which connects two distant regions of the same asymptotically flat universe. This figure was constructed by a similar figure  from \cite{mtw}.
(b) An inter-universe wormhole which connects two distinct asymptotically flat universes.}
\end{figure}

\section{Wormholes in General Relativity}

A wormhole is a solution of the Einstein’s field equations which has the property to connect two distant regions in spacetime \cite{Visser:1995cc}. A wormhole may connect
two distant regions of our universe (intra-universe wormhole) or
two different universes (inter-universe wormhole).  
The difference between the two kinds of wormholes is depicted in Fig. \ref{fig1b}. However, this difference is  topological. An observer who makes measurements near the wormhole cannot identify the class of the wormhole.
The first wormhole solution, known as Einstein-Rosen Bridge, was found by  Einstein and Rosen at 1935 \cite{erbr}.
The Einstein-Rosen Bridge was constructed as a geometrical model for the electron and other elementary particles. This wormhole solution  may be easily constructed from the Schwarzschild black hole  
\begin{equation}
ds^2= -\left( 1-\frac{2M}{r}\right) dt^2 + \left( 1-\frac{2M}{r}\right)^{-1} dr^2 + r^2\left(d\theta^2 +\sin^2\theta \,d\varphi^2 \right).
\end{equation}
By making the coordinate change $u^2=r-2M$ the Schwarzschild solution takes the form:
\begin{equation}
ds^2=- \frac{u^2}{u^2+2M} dt^2 +4 \left( u^2+2M \right)  du^2 + \left( u^2+2M \right)^2\left(d\theta^2 +\sin^2\theta \,d\varphi^2 \right).
\end{equation}
The allowed region of the new coordinate $u$ is   $-\infty<u<+\infty$. 
The circumferential radius, i.e. the coefficient in front of the angular part of the metric, is $R_c=\left( u^2+2M \right)$ and has a minimum at $u=0$ with  $R_c(0)=2M$. The throat is a property of a wormhole  and is located at the minimum of the circumferential radius.  

The Einstein-Rosen wormhole is constructed by a coordinate transformation. However, the coordinate $u$ is a bad coordinate which covers only a part of the black-hole geometry. Actually, the coordinate transformation discards the interior region of the black hole $r\in[0,2M)$ and double covers the exterior region $r\in[2M,+\infty)$. Therefore, the Einstein-Rosen Bridge is not an actual wormhole, instead it is a ``disguised'' black hole.   There are some serious  problems with the  Einstein-Rosen wormholes:  First of all, Einstein-Rosen bridges are one-way wormholes. The throat of the wormhole is located at $r=2M$ which is also the location of the horizon of the black hole. However in order to travel through the wormhole we need to cross the horizon which is an one-way surface. A possible exit would be a white hole but we know that white holes are unstable objects; under a small perturbation they collapse to black holes \cite{Eardley:1974zz}. Thus, if one crosses the horizon, he is trapped inside the black hole and is forced to hit the singularity.  Actually, the aspiring traveler should not worry so much about the interior of the black hole. The huge tidal forces near the wormhole would rip apart the traveler way before he  reaches the throat.  As Matt Visser states \textit{``If you discover an Einstein-Rosen bridge, do not attempt to cross it, you will die. You will die because you are jumping into a black hole"}\cite{Visser:1995cc}. Finally, the interior of the black hole is dynamic. Even if we suppose that the    Einstein-Rosen bridge exists, its throat opens and closes so fast that even a photon could not cross it. The second work in this direction was made by Wheeler in 1955 with the construction of the geons  (gravitational-electromagnetic entities) \cite{geon1, geon2}   which are solutions of the
Einstein-Maxwell coupled system. The metric for a geon is everywhere flat except for two widely separated regions which are connected with a wormhole-like structure. The geons were constructed as models for electrical charges and particle-like entities. In modern language, the geon is an unstable gravitational-electromagnetic quasi-soliton i.e. a particle-like solution. The main problem with the Wheeler wormholes, apart from their stability, is their size. A typical size for a geon is of the order of the Planck length.

All the aforementioned wormholes have serious problems which made them unrealistic. However, a work by Morris and Thorne established the foundations for  the modern wormhole physics \cite{Morris:1988cz}. The usual way for the derivation of a solution in General Relativity is to choose a form for the energy-momentum tensor and then to solve the differential set of the field equations in order to find the metric. Morris and Thorne instead, adapted an engineering-like technique. They started from a metric which describes a traversable wormhole and by solving the Einstein field equations in the reverse direction they determined the form of the associated energy-momentum tensor. Then, if one knows  the form of the matter distribution which supports the solution and the characteristics of the metric, one could theoretically build the wormhole or search in the sky for astrophysical wormholes. For simplicity, Morris and Thorne also assumed that the wormholes are static, non-rotating and spherically symmetric. In order to construct traversable wormholes they used the following metric: 
\begin{equation}
ds^2=-e^{2\f}dt^2+\left(1-\frac{b(r)}{r}\right)^{-1}dr^2+r^2\left(d\theta^2 +\sin^2\theta \,d\varphi^2 \right).\label{metw}
\end{equation}
The above metric must obey everywhere the Einstein's field equations. Since every wormhole possesses at least one throat, the first condition we should impose in the above metric is the existence of the throat. In this coordinate system the location of the throat is determined from the equation $1/g_{rr}=0$, thus $\exists\, r_0 : b(r_0)=r_0$. This definition of the throat is very similar with the definition of the Schwarzschild radius and the horizon of a black hole. In order to ensure that our wormholes do not possess horizons nor singularities and thus are traversable in both directions we demand that   $g_{tt}$ is everywhere non-vanishing and finite or equivalently that the metric function $\f(r)$ is everywhere finite. The coordinate system covers two asymptotically flat regions $-\infty<r_-\leq -r_0$ and $r_0\leq r_+<\infty$ which are joined at the throat $r_0$. In addition, in order to ensure   asymptotic  flatness we demand: 
\begin{equation}
\lim_{r\to\ \pm\infty} \f(r)\,\,=\lim_{r\to\ \pm\infty} b(r) =0.
\end{equation}
Since the wormhole is asymptotically flat, we may find its mass from the expansion at infinity. From the weak field analysis we found that $|g_{tt}|=1+2\F$, thus at infinity $|g_{tt}|=1-2M/r$, where $M$ is the mass of the wormhole. 
We may easily eliminate the divergence of the $g_{rr}$ component by using a coordinate transformation. For example using the transformation $r^2=\eta^2+r_0^2$ which covers both asymptotic regions $\eta\in(-\infty,\infty)$ the metric takes the form: 
\begin{equation}
ds^2=-e^{2\text{v(\eta)}}dt^2+f(\eta)d\eta^2+\left(\eta^2+r_0^2\right)\left(d\theta^2 +\sin^2\theta \,d\varphi^2 \right).
\end{equation}
Alternatively we may use the proper distance 
\begin{equation}
l(r)=\int_{r_0}^r \sqrt{g_{rr}(r')}dr'=\int_{r_0}^r \frac{dr'}{\sqrt{1-b(r')/r'}} \4\Leftrightarrow\4 \frac{dr}{dl}=\sqrt{1-\frac{b(r)}{r}},
\end{equation}
 as a radial coordinate, in terms of which the metric takes the form: 
\begin{equation}
ds^2=-e^{2\f(l)}dt^2+ dl^2+r^2(l)\left(d\theta^2 +\sin^2\theta \,d\varphi^2 \right).
\end{equation}
In the above metrics the functions $\text{v}(\eta)$, $f(\eta)$ and $\f(l)$ are everywhere finite. Also, in both metrics the throat is located at the minimum of the circumferential radius  
\begin{equation}
\frac{d}{d\eta}\sqrt{\eta^2+r_0^2}=\frac{d r(l)}{dl}=0. 
\end{equation}
As a final step,  Morris and Thorne suggested that we should adjust the parameters of the wormhole in order to made it traversable by humans. An observer should be able to travel through the wormhole and safely return. The wormhole should be stable under perturbations otherwise it would collapse as a traveler tries to cross it. Also the acceleration that a traveler would feel should be comparable to the Earth's gravity acceleration and the tidal forces should be small. Finally, a   traveler should be able to cross the wormhole in a finite and small proper time (Morris and Thorne demanded less than a year).

In order to find the form of the energy momentum tensor we have to calculate the components of the Einstein's tensor. Using the metric (\ref{metw}) we find: 
\begin{align}
G_{tt}&=\frac{e^{2\f}b'}{r^2},\\[4mm]
G_{rr}&=\frac{2 r^2 \phi '-b \left(2 r \phi '+1\right)}{r^2 (r-b)},\\[4mm]
G_{\theta\theta}&=\sin^2\theta \,G_{\varphi\varphi}=\frac{r \left(2 r \left(r \phi ''+r  \phi '^2+\phi '\right)-b'  \left(r \phi '+1\right)\right)-b \left(2 r^2
   \phi ''+2 r^2  \phi '^2+r \phi '-1\right)}{2 r}.
\end{align}
The above equations are rather complicated. However, we may simplify our analysis by changing the reference frame as $\vec{e}_{\hat \a}=L_{\hat{\a}}^{\,\,\m}\vec{e}_\m$, with 
\begin{equation}
L_{\hat{\a}}^{\,\,\m} =\text{diag}\left(e^{-\f},\,\,\sqrt{1-\frac{b}{r}},\,\,\frac{1}{r},\,\,\frac{1}{r \,\sin\theta}\right). 
\end{equation}
In terms of the above transformation the metric tensor assumes a Minkowski-like form $g_{\hat{\a}\hat{\b}}=L_{\hat{\a}}^{\,\,\m}L_{\hat{\b}}^{\,\,\n}g_{\m\n}=\eta_{\hat{\a}\hat{\b}}$, while the components of the Einstein's tensor $G_{\hat{\a}\hat{\b}}=L_{\hat{\a}}^{\,\,\m}L_{\hat{\b}}^{\,\,\n}G_{\m\n}$  assume a simplified form: 
\begin{align}
G_{\hat{t}\hat{t}}&=\frac{b'}{r^2},\\[4mm]
G_{\hat{r}\hat{r}}&= -\frac{b}{r^3}+2\left(1-\frac{b}{r}\right)\frac{\f'}{r} ,\\[4mm]
G_{\hat{\theta}\hat{\theta}}&=\,G_{\hat{\varphi}\hat{\varphi}}= \left(1-\frac{b}{r}\right)\left[\f''+\f'\left(\f'+\frac{1}{r}\right)\right]-\frac{1}{2r^2}[b'r-b]\left(\f'+\frac{1}{r}\right).
\end{align}
For a spherically symmetric spacetime, the more general form of the energy momentum tensor is $T_{\hat{t}\hat{t}}=\r,$ $T_{\hat{r}\hat{r}}=-\tau,$ and $T_{\hat{\theta}\hat{\theta}} =\,T_{\hat{\varphi}\hat{\varphi}}=p.$ With $\r$ we denote the energy density, with $\tau$ the radial tension and finally with $p$ the transverse pressure. Substituting in the Einstein's field equations $G_{\m\n}=8\p T_{\m\n}$ we may find the form for each one of the above quantities. However, the fact that we have found the form of the energy momentum tensor does not mean that it is indeed physically acceptable. By taking the combination of the $(t,t)$ and $(r,r)$ equations we find: 
\begin{equation}
8\p(\r-\tau)=-\frac{e^{2\f}}{r}\left[e^{-2\f}\left(1-\frac{b}{r}\right)\right]'. \label{nullw}
\end{equation}
Since the circumferential radius $r(l)$ has a minimum at the throat its second derivative near the throat should be positive, thus
\begin{equation}
\frac{d^2r}{dl^2}=\frac{d}{dl}\left(\frac{dr}{dl}\right)=\frac{dr}{dl}\frac{d}{dr}\left(\frac{dr}{dl}\right)=\frac{1}{2}\frac{d}{dr}\left[\left(\frac{dr}{dl}\right)^2\right]=\frac{1}{2}\left(1-\frac{b}{r}\right)'>0.
\end{equation}
Therefore, from Eq. (\ref{nullw}) we deduce that near the throat of the wormhole  
\begin{equation}
\r-\tau<0. \label{nulle}
\end{equation}
However this combination appears in the Null Energy Condition   which states that $T_{\m\n}n^\m n^\n\geq 0$ for any null vector  $n^\m$. A null vector is defined as the vector which obeys the equation $g_{\m\n}n^\m n^\n=0$. For a spherically symmetric energy momentum tensor  the Null Energy Condition takes the explicit form $\r-\tau \geq 0$ and  $\r+p \geq 0$. Therefore from Eq. (\ref{nulle}) we conclude that near the throat of a traversable wormhole the Null Energy Condition is violated. The only way to construct a traversable wormhole and to keep its throat open is to use some kind of exotic matter, i.e. matter which violates the energy conditions, near the throat. In other words, one need to seek for wormholes in theories beyond General Relativity. 

\section{Modified Gravitational Theories}

Although General Relativity is well tested, it is known that is not a perfect theory. The
 motivations for the modifications of the General Relativity are mainly cosmological. The Standard Model for Cosmology has many open problems, like the nature of the dark energy and dark matter, the accurate model for the inflation or the initial singularity problem. All the above, together with theoretical reasons such as the non-renormalizability of the General Relativity, motivate the need for modified gravitational theories.

\begin{figure}[t!] 
\begin{center}
\hspace{0.0cm} \hspace{-0.6cm}
\includegraphics[height=.64\textheight, angle =0]{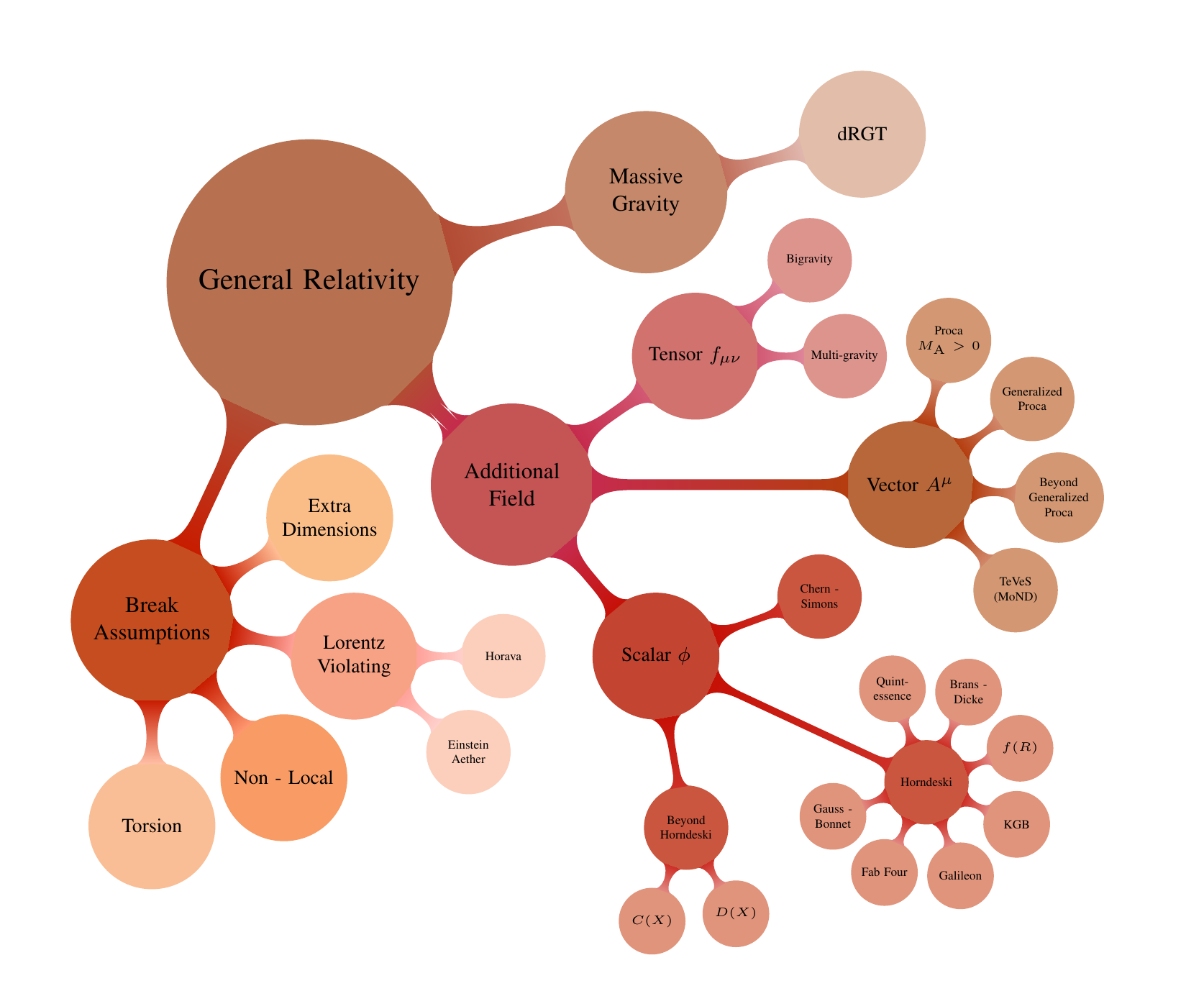}
\end{center}
\caption{A roadway to the  modified gravitational theories. This figure is inspired from a similar figure in \cite{road}. 
\label{fig02}}
\end{figure} 
The usual way to modify gravity is through the modification of the Einstein-Hilbert action
\begin{equation}
S=\int d^4x\sqrt{-g}\left[  \frac{1}{16\p}\left[R(g_{\m\n})-2\L\right]+\mathcal{L}_M(g_{\m\n},\psi_i)  \right],\label{eha}
\end{equation}
where with $\psi_i$ we denote the matter fields. The first thing that one would  do in order to modify gravity is to allow the graviton to has a mass $m_g$. These theories are known as Massive Gravitational theories. The first Massive Gravitational theory was presented by Fierz and Pauli in 1939 \cite{fp}. This theory   failed to reproduce the General Relativity as a $m_g \rightarrow 0$ limit and also was containing unstable problematic degrees of freedom known as ghosts \cite{fp1, fp2}. Many other  Massive Gravitational theories have been proposed after the work of Fierz and Pauli \cite{fp3, fp4, fp5}. However, the first Ghost Free Massive Gravitational theory was proposed in 2011 by de Rham, Gabadadze and Tolley, a theory  known as  dRGT theory \cite{dr1, dr2, dr3, dr4, dr5, dr6, dr7}. Another way towards the modification of   gravity is to break the assumptions of   General Relativity. The most known theories in this direction are the theories with extra spacetime dimensions like the Kaluza-Klein theory, the Lovelock theory or the String theory \cite{love, klei1, klei2, ss1, ss2, ss3, ss31, ss32, ss4, ss5, ss6, ss7}. Also the General Relativity may be modified by breaking the Lorentz invariance \cite{l1, l2, l3, l4, l5, l6, l7} or by non-local theories which include inverse powers of the Laplacian operator $(\nabla^2=g^{\m\n}\nabla_\m\nabla_\n)$ in their action like the term $R \frac{1}{\nabla^2} R$ \cite{n1, n2, n3, n4, n5, n6, n7, n8}. Furthermore, a class of theories which are very popular the last years are the torsion theories in which the field equations of General Relativity may be described using the torsion instead of the curvature \cite{tg1, tg2}. The last way to modify gravity is the addition of extra fields in the theory. Since all the fields that carry energy couple minimally with gravity (through the term $\mathcal{L}_M$ in the action (\ref{eha})), an extra field which modifies gravity should result to extra propagating degrees of freedom in the spectrum of the gravitational waves; otherwhise it is just a matter field. The extra fields may be tensors, vectors or scalars \cite{addfields}. In tensor theories there are two (Bigravity) \cite{bigrav} or more (Multi-Gravity) \cite{multigrav} interacting metric tensors in the theory resulting to the existence of two or more spin-$2$ particles. The vector theories, also known as vector-tensor theories, contain one or more vector fields coupled with the gravitational field. The most simple vector theory is known as Proca theory \cite{pr} and contains the kinetic term of a vector field $F_{\m\n}F^{\m\n}$ defined as $F_{\m\n}=\partial_\m A_\n -\partial_\n A_\m$ and a mass term for the field $M_A^2 A_\m^2$. Also the most general vector theory with second order equations of motion is known as Generalized Proca theory \cite{gpr1, gpr2, gpr3, gpr4} while the theories with higher order equations are known as Beyond Generalized Proca \cite{bpr1, bpr2}. The scalar theories, also known as scalar-tensor theories, contain scalar fields interacting with the gravitational field. Like the vector theories we have two major classes of scalar theories, the Horndeski theory \cite{Horndeski} which is the most general scalar theory with second order equations of motion and the Beyond Horndeski theories with higher order derivatives \cite{beyondhorn1, beyondhorn2}, also known as  Degenerate Higher Order Scalar Tensor (DHOST) theories.  There also exist theories which contain both scalars and vectors. These theories are known as Tensor-Vector-Scalar theories (TeVeS) \cite{teve1, teve2}.  In Fig. (\ref{fig02}) we depict the roadway for the modern modified gravitational theories.

The most simple and most studied modified theories of gravity are the scalar tensor theories. One of the  first scalar tensor theories that appeared in the literature  is the Brans-Dicke  theory \cite{bd} which consists of a scalar field non minimally coupled with the gravitational field: 
\begin{equation}
S=\frac{1}{16\p}\int d^4x\,\sqrt{-g}\left(\f_\p R-\frac{\omega}{\f_\p}\partial_\m\f_\p\partial^\m\f_\p-U(\f_\p)\right),
\end{equation}
where $\f_\p$ is the scalar field, $\omega$ is a coupling constant and $U(\f_\p)$ is the potential of the scalar field. In this theory, the scalar field generates an effective and varying Newton's constant through the relation $G(\f_\p)=1/\f_\p$. From the variation of the above action with respect to the metric $g_{\m\n}$ we may derive the field equations while by varying with respect to the scalar field we may find the scalar equation. However, we may simplify our analysis by performing a conformal transformation which removes the non-minimal coupling \cite{carroll, conformal}. Using the following transformation of the metric $\tilde{g}_{\m\n}=\f_\p g_{\m\n}$ the  action takes the form: 
\begin{equation}
S=\frac{1}{16\p}\int d^4x \sqrt{-\tilde{g}}\left(\tilde R  -\frac{1}{2}\left[\frac{1}{\f_\p^2}\left(\frac{\omega}{2}+3\right)  \right]\tilde{\nabla}_\m\f_\p\tilde{\nabla}^\m\f_\p -\frac{U(\f_\p)}{\f_\p^2}     \right).
\end{equation}
By redefining the scalar field as $d\f=\left[\frac{1}{\f_\p^2}\left(\frac{\omega}{2}+3\right)  \right]^{1/2}d\f_\p$, the action takes the simpler form
\begin{equation}
S=\frac{1}{16\p}\int d^4x \sqrt{-\tilde{g}}\left(\tilde R  -\frac{1}{2}\tilde{\nabla}_\m\f\tilde{\nabla}^\m\f-V(\f)    \right),
\end{equation}
where we define the new potential as $V=U/\f_\p^2$. Since the Einstein's equations take on their conventional form in the above frame, it is usually called the Einstein frame. The initial frame with the non-minimal coupling and the metric $g_{\m\n}$ is called Jordan frame. Note that the above action does not contain  a matter term. If we want to include a matter term in the Einstein frame we should take into consideration the conformal transformation in the calculation of the energy-momentum tensor: 
\begin{equation}
\tilde{T}^M_{\m\n}=-2\frac{1}{\sqrt{-\tilde{g}}}\frac{\d S_M}{\d \tilde{g}^{\m\n}}=-2\frac{1}{\sqrt{-\tilde{g}}}\frac{\partial g^{\a\b}}{\partial \tilde{g}^{\m\n}}\frac{\d S_M}{\d g^{\a\b}}=T^M_{\m\n}/\f_\p. 
\end{equation}
Also, using a similar method we find the contribution of the matter term in the scalar equation: 
\begin{equation}
\frac{\d S_M}{\d \f}=\frac{\partial g^{\a\b}}{\partial \phi}\frac{\d S_M}{\d g^{\a\b}}=-\frac{1}{2 \f}\frac{d \f_\p}{d \f}\sqrt{-\tilde{g}} \tilde{T}^M,
\end{equation}
where $\tilde{T}^M$ is the trace of the energy momentum tensor. Finally, the variation of the full action with respect both to the metric and to the scalar field leads to the following field equations: 
\begin{equation}
\tilde{G}_{\m\n}=\tilde{T}^\f_{\m\n}+8\p\tilde{T}^M_{\m\n},
\end{equation}
where the effective energy-momentum tensor is defined as
\begin{equation}
\tilde{T}^\f_{\m\n}=\frac{1}{2}\tilde{\nabla}_\m\f\tilde{\nabla}_\n\f-\frac{1}{2}\tilde{g}_{\m\n}\left(\frac{1}{2}\tilde{\nabla}_\r\f\tilde{\nabla}^\r\f+V(\f)\right).
\end{equation}
The scalar field equation has the following form: 
\begin{equation}
\tilde{\nabla}^2\f-\frac{d V}{d \f}=\frac{1}{2\f}\frac{d\f_\p}{d \f}\tilde{T}^M.
\end{equation}
While the field equations in the Einstein frame look  like the classical General Relativity sourced by a scalar field, we should not forget that we started from a modified gravitational theory. A test particle will move along geodesics of the Jordan frame metric $g_{\m\n}$ which in general are different from the geodesics of the conformal metric $\tilde{g}_{\m\n}$ on the Einstein frame. The Brans-Dicke theory is very useful in the investigation of the inflation and as a dark energy model as well \cite{bdc1, bdc2, bdc3, bdc4, bdc5}. The first Brans-Dicke theory that appeared in literature which had no potential $U=0$ is experimentally  excluded as $\omega>40000$ \cite{General1, Bertotti:2003rm}. However, theories with a potential or even with a cosmological constant seem to still be valid \cite{bdc1,Yadav:2019fjx,Sola:2019jek,Sola:2020lba,Babichev:2013usa}.   Also, it is useful in the understanding of the $f(R)$ theories since by doing an appropriate conformal transformation the $f(R)$ theory reduces to an equivalent Brans-Dicke  theory \cite{fr1, fr2, fr3, fr4}. 
 
Finally, we have already mentioned that the most general scalar-tensor theory in four dimensions and with second order equations of motion is the Horndeski theory \cite{Horndeski}. This theory was constructed in 1974 by Horndeski and rediscovered in 2011 in the framework of the Generalized Galileon theory \cite{gg1, gg2, Kobayashi:2011nu,horth1}. As shown by Ostrogradski \cite{os1}, the significance of the construction of a theory with second order equation of motion is that it does not contain problematic ghost states \cite{os2, os3}. Using the modern formulation of the Generalized Galileons, the Horndeski theory may be expressed as: 
\begin{equation}
S=\int d^4x\,\sqrt{-g}\left[\sum_{i=2}^5 \mathcal{L}_i(g_{\m\n},\f)+\mathcal{L}_M(g_{\m\n},\psi_i)\right], 
\end{equation}
where we assume that as usual the matter field are coupled minimally with the gravitational field. The four Lagrangians in the sum have the following form
\begin{align}
\mathcal{L}_2=& G_2(\f, X),\\[3mm]
\mathcal{L}_3=& G_3(\f, X)\nabla^2\,\f,\\[3mm]
\mathcal{L}_4=& G_4(\f, X)R+G_{4,X}\left[(\nabla^2\f)^2-(\nabla_\m\,\nabla_\n \,\f)^2 \right],\\[2mm]
\mathcal{L}_5=&G_5(\f, X)G_{\m\n}\nabla^\m\,\nabla^\n \f-\frac{1}{6}G_{5,X} \big[ (\nabla^2\f)^3 -3(\nabla^2\f)(\nabla_\m\,\nabla_\n \,\f)^2 \nn\\[3mm]
&+2 (\nabla_\n\,\nabla^\m \,\f)\,\,(\nabla_\a\,\nabla^\n \,\f)\,\,(\nabla_\m\,\nabla^\a \,\f)   \big],
\end{align}
where $G_i$ are functions of $X$ and $\f$, $X=\nabla^\r\f\nabla_\r\f/2$ and $G_{i,X}=\partial G_i/\partial X$. Horndeski theory includes the General Relativity ($G_4=1$, $G_2=G_3=G_5=0$), and the Brans-Dicke-$f(R)$ theories ($G_2=-\omega X/\f-V(\f)$, $G_4=\f$, $G_3=G_5=0$)   as well as  many other scalar-tensor theories as special cases \cite{sst1, sst2, sst3, sst4, fab1, fab2, fab3}. Except of the Horndeski theories, the last years have appeared many other ghost-free theories like the Degenerate Higher-Order Scalar-Tensor (DHOST) theories \cite{Langlois:2015cwa,BenAchour:2016fzp
} or the Extended Scalar-Tensor (EST) theories  \cite{Crisostomi:2016czh}.

\section{The No-Scalar-Hair theorem}\label{nohhhh}

One of the most interesting properties of the black-hole solutions in General Relativity  is their uniqueness and simplicity: black holes are uniquely determined only by three physical quantities (mass, electromagnetic charge and angular momentum). No-Hair theorems, which forbid the association of the GR black holes with any other type of conserved  ``charge", were formulated quite early on. The first No-Hair type theorem that appeared was   Birkhoff's theorem \cite{birk, birk1, birk2, carroll, inverno, mtw, hartle}  which states that the Schwarzschild solution is unique. In other words, the only  spherically symmetric black hole in vacuum  is the Schwarzschild one  and the only conserved ``charge'' of this solution is the mass. The  Birkhoff's theorem may be extended  in the case of the presence of an electromagnetic field. In this case, the only spherically symmetric black hole is the Reissner–Nordström one and the associated conserved charges are the mass and the electromagnetic charge. Finally, in the case of rotating black holes no-hair theorems were formulated by Carter and Robinson \cite{car1, car2, car3, car4} and also by Hawking \cite{hok}. These theorems state that the only axisymmetric black hole is the Kerr-Newmann\footnote{Rotating black holes which do not have electromagnetic charge are known as Kerr black holes.} one which has the mass, the electromagnetic charge and the angular momentum as conserved charges. All the other information about the object which collapsed to a black hole (or about an object falling into the black hole) disappears permanently behind the horizon. 

Even in the case of the scalar-tensor theories, No-scalar Hair theorems were proposed which made difficult the derivation of new black-hole solutions in these theories. The first No-Scalar-Hair theorem was developed by Bekenstein in 1972 \cite{NH-scalar1} \cite{NH-scalar2} in the framework of a scalar-tensor theory with a non-minimally coupled scalar field
 \begin{equation}
S=\frac{1}{16\p}\int d^4x \sqrt{-g}\left( R  -\frac{1}{2}\nabla_\m\f\nabla^\m\f-V(\f)    \right).
\end{equation}
In order to derive the No-Scalar-Hair theorem we use the scalar field equation
\begin{equation}
\nabla^2\f-\dot V=0,
\end{equation}
where $\dot V = dV/d\f$. We assume that there exists a  spherically symmetric and asymptotically flat black hole in the theory which possesses scalar hair. We also assume that the scalar field has the same symmetries as the spacetime i.e. $\f(x^\m)=\f(r)$. Multiplying the scalar equation with $\f$ and integrating from $r_h$ up to infinity we get
\begin{equation}
\int_{r_h}^\infty \sqrt{-g} \left(\nabla_\m\f\,\nabla^\m\f +\f\dot V \right)-\left[\sqrt{-g}\f\nabla^\m\f\right]_{r_h}^\infty=0.
\end{equation}
The boundary term may be written as $\sqrt{-g}g^{rr}\f\f'$ and vanishes in both boundaries since $g^{rr}=0$ at the horizon and for an asymptotically flat spacetime $\f(r)=constant$ at infinity. The first term in the integral is   $ \nabla_\m\f\,\nabla^\m\f=g^{rr}\f'^2 $ and is positive definite since $g^{rr}$ is positive everywhere outside the horizon. Thus     the integral in the above equation  vanishes only if
\begin{equation}
 \f\dot V<0.
\end{equation}
However, for many realistic potentials like a mass term $V=m^2 \f^2/2 $ the combination $\f\dot V=m^2\f^2$ is positive. Therefore in these cases, the only spherically symmetric solution is the Schwarzschild which is accompanied with a trivial (constant everywhere) scalar field. In the previous section we showed that the Brans-Dicke theory and the $f(R)$ theory may be reduced to a scalar-tensor theory like the above using a conformal transformation. Thus the No-Scalar-Hair theorem may be easily generalized in order to cover both the Brans-Dicke theory and the $f(R)$ theories \cite{frh1, frh2}. The above theorem is known as the   Bekenstein's  old No-Scalar-Hair theorem.

The old No-Hair theorem does not exclude all   possible potentials. As a result, soon after its derivation new solutions for black holes were found in scalar-tensor theories \cite{YM1, YM2, YM3, YM4, Skyrmions1, Skyrmions2, Conformal1, Conformal2}. Therefore, in order to construct a more general theorem, Bekenstein developed the novel No-Scalar-Hair theorem in 1995 \cite{Bekenstein}. This novel theorem was constructed in the framework of the following scalar-tensor theory: 
\begin{equation}
S=\int d^4x\sqrt{-g}\left[R+\mathcal{E}(\mathcal{J},\mathcal{F,\mathcal{K}})\right],
\end{equation}
where the quantities $\mJ$, $\mF$ and $\mK$ are defined in terms of two scalar fields $\chi$ and $\f$
\begin{equation}
\mJ=\aa_\m\chi\,\aa^\m\chi,\4\mF=\aa_\m\f\,\aa^\m\f\4\text{and}\4 \mK=\aa_\m\chi\,\aa^\m\f. 
\end{equation}
The energy momentum tensor for the above theory is 
\begin{equation}
T_\m^\n=-\mE\,\d_\m^\n+2\frac{\aa\mE}{\aa\mJ}\aa_\m\f\aa^\n\f+2\frac{\aa\mE}{\aa\mF}\aa_\m\chi\aa^\n\chi+ \frac{\aa\mE}{\aa\mK}\left( \aa_\m\f\aa^\n\chi+\aa_\m\chi\aa^\n\f  \right).
\end{equation}
Bekenstein assumed again that an asymptotically flat and spherically symmetric black hole exists in the theory and also that the scalar field has the same symmetries with the spacetime. Also he assumed that the quantity $\mE$ is identified with the local energy density as measured by an observer which moves in a geodesic curve and thus it should be positive.  Then, by examining the sign of the $T_r^r$ component of the energy momentum tensor as well as the $(T_r^r)'$ using the conservation equation $\nabla_\m T^{\m\n}=0$ he found that
\begin{itemize}
\item near the horizon of the black hole $T_r^r<0$ and $(T_r^r)'<0,$
\item while at infinity $T_r^r>0$ and $(T_r^r)'<0.$
\end{itemize}
The above behavior indicates that there is a region $[r_a,r_c]$ --between $[r_h,\infty)\,\,$-- in which the $T_r^r$ component changes sign with $(T_r^r)'>0$. However, using the Einstein's field equations, Bekenstein showed that both $T_r^r$ and $(T_r^r)'$ remain negative in the whole  $[r_h,\infty)$ region. Thus, there is no way for the horizon and an  asymptotically flat region to be  smoothly connected. The only way to avoid this contradiction is the trivial vanishing of the scalar fields $\f$ and $\chi$. Therefore, even in this case, the only black hole solution is the Schwarzschild one. As we will see in the next chapter, the novel No-scalar-Hair theorem may be also evaded in the context of specific modified gravitational theories.

\section{The Einstein-scalar-Gauss-Bonnet theory}\label{esgbi}

In this work, we will focus on a Scalar Tensor theory which is known as the Einstein-scalar-Gauss-Bonnet  theory. In this theory, the scalar field   has a general coupling function  $f(\f)$ with the quadratic gravitational Gauss-Bonnet term defined as: 
\begin{equation}
R^2_{GB}=R_{\mu\nu\rho\sigma}R^{\mu\nu\rho\sigma}-4R_{\mu\nu}R^{\mu\nu}+R^2\,,\label{gabt}
\end{equation}
in terms of the Riemann tensor $R_{\mu\nu\rho\sigma}$, the Ricci tensor $R_{\mu\nu}$
and the Ricci scalar $R$.
It may be shown that the Gauss-Bonnet term in four dimensions is a total derivative. Therefore, the only way to make it contribute to the field equations is through a coupling with an additional field. The Einstein-scalar-Gauss-Bonnet  theory is thus described by the following action functional: 
\begin{equation} 
S=\frac{1}{16\pi}\int{d^4x \sqrt{-g}\left[R-\frac{1}{2}\,\partial_{\mu}\phi\,\partial^{\mu}\phi+f(\phi)R^2_{GB}\right]}.
\end{equation}
The most attractive feature of this theory is that it belongs to the family of Horndeski theories; the field equations are of second order and as a result, it avoids Ostrogradsky instabilities or problematic ghost states. After some tedious and non-trivial calculations we may show that the Horndeski theory with the following $G$ functions
\begin{align}
G_2=&-X+8f^{(4)} X^2(3-\ln X),\\[3mm]
G_3=&4f^{(3)} X (7-3\ln X),\\[3mm]
G_4=&1+4\ddot f  X(2-\ln X),\\[3mm]
G_5=&-4\dot f \ln X,
\end{align}
reduces to the Einstein-scalar-Gauss-Bonnet  theory \cite{Kobayashi:2011nu, DeFelice}. In the above equations with the dots and the numbers at the superscripts we denote differentiation with respect to the scalar field. Another attractive feature of the theory is that the special case with $f(\f)=\a\exp (\l\f)$, where $\a$ is a coupling constant and $\l=\pm 1$ is a part of the low-energy limit of theories with extra dimensions like the heterotic superstring theory \cite{metsa}. The special case with the exponential coupling function is also part of  effective theories constructed from the Lovelock theory in four dimensions \cite{lovetohor}. This special case, which was the first that was studied, is known as Einstein-dilaton-Gauss-Bonnet theory since the scalar field is identified with the dilaton field. Also, the special case with the linear coupling $f(\f)=\a\f$ constitutes a shift symmetric Galileon theory since the action is invariant under the shift $\f\rightarrow\f+c$, where $c$ is a constant \cite{gg1, gg2, Kobayashi:2011nu}. Finally, the  Einstein-scalar-Gauss-Bonnet  theory represents the Ringo  term ($\mathcal{L}_{ringo}=f(\f)R^2_{GB}$) of the Fab Four theory which is the most general scalar tensor theory that is capable of self-tuning (as a dark energy model)\cite{fab1, fab2}. 

Over the last twenty years, many works have been done in the framework of the Einstein-scalar-Gauss-Bonnet theory. First of all,  in the case of  the Einstein-scalar-Gauss-Bonnet theory the No-scalar-Hair theorem is evaded. The first derivation of this evasion was formulated  in the framework of the Einstein-dilaton-Gauss-Bonnet theory \cite{DBH1} only a year after the proposal of the No-scalar-Hair theorem. A few years ago, a new derivation was proposed, this time for the linear coupling \cite{Babichev, SZ, Benkel1, Benkel2, Benkel3}. In both cases   new solutions for asymptotically flat black holes were found \cite{DBH1, DBH2, SZ}. Also, the last twenty years many other solutions for black holes have been found in the framework of the Einstein-scalar-Gauss-Bonnet theory \cite{Lee, konopap, konoko,pertt1,pertt2,pertt3}, including solutions with rotation\cite{Kleihaus1, Kleihaus2, Pani1, Pani2, Herdeiro, Ayzenberg,Maselli:2015tta,Kleihaus:2014lba,Kunz,Ayzen,Pani:2009wy, Cunha:2019dwb}, solutions in the presence of an electromagnetic field \cite{Doneva-Papa, Hartmann2, Brihaye} or even solutions with a cosmological constant \cite{Brihaye:2019dck, Brihaye:2019gla, Hartmann1}. 

 While General Relativity predicts the existence of three families of black hole solutions, it does not admit stable solutions for other compact objects, like wormholes or particle-like solutions also known as solitons. In General Relativity, solitons are unstable \cite{geon1, geon2} while wormholes demand exotic matter near the throat \cite{Morris:1988cz,bbwh1,bbwh2,bbwh3,bbwh4,bbwh5,bbwh6,bbwh7,bbwh8} as discussed previously. However, we know that ultra compact objects which are not black holes nor neutron stars exist in the universe, like the X-ray transient GROJ0422+32 \cite{compact}. These objects have radius between $r_s<r<3r_s/2$, where $r_s$ is the Schwarzschild radius, and as a result do not have a horizon but they possess a light ring.  A soliton, and even a wormhole would be a good model for such objects. On the other side, it is known that the Einstein-dilaton-Gauss-Bonnet theory admits stable traversable wormhole solutions with no need for exotic matter \cite{Kanti:2011jz, Kanti:2011yv} and solutions for solitons as well \cite{Kleihaus:2019rbg,Kleihaus:2020qwo}. These solutions are characterized by a regular spacetime, a regular energy-momentum tensor and may possess throats as well. Thus, the Einstein-scalar-Gauss-Bonnet theory not only predicts new black-hole solutions but may provide reliable models for ultra compact objects and realistic traversable wormholes. 

In addition, in the case of the Einstein-scalar-Gauss-Bonnet theory many cosmological solutions have been produced\cite{gbcc1,gbcc2,gbcc3,gbcc4,gbcc5,gbcc6,gbcc7,gbcc8,gbcc9,gbcc10}. However, as in many modified gravitational theories, the prediction for the speed of the gravitational waves is different than the speed of light \cite{DeFelice}. After the detection of the GW170817 gravitational wave event and its GRB170817A gamma-ray burst counterpart \cite{gw2, gw3, gw4} many modified gravitational theories were constrained 
\cite{ddr1, ddr2}. However, there have been certain criticisms on these constraints \cite{deRham:2018red}. The Einstein-scalar-Gauss-Bonnet theory was one of them and it is widely accepted today that we cannot use this theory as a model for dark energy. However, we can still use the Einstein-scalar-Gauss-Bonnet theory as a model for early time cosmology \cite{Odintsov:2019clh}  or as a model for local gravitational solutions \cite{Nair:2019iur}. In the framework of the Einstein-scalar-Gauss-Bonnet theory  cosmological solutions have been found without initial singularities \cite{cos1, cos2, cos33, cos4, cos5, cos6, cos7, cos8, cos9} and with inflationary solutions \cite{cos9, inf1, inf2, inf3, inf4, inf5, inf6, inf7, inf8, inf9, inf10, inf11}. In addition, early time cosmological solutions have also been found in the framework of the pure scalar-Gauss-Bonnet theory \cite{pgb1, pgb2, pgb3, bakoppgb}. This theory is an approximation in which in regions where the curvature is large and the Gauss-Bonnet term is larger than the Ricci scalar we may remove the Ricci scalar from the EsGB theory. 

In this thesis we will focus only on local solutions in Einstein-scalar-Gauss-Bonnet theory. In order to keep our analysis as general as possible we will keep arbitrary the form of the coupling function $f(\f)$. At first, we will examine whether or not the Einstein-scalar-Gauss-Bonnet theory with general coupling function evade the No-scalar-Hair theorem and we will produce novel solutions for spherically symmetric asymptotically flat black holes. Then, we will construct solutions for black holes in the presence of a cosmological constant. Finally, we will produce new solutions for traversable wormholes.

\clearpage
\thispagestyle{empty}


\chapter{Evasion of the No-Scalar-Hair theorem and new black hole solutions in Einstein-Scalar- Gauss-Bonnet theory}\label{3}

\section{Introduction}

 The Scalar-Tensor theories are the most popular and at the same time the most studied modified gravitational theories. While many cosmological solutions have been produced in scalar-tensor theories over a period of many decades, local gravitational solutions, like black holes, are limited. 
Thus, the existence or not of black holes associated with a non-trivial scalar field in the exterior region has attracted the attention of researchers. Early on,
a {\it no-scalar-hair theorem} \cite{NH-scalar1, NH-scalar2}  appeared, which excluded static black holes with a
scalar field, but this was soon outdated by the discovery of black holes with Yang-Mills
\cite{YM1, YM2, YM3, YM4} or Skyrme fields \cite{Skyrmions1, Skyrmions2}. The emergence of additional black-hole solutions where
the scalar field had a conformal coupling to gravity \cite{Conformal1, Conformal2} led to the
formulation of a novel no-scalar-hair theorem \cite{Bekenstein} (for a review, see 
\cite{Herdeiro}). In more recent years, this argument
was extended to the case of standard scalar-tensor theories \cite{SF}, and also, a new
form was proposed, which covers the case of Galileon fields \cite{HN}. 

However, both novel forms of the no-hair theorem \cite{Bekenstein, HN} were shown
to be evaded: the former in the context of the Einstein-Dilaton-Gauss-Bonnet theory
\cite{DBH1, DBH2} and the latter in a special case of shift-symmetric Galileon theories
\cite{Babichev, SZ, Benkel1, Benkel2, Benkel3}.
The common feature of the above theories was the presence of a non minimal coupling between the scalar field and the gravitational field through
the quadratic Gauss-Bonnet (GB) term defined in Eq.(\ref{gabt}). In the first theory the coupling function was of an exponential form while in the second one the coupling function was linear in the scalar field. In both cases, basic requirements of the no-scalar-hair theorems
were invalidated, and this paved the way for the construction of
new black-hole solutions. 

Here, we consider the most general class of scalar-GB theories,
of which the cases \cite{DBH1, DBH2, SZ} constitute particular examples. We demonstrate
that black-hole solutions, with a regular horizon, an asymptotically-flat limit and a non trivial scalar field 
may in fact be constructed for a large class of such theories under mild only
constraints on the parameters of the theory. 
We address the requirements of both the old and novel no-hair theorems, and we
show that they are not applicable for this specific class of theories. In accordance to the
above,
we then present a large number of exact, regular black-hole solutions with scalar hair
for a variety of forms  of the coupling function. The analysis of this chapter is based on \cite{ABK1}.

The outline of this chapter is as follows: in Section \ref{evthf}, we present our theoretical framework.
In section \ref{sap1} we perform an  analytic study of the near-horizon and far-away radial regimes. In Section \ref{eva33}, we examine the evasion of the No-scalar-Hair theorem in the framework of the EsGB theory. The spontaneous scalarisation effect is discussed in Section \ref{evaspo} while in Section \ref{evapure} we investigate the existence of black-hole solutions in the framework of the pure-Gauss-Bonnet theory. We finish with our conclusions in Section \ref{evadis}. The analysis of this chapter is based on \cite{ABK1}.


\section{The theoretical framework}\label{evthf}

We start our analysis using the following action functional
\begin{equation}\label{action}
S=\frac{1}{16\pi}\int{d^4x \sqrt{-g}\left[R-\frac{1}{2}\,\partial_{\mu}\phi\,\partial^{\mu}\phi+f(\phi)R^2_{GB}\right]},
\end{equation}
which describes a  modified gravitational theory where the Einstein's term (the Ricci scalar curvature $R$) is accompanied by a real scalar field $\phi$ non minimally coupled with the quadratic gravitational Gauss-Bonnet term $R^2_{GB}$. The latter in four dimensions is a topological invariant i.e. a total derivative, thus the only way to keep it in the theory is  through the coupling  with the scalar field $f(\phi)$. 


The variation of  the action (\ref{action}) with respect to the metric tensor $g_{\mu\nu}$
and the scalar field $\phi$ as well  leads to the derivation of the gravitational Einstein's field equations and the
equation for the scalar field, respectively. These equations have the following form:
\begin{eqnarray}
& G_{\mu\nu}=T_{\mu\nu}\,,& \label{Einstein-eqs}\\[1mm]
&\nabla^2 \phi+\dot{f}(\phi)R^2_{GB}=0\,,& \label{scalar-eq}
\end{eqnarray}
where $G_{\mu\nu}$ is the Einstein tensor and the effective energy-momentum tensor is defined as 
\begin{align}\label{f}
T_{\mu\nu}=-\frac{1}{4}g_{\mu\nu}\partial_{\rho}\phi\partial^{\rho}\phi+\frac{1}{2}\partial_{\mu}\phi\partial_{\nu}\phi
-
\frac{1}{2}(g_{\rho\mu}g_{\lambda\nu}+g_{\lambda\mu}g_{\rho\nu})
\eta^{\kappa\lambda\alpha\beta}\tilde{R}^{\rho\gamma}_{\quad\alpha\beta}
\nabla_{\gamma}\partial_{\kappa}f.
\end{align}
In the above equation $\tilde{R}^{\rho\gamma}_{\quad\alpha\beta}$ is the dual of the Riemann tensor, defined as $\tilde{R}^{\rho\gamma}_{\quad\alpha\beta}=\eta^{\rho\gamma\sigma\tau}
R_{\sigma\tau\alpha\beta}=\varepsilon^{\rho\gamma\sigma\tau}
R_{\sigma\tau\alpha\beta}/\sqrt{-g}$. We can write the energy-momentum tensor in the form:
\begin{align}
T_{\m\n}=T^{(\f)}_{\m\n}+T^{\rm(int)}_{\m\n},
\end{align}
where
\begin{align}
T^{(\f)}_{\m\n}&=-\frac{1}{4}g_{\mu\nu}\partial_{\rho}\phi\partial^{\rho}\phi+\frac{1}{2}\partial_{\mu}\phi\partial_{\nu}\phi\\[2mm]
T^{\rm(int)}_{\m\n}&=-
\frac{1}{2}(g_{\rho\mu}g_{\lambda\nu}+g_{\lambda\mu}g_{\rho\nu})
\eta^{\kappa\lambda\alpha\beta}\tilde{R}^{\rho\gamma}_{\quad\alpha\beta}
\nabla_{\gamma}\partial_{\kappa}f.
\end{align}
The energy-momentum tensor
$T_{\mu\nu}$ receives contributions from both the kinetic term of the scalar field $T^{(\f)}_{\m\n}$ and the interaction with the
Gauss-Bonnet term $T^{\rm(int)}_{\m\n}$. Also from Eq. (\ref{scalar-eq}) it is clear that the interaction with the Gauss-Bonnet acts as a potential for the scalar field. Finally, from now on, the dot over a function denotes derivation with respect to the scalar field i.e $\dot f = df/d\f$.

Our aim is to find spherically symmetric solutions for black holes in the context of the EsGB theory. In order to do that, we start
with a line-element of the form
\begin{equation}\label{metric}
{ds}^2=-e^{A(r)}{dt}^2+e^{B(r)}{dr}^2+r^2({d\theta}^2+\sin^2\theta\,{d\varphi}^2)\,,
\end{equation}
that describes a spherically symmetric spacetime. We also assume that the scalar field has the same symmetries as the spacetime i.e. $\phi(x^\m)=\phi(r)$. We seek regular, static and asymptotically-flat black holes with a non trivial scalar field. Our analysis in this chapter
will  investigate whether or not a black hole may be constructed in the framework of the EsGB theory. We  keep general the form of the coupling function $f(\phi)$ and we find the general constraints it needs to obey in order for these solutions to arise. 

Using the above line-element (\ref{metric}), the Einstein's field equations take
the explicit form
\bea
4e^B(e^{B}+rB'-1)&=&\phi'^2\bigl[r^2e^B+16\ddot{f}(e^B-1)\bigr] \nonumber\\
&& \hspace*{-2.5cm}-8\dot{f}\left[B'\phi'(e^B-3)-2\phi''(e^B-1)\right], \label{tt-eq}
\eea
\vskip -0.6cm
\beq
4e^B(e^B- r A'-1)=-\phi'^2 r^2 e^B +8\left(e^B-3\right)\dot{f}A'\phi',
\label{rr-eq}
\eeq
\vskip -0.6cm
\begin{eqnarray}
&&\hspace*{-0.7cm} e^B\bigl[r{A'}^2-2B'+A'(2-rB')+2rA''\bigr]= -\phi'^2 r e^B
\nonumber \\
&&\hspace*{-0.8cm} +8 \phi'^2 \ddot{f}A'+ 
4\dot{f}[\phi'(A'^2+2A'')+A'(2\phi''-3B'\phi')], \label{thth-eq}
\end{eqnarray}
while the scalar equation has the following form:
\beq
 2r\phi''+(4+rA'-rB')\,\phi'+
\frac{4\dot{f}e^{-B}}{r}\bigl[(e^B-3)A'B' 
 -(e^B-1)(2A''+A'^2)\bigr]=0\,. \label{phi-eq}
\eeq
Throughout 
this work, the prime denotes differentiation with respect to the radial coordinate $r$\footnote{In some cases, in the following chapters, we will use different symbols for the radial coordinate. However the prime will still denote differentiation with respect to the radial coordinate, independently of the symbol we use.}.

The $(rr)$-component of the Einstein's field equations  (\ref{rr-eq})  takes the form of a second-order polynomial
with respect to $e^B$ i.e. $e^{2B}+\beta e^B + \gamma=0$, which then, may be solved to give 
\begin{equation}\label{Bfunction}
e^B=\frac{-\beta\pm\sqrt{\beta^2-4\gamma}}{2},
\end{equation}
with
\beq
\beta=\frac{r^2{\phi'}^2}{4}-(2\dot{f}\phi'+r) A'-1\,, \qquad 
\gamma=6\dot{f}\phi'A'\label{bg}\,.
\eeq
Then, we can also eliminate the $B'$ from the remaining equations by taking the derivative of Eq. (\ref{Bfunction}) with respect to the radial coordinate. We easily find
\begin{equation}
B'=\frac{\gamma'+\beta'e^B}{2e^{2B}+\beta e^B}.\label{bdd}
\end{equation}

As a result, once the solutions for the scalar
field $\phi(r)$ and the metric function $A(r)$ are determined, the metric function $B$ follows from Eq. (\ref{Bfunction}). Of the set of the remaining three, Eqs. (\ref{tt-eq}),
(\ref{thth-eq}) and (\ref{phi-eq}), the $e^B$ may be eliminated using the above equations and only two of them are independent. Then we can choose any two of these equations to form a system of two independent,
ordinary differential equations of second order for the metric function $A$ and the scalar field
$\phi$:
\begin{align}
A''&=\frac{P}{S}\,,\label{Aphi1}\\[2mm]
\phi''&=\frac{Q}{S}\,, \label{Aphi-system}
\end{align}
where the three functions $P$, $Q$ and $S$ are lengthy expressions of $(r, \phi', A', \dot f,
\ddot f)$ which are given in the Appendix \ref{apa1}. For simplicity, in these expressions we have eliminated $B'$ using Eq. (\ref{bdd}) but we have kept the dependence on $e^B$.


\section{Approximate solutions at the boundaries}\label{sap1}
\subsection{Solutions near the black-hole horizon}

We will demonstrate now  that the set of our equations, with a general form for the coupling function $f(\phi)$,
may allow   the construction of a solution with a regular black-hole horizon,  provided that the coupling function
$f$ satisfies certain constraints. 
For a spherically-symmetric spacetime, the conditions for the presence of
a black hole horizon at $r=r_h$ are:
\begin{equation}
g_{tt}|_{r=r_h}\rightarrow 0 \4 \2 \text{and} \4\2 g_{rr}|_{r=r_h}\rightarrow \infty.\label{cobbb}
\end{equation}
In our case, the metric function $B$ is a dependent function through Eq. (\ref{Bfunction}), and the above constraints translate to $e^A \rightarrow 0$ near the horizon $r \rightarrow r_h$. Equivalently near the horizon of the black hole, we have
$A' \rightarrow \infty$ -- in our analysis, we will use that as an assumption
but, from the general near-horizon solution that we will construct, this assumption will be shown to follow from the first condition of Eq. (\ref{cobbb}).
 Finally,   we search for solutions with regular black-hole
 horizon; this demands that
$\phi$, $\phi'$ and $\phi''$ should be finite in the near-horizon limit $r \rightarrow r_h$, otherwise the trace of the energy momentum tensor would diverge near the horizon.

Starting with Eq. (\ref{Bfunction})
and by assuming that $A' \rightarrow \infty$ and $\phi'$ remains finite, we expand its right-hand-side and we easily obtain  the result
\begin{equation}\label{expbbh}
e^B=(2\f'\dot f +r)A'-\frac{4r^2\f'^3\dot f +2 r^3 \f'^2+32\f'\dot f-8r}{8(2\f'\dot f +r)}+\mathcal{O}\left( \frac{1}{A'}\right).
\end{equation}
As we expected, the $g_{rr}$ component of the metric $e^B$ diverges at the horizon. Note that, in the expression above, we
have kept only the ($+$) sign in front of the square root in Eq. (\ref{Bfunction}), as the ($-$) sign, under the same assumptions, leads to the behavior $e^B \simeq {\cal O}(1)$, which
is not a solution for a black-hole. Employing the above result Eq. (\ref{expbbh}) together with the aforementioned behavior of the functions $A'$, $\f$, $\f'$ and $\phi''$ near the horizon of the black hole $r\rightarrow r_h$, Eqs. (\ref{Aphi1})-(\ref{Aphi-system})
take the following approximate forms
\bea
&&\hspace*{-1cm}A''=-\frac{r^4+4r^3\phi'\dot{f}+4r^2\phi'^2\dot{f}^2-24\dot{f}^2}{r^4+2r^3\phi'\dot{f}-48\dot{f}^2}A'^2+...\label{A-approx-h}\\
&&\hspace*{-1cm}\phi''=-\frac{(2\phi'\dot{f}+r)(r^3\phi'+12\dot{f}+2r^2\phi'^2\dot{f})}{r^4+2r^3\phi'\dot{f}-48\dot{f}^2}A'+...\label{phi-approx-h}
\eea
Focusing on the second of the above equations, we observe that, if the coupling function vanishes, $\phi''$ diverges at
the horizon of the black hole. Also, the same happens if the coupling function is  left unconstrained. Therefore, the only way for $\phi''$ to
be rendered finite is if either one of the two expressions in the numerator of 
Eq. (\ref{phi-approx-h}) is zero close to the horizon of the black hole.

Let us examine now the two cases. First we assume  that $(2\phi'\dot{f}+r)=0$ near the horizon; then a more careful
inspection of our Eqs. (\ref{Bfunction})-(\ref{bg})  reveals that the coefficient $\b$ remains finite whereas $\g$ is the one that diverges. In that case, we find that $e^B\simeq\sqrt{A'}$. Using this result in Eq. (\ref{Aphi-system}), we find that  $\phi'' \simeq \sqrt{A'}/\dot f$.
Therefore, the only way for $\phi''$ to remain finite is to demand that $\dot f \rightarrow \pm\infty$, 
near the horizon of the black hole $r\rightarrow r_h$. For a polynomial coupling function $f=\alpha \f^n$ with $n>0$, or an exponential coupling function $f=\a e^{\l\f}$, with $\l$ a real constant, this is achieved only if the scalar field diverges near the horizon, $\phi\rightarrow\pm\infty$, as $r\rightarrow r_h$. But this breaks the assumption for a regular black hole horizon. If we use an inverse polynomial coupling function $f=\alpha \f^n$ with $n<0$, or a logarithmic one $f=\ln\f$, the $\f''$ vanishes only for $\f\rightarrow 0 $ near the horizon. However since $(2\phi'\dot{f}+r)=0$, $\f'$ should also vanish for the combination $\phi'\dot{f} $ to remain finite. But if both the scalar field $\phi$ and its first derivative $\phi'$ vanish at the horizon of the black hole,  the only solution for the scalar field is the trivial one.

Thus, the only way to construct  a regular black hole horizon in the presence of a non-trivial
scalar field, is  to consider the second choice: 
\begin{equation}\label{constraint1}
r_h^3\phi_h'+12\dot{f}_h+2r_h^2{\phi'_h}^2\dot{f}_h=0,
\end{equation}
where all quantities have been evaluated at the horizon of the black hole $r_h$. 
The above equation may be
easily solved to yield:
\begin{equation}\label{con-phi'}
\phi'_h=\frac{r_h}{4\dot{f}_h}\left(-1\pm\sqrt{1-\frac{96\dot{f}_h^2}{r_h^4}}\right).
\end{equation}
In order to ensure that the above solution for  $\phi'_h$ is real, we need to impose the following constraint
\begin{equation}\label{con-f}
\dot{f}_h^2<\frac{r_h^4}{96}\,.
\end{equation}
Note that the above constraint is a constraint on the parameters of the theory and not on the form of the coupling function. As we will see in the following chapter, for every form of the coupling function, there is always a  range of values for the parameters which validate the above constraint. Also the above constraint imposes a lower-bound for the radius and consequently the mass of the Gauss-Bonnet black holes.

If we use  Eq. (\ref{con-phi'}), we can easily obtain
\begin{equation}\label{zz}
2 \phi'_h\dot{f}_h +r_h=\frac{r_h}{2}\left(1\pm\sqrt{1-\frac{96\dot{f}_h^2}{r_h^4}}\right).
\end{equation}
For a non-trivial coupling function $f$, the above is always non-zero and positive, in accordance to the discussion of the previous paragraph. The positivity of the above relation also ensures the positive sign of the $g_{rr}$ component of the metric $e^B$ close to, but outside of course, the horizon according to Eq. (\ref{expbbh}) (we remind the reader that, for a spherically symmetric and static black hole solution the metric function $e^A$ always increases from zero (close to $r_h$) to unity as  $r\rightarrow \infty$ so $A'>0$ over the whole radial regime $r_h<r<\infty$).

Finally, we turn to Eq. (\ref{A-approx-h}), and by using the constraint for the scalar field (\ref{con-phi'}),
we easily find that this  simplifies to 
\begin{equation}
A''=-A'^2+\mathcal{O}(A').
\end{equation}
Then, the above equation can be easily integrated once, with respect to $r$, to give
\begin{equation}\label{add1}
 A'=\frac{1}{(r-r_h)}+\mathcal{O}(1),
\end{equation}
which, in agreement to our initial assumption, diverges close to the horizon of the black hole. 
Now if we integrate once more and put  everything together, we can write the 
near-horizon solutions as
\bea
&&e^{A}=a_1 (r-r_h) + ... \,, \label{A-rh}\\[1mm]
&&e^{B}=b_0+\frac{b_1}{(r-r_h)} + ... \,, \label{B-rh}\\[1mm]
&&\phi =\phi_h + \phi_h'(r-r_h)+ \phi''_h (r-r_h)^2+ ... \,. \label{phi-rh}
\eea
From Eqs. (\ref{expbbh}), (\ref{B-rh})  and by using Eqs. (\ref{con-phi'}) and (\ref{add1}) we can easily find that: 
\begin{align}
b_0=4  \4\2 \text{and} \4\2 b_1=2 \f'_h\dot f_h +r_h.
\end{align}
The above solutions describe a regular black-hole horizon in the presence of a non trivial scalar
field provided that $\phi'$ and the coupling function $f$ satisfy the constraints
(\ref{con-phi'})-(\ref{con-f}).


\subsection{Asymptotic solutions at infinity}

We move now to the other asymptotic region. We will show that    for any form of  the  coupling function $f(\phi)$, an asymptotically flat limit for the spacetime (\ref{metric}) may be always constructed. We will start by
assuming that, in the limit $r \rightarrow \infty$, the expressions for the metric functions and
scalar field have the following power-law forms
\begin{align}
e^{A}&=1+\sum_{n=1}^\infty{\frac{p_n}{r}}\,,\\
e^B&=1+\sum_{n=1}^\infty{\frac{q_n}{r}}\,,\\
\phi &=\phi_{\infty}+\sum_{n=1}^{\infty}{\frac{d_n}{r}}\,.
\end{align}
Then, substituting in the field equations, we can determine the arbitrary
coefficients $(p_n, q_n,  d_n)$ from the expansion. In fact, the coefficients $p_1$ and $d_1$
remain arbitrary, and we can associate them with the ADM mass and scalar charge of the black hole,
respectively: $p_1 \equiv -2M$ and $d_1=D$. Finally, the asymptotic form of the solutions
for the two metric functions and the scalar field read:
\begin{align}
e^A=&\; 1-\frac{2M}{r}+\frac{MD^2}{12r^3}+\frac{24MD\dot{f}+M^2D^2}{6r^4}\nonumber\\
&-\frac{96M^3D-3MD^3+512M^2\dot{f}-64D^2\dot{f}+128MD^2\ddot{f}}{90r^5}+\mathcal{O}(1/r^6)\,,\label{Afar}\\
e^B=&\; 1+\frac{2M}{r}+\frac{16M^2-D^2}{4r^2}+\frac{32M^3-5MD^2}{4r^3}\nonumber\\
&+\frac{6144M^5-2464M^3D^2+97MD^4-6144M^2D\dot{f}+192D^3\dot{f}-384MD^2\ddot{f}}{192r^5}\nonumber\\
&+\frac{768M^4-208M^2D^2-384MD\dot{f}+3D^4}{48r^4} +\mathcal{O}(1/r^6)\,,\label{Bfar}\\
\phi=&\; \phi_{\infty}+\frac{D}{r}+\frac{MD}{r^2}+\frac{32M^2D-D^3}{24r^3}+\frac{12M^3D-24M^2\dot{f}-MD^3}{6r^4}\nonumber\\
&+\frac{6144M^4D-928M^2D^3+9D^5
-12288M^3\dot{f}-1536MD^2\dot{f}}{1920r^5}+\mathcal{O}(1/r^6)\,.\label{phifar}
\end{align}
As we observe from the above expressions, the explicit form of the coupling function
$f(\phi)$ makes its first appearance not earlier than in the order $\mathcal{O}(1/r^4)$. In general, the quadratic higher-curvature Gauss-Bonnet term
is expected to have a minor contribution at   distances  far away from the black-hole horizon where the curvature is extremely
small. The Gauss-Bonnet term, through the coupling function $f$, works mainly in the near horizon regime to create a black hole topology with a non-trivial scalar field. In the far-away regime, this action is reflected in the modified metric functions, in $\mathcal{O}(1/r^2)$ compared to the Schwarzschild case, but more importantly in the non-trivial scalar charge $D$ whose vanishing would have trivialised the scalar field $\f$. The appearance of the explicit coupling function at a much higher order signifies the fact that the existence of regular black hole solutions is a generic feature of the theory (\ref{action}).


\section{Evasion of the No-Scalar-Hair Theorems}\label{eva33}

\subsection{Evasion of the Old No-Scalar-Hair Theorem}\label{seves}

Now let us turn our attention to the no-scalar-hair theorems, which forbid the existence of new black-hole solutions in  the framework of Scalar-Tensor theories, i.e. the existence of solutions that smoothly
connect the two asymptotic solutions --the near-horizon and far-away  solutions-- found in the previous two subsections. We will demonstrate that the presence of the Gauss-Bonnet term, coupled to the
scalar field through a general coupling function $f(\f)$, causes the evasion of those theorems under a sole constraint on the coupling function.

We will start with the older version of  Bekenstein's no-scalar-hair theorem \cite{NH-scalar1} \cite{NH-scalar2}.
As explained in Section \ref{nohhhh} this argument uses the scalar equation of motion Eq. (\ref{scalar-eq}) and relies on the sign of $V'(\f)$, where $V(\f)$ is the potential for the scalar field. Let us investigate now how this theorem develops in our case. We start from the scalar equation (\ref{scalar-eq}), we multiply  it with $f(\phi)$ and by integrating over the exterior region of the black-hole, we
obtain the following integral constraint
\beq
\int d^4x \sqrt{-g} \,f(\phi) \left[\nabla^2 \phi + \dot f(\phi) R^2_{GB} \right] =0\,.
\eeq
Integrating by parts the first term, the above equation takes the form
\beq
\int  d^4x \sqrt{-g}\,\dot f(\phi) \left[\partial_\mu \phi\,\partial^\mu \phi  - 
f(\phi)  R^2_{GB}\right] =0\,. \label{old-con}
\eeq
Note that the boundary term $[\sqrt{-g} f(\phi) \partial^\mu \phi]_{r_h}^\infty$
vanishes at the horizon of the black-hole and at infinity as well. At infinity  it vanishes due to the fact that the scalar field asymptotes to a constant value while near the black-hole horizon it does so due to the combination  $e^{(A-B)/2}$ which appears there.
Returning now to  Eq. (\ref{old-con}), one may easily see that, due to the staticity and spherical symmetry of our solutions, 
the first term  gives
$\partial_\mu \phi\,\partial^\mu \phi = g^{rr}(\partial_r \phi)^2>0$ 
throughout the exterior region of the black hole. Therefore Eq. (\ref{old-con}) may be satisfied only if the second term $(f(\f)R^2_{GB})$ is positive. For the metric (\ref{metric}),
the GB term has the following explicit form
\beq
R^2_{GB} =\frac{2 e^{-2B}}{r^2}\bigl[(e^B-3)\,A' B' -(e^B-1)\,(2A''+A'^2)
\bigr]. \label{GB}
\eeq
If we employ  the asymptotic solutions near the horizon (\ref{A-rh})-(\ref{B-rh})
and at   infinity (\ref{Afar})-(\ref{phifar}), we may easily see that the GB term
takes a positive value at both asymptotic regimes since 
\begin{align}
R^2_{GB}=\rightarrow\frac{2(2+b_0)}{b_1^2r_h^2} \4\text{as}\4 r\rightarrow r_h, \4\2\text{and}\4\2 R^2_{GB}\rightarrow\frac{48M^2}{r^6}\4\text{as}\4r\rightarrow \infty.
\end{align}
The above result points toward a monotonic behaviour of the GB term in the exterior region -- the explicit numerical construction of the solutions will demonstrate that this is indeed the case. As a result, the old no-scalar-hair theorem may be evaded if the coupling function merely satisfies the following constraint: 
\begin{equation}
f(\f)>0.
\end{equation}
Thus, for a positive coupling function  we can always find  solutions for black holes with a non-trivial scalar field. 

Although we have managed to show that the No-Hair theorem is evaded and to derive a very simple constraint, there are some indications that this integral method is not reliable and the above  is not a true constraint. We started our analysis multiplying the scalar field equation with $ f$ and integrating over the exterior region of the black hole. However with a   slightly different manipulation we can find new constraints. For example multiplying 
\begin{itemize}
\item with $\dot f$, we get the constraint: $\ddot f>0$,
\item or with $\f$, we get the constraint: $\f \dot{f} >0.$
\end{itemize}
Using the same procedure we may end up with a large number of constraints. However, which one of those is the fundamental constraint? Furthermore, using our numerical code, we can find regular solutions for black holes which invalidate many of the above constraints. For more details see the discussion below Figs. \ref{Phi_phi3} and \ref{nesc} in the following chapter. Since the method of the old no-scalar-hair theorem is still used in many works we conclude that we should be extremely careful when we use it. In order to find if the no-scalar-hair theorem is indeed evaded and to derive a reliable constraint, we use the method of  Bekenstein's Novel No-Scalar-Hair theorem. 

\subsection{Evasion of the Novel No-Scalar-Hair Theorem}

We will continue now with the `novel' no-hair theorem which was developed by Bekenstein \cite{Bekenstein}. The basic  argument of this theorem
was built around the behavior  of the $T^r_{\;\,r}$ component of the energy-momentum tensor of
the theory in terms of the radial coordinate $r$. Bekenstein demonstrated that, if we make use of the assumptions of the energy conditions from General Relativity --in other words if we demand the positivity and conservation of the energy-- then the profiles of $T^r_{\;\,r}$ component near the horizon
and at asymptotic infinity could never be smoothly matched. 
This argument proved   
beyond doubt that, for
a large class of physically interesting theories, there were no black-hole solutions associated with a non trivial scalar field. However, that class of theories contained
by definition only minimally-coupled-to-gravity scalar fields. Concluding, the Bekenstein's no-scalar hair theorem states that for a scalar-tensor theory, under the assumptions of asymptotic flatness, staticity and minimal  coupling we cannot construct hairy black hole solutions. In other words the no-scalar hair theorem is a uniqueness theorem which states that in the above theories the only black hole solution we may construct is the Schwarzschild one.

The first theory which was shown to evade Bekenstein’s no-scalar-hair theorem was the
Einstein-Dilaton-Gauss-Bonnet (EdGB) theory  \cite{DBH1, DBH2}, inspired by the low-energy approximation of the superstring theory,
in which the GB term was coupled to the dilaton scalar field through an exponential coupling function.
There, it was shown that the structure of the EdGB theory was such that it invalidated certain
requirements of the no-scalar-hair theorem, which therefore was not applicable any more. In the framework of the EsGB theory the second case was the theory with a linear coupling function which is also a shift symmetric Gallileon theory \cite{SZ}. In this work, the authors used a specific No-Scalar-Hair theorem formulated for the Gallileon theories but as we will see the theory also evades  Bekenstein's novel no-hair theorem.  Here, we will show that the coupling of the scalar field to
the quadratic GB term causes the complete evasion of Bekenstein's theorem. This may
be in fact realised for a large class of scalar field theories, with the
previously-studied exponential  \cite{DBH1, DBH2} and linear \cite{SZ} GB couplings comprising special cases of our present argument.

We start with  the  
conservation equation $\nabla_\mu T^\mu_{\;\,\nu}=0$ which satisfies the energy-momentum tensor $T_{\mu\nu}$  due to the invariance of the action
(\ref{action}) under coordinate transformations. The $r$-component of the above equation may take
the explicit form 
\beq
\left(e^{A/2} r^2 T^r_{\;\,r}\right)'=\frac{1}{2}\,e^{A/2} r^2 \left(A'\,T^t_{\;\,t} +
\frac{4}{r}\,T^\theta_{\;\,\theta}\right),
\label{Trr-conserv}
\eeq
where we have used  the relation $T^\theta_\theta=T^\varphi_\varphi$   due
to the spherical symmetry of our solutions. Then carrying out the derivative on the left-hand-side, we may solve the above equation for $(T^r_{\;\,r})'$ to obtain
\beq
(T^r_{\;\,r})'=\frac{A'}{2}\,(T^t_{\;\,t} -T^r_{\;\,r}) +
\frac{2}{r}\,(T^\theta_{\;\,\theta}-T^r_{\;\,r})\,.
\label{Trr-deriv}
\eeq
 The non-trivial components of the energy momentum
tensor $T_{\mu\nu}$ for our theory (\ref{action}) with a generic coupling function
$f$ are:
\begin{align} 
T^t_{\;\,t}&=-\frac{e^{-2B}}{4r^2}\left\{\phi'^2\bigl[r^2e^B+16\ddot{f}(e^B-1)\bigr]\right.
-\left.8\dot{f}\bigl[(B'\phi'(e^B-3)-2\phi''(e^B-1)\bigr]\right\},\label{Ttt}\\[2mm]
T^r_{\;\,r}&=\frac{e^{-B}\phi'}{4}\Bigl[\phi'-
\frac{8e^{-B}\left(e^B-3\right)\dot{f}A'}{r^2}\Bigr],\label{Trr}\\[2mm]
\hspace*{-0.5cm}T^{\theta}_{\;\,\theta}&=
-\frac{e^{-2B}}{4 r}\left\{\phi'^2\bigl(re^B-8\ddot{f}A'\bigr)\right.  
 - \left. 4\dot{f}\bigl[\phi' (A'^2+2A'') +
A'(2\phi''-3B'\phi')\bigr]\right\}. \label{Tthth}
\end{align}

Using the above equation we may investigate the profile of $T^r_{\;\,r}$ component close to and far away from the
horizon of the black hole. First, we start from asymptotic infinity: using the asymptotic solutions at infinity (\ref{Afar})-(\ref{phifar}), we easily find that
\begin{equation}
T^t_{\;\,t} \simeq- \frac{1}{4}\phi'^2
+ {\mathcal O}\left(\frac{1}{r^6}\right),\4   T^r_{\;\,r} \simeq  \frac{1}{4}\phi'^2
+ {\mathcal O}\left(\frac{1}{r^6}\right),  \4 T^\theta_{\;\,\theta} \simeq- \frac{1}{4}\phi'^2
+ {\mathcal O}\left(\frac{1}{r^6}\right).
\end{equation}
Taking into account the fact that, as the radial coordinate approaches infinity, the metric function $e^A$  adopts a constant value ($e^A \rightarrow 1$),
we conclude that the dominant contribution to the right-hand-side of Eq. (\ref{Trr-deriv}) at asymptotic infinity is
\beq
(T^r_r)' \simeq \frac{2}{r}\,(T^\theta_\theta-T^r_r) \simeq 
-\frac{1}{r}\,\phi'^2 +....\,.
\eeq
According to the results of the above equation, far away from the  horizon of the black hole, the $T^r_{\;\,r}$ component is positive and decreasing.
This was also the result obtained in the novel no-scalar-hair theorem  \cite{Bekenstein} since the higher derivative GB term is not expected to have any significant contribution in this regime.

We continue now with the near-horizon regime. In the limit $r \rightarrow r_h$, the $T^r_{\;\,r}$ component (\ref{Trr})
takes the following approximate form:
\beq
T^r_{\;\,r}=-\frac{2e^{-B}}{r^2}A'\phi'\dot{f}+\mathcal{O}(r-r_h)\,.\label{Trr-rh}
\eeq
The dominant behavior of the metric functions $A$, $B$ and the scalar field $\phi$ can be read from the near-horizon solutions (\ref{A-rh})-(\ref{phi-rh}) and we can easily verify that it is such that it renders the above combination finite. In the novel no-scalar-hair theorem \cite{Bekenstein}, the $T^r_{\;\,r}$  component was
strictly negative-definite which, in conjunction with its positivity at large distances,
demanded a non-monotonic behavior for this component that at the end could not be
supported. Here, the $T^r_{\;\,r}$ component can be instead made positive under the simple
assumption that, close to the black-hole horizon, the condition
\begin{equation}\label{constr1}
\dot f \phi'<0
\end{equation}
holds. However, rearranging Eq. (\ref{con-phi'}), we observe that
\begin{equation}\label{condev}
\dot{f}_h\phi'_h=\frac{r_h}{4}\left(-1\pm\sqrt{1-\frac{96\dot{f}_h^2}{r_h^4}}\right)<0,
\end{equation}
 and the   constraint (\ref{constr1}) is always satisfied. 
Therefore, the requirement of the existence of a regular black-hole horizon in the context of the
theory (\ref{action}) automatically evades one of the two requirements of the
novel no-hair theorem.

Using  the components  of the energy momentum
(\ref{Ttt})-(\ref{Tthth}) in Eq. (\ref{Trr-deriv}), we  may obtain  that in the same regime the
 expression for  $(T^r_r)'$  takes the following form:
\begin{align}
(T^r_r)' =& -\frac{e^{-B} A'\phi'^2}{4} + \frac{2 e^{-B}}{r^2}\left\{
-A'\,(\ddot f \phi'^2 +\dot f \phi'')\right. \nonumber \\[2mm]
&+ \left.\dot f \phi'\left[\frac{A'}{2}\,(A'+ B') + 2e^{-B} A'' +
e^{-B} A'\,(A'-3B') + \frac{2 A'}{r}\right]\right\} + {\mathcal O}(r-r_h)\,.
\label{Trr'-inter}
\end{align}
Let us now work a bit more on the above equation: if we take the sum of the  $(tt)$ and $(rr)$ components of
Einstein’s equations (\ref{tt-eq})-(\ref{rr-eq}) in the limit $r \rightarrow r_h$, we find
\begin{equation}
A'+B'\simeq\frac{1}{Z}\left[\frac{r^2\f'^2}{2}+4\left( \ddot f \f'^2+\dot f \f''  \right)\right]+\mathcal{O}(r-r_h),
\end{equation}
where we have defined $Z \equiv r+2 \dot f \phi'$. As shown in Eq. (\ref{zz}), this latter combination is not allowed to vanish and is in fact always positive if Eq. (\ref{con-phi'}) holds. From the $(\theta\theta)$ component (\ref{thth-eq}) finally, we find
\begin{equation}
A''=-\frac{A'}{2}(A'-B')+\mathcal{O}(r-r_h)^{-1}.
\end{equation}
Substituting the above relation into Eq. (\ref{Trr'-inter}) and performing some simple algebra, we take the
final result for $(T^r_r)'$ close to the black-hole horizon
\beq  
(T^r_r)' =  e^{-B} A'\Bigl[-\frac{r \phi'^2}{4Z} -\frac{2 (\ddot f \phi'^2
+\dot f \phi'')}{r Z} + \frac{4 \dot f \phi'}{r^2}\,\Bigl(\frac{1}{r}-e^{-B} B'\Bigr)
\Bigr]+ {\mathcal O}(r-r_h)\,.
\label{Trr'-rh}
\eeq
 Given that, close to the black-hole
horizon, $A'>0$, $B'<0$ and that $Z>0$ and $\dot f \phi'<0$, as dictated by Eq. (\ref{condev}), we conclude
that $(T^r_r)'$ is negative in the near-horizon regime if a sole additional
constraint, namely $\ddot f \phi'^2+\dot f \phi''>0$, is satisfied. This may
be alternatively written as $\partial_r (\dot f \phi')|_{r_h}>0$, and merely
demands that the negative value of the quantity $(\dot f \phi')|_{r_h}$,
 that is necessary for the existence of a regular black-hole horizon, should be 
constrained far away from the horizon. This is in fact the only way for the
two asymptotic solutions (\ref{A-rh})-(\ref{phi-rh}) and
(\ref{Afar})-(\ref{phifar}) to be smoothly connected. Therefore, for $\partial_r (\dot f \phi')|_{r_h}>0$,
the $T^r_r$ component is always positive and monotonically decreasing.
 As a result, both requirements of   Bekenstein's novel no-scalar-hair theorem 
\cite{Bekenstein} do not apply in this theory, and thus they can be evaded.

\subsection{Derivation of new black-hole solutions}\label{nbhsww}

\begin{figure}[t!] 
\begin{center}
\hspace{0.0cm} \hspace{-0.6cm}
\includegraphics[height=.25\textheight, angle =0]{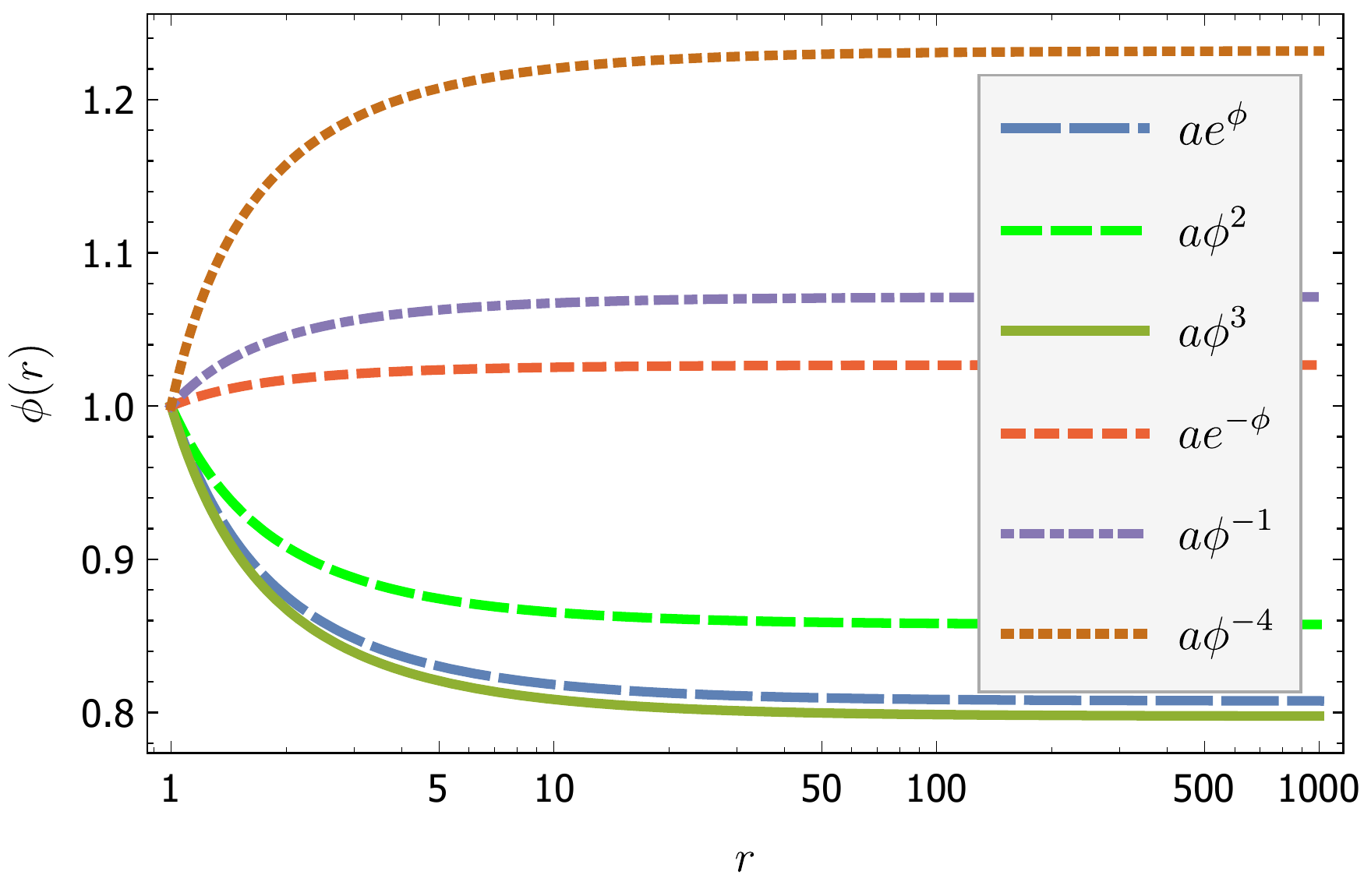}
\hspace{0.52cm} \hspace{-0.6cm}
\includegraphics[height=.25\textheight, angle =0]{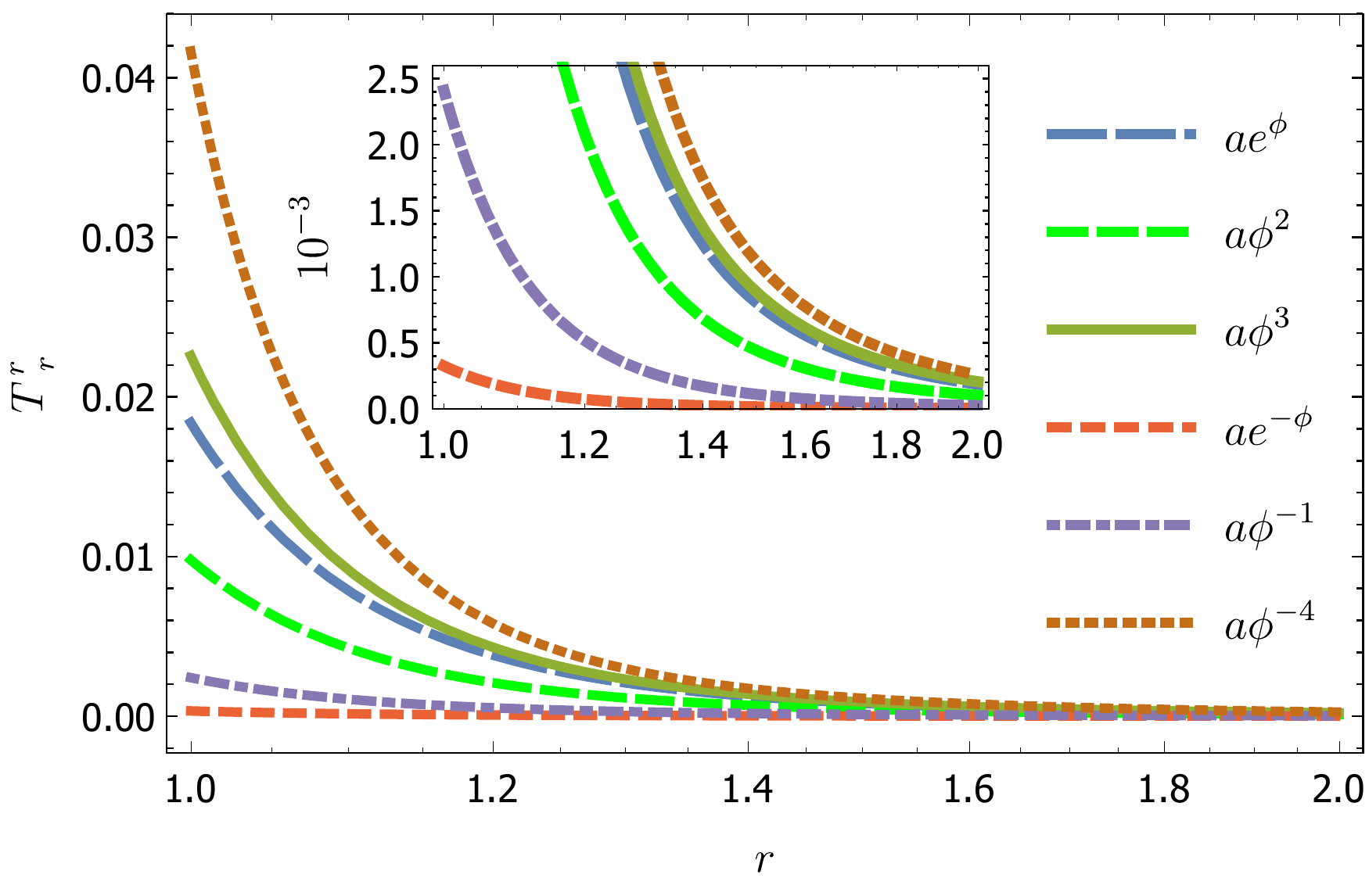}
\\
\hspace*{0.7cm} {(a)} \hspace*{7.5cm} {(b)}  \vspace*{-0.5cm}
\end{center}
\caption{(a) The scalar field $\phi$, and (b) the $T^r_{\,\,r}$ component of the energy-momentum tensor
  in terms of the radial coordinate $r$, for different coupling functions $f(\phi)$,
  for $a=0.01$ and $\phi_h=1$.}
  \label{Fig_phi}
\end{figure}

In order to demonstrate the validity of our arguments regarding the evasion of the No-Scalar-Hair theorem, we
have numerically solved the system of the field equations (\ref{Aphi-system}), and we have
produced a large number of black-hole solutions with  a non trivial scalar field. The numerical analysis will be given in detail in the following chapter.
In agreement with our results from this section, we may find regular numerical black hole solutions for  every form of the coupling function.
 In Fig. \ref{Fig_phi}(a) we depict   the solutions
for the scalar field  for many different  
forms  of the coupling function $f(\phi)$: we use  exponential, odd and even power-law,
odd and even inverse-power-law functions. These forms are all simple, natural choices to
keep the GB term in the 4-dimensional theory inspired from previous works on local or cosmological solutions. For easy comparison between the solutions, the coupling
constant in all cases has been set to $\alpha/r_h^2 =0.01$\footnote{As we wrote the action \ref{action} the value of $\f_h$ is dimensionless but not that of $\a$ which has units of square length $[r^2]$. For our numerical analysis, it is convenient to define the dimensionless quantity $\a/r_h^2$. However, since we set $r_h=1$ everywhere, sometimes instead of $\a/r_h^2$ we simply write $\a$.}, the near-horizon value
of the scalar field to $\phi_h=1$ and finally the horizon radius to $r_h=1$. We have chosen three coupling functions  $f(\phi)=(\alpha e^\phi, \alpha \phi^2, 
\alpha \phi^3)$, that all have $f'_h>0$, thus our constraint (\ref{con-phi'}) leads
to a negative sign for the first derivative of the scalar field near the horizon $\phi'_h$; and three functions
$f(\phi)=(\alpha e^{-\phi}, \alpha \phi^{-1}, \alpha \phi^{-4})$, that
have $f'_h<0$, thus, Eq. (\ref{con-phi'}) demands a positive sign for the first derivative $\phi'_h$. The sign of the derivative  determines   the
decreasing or increasing, respectively, profile of the solutions for the scalar field which are clearly depicted
in Fig. \ref{Fig_phi}(a). In all cases, for a given value of the $\phi_h$ parameter,
Eq. (\ref{con-phi'}) uniquely determines the value of the quantity $\phi'_h$.
The integration of the system (\ref{Aphi-system}) then, with initial conditions
$(\phi_h, \phi'_h)$  leads to the presented solutions.
The positivity and also the decreasing profile of the $T^r_{\,\,r}$ component of the energy momentum tensor near the horizon, 
 which are necessary features for the evasion of the novel no-hair theorem, are clearly
seen in Fig. \ref{Fig_phi}(b). It is worth mentioning that the second constraint,
$\partial_r (\dot f \phi')|_{r_h}>0$, for the evasion of the no-hair theorem is automatically satisfied for every
solution found and does not  require any further action or any fine-tuning of the
free parameters. Finally we note that  for all the forms
of $f(\phi)$ the above solutions satisfy also the constraint $f(\phi)>0$, which was derived   from
the evasion of the old no-hair theorem.


\section{The Spontaneous Scalarisation Mechanism}\label{evaspo}

After our paper with the results for the evasion of the No-Scalar-Hair theorem was submitted for
publication \cite{ABK1}, two additional works appeared 
where particular solutions of our general theory was
discussed \cite{Doneva, Silva}.  In these two works, the spontaneous scalarization effect for black holes was first introduced. In the spontaneous scalarization effect, the Schwarzschild solution  becomes unstable for some values of the coupling constant and a scalarized solution bifurcates from the Schwarzschild one. Here, the term scalarized solution stands for a black-hole solution with a non trivial scalar field. As we will see, the spontaneous scalarized solutions have nothing special compared with other solutions found in the framework of the EsGB theory --like the solutions in Fig. \ref{Fig_phi}. However, we will introduce the term ``naturally scalarized" in order to separate all the other solutions of the theory from the spontaneously scalarized ones.  Finally, the spontaneous scalarization of black holes is quite similar with the spontaneous  scalarization of neutron stars \cite{neutronsc} but it is triggered by the curvature of the space-time itself instead of the presence of some kind of matter. 

We start our analysis from the Schwarzschild solution
\begin{equation}
{ds}^2=-h(r){dt}^2+\frac{1}{h(r)}{dr}^2+r^2({d\theta}^2+\sin^2\theta{d\varphi}^2),
\end{equation}
with $h(r)=1-\frac{2M}{r}.$ 
Substituting the above solution to the field equations  (\ref{tt-eq})-(\ref{scalar-eq}) we find that the Schwarzschild solution is a solution of the EsGB theory only if
\begin{align}
\phi(r)&=\phi_0,\\[2mm]
\dot f (\phi_0)&=0,
\end{align}
where $\f_0$ is a constant. In order to examine the stability of the system, we perturb  the above solution i.e. $g_{\m\n}(x^\m)\rightarrow g_{\m\n}(r)+\varepsilon {h_1}_{\m\n}(x^\m)$ and $\phi(x^\m)\rightarrow \phi_0+\varepsilon \phi_1(x^\mu)$, where ${h_1}_{\m\n}$ and $\f_1$ are small perturbations and $\varepsilon$ is an index for the expansion.  Then, we easily find that --in the system of the field equations-- the two perturbations decouple  due to the simplicity of the Schwarzschild solution. The system of the Einstein's equations takes the form of the well studied Schwarzschild perturbations while the scalar equation has the following form: 
\begin{equation}
\frac{1}{\sqrt{-g}}\partial_\m\left( \sqrt{-g}\partial^\mu \phi_1(x^\r)    \right)+\ddot f(\f_0) {R^2_0}_{GB} \f_1(x^\r)=0,\label{se1}
\end{equation}
with ${R^2_0}_{GB}=\frac{48M^2}{r^6}.$
 Using the method of the   separation of variables, we may easily find the solutions for the angular and the \textit{time} part of the perturbation. Then, the scalar perturbation may be written in the following form:
 \begin{equation}
 \f_1(x^\m)=\frac{u(r)}{r}e^{i\omega t }Y_\ell^m(\theta,\varphi),
 \end{equation}
where $Y_\ell^m(\theta,\varphi)$ are the spherical harmonics and $\omega$ a constant. Substituting the above equation in  Eq. (\ref{se1})  we find that its radial part becomes:
\begin{equation}
\frac{h(r)}{r}\frac{d}{dr}\left( r^2 h(r)\frac{d}{dr}\left(\frac{u(r)}{r}\right)\right)+\left(\omega^2+h(r)\left(\ddot f(\f_0) {R^2_0}_{GB}-\frac{\ell(\ell+1)}{r^2}\right)\right)=0.
\end{equation}
Using the \textit{tortoise} coordinate $dr_*=dr/h(r)$ --which maps the domain $r\in[2M,+\infty)$ to $r_*\in(-\infty,+\infty)$-- we may eliminate the first derivative $u'$ that appears in the above equation which then  takes the following Schrödinger-like form:
\begin{equation}
\frac{d^2}{dr_*^{2}}u+\left(\omega^2-V(r)\right)u=0,
\end{equation}
where the potential is defined as
\begin{equation}
V(r)=h(r)\left(\frac{h'(r)}{r}+\frac{\ell(\ell+1)}{r^2}-\ddot f(\f_0) {R^2_0}_{GB}\right).
\end{equation}
Then, the condition for an  unstable mode is:
\begin{equation}
\int_{-\infty}^{+\infty}V dr_*=\int_{2M}^{\infty}\frac{V(r)}{h(r)}dr\,=\,\frac{ 5M^2+10 M^2 \,\ell(\ell+1) -6 \ddot f(\f_0)}{20 M^3}<0.
\end{equation}

Since $M>0$ and $\ell>0$ always, the only way for the above quantity to be negative is to assume that the second derivative of the coupling function is also positive i.e. $\ddot f(\f_0)>0$. Then, the constraint takes the form
\begin{align}
M^2<\frac{6 \ddot f(\f_0)}{5+10\ell(\ell+1)}.
\end{align}
Thus, in the framework of the EsGB theory, the Schwarzschild solution with a mass $M$, which validates the above constraint is unstable. As the mass of the black hole decreases the horizon radius also decreases and this results to an increasing of the horizon curvature. So, the value of the GB term near the horizon also increases. When the value  of the GB at the horizon exceeds the critical value ${{R^2_0}_{GB}}_{\rm c}=3/(4M_c^4)$, with $M_c^2=\frac{6 \ddot f(\f_0)}{5+10\ell(\ell+1)}$, the Schwarzschild solution becomes unstable and then a  spontaneously scalarized   black-hole solution may arise. 

Although the spontaneous scalarization effect may provide an explanation on how these solutions may arise it is not an existence theorem. The instability of the Schwarzs- child solution does not guarantee that a scalarized solution exists. Also the spontaneous scalarization mechanism is over-restrictive and applies only on specific forms of the coupling function i.e. for coupling functions with $\dot f(\f_0)=0$ and $\ddot f(\f_0)>0.$ On the other hand, in the previous section,  we proved that  for any form of the  coupling function in the EsGB theory, scalarized solutions exist provided that the parameters of the theory respect the constraint (\ref{con-f}). As we have already mentioned, the spontaneous scalarized solutions have nothing special compared to the naturally scalarized solutions. Once a spontaneously scalarized solution is produced there is no way to distinguish it from a naturally scalarized one. However, in the spontaneous scalarization the Schwarzschild solution co-exists with the scalarized solutions. In the more general case of natural scalarization,   the Schwarzschild solution may not exist in the configuration space of the theory. Concluding, the spontaneously scalarized solutions comprise only a subgroup of our more general naturally scalarized solutions. 

\section{The Scalar-GB Theory}\label{evapure}

In the previous section, we showed that the synergy between the scalar field $\f$ and the
GB term is in a position to support asymptotic solutions for regular black holes, and no
no-hair theorem can exclude the existence of a complete solution that smoothly connects
them. In the light of previous studies in a cosmological context, where it was shown
that, in a regime of large curvature, the quadratic GB term dominates over the linear
Ricci term \cite{pgb1, pgb2, pgb3, bakoppgb}, here, we want to investigate whether the black-hole horizon could owe its existence
exclusively on the GB term. If that is the case, then in the near-horizon regime one could
ignore all terms in the field equations coming from the Ricci term and attempt to find a
more sophisticated, analytic near-horizon solution.

In what follows, we will ignore the Ricci term from Eqs. (\ref{tt-eq})-(\ref{phi-eq}) and pose anew the
question of the existence of regular black-hole solutions. We first note that the discussion around the old no-hair theorem remains unchanged: due to the absence of a direct
coupling between $\f$ and $R$, the scalar equation is insensitive to the presence or absence
of $R$. The novel no-hair theorem, on the other hand, is affected as all components of
the energy-momentum tensor should be formally zero in the near-horizon regime. If we
accept for the moment that the rest of the analysis remains unchanged, that means that
a complete black-hole solution would need to match a zero and decreasing $T^r_{\,\,r}$ at the
near-horizon regime with a positive and decreasing $T^r_{\,\,r}$ at asymptotic infinity, a task
that is far from trivial. However, a problem appears at a more fundamental level: at the
construction of an asymptotic near-horizon solution analogous to that of Eqs. (\ref{A-rh})-(\ref{phi-rh}).

Starting from Eq. (\ref{Trr}) and demanding this to be zero, we easily obtain the result:
\begin{equation}
e^B=\frac{24A'\dot f}{8A'\dot f -r^2 \f'}.
\end{equation}
  If we demand again that, as $r \rightarrow r_h$, $\phi'$ remains finite while $A'$ diverges, then we obtain
$e^B \simeq 3 +{\cal O}(1/A')$. Forming a system similar to that of Eqs. (\ref{Aphi1})-(\ref{Aphi-system}) and using the
above result for $e^B$, one may easily find that, in the same limit,
\begin{align}
\f''=&\,\,\mathcal{O}(1),\\[2mm]
A''=&-\frac{1}{2}A'^2+\mathcal{O}(A').
\end{align}
The above leads to a solution with a regular scalar field $\f$, a finite metric function $e^B$
and a vanishing $e^A=a_2(r-r_h)+...,$  as $r\rightarrow r_h$. Clearly, the above is not a black-hole
solution and its properties and physical interpretation will be investigated elsewhere.

Let us change approach and demand directly the divergence of $g_{rr}$ as $r\rightarrow r_h$. Then,
Eq. (\ref{Trr}) can now be solved for $A'$ to give:
\begin{equation}
A'=\frac{r^2 e^B \f'}{8\left(e^B-3\right)}\simeq\frac{r^2\f'}{8\dot f}+\mathcal{O}\left(e^{-B}\right),
\end{equation}
where we have excluded the case $e^B=3$ found above, and demanded instead that
$e^B\rightarrow\infty$. Using the above and the remaining components of the energy-momentum tensor
(\ref{Ttt}) and (\ref{Tthth}), we may form again a system of two independent differential equations,
this time for $B$ and $\f$. In the limit $r\rightarrow r_h$, this leads to the behavior
\bea
B'&=&-\frac{2}{r}\,e^B+\mathcal{O}\left(e^{-B}\right),\label{106}\\
\phi''&=&-\frac{e^B}{r}\,\phi'+\mathcal{O}\left(e^{-B}\right).\label{107}
\eea
Solving Eq. (\ref{106}) we obtain the solution: 
\begin{equation}
e^{-B}=2 \ln\left(r/r_h\right)
\end{equation}
which does point towards the existence of a horizon. However, for this
horizon to be regular, Eq. (\ref{107}) demands that $\phi'(r_h)=0$. But in this case, $A'$ vanishes too, and the
metric function $e^A$ assumes a constant value. In addition, from Eq. (\ref{107}) it follows that
$\phi''(r_h) = 0$, too, which in conjunction with $\phi'(r_h)=0$, does not allow the scalar field to
deviate from its constant horizon value $\f_h$. As a result, the scalar field remains constant
and the same holds for the coupling function $f$; but, then the GB term does not contribute
to the field equations and the above solution disappears.

The emergence of black-hole solutions in the absence of the Ricci term was extensively studied in a follow-up work \cite{bakoppgb}. In this work  we further  investigated the above results and also we extended them. In the quest of constructing black-hole solutions in the framework of the pure Gauss-Bonnet theory, we used more general ansatzes for the metric: 
\begin{equation}
ds^2=-e^{A(r)}dt^2+e^{B(r)}dr^2+H^2(r)\left(d\theta^2+\sin^2\theta d\varphi\right),
\end{equation}
and we even  considered solutions of the system (\ref{106}-\ref{107}) with an irregular scalar field. In addition we searched for pure scalar-GB black hole solutions in the presence of a cosmological constant $\L$. Despite the use of a general form of the spacetime line-element, no black-hole solutions were found.  In contrast, solutions that resemble irregular particle-like solutions or completely regular gravitational solutions with a finite energy-momentum tensor do emerge. In addition, in the presence of a cosmological constant, solutions with a horizon also emerge, however,  the  latter  corresponds  to  a  cosmological  rather  than  to  a  black-hole  horizon.

In order to investigate the role of the GB term, we employ the near-horizon expansion and obtain the  dominant contributions to the effective energy density and radial pressure of the  system: 
\begin{align}
\r=&-T^t_t=\frac{2e^{-B}}{r^2}B'\f'\dot f,\\[3mm]
p_r=&-T^r_r=\frac{2e^{-B}}{r^2}A'\f'\dot f.
\end{align}
From Eqs. (\ref{A-rh})-(\ref{B-rh}) and (\ref{condev}) we recall  that near the black-hole horizon $\f'\dot f<0$ and $A'\approx-B'>0$. Thus, near the horizon of the black hole the effective energy density is negative $\r<0$, while the radial pressure  is positive $p_r>0$. This behavior of the energy density components, which is essential for the evasion of the no-scalar-hair theorem, indicates also the repulsive role of the Gauss-Bonnet term. Therefore, the presence of the Ricci term is necessary in order to provide the attractive force that will create the positively-curved topology around the formed black hole. Concluding,  the  Gauss-Bonnet  term  makes  the  formation  of  a  black  hole  more  difficult.   What  it does  facilitate  is  the  dressing  of  the  black-hole  solution  with  a  non-trivial  scalar  field,  a feature that would have been forbidden in its absence.  As the value of the coupling constant increases, the weight of the Gauss-Bonnet term in the theory gradually increases, too.  The same holds for its repulsive effect.  Beyond the maximum value of the Gauss-Bonnet coupling parameter, no black-hole horizon can be formed – or sustained – any more. This also explains the existence of a minimum mass in the EsGB theory given in Eq. (\ref{con-f}).  We may easily then justify the fact that a pure scalar-GB theory can not, in the absence of the Ricci scalar, create by itself a black-hole solution.


\section{Discussion}\label{evadis}

In the context of the general Einstein-scalar-Gauss-Bonnet theory with an arbitrary coupling
function $f(\phi)$, we have demonstrated that the emergence of regular black-hole
solutions is a generic feature of the theory: the explicit
form of $f(\phi)$ affects very little the asymptotically-flat limit at infinity,
while a regular horizon is formed provided that $\phi'_h$ and $f(\phi)$ satisfy
the constraints (\ref{con-phi'})-(\ref{con-f}). 

The existing no-scalar-hair theorems were shown to be evaded under mild assumptions
on the values of the parameters of the theory. The old no-hair theorem \cite{NH-scalar1, NH-scalar2}  is easily evaded for 
$f(\phi)>0$ while the novel no-hair theorem \cite{Bekenstein} is non-applicable
if the same constraint (\ref{con-phi'}) holds. Based on this, we have produced
a large number of regular black-hole solutions with non-trivial scalar hair for
arbitrary forms of the coupling function $f(\phi)$. They are all characterised
by a minimum black-hole radius and mass, and their near-horizon strong dynamics
is expected to leave its imprint on a number of observables. The obtained
solutions survive only when the synergy of $\phi$ with the Gauss-Bonnet term is 
supplemented by the linear Ricci term.


\clearpage
\thispagestyle{empty}

\chapter{Properties of Black-Hole Solutions with Scalar Hair in Einstein-Scalar-Gauss-Bonnet-Theories}\label{4}

\section{Introduction}

A common feature of the theories which evaded the no-hair theorems was
the presence of higher-curvature terms, such as the quadratic Gauss-Bonnet
(GB) term inspired by the string theory \cite{Metsaev} or Horndeski 
theory \cite{Horndeski}. It is the presence of such terms that invalidate basic
requirements of the no-hair theorems, and open the way for the construction
of black-hole solutions with scalar hair. In the previous chapter,
we considered a general class of Einstein-scalar-GB theories \cite{ABK1}, of which the cases
\cite{DBH1,DBH2, SZ} constitute particular examples. We demonstrated that, under
certain constraints on the form of the coupling function between the scalar
field and the Gauss-Bonnet term, and in conjunction with the profile of
the scalar field itself, a regular black-hole horizon regime and an
asymptotically-flat regime may be smoothly connected, and thus the
no-hair theorems may be evaded. A number of novel black-hole solutions
with scalar hair was thus determined and briefly presented \cite{ABK1}.

In this chapter, we provide additional support to the arguments presented
in \cite{ABK1}. We consider the same general gravitational theory containing
the Ricci scalar, a scalar field and the GB term, with the latter two
quantities being coupled together through a coupling function $f(\phi)$.
Guided by the findings of our previous work \cite{ABK1}, we impose the
aforementioned constraints on the coupling function $f$ and the scalar
field $\phi$, and find a large number of regular, black-hole
solutions with a non-trivial scalar hair  for a variety of forms for the coupling function:
exponential, polynomial (even and odd), inverse polynomial (even and
odd) and logarithmic. In all cases, the solutions for the metric
components, scalar field, curvature invariant quantities and components
of the energy-momentum tensor are derived and discussed. Further
characteristics of the produced solutions, such as the scalar charge,
horizon area and entropy, are also determined, studied in detail and
compared to the corresponding Schwarzschild values.

The outline of this chapter is as follows: in Section \ref{thof0}, we briefly review our theoretical framework
and   analytic study of the near and far-away radial regimes. In Section \ref{nums3}, we present extensive numerical results for the obtained solutions and their properties. We finish with our conclusions in Section \ref{dis3}. The analysis of this chapter is based on \cite{ABK2}.


\section{The Theoretical Framework}\label{thof0}

The theoretical framework is the same with the previous chapter (see section \ref{evthf}). Here we will give a brief overview. 
We start  with the action functional that describes the aforementioned general
class of higher-curvature gravitational theories \cite{ABK1}:
\begin{equation}
S=\frac{1}{16\pi}\int{d^4x \sqrt{-g}\left[R-\frac{1}{2}\,\partial_{\mu}\phi\partial^{\mu}\phi+f(\phi)R^2_{GB}\right]}.\label{act2}
\end{equation}

Our aim is to find solutions of the set of Eqs. (\ref{Einstein-eqs}-\ref{scalar-eq}) 
that describe regular, static, asymptotically-flat black-hole solutions with a non-trivial
scalar field. In particular, we will assume again that the line-element takes the following
spherically-symmetric form
\begin{equation}\label{metric1}
{ds}^2=-e^{A(r)}{dt}^2+e^{B(r)}{dr}^2+r^2({d\theta}^2+\sin^2\theta\,{d\varphi}^2)\,,
\end{equation}
and that the scalar field is also static and spherically-symmetric, $\phi=\phi(r)$.
In our quest for the aforementioned solutions, we will consider a variety of
coupling functions $f(\phi)$, that will however need to obey certain constraints
\cite{ABK1}.  
By employing the line-element (\ref{metric1}), we may obtain the explicit forms of the Einstein's field equations 
\begin{align}
&4e^B(e^{B}+rB'-1)=\phi'^2\bigl[r^2e^B+16\ddot{f}(e^B-1)\bigr] 
-8\dot{f}\left[B'\phi'(e^B-3)-2\phi''(e^B-1)\right], \label{tt-eq22}\\
&4e^B(e^B- r A'-1)=-\phi'^2 r^2 e^B +8\left(e^B-3\right)\dot{f}A'\phi',
\label{rr-eq22}\\[4mm]
&e^B\bigl[r{A'}^2-2B'+A'(2-rB')+2rA''\bigr]= -\phi'^2 r e^B+8 \phi'^2 \ddot{f}A'+ 4\dot{f}[\phi'(A'^2+2A'')\nn\\[2mm]&\hspace{1cm} +A'(2\phi''-3B'\phi')], \label{thth-eq22}
\end{align}
while the scalar equation has the following form:
\beq
 2r\phi''+(4+rA'-rB')\,\phi'+
\frac{4\dot{f}e^{-B}}{r}\bigl[(e^B-3)A'B' 
 -(e^B-1)(2A''+A'^2)\bigr]=0\,. \label{phi-eq22}
\eeq
Before proceeding to numerically integrate the differential system of the field equations let us also briefly review the form of the solutions at the two asymptotic regimes of spacetime.

\subsection{Asymptotic Solution at Black-Hole Horizon}\label{horr0}

We will start our quest for black-hole solutions with a non trivial scalar hair by
determining first the asymptotic solutions of the set of Eqs. (\ref{tt-eq22})-(\ref{phi-eq22})
near the black-hole horizon and at asymptotic infinity. These solutions will serve as
boundary conditions for our numerical integration but will also provide important
constraints on our theory (\ref{act2}). Near the black-hole horizon $r_h$, it is
usually assumed that the metric functions and the scalar field may be expanded as
\begin{align}
e^{A}&=\sum_{n=1}^\infty{a_n(r-r_h)^n}\,,\label{expA_rh}\\
e^{-B}&=\sum_{n=1}^\infty{b_n(r-r_h)^n}\,,\label{expB_rh}\\
\phi &=\sum_{n=0}^{\infty}\frac{\phi^{(n)}(r_h)}{n!}\,(r-r_h)^n\,, \label{phi_rh}
\end{align}
where $(a_n, b_n)$ are constant coefficients and $\phi^{(n)}(r_h)$ denotes the ($n$th)-derivative
of the scalar field evaluated at the black-hole horizon. Equations (\ref{expA_rh})-(\ref{expB_rh})
reflect the expected behaviour of the metric tensor near the horizon of a spherically-symmetric
black hole with the solution being regular if the scalar coefficients $\phi^{(n)}(r_h)$ in
Eq. (\ref{phi_rh}) remain finite at the same regime. 

In the previous chapter, we have followed instead the alternative approach 
of assuming merely that, near the horizon, the metric function $A(r)$ diverges, in 
accordance to Eq. (\ref{expA_rh}). Then, the system of differential equations was evaluated
in the limit $r \rightarrow r_h$, and the finiteness of the quantity $\phi''$ was demanded.
This approach was followed in \cite{DBH1,DBH2} where an exponential coupling function was
assumed between the scalar field and the GB term. In the previous chapter, the form of the
coupling function $f(\phi)$ was left arbitrary, and the requirement of the finiteness of
$\phi''_h$ was shown to be satisfied only under the constraint \cite{ABK1}

\begin{equation}\label{constraint11}
r_h^3\phi'_h+12\dot{f}_h+2r_h^2{\phi'}_h^2\dot{f}_h=0\,,
\end{equation}
where all quantities have been evaluated at the horizon $r_h$. The above is a 
second-order polynomial with respect to $\phi'_h$, and may be easily solved to yield
the solutions:
\begin{equation}\label{con-phi'1}
\phi'_h=\frac{r_h}{4\dot{f}_h}\left(-1\pm\sqrt{1-\frac{96\dot{f}_h^2}{r_h^4}}\right).
\end{equation}
The above ensures that an asymptotic black-hole solution with a regular scalar field 
exists for a general class of theories of the form (\ref{act2}). The only constraint
on the form of the coupling function arises from the demand that the first derivative
of the scalar field on the horizon must be real, which translates to the inequality
\begin{equation}\label{con-f1}
\dot{f}_h^2<\frac{r_h^4}{96}\,.
\end{equation}
Assuming the validity of the constraint (\ref{constraint11}), Eq. (\ref{A-approx-h}) then uniquely
determines the form of the metric function $A$ in the near-horizon regime; through
Eq. (\ref{Bfunction}), the metric function $B$ is also determined. Therefore, the
asymptotic solution for the metric functions and the scalar field in the
limit $r \rightarrow r_h$, is given by the expressions
\bea
&&e^{A}=a_1 (r-r_h) + ... \,, \label{A-rh1}\\[1mm]
&&e^{-B}=b_1 (r-r_h) + ... \,, \label{B-rh1}\\[1mm]
&&\phi =\phi_h + \phi_h'(r-r_h)+ \phi_h'' (r-r_h)^2+ ... \,, \label{phi-rh1}
\eea
and describes, by construction, a black-hole horizon with a regular scalar field
provided that $\phi'$ obeys the constraint (\ref{con-phi'1}) and the coupling
function $f$ satisfies Eq. (\ref{con-f1}). We note that the desired form of the
asymptotic solution was derived only for the choice of the $(+)$-sign in
 Eq. (\ref{Bfunction}) as the $(-)$-sign fails to lead to a black-hole solution \cite{ABK1}.

The aforementioned regularity of the near-horizon solution should be reflected 
to the components of the energy-momentum tensor $T_{\mu\nu}$ as well as to the scalar 
invariant quantities of the theory.
Employing the asymptotic behaviour given in Eqs. (\ref{A-rh1})-(\ref{phi-rh1}), near the black-hole horizon, the non-vanishing components of the energy momentum
tensor (\ref{Ttt})-(\ref{Tthth}) assume the following approximate behaviour:
\begin{align}
T^t_{\;\,t}=&+\frac{2e^{-B}}{r^2}B'\phi'\dot{f}+\mathcal{O}(r-r_h)\,,\label{Ttt-rh}\\
T^r_{\;\,r}=&-\frac{2e^{-B}}{r^2}A'\phi'\dot{f}+\mathcal{O}(r-r_h)\,,\label{Trr-rh1}\\
T^{\theta}_{\;\,\theta}=&T^{\varphi}_{\;\,\varphi}=\frac{e^{-2B}}{r}\phi' \dot{f}\left(2A''+A'^2-3A'B'\right)
+\mathcal{O}(r-r_h)\,.\label{Tthth-rh}
\end{align}
The above expressions, in the limit $r \rightarrow r_h$, lead to constant values for all
components of the energy-momentum tensor. Similarly, one may see that the scalar
invariant quantities
$R$, $R_{\mu\nu}R^{\mu\nu}$ and $R_{\mu\nu\rho\sigma} R^{\mu\nu\rho\sigma}$,
whose exact expressions are listed in the Appendix \ref{apa2}, reduce to the approximate forms
\begin{align}
R = &+\frac{2 e^{-B}}{r^2} \left(e^{B}-2 r A'\right) +\mathcal{O}(r-r_h)\,,\\
R_{\mu\nu}R^{\mu\nu}= &+\frac{2e^{-2 B}}{r^4} \left(e^{B} -r A'\right)^2+\mathcal{O}(r-r_h)\,,\\
R_{\mu\nu\rho\sigma}R^{\mu\nu\rho\sigma} =&+\frac{4 e^{-2 B}}{r^4} \left( e^{2B} + 
r^2 A'^2\right)+\mathcal{O}(r-r_h)\,.
\end{align}
In the above, we have used that, near the horizon, $A' \approx -B' \approx 1/(r-r_h)$
and $A'^2 \approx -A''$,
as dictated by Eqs. (\ref{A-rh1})-(\ref{B-rh1}). Again, the dominant term in each curvature
invariant adopts a constant, finite value in the limit $r \rightarrow r_h$. Subsequently,
the GB term also comes out to be finite, in the same limit, and given by
\beq
R^2_{GB} = +\frac{12  e^{-2 B}}{r^4}A'^2 +\mathcal{O}(r-r_h)\,. \label{GB-rh}
\eeq


\subsection{Asymptotic Solution at Infinity}

At the other asymptotic regime, that of radial infinity ($r \rightarrow \infty$), the 
metric functions and the scalar field may be again expanded in power series,
this time in terms of $1/r$. Demanding that the metric components reduce to
those of the asymptotically-flat Minkowski space-time while the scalar field assumes
a constant value, we find in brief:
\begin{align}
e^A=&\; 1-\frac{2M}{r}+\mathcal{O}(1/r^3)\,,\label{Afar1}\\
e^B=&\; 1+\frac{2M}{r}+\mathcal{O}(1/r^2)\,,\label{Bfar1}\\
\phi=&\; \phi_{\infty}+\frac{D}{r}+\mathcal{O}(1/r^2)\,\label{phifar1},
\end{align}
where with $M$ we denote the Arnowitt-Deser-Misner (ADM) mass and with $D$ the scalar charge. The remaining coefficients may be calculated to an arbitrary order: in the previous chapter we have performed
this calculation up to order $\mathcal{O}(1/r^6)$, see Eqs. (\ref{Afar}-\ref{phifar}). We found that the scalar charge $D$ modifies significantly the expansion
of the metric functions at order $\mathcal{O}(1/r^2)$ and higher. The
existence itself of $D$, and thus of a non-trivial form for the scalar field,
is caused by the presence of the GB term in the theory. The exact form,
however, of the coupling function does not enter in the above expansions
earlier than the order $\mathcal{O}(1/r^4)$. This shows that an asymptotically
flat solution of Eqs. (\ref{tt-eq22}-\ref{phi-eq22}), with a constant scalar field does not require a
specific coupling function and, in fact, arises for an arbitrary form of this function. 

The asymptotic solution at infinity, given by Eqs. (\ref{Afar1})-(\ref{phifar1}), is also
characterised by regular components of $T_{\mu\nu}$ and curvature invariants.
Employing the facts that, as $r \rightarrow \infty$, $(e^A,\,e^B,\,\phi) \approx \mathcal{O}(1)$
while $(A',\,B',\,\phi') \approx \mathcal{O}\left(1/r^2\right)$, we find
for the components of the energy-momentum tensor the asymptotic behaviour:
\beq
T^t_{\;\,t} \simeq -T^r_{\;\,r} \simeq 
T^\theta_{\;\,\theta} \simeq -\frac{1}{4}\,\phi'^2 + {\mathcal O}\left(\frac{1}{r^6}\right).
\label{Tmn-far}
\eeq
Clearly, all of the above components go to zero, as expected. A similar behaviour is
exhibited by all curvature invariants and the GB term, in accordance to the 
asymptotically-flat limit derived above. In particular, for the GB term, we obtain
\beq
R^2_{GB} \approx \frac{48 M^2}{r^6}\,, \label{GB-far}
\eeq
which quickly reduces to zero at asymptotic infinity as anticipated.



\section{Numerical Solutions}\label{nums3}

\begin{figure}[t!] 
\begin{center}
\hspace{0.0cm} \hspace{-0.6cm}
\includegraphics[height=.24\textheight, angle =0]{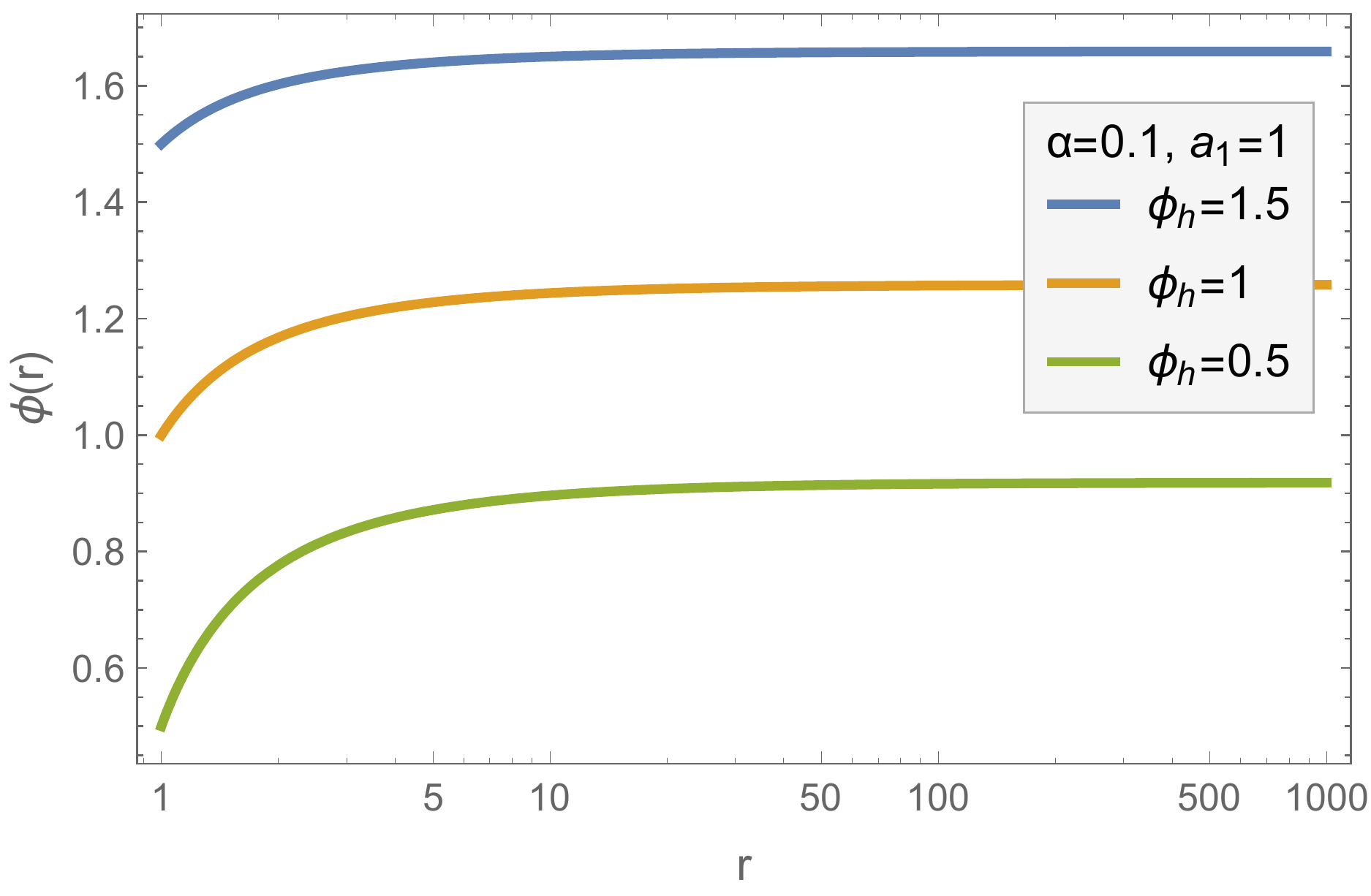}
\hspace{0.52cm} \hspace{-0.6cm}
\includegraphics[height=.24\textheight, angle =0]{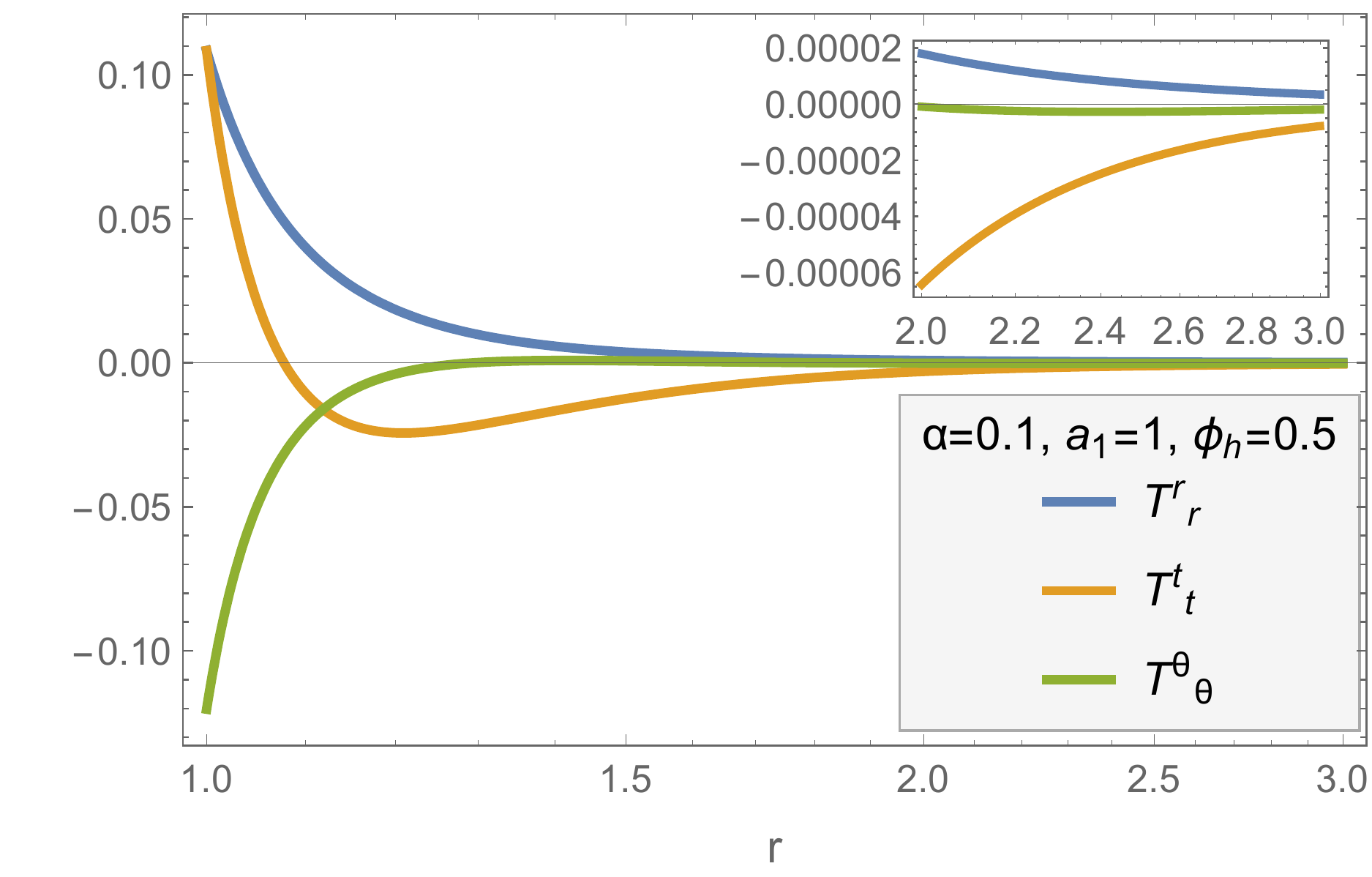}
\\
\hspace*{0.7cm} {(a)} \hspace*{7.5cm} {(b)}  \vspace*{-0.5cm}
\end{center}
\caption{(a) The scalar field $\phi$, and (b) the energy-momentum tensor
$T_{\mu\nu}$ in terms of the radial coordinate $r$, for $f(\phi)=\alpha e^{-\phi}$.}
  \label{Phi_Exp}
\end{figure} 

In the previous two subsections, we have constructed a near-horizon solution
with a regular scalar field and an asymptotically-flat solution with a constant
scalar field -- that was achieved under mild constraints on the form of
the coupling function $f(\phi)$. As demonstrated in the previous chapter, both the old and the novel no-scalar-hair theorems can be evaded in the context of the EsGB theory, and the asymptotic solutions at black-hole horizon and radial infinity may be smoothly matched. However, given the complexity of the
equations of the theory, it is the numerical integration of the   field equations
(\ref{tt-eq22})-(\ref{phi-eq22}) that will reveal whether  a black-hole solution with a non-trivial scalar hair
valid over the entire radial domain can be constructed.

We therefore proceed to numerically solve ours equations (\ref{tt-eq22}-\ref{phi-eq22}). As described in the previous chapter, the set of equations reduces to a set of two independent equations for $A$ and $\phi$
\begin{align} \label{A}
A''=\frac{P}{S}\,, \4\4\text{and}\4\4\phi''=\frac{Q}{S}\,, 
\end{align} 
while the metric function $B$ may be expressed in terms of the functions $A$ and $\phi$, see Eq. (\ref{Bfunction}).
In the above equations, $P$, $Q$ and $S$ are complicated expressions of
$(r, e^B, \phi', A', \dot f, \ddot f)$ that are given in the Appendix \ref{apa1}. 
 Our numerical code is written in Mathematica software. The integration
starts at a distance very close to the horizon of the black hole, i.e. at
$r\approx r_h+\mathcal{O}(10^{-5})$ (for simplicity, we set $r_h=1$). There,
we use as boundary conditions the asymptotic solution (\ref{A-rh1})-(\ref{phi-rh1})
together with Eq. (\ref{con-phi'1}) for $\phi'_h$ upon choosing a particular coupling
function $f(\phi)$. The integration proceeds towards large values of the radial
coordinate until the form of the derived solution matches the asymptotic solution
(\ref{Afar1})-(\ref{phifar1}). In the next sub-sections, we present a variety of
regular black-hole solutions with scalar hair for different choices of the coupling
function $f(\phi)$. 

\begin{figure}[t!] 
\begin{center}
\hspace{0.0cm} \hspace{-0.6cm}
\includegraphics[height=.25\textheight, angle =0]{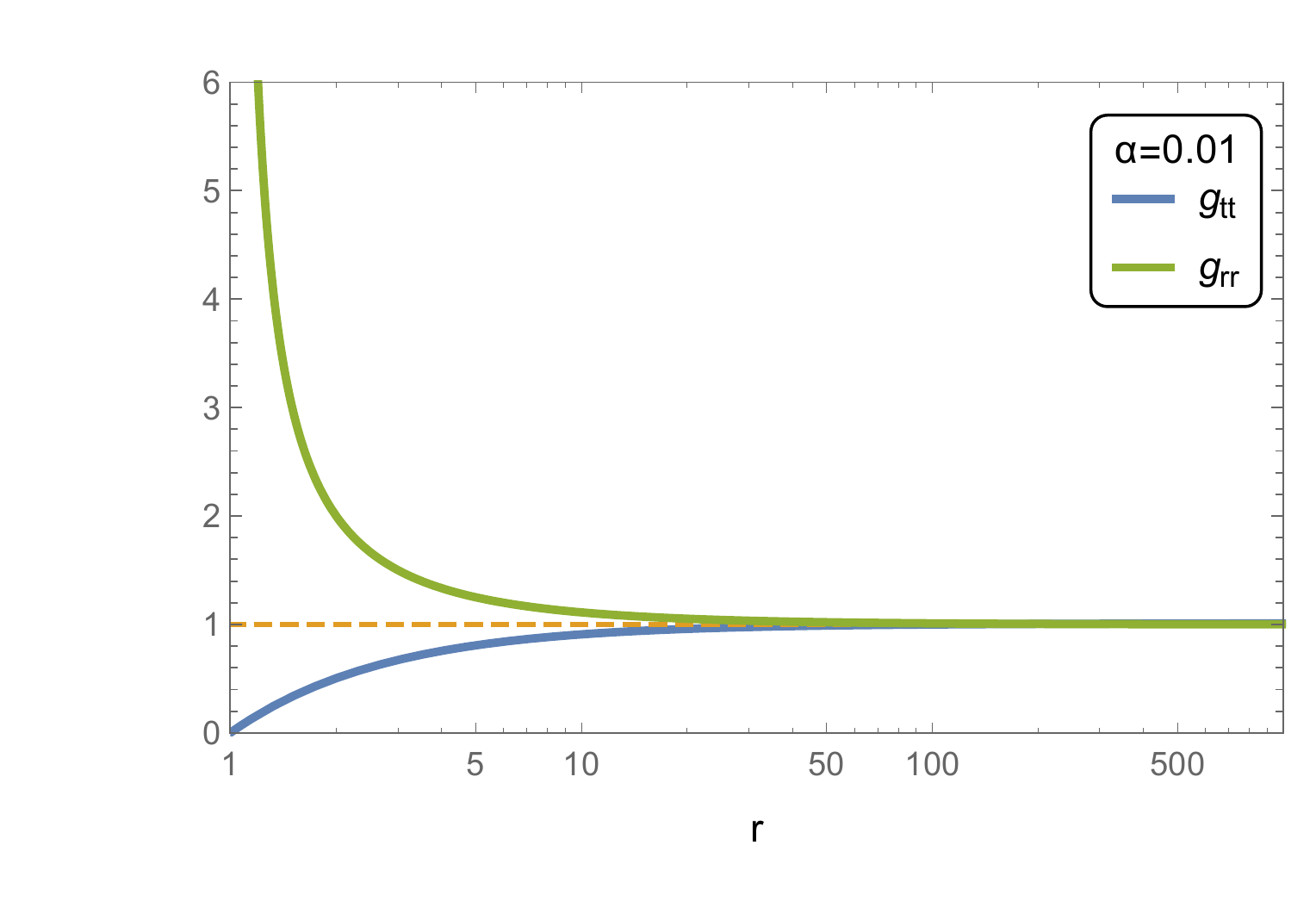}
\hspace{0.52cm} \hspace{-0.6cm}
\includegraphics[height=.25\textheight, angle =0]{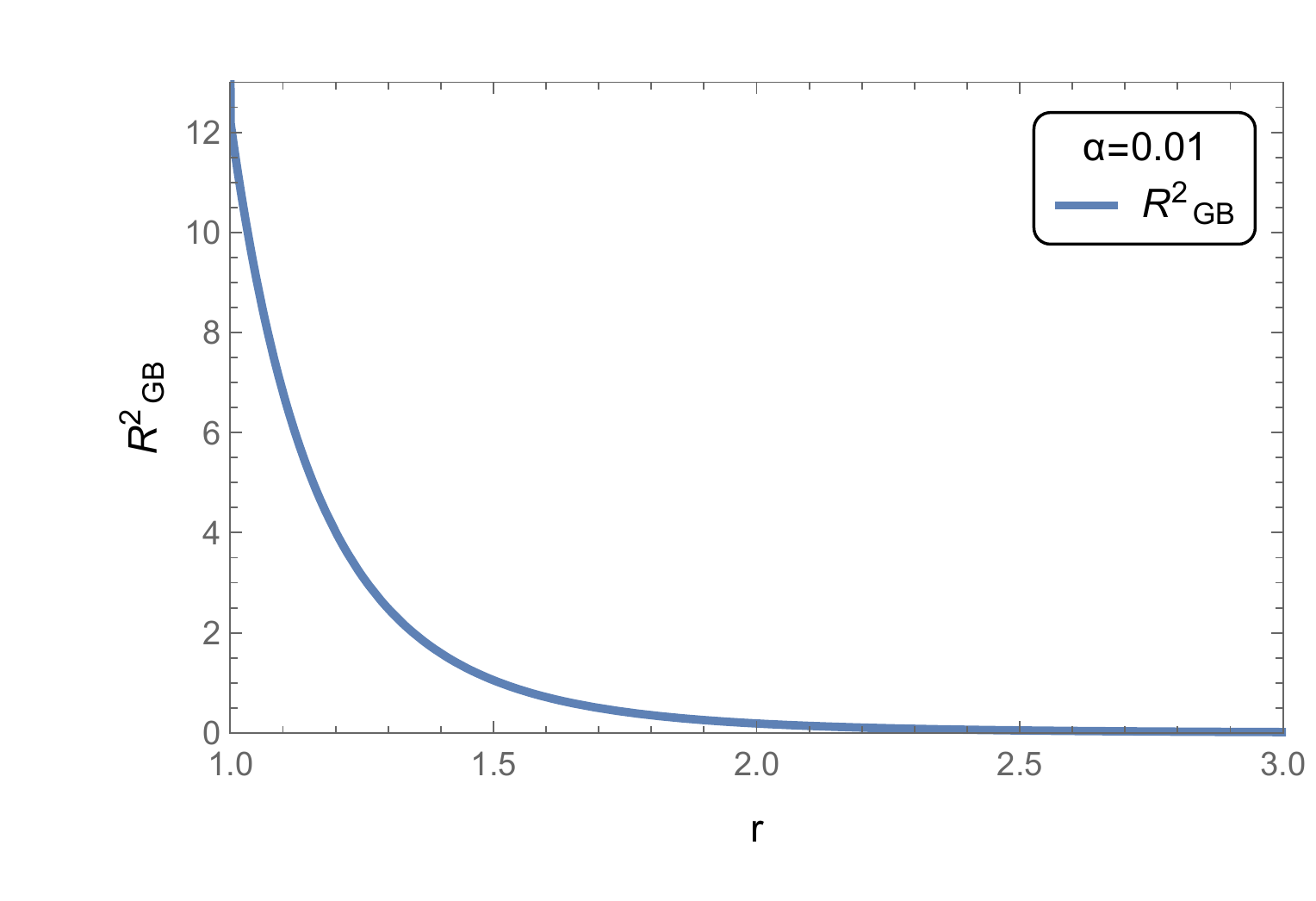}
\\
\hspace*{0.7cm} {(a)} \hspace*{7.5cm} {(b)}  \vspace*{-0.5cm}
\end{center}
\caption{(a) The metric components $g_{tt}$ and $g_{rr}$, and (b)  the Gauss-Bonnet
term $R^2_{GB}$  in terms of the radial coordinate $r$, for $f(\phi)=\alpha e^{-\phi}$.}
  \label{Metric_GB}
\end{figure} 

\subsection{Exponential Coupling Function}

First, we consider the case where $f(\phi)=\alpha e^{\kappa \phi}$. According to
the arguments presented  in section \ref{seves}, the evasion of the  old no-hair theorem
is ensured for $f(\phi)>0$; therefore, we focus on the case with $\alpha>0$. As
the exponential function is always positive-definite, the sign of 
$\dot f= \alpha \kappa e^{\kappa \phi}$ is then determined by the sign of $\kappa$.
In order to evade also the novel no-hair theorem and allow for regular black-hole
solutions to emerge, we should satisfy the constraint $\dot f \phi'<0$ according to Eq. (\ref{constr1}),
or equivalently $\kappa\,\phi'<0$, near the horizon. Therefore, for $\kappa>0$,
we should have $\phi'_h<0$, which causes the decreasing of the scalar field as
we move away from the black-hole horizon. The situation is reversed for
$\kappa<0$ when $\phi_h'>0$ and the scalar field increases with $r$. 

The case of $f(\phi)=\alpha e^\phi$, with $\alpha>0$, was studied in \cite{DBH1,DBH2}
and led to the well-known family of Dilatonic Black Holes. The solutions were
indeed regular and asymptotically-flat with the scalar field decreasing away
from the horizon, in agreement with the above discussion. Here, we present the
complimentary case with $f(\phi)=\alpha e^{-\phi}$ (the exact value
of $\kappa$ does not alter the physical characteristics of the solution and,
here, we set it to $\kappa=-1$). 
In   Fig. \ref{Phi_Exp}(a), we present a family of solutions for
the scalar field $\phi$ for different initial values $\phi_h$: for $\kappa=-1<0$,
the scalar field must necessarily have $\phi_h'>0$, and therefore increases as
$r$ increases. The value of the coupling constant $\alpha$, once the form of
the coupling function and the asymptotic value $\phi_h$ are chosen, is
restricted by the inequality (\ref{con-f1}) - here,
we present solutions for indicative allowed values of $\alpha$.

\begin{figure}[t!] 
\begin{center}
\hspace{0.0cm} \hspace{-0.6cm}
\includegraphics[height=.24\textheight, angle =0]{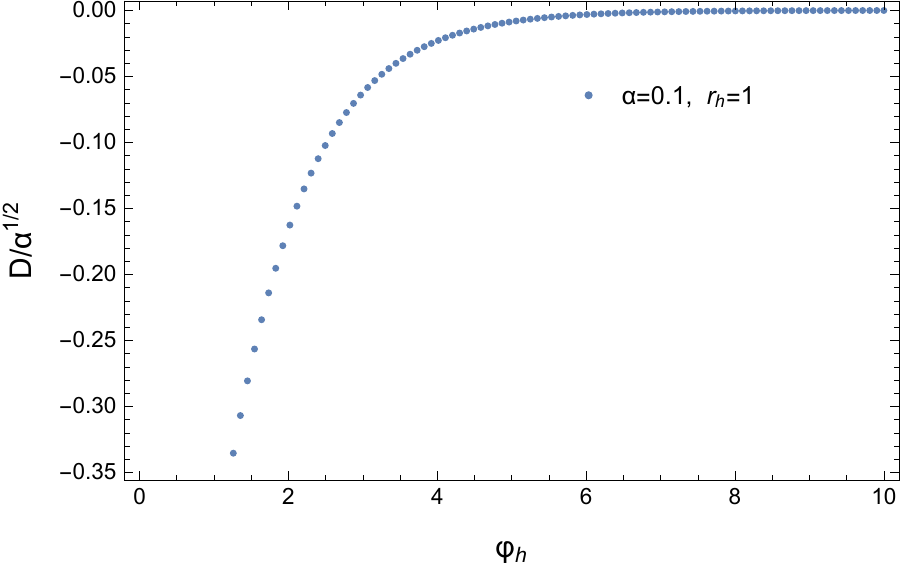}
\hspace{0.52cm} \hspace{-0.6cm}
\includegraphics[height=.24\textheight, angle =0]{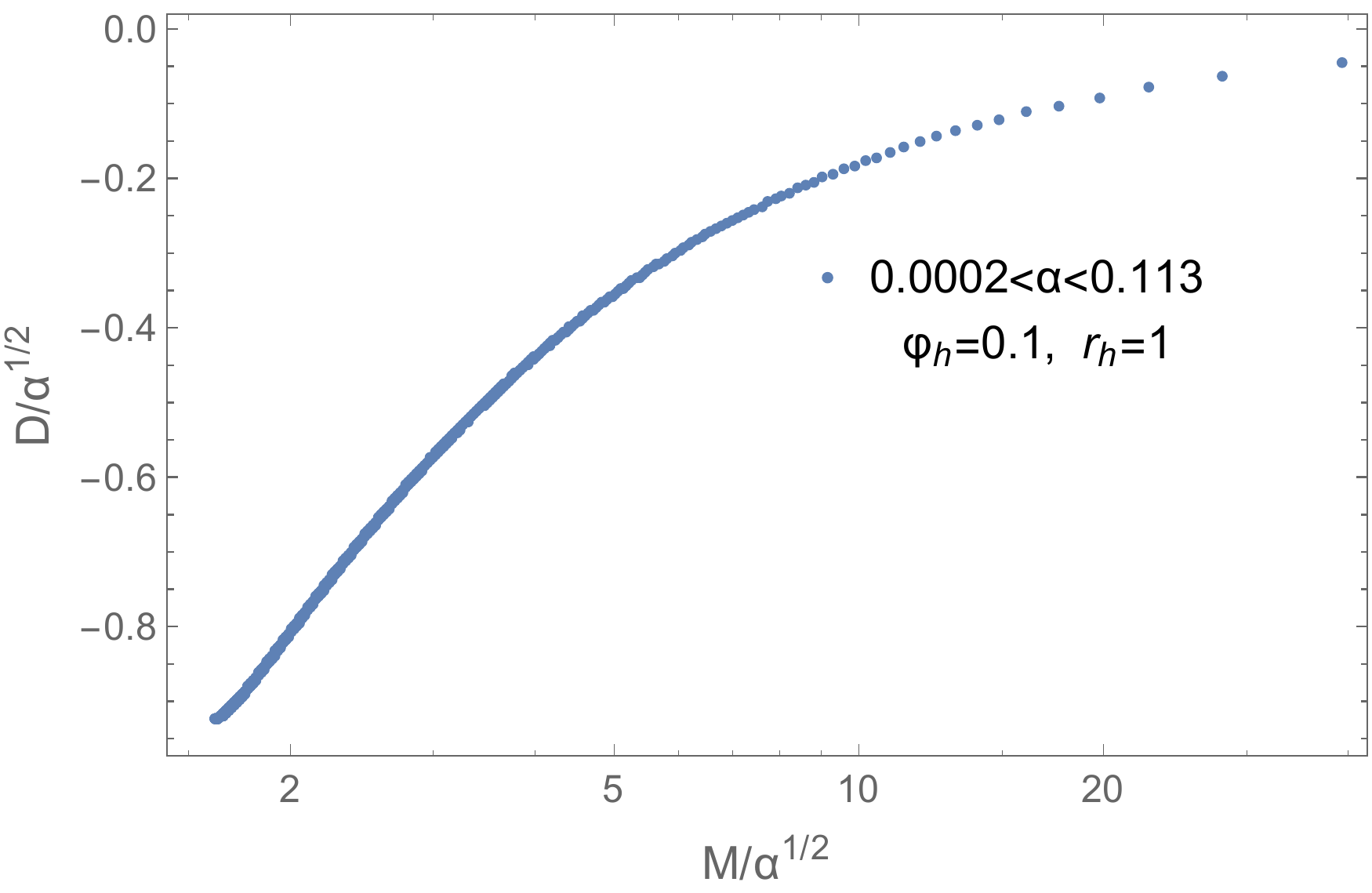}
\\
\hspace*{0.7cm} {(a)} \hspace*{7.5cm} {(b)}  \vspace*{-0.5cm}
\end{center}
\caption{(a) The scalar charge $D$ as a function of the near-horizon value $\phi_h$, and (b) the scalar charge $D$ as a function  of its mass $M$, for $f(\phi)=\alpha e^{-\phi}$.}
  \label{scalarD-exp}
\end{figure} 

In Fig. \ref{Phi_Exp}(b), we present
the energy-momentum tensor components for an indicative solution of this family:
clearly, all components remain finite over the whole radial regime. In particular,
$T^r_{\,\,r}$ remains positive and monotonically decreases towards infinity,
exactly the behaviour that ensures the evasion of the novel no-hair theorem.
As discussed in the previous chapter, apart from choosing the input value $\phi'_h$
for our numerical integration in accordance with Eq. (\ref{con-phi'1}), no
other fine-tuning is necessary; the second constraint for the evasion of the novel
no-hair theorem, i.e. $\dot f \phi'' + \ddot f \phi'^2>0$, is automatically satisfied
without any further action, and this is reflected in the decreasing behaviour of $T^r_{\,\,r}$
component near the horizon.

At the left and right plots of Fig. \ref{Metric_GB}, we also present the solution
for the two metric components ($|g_{tt}|,g_{rr}$) and the GB term $R^2_{GB}$, respectively. 
The metric components exhibit the expected behaviour near the black-hole horizon
with $g_{tt}$ vanishing and $g_{rr}$ diverging at $r_h=1$. In order to ensure
asymptotic flatness at radial infinity, the free parameter $a_1$ appearing in the
near-horizon solution (\ref{A-rh1}) is appropriately chosen. On the other hand,
the GB term remains finite and
positive-definite over the entire radial domain - in fact it displays the monotonic
behaviour, hinted by its two asymptotic limits (\ref{GB-rh}) and (\ref{GB-far}),
that causes the evasion of the old no-hair theorem. As expected, it contributes
significantly near the horizon, where the curvature is large, and quickly fades
away as we move towards larger distances. The profile of the metric
components and GB term exhibit the same qualitative behaviour in all families
of black-hole solutions presented in this work, so we refrain from giving
additional plots of these two quantities in the next sub-sections. 

The profile of the scalar charge $D$ as a function of the near-horizon value $\phi_h$
and of the mass $M$ is given at the left and right plot, respectively, of Fig.
\ref{scalarD-exp} (each dot in these, and subsequent, plots stand for a different
black-hole solution). 
For the exponential coupling function $f(\phi)=\alpha e^{-\phi}$, and for fixed $\alpha$
and $r_h$, the scalar field near the horizon $\phi_h$ may range from a
minimum value, dictated by Eq. (\ref{con-f1}), up to infinity. As the left plot reveals,
as $\phi_h \rightarrow \infty$, the coupling of the scalar field to the GB term vanishes,
and we recover the Schwarzschild case with a trivial scalar field and a vanishing charge.
On the other hand, in the right plot, we observe that as the mass of the black-hole
increases, the scalar charge decreases in absolute value, and thus larger black holes
tend to have smaller charges.

\begin{figure}[t!] 
\begin{center}
\hspace{0.0cm} \hspace{-0.6cm}
\includegraphics[height=.24\textheight, angle =0]{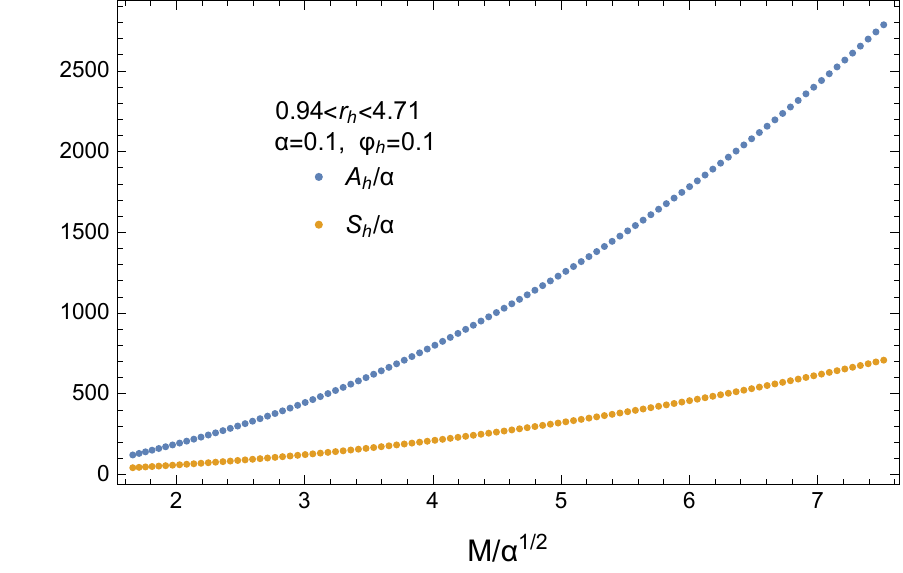}
\hspace{0.52cm} \hspace{-0.6cm}
\includegraphics[height=.24\textheight, angle =0]{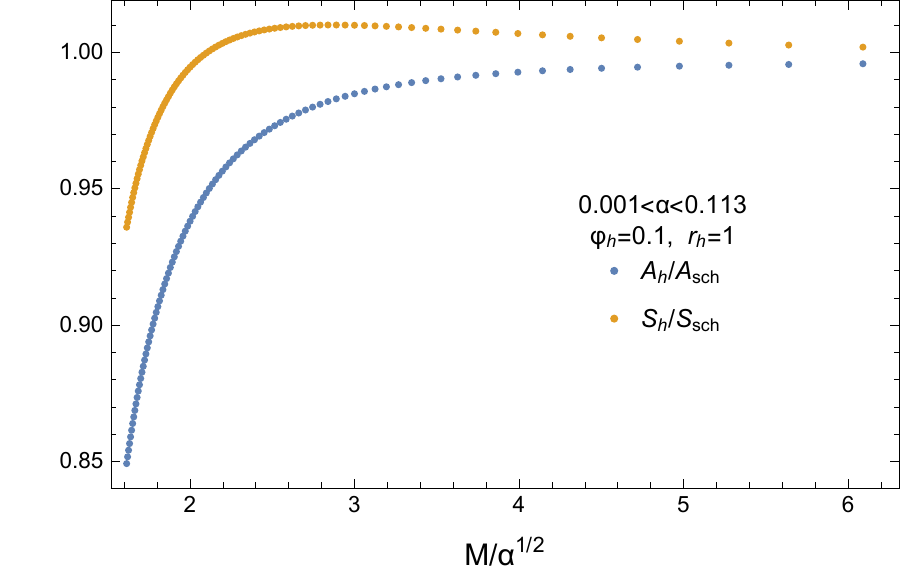}
\\
\hspace*{0.7cm} {(a)} \hspace*{7.5cm} {(b)}  \vspace*{-0.5cm}
\end{center}
\caption{(a) The horizon area $A_h$ and entropy $S_h$ of the black hole, and (b) their ratios to the corresponding Schwarzschild values (right plot) in terms of the
mass $M$ for $f(\phi)=\alpha e^{-\phi}$.}
  \label{AS-exp}
\end{figure} 

It is also interesting to study the profiles of the area of the black-hole horizon, 
$A_h=4 \pi r_h^2$, and of the entropy $S_h$ of this class of solutions. 
The entropy is defined through the relation \cite{GH}
\beq
S_h=\beta\left[\frac{\partial (\beta F)}{\partial \beta} -F\right],
\label{entropy-def}
\eeq
where $F=I_E/\beta$ is the Helmholtz free-energy of the system given in terms of the
Euclidean version of the action $I_E$. Also, $\beta=1/(k_B T)$ with the temperature 
following easily from the definition \cite{York, GK}
\beq 
T=\frac{k}{2\pi}=\frac{1}{4\pi}\,\left(\frac{1}{\sqrt{|g_{tt} g_{rr}|}}\,
\left|\frac{dg_{tt}}{dr}\right|\right)_{r_h}=\frac{\sqrt{a_1 b_1}}{4\pi}\,.
\label{Temp-def}
\eeq
The calculation of the temperature and entropy of the dilatonic black hole, with
an exponential coupling function of the form $f(\phi)=\alpha e^\phi$, was performed
in detail in \cite{KT}. By closely repeating the analysis, we find the following expressions
for the temperature
\beq 
T=\frac{1}{4\pi}\,\frac{(2M+D)}{r_h^2+4 f(\phi_h)}\,,
\label{Temp-gen}
\eeq
and entropy 
\beq
S_h=\frac{A_h}{4} +4 \pi f(\phi_h)
\label{entropy}
\eeq
of a GB black-hole arising in the context of our theory (\ref{act2}) with a general
coupling function $f(\phi)$ between the scalar field and the GB term. We easily confirm
that, in the absence of the coupling function, the above quantities reduce to the
corresponding Schwarzschild ones, $T=1/(4\pi r_h)$ and $S_h=A_h/4$, respectively.

By employing the expressions for $A_h$ and $S_h$ as given above, we depict the horizon
area and entropy, in terms of the mass of the black hole, in   Fig. \ref{AS-exp}(a)
we observe that both quantities increase fast as the mass increases.
The plot in Fig. \ref{AS-exp}(b) allows us to compare more effectively our solutions to the Schwarzschild
one. The lower curve depicts the ratio of $A_h$ to the area of the Schwarzschild
solution $A_{Sch}=16 \pi M^2$, as a function of $M$; we observe that, for large black-hole
masses, the ratio $A_h/A_{Sch}$ approaches unity, therefore, large GB black holes are
not expected to deviate in their characteristics from the Schwarzschild solution of the
same mass. On the other hand, in the small-mass limit, the ratio $A_h/A_{Sch}$
significantly deviates from unity; in addition, a lower bound appears for the black-hole
radius, and thus of the mass of the
black hole, due to the constraint (\ref{con-f1}) not present in the Schwarzschild case --
this feature has been noted before in the case of the dilatonic black holes \cite{DBH1,DBH2,
Blazquez}. It is worth noting that the GB term, as an extra gravitational effect, causes
the shrinking of the size of the black hole as the ratio $A_h/A_{Sch}$ remains for
all solutions below unity. 

\begin{figure}[t!] 
\begin{center}
\hspace{0.0cm} \hspace{-0.6cm}
\includegraphics[height=.24\textheight, angle =0]{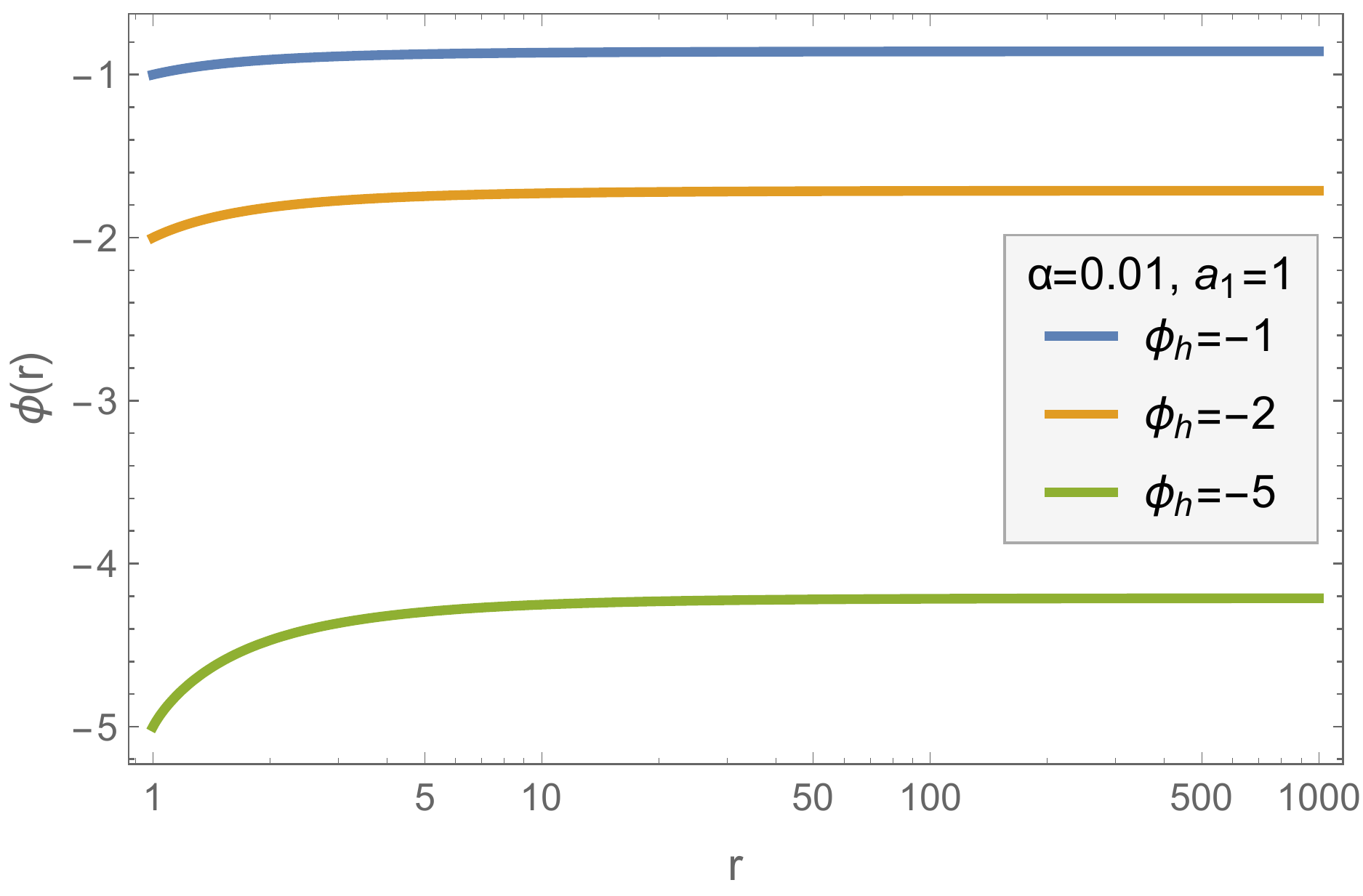}
\hspace{0.52cm} \hspace{-0.6cm}
\includegraphics[height=.24\textheight, angle =0]{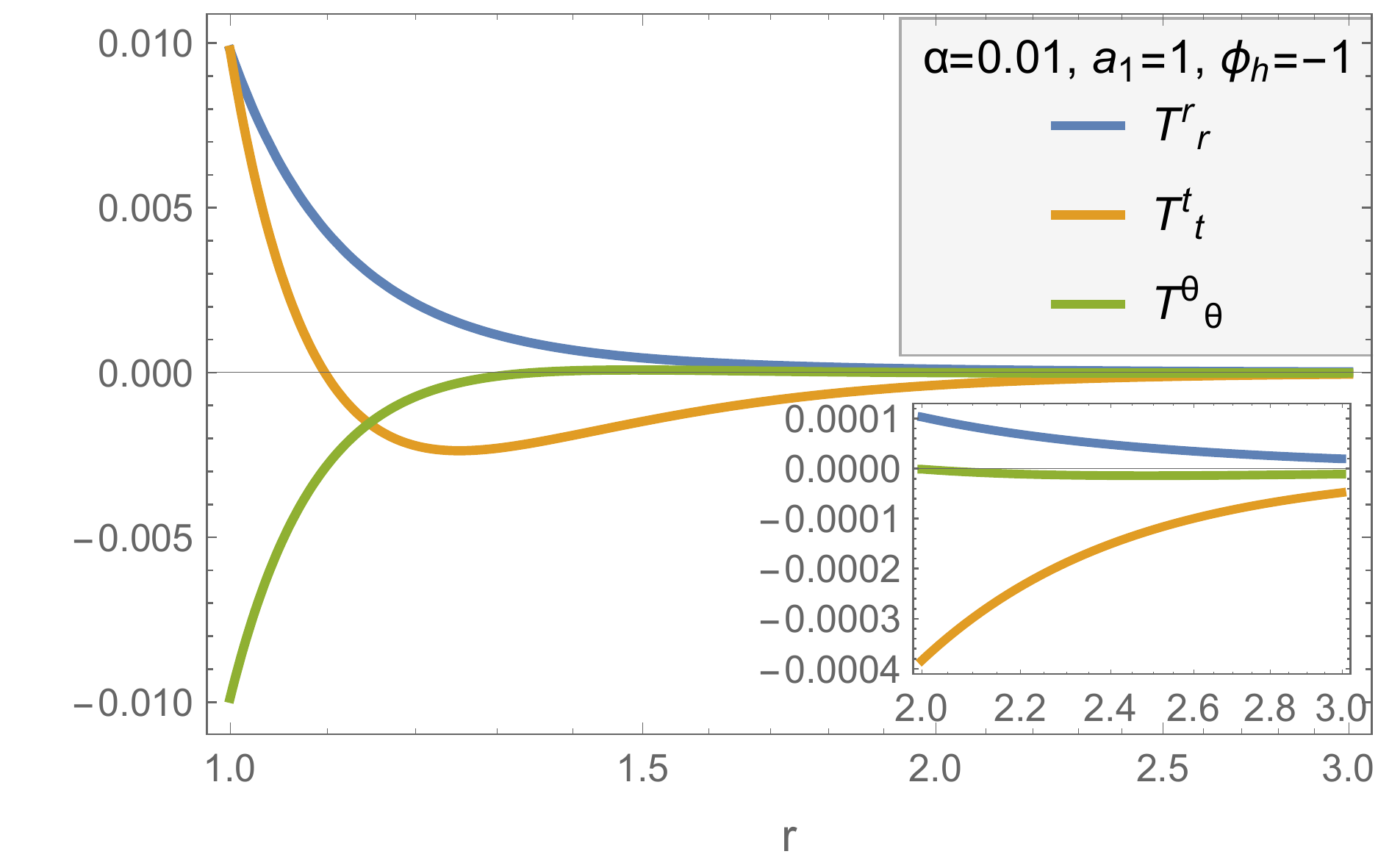}
\\
\hspace*{0.7cm} {(a)} \hspace*{7.5cm} {(b)}  \vspace*{-0.5cm}
\end{center}
\caption{(a) The scalar field $\phi$, and (b) the energy-momentum tensor
$T_{\mu\nu}$  in terms of the radial coordinate $r$, for $f(\phi)=\alpha \phi^2$.}
  \label{Phi_phi2}
\end{figure} 

Turning finally to the entropy $S_h$ of our black-hole solutions, we find a similar
pattern: very small GB black holes differ themselves from the Schwarzschild solution by
having a lower entropy whereas large GB black holes tend to acquire, among other
characteristics, and the entropy of the Schwarzschild solution. We observe however
that, apart from the very small-mass regime close to the minimum value, the 
`exponential' GB
black holes have in general a higher entropy than the Schwarzschild solution, a 
characteristic that points to the thermodynamical stability of these solutions. In
fact, the dilatonic GB black holes \cite{DBH1,DBH2}, which comprise a subclass of this
family of solutions with a coupling function of the form $f(\phi)=\alpha e^\phi$,
have an identical entropy pattern, and were shown to be linearly stable under small
perturbations more than twenty years ago.


\subsection{Even Polynomial Function}

\begin{figure}[t!] 
\begin{center}
\hspace{0.0cm} \hspace{-0.6cm}
\includegraphics[height=.24\textheight, angle =0]{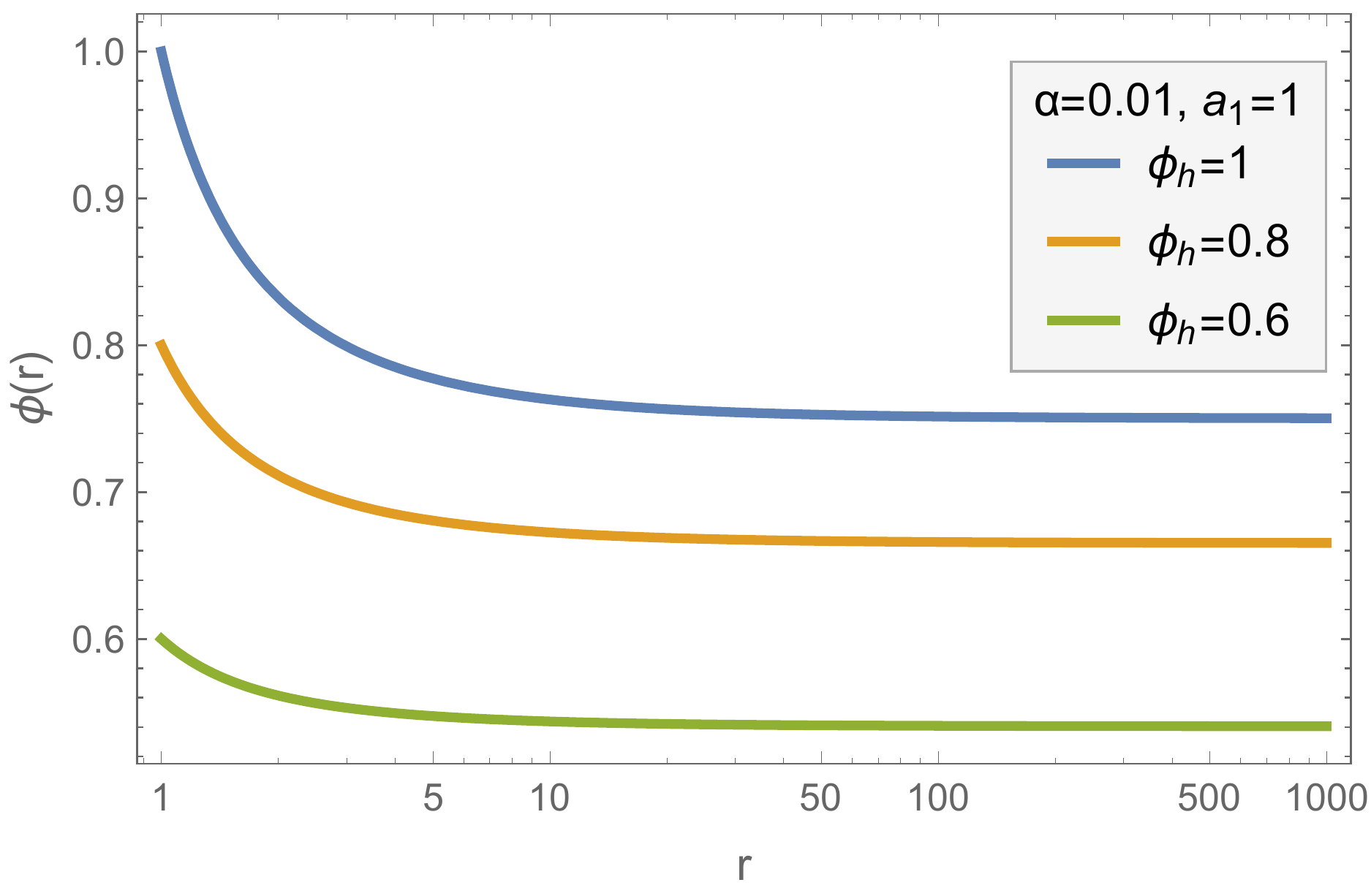}
\hspace{0.52cm} \hspace{-0.6cm}
\includegraphics[height=.24\textheight, angle =0]{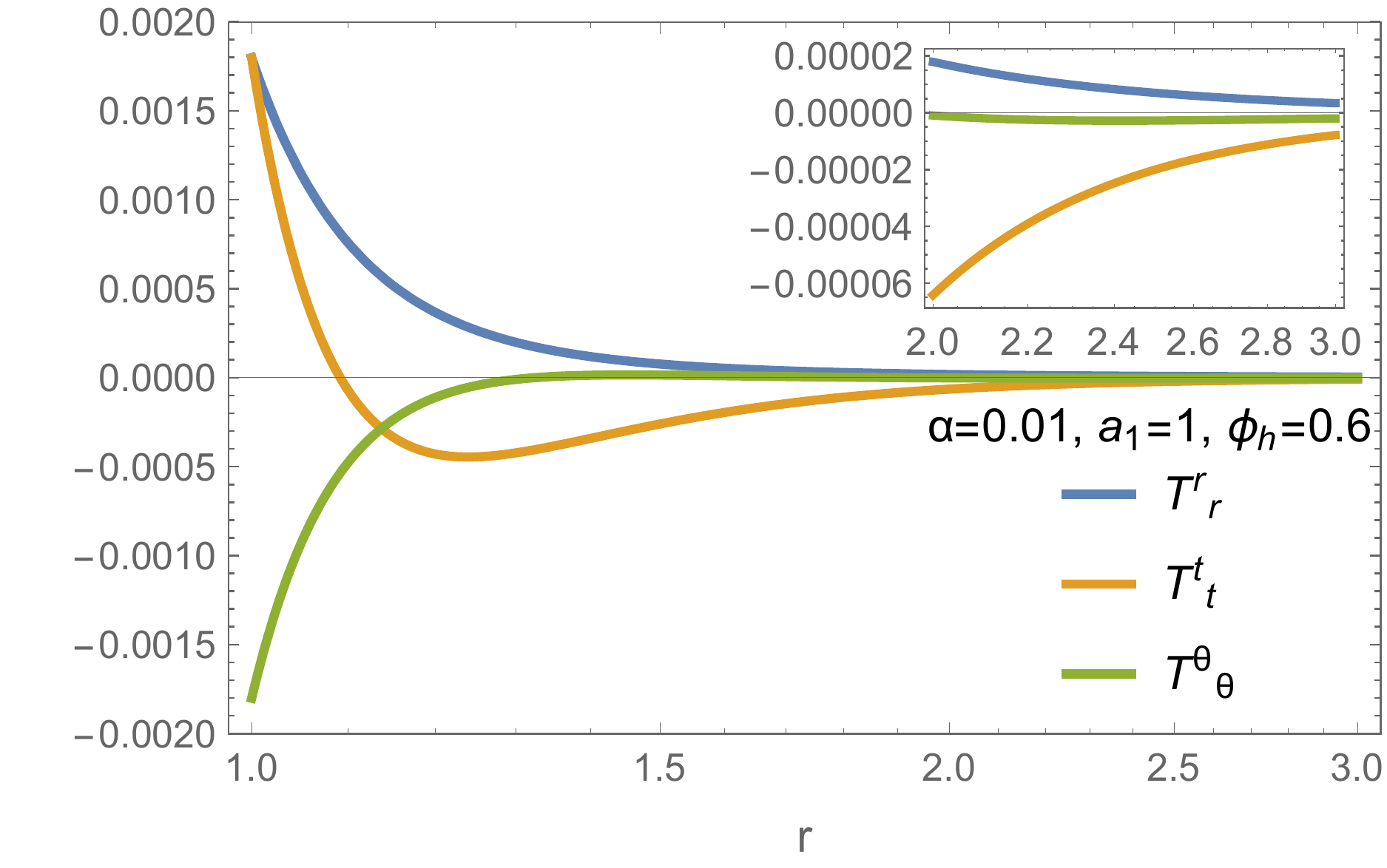}
\\
\hspace*{0.7cm} {(a)} \hspace*{7.5cm} {(b)}  \vspace*{-0.5cm}
\end{center}
\caption{(a) The scalar field $\phi$, and (b) the energy-momentum tensor
$T_{\mu\nu}$  in terms of the radial coordinate $r$, for $f(\phi)=\alpha \phi^4$.}
  \label{Phi_phi4}
\end{figure}

Next, we consider the case where $f(\phi)=\alpha \phi^{2n}$, with $n \geq 1$. Since the
coupling function must be positive-definite, we assume again that $\alpha>0$. The first
constraint for the evasion of the novel no-hair theorem, $\dot f \phi' <0$ near the horizon,
now translates to $\phi_h \phi'_h<0$. Therefore, two classes of solutions appear for each value
of $n$: one for $\phi_h>0$, where $\phi'_h<0$ and the solution for the scalar field decreases
with $r$, and one for $\phi_h<0$, where $\phi'_h>0$ and the scalar field increases away
from the black-hole horizon. In  Fig. \ref{Phi_phi2}(a), we depict the first family
of solutions with $\phi_h<0$ and $\phi'_h>0$ for the choice $f(\phi)=\alpha \phi^2$ while,
in Fig. \ref{Phi_phi4}(a), we depict the second class with $\phi_h>0$ and
$\phi'_h<0$ for the choice $f(\phi)=\alpha \phi^4$. The complimentary classes of solutions
may be easily derived in each case by reversing the signs of $\phi_h$ and $\phi'_h$.

The form of the energy-momentum tensor components for the two choices  $f(\phi)=\alpha \phi^2$
and  $f(\phi)=\alpha \phi^4$, and for two indicative solutions, are depicted in   Figs. \ref{Phi_phi2}(b) and \ref{Phi_phi4}(b), respectively. We observe that the qualitative
behaviour of the three components largely remain the same despite the change in the form
of the coupling function $f(\phi)$ (note also the resemblance with the behaviour
depicted in   Fig. \ref{Phi_Exp}(b)). In fact, the asymptotic behaviour of
$T_{\mu\nu}$ near the black-hole horizon and radial infinity is fixed, according to 
Eqs. (\ref{Ttt-rh})-(\ref{Tthth-rh}) and (\ref{Tmn-far}), respectively. Independently
of the form of the coupling function $f(\phi)$, at asymptotic infinity $T^r_{\,\,r}$
approaches zero from the positive side, while $T^t_{\,\,t}$ and $T^{\theta}_{\,\,\theta}$
do the same from the negative side. In the near-horizon regime, 
Eqs. (\ref{Ttt-rh})-(\ref{Tthth-rh}) dictate that 

 \begin{align}
 \sign\left(T^t_{\;\,t}\right)_h,\;\sign\left(T^r_{\;\,r}\right)_h\sim & -\sign\left(\phi'_h\dot{f}_h\right),\\
 \sign\left(T^\theta_{\;\,\theta}\right)_h\sim & +\sign\left(\phi'_h\dot{f}_h\right).
 \end{align}
 
Using that $\dot f_h \phi'_h<0$ and the scaling behaviour of the metric functions near
the horizon, we may easily derive that  $(T^t_{\,\,t})_h$ and $(T^r_{\,\,r})_h$ always
assume positive values while ($T^\theta_{\,\,\theta})_h$ assumes a negative one.
The form of $f(\phi)$ merely changes the magnitude of these
asymptotic values: in the case of a polynomial coupling function, the higher the degree is,
the larger the asymptotic values near the horizon are.

\begin{figure}[t!] 
\begin{center}
\hspace{0.0cm} \hspace{-0.6cm}
\includegraphics[height=.24\textheight, angle =0]{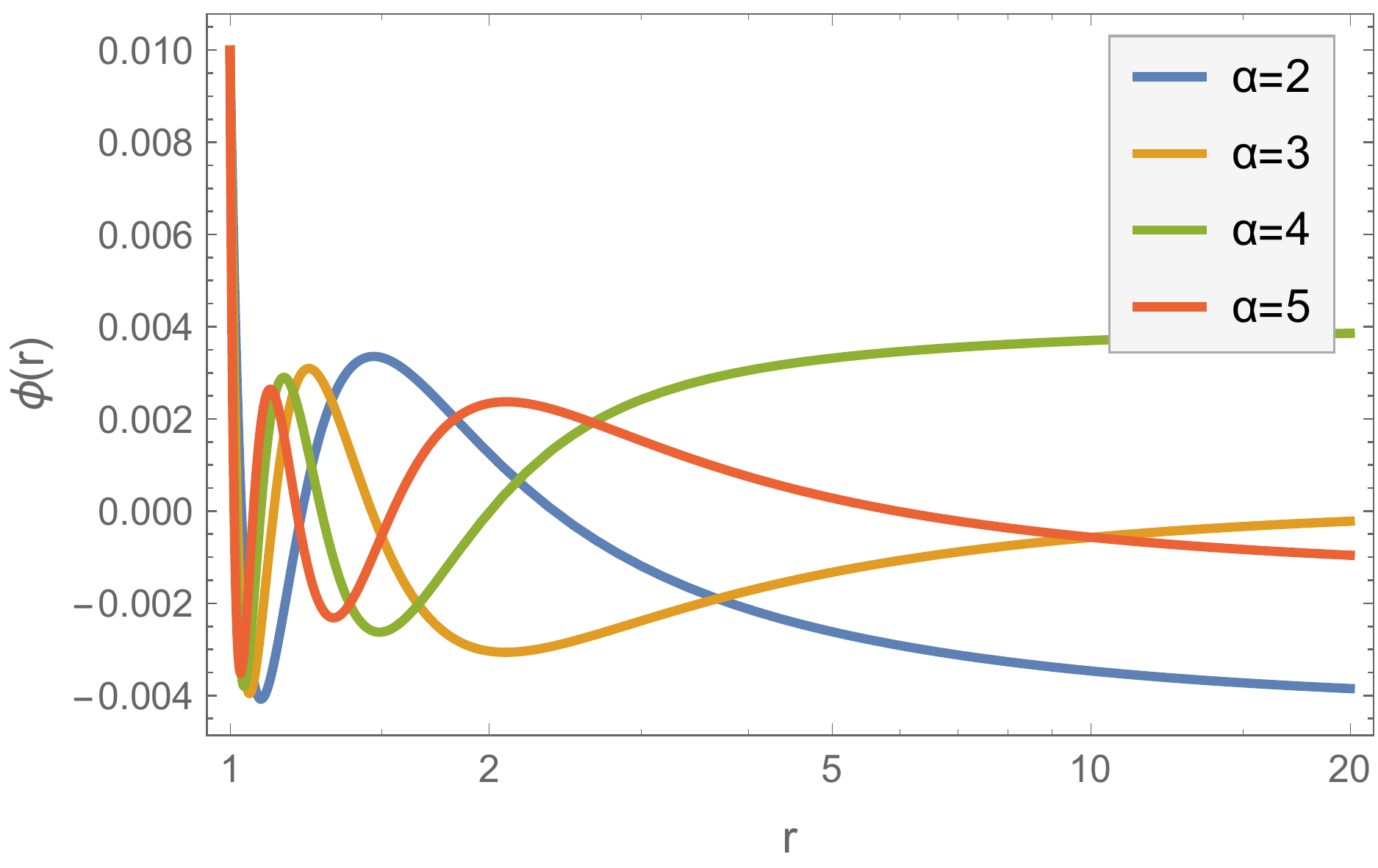}
\hspace{0.52cm} \hspace{-0.6cm}
\includegraphics[height=.24\textheight, angle =0]{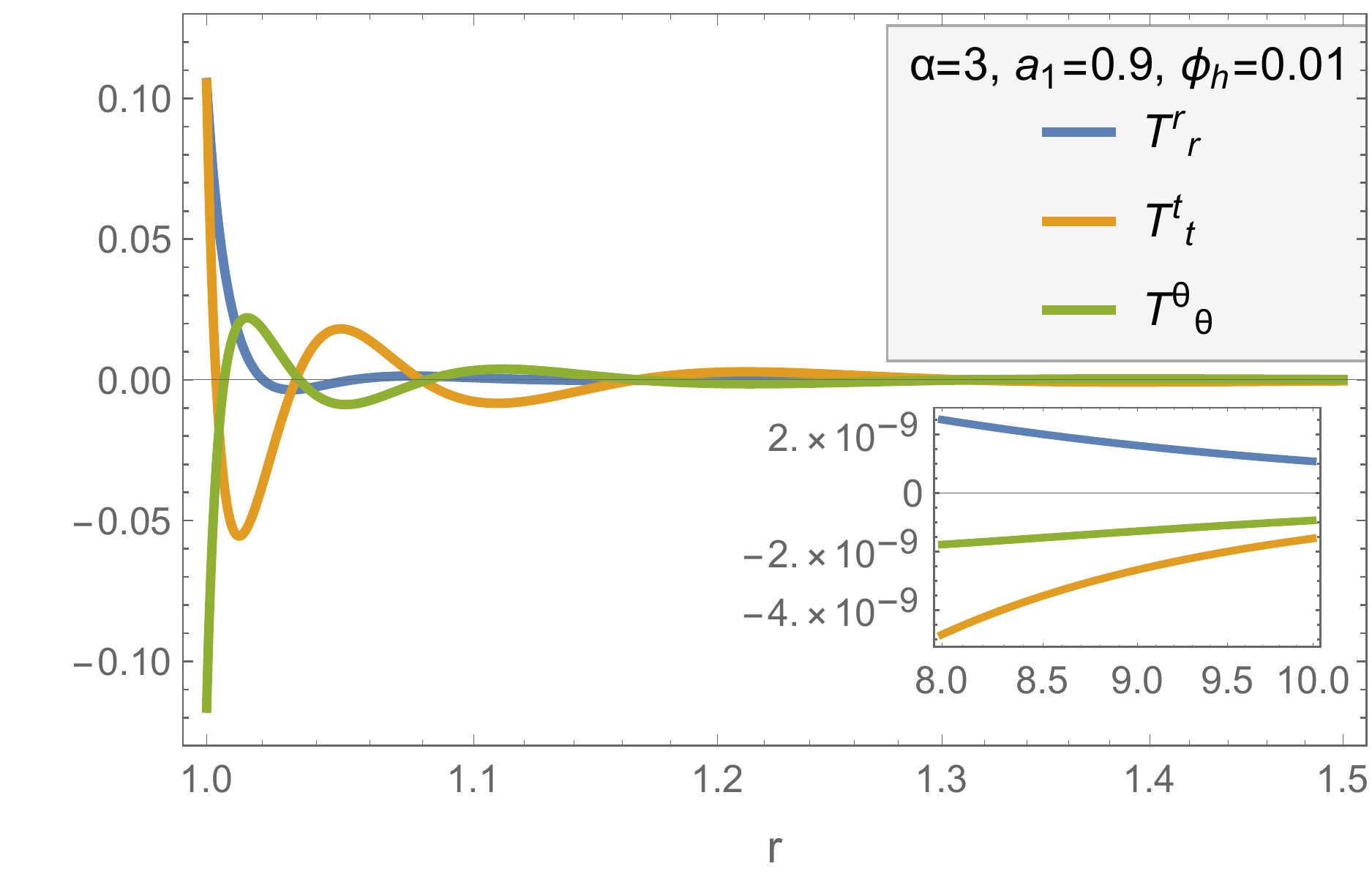}
\\
\hspace*{0.7cm} {(a)} \hspace*{7.5cm} {(b)}  \vspace*{-0.5cm}
\end{center}
\caption{(a) The scalar field $\phi$, and (b) the energy-momentum tensor
$T_{\mu\nu}$  in terms of the radial coordinate $r$, for $f(\phi)=\alpha \phi^2$ and various values of the coupling constant $\alpha$.}
  \label{Phi_alpha_phi2}
\end{figure} 

One could assume that the intermediate behaviour of the scalar field and the energy-momentum
tensor always remains qualitatively the same. In fact, this is not so. Let us fix for simplicity
the values of $r_h$ and $\phi_h$, and gradually increase the value of the coupling
parameter $\alpha$; this has the effect of increasing the magnitude of the GB source-term
appearing in the equation of motion (\ref{phi-eq22}) for $\phi$. As an indicative case,
in   Fig. \ref{Phi_alpha_phi2}(a), we
depict the behaviour of $\phi$ for four large values of $\alpha$, and, in Fig. \ref{Phi_alpha_phi2}(b), the $T_{\mu\nu}$ for one of these solutions. We observe that all of these quantities
are not monotonic any more; they go through a number of maxima or minima -- with
that number increasing with the value of $\alpha$ -- before reaching their asymptotic
values at infinity. Note that the near-horizon behaviour of both $\phi$ and $T_{\mu\nu}$
is still the one that guarantees the evasion of the no-hair theorem. We may thus
conclude that the presence of the GB term in the theory not only ensures that
the asymptotic solutions (\ref{A-rh1})-(\ref{phi-rh1}) and (\ref{Afar1})-(\ref{phifar1}) may be
smoothly connected to create a regular black hole but it allows for this to happen
even in a non-monotonic way.

\begin{figure}[t!] 
\begin{center}
\hspace{0.0cm} \hspace{-0.6cm}
\includegraphics[height=.24\textheight, angle =0]{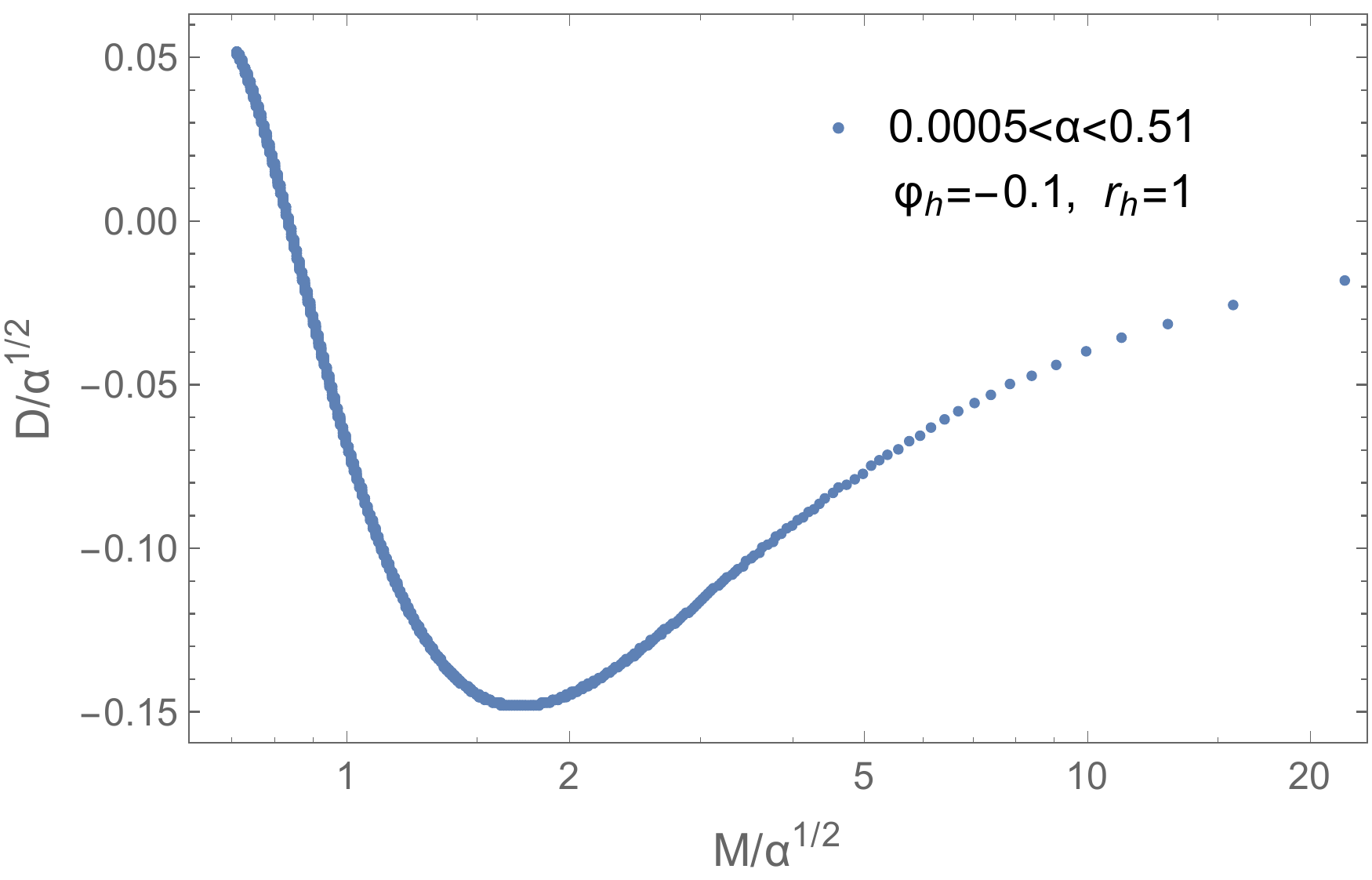}
\hspace{0.52cm} \hspace{-0.6cm}
\includegraphics[height=.24\textheight, angle =0]{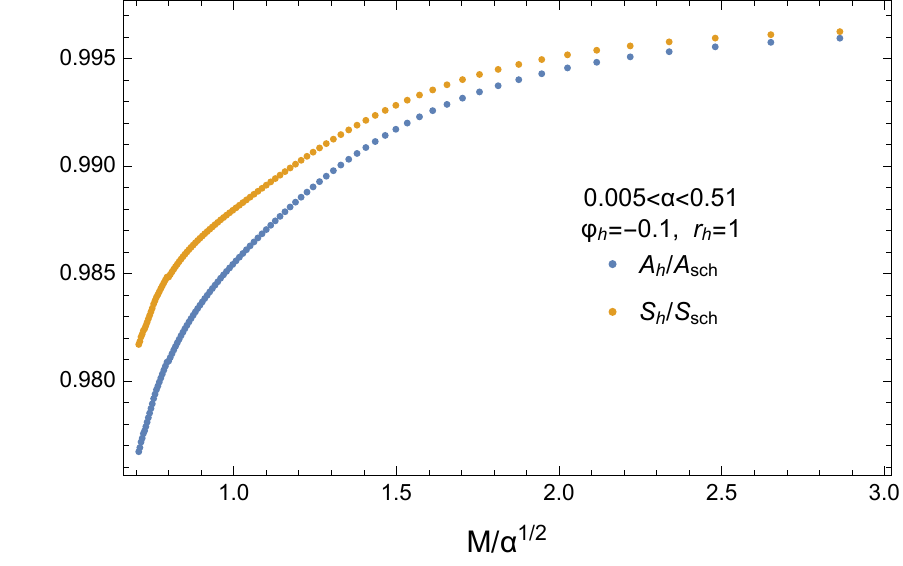}
\\
\hspace*{0.7cm} {(a)} \hspace*{7.5cm} {(b)}  \vspace*{-0.5cm}
\end{center}
\caption{(a) The scalar charge $D$, and (b) the ratios $A_h/A_{Sch}$ and $S_h/S_{Sch}$ 
 in terms of the mass  $M$, for $f(\phi)=\alpha \phi^2$.}
\label{D-AS-even}
\end{figure} 

Let us also study the characteristics of this class of black-hole solutions arising for an
even polynomial coupling function. In Fig. \ref{D-AS-even}(a), we depict the
scalar charge $D$ in terms of the mass $M$ of the black hole, for the quadratic coupling 
function $f(\phi)=\alpha \phi^2$: we observe that, in this case, the function $D(M)$ is
not monotonic in the small-mass regime but it tends again to zero for large values of
its mass. In terms of the near-horizon value $\phi_h$,
the scalar charge exhibits the expected behaviour: for large values of $\phi_h$, the
effect of the GB term becomes important and $D$ increases; on the other hand, for
vanishing $\phi_h$, i.e. a vanishing coupling function, the scalar charge also vanishes
-- in order to minimise the number of figures, we refrain from showing plots depicting
the anticipated behaviour; in the same spirit, we present no new plots for the quartic
coupling function as it leads to exactly the same qualitative behaviour. 

Turning to the horizon area $A_h$ and entropy $S_h$ of these black-hole solutions, we find
that, in terms of the mass $M$, they both quickly increase, showing a profile similar to
that of Fig. \ref{AS-exp}(a) for the exponential case. The ratio $A_h/A_{Sch}$
remains again below unity over the whole mass regime, and interpolates between a
value corresponding to the lowest allowed value of the mass, according to Eq. (\ref{con-f1}),
and the asymptotic Schwarzschild value at the large-mass limit. The entropy ratio
$S_h/S_{Sch}$, on the other hand, is found to have a different profile by remaining
now always below unity - this feature points perhaps towards a thermodynamic instability
of the `even polynomial' GB black holes compared to the Schwarzschild solution.


\subsection{Odd Polynomial Function}

\begin{figure}[t!] 
\begin{center}
\hspace{0.0cm} \hspace{-0.6cm}
\includegraphics[height=.24\textheight, angle =0]{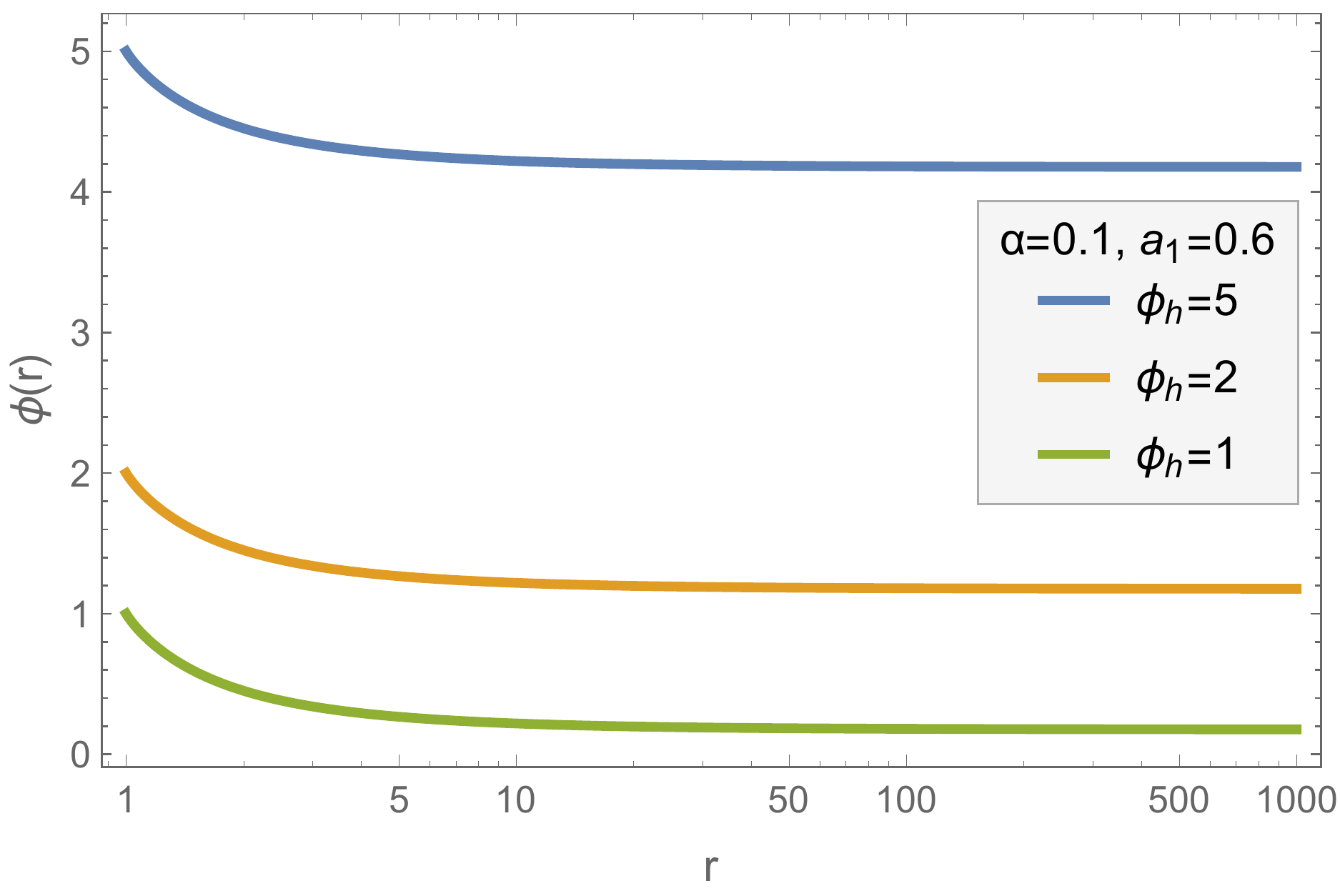}
\hspace{0.52cm} \hspace{-0.6cm}
\includegraphics[height=.24\textheight, angle =0]{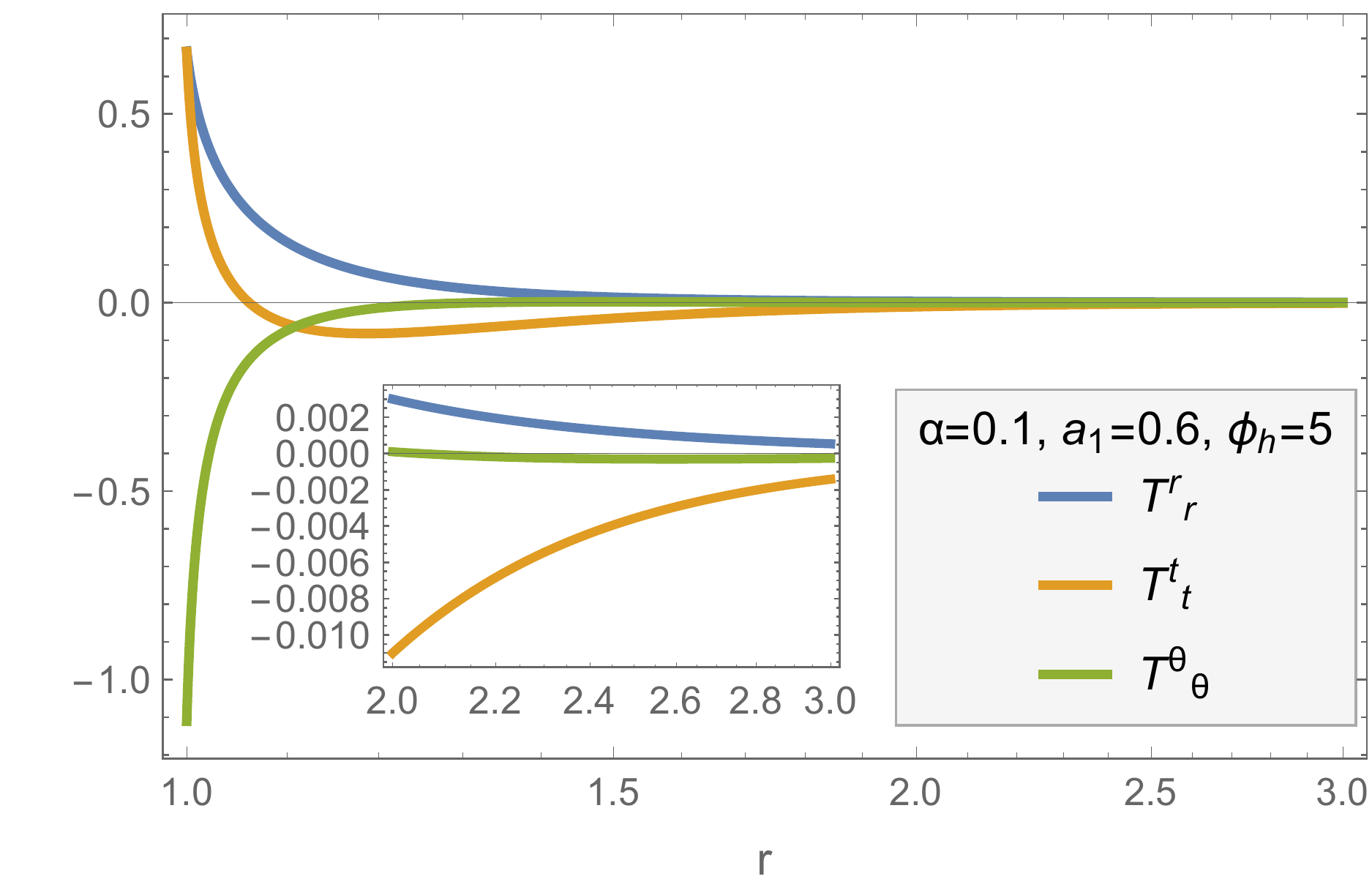}
\\
\hspace*{0.7cm} {(a)} \hspace*{7.5cm} {(b)}  \vspace*{-0.5cm}
\end{center}
\caption{(a) The scalar field $\phi$, and (b) the energy-momentum tensor
$T_{\mu\nu}$ 
 in terms of the radial coordinate $r$, for $f(\phi)=\alpha \phi$.}
\label{Phi_phi}
\end{figure} 

We now consider the case of $f(\phi)=\alpha\,\phi^{2n+1}$, with $n \geq 0$
and $\alpha>0$. Here, the constraint $\dot f \phi'<0$ translates to 
$\phi^{2n} \phi' <0$, or simply to $\phi'_h<0$ for all solutions. 
In Fig. \ref{Phi_phi} (a) and (b), we have chosen the linear case, i.e. $f(\phi)=\alpha \phi$,
and presented an indicative family of solutions for the scalar field (left plot)
and the components of the energy-momentum tensor for one of them (right plot).
The decreasing profile of $\phi $ for all solutions, as we move away from the
black-hole horizon, is evident and in agreement with the above constraint.
The energy-momentum tensor clearly satisfies the analytically predicted behaviour
at the two asymptotic regimes, that once again ensures the evasion of the novel
no-hair theorem.

\begin{figure}[t!] 
\begin{center}
\hspace{0.0cm} \hspace{-0.6cm}
\includegraphics[height=.24\textheight, angle =0]{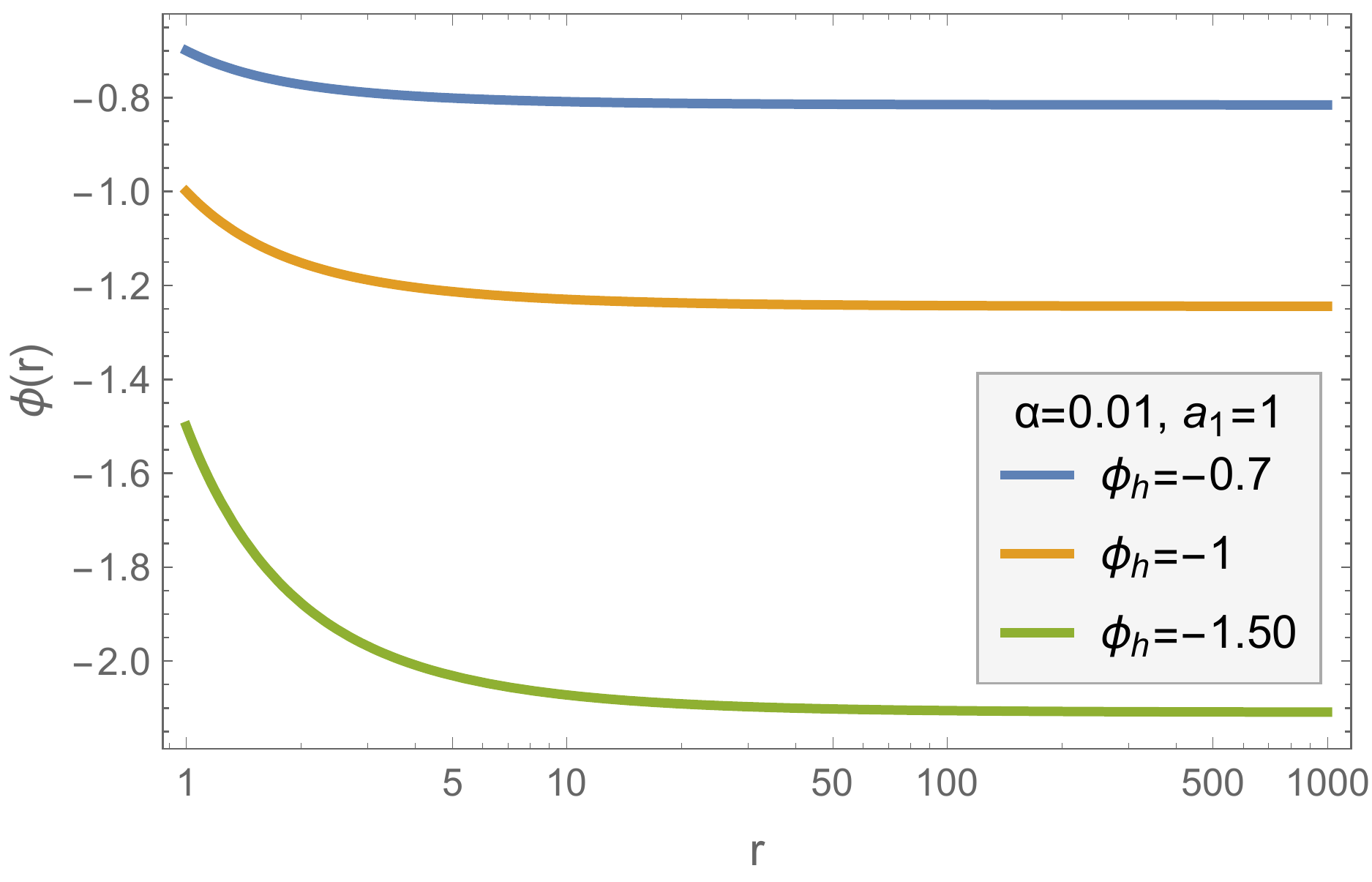}
\hspace{0.52cm} \hspace{-0.6cm}
\includegraphics[height=.24\textheight, angle =0]{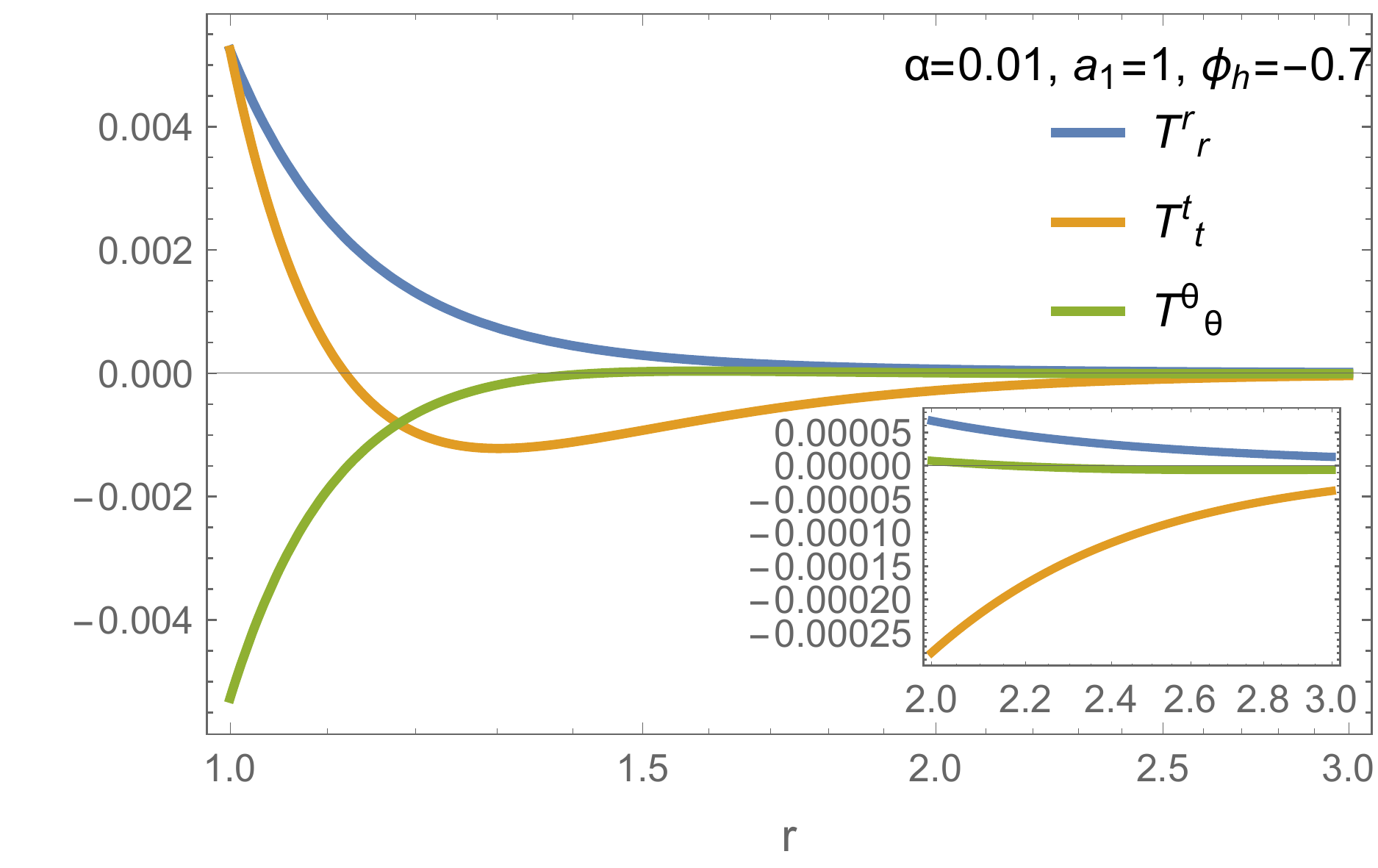}
\\
\hspace*{0.7cm} {(a)} \hspace*{7.5cm} {(b)}  \vspace*{-0.5cm}
\end{center}
\caption{(a) The scalar field $\phi$, and (b) the energy-momentum tensor
$T_{\mu\nu}$ 
 in terms of the radial coordinate $r$, for $f(\phi)=\alpha \phi^3$.}
\label{Phi_phi3}
\end{figure} 

\begin{figure}[b!] 
\begin{center}
\hspace{0.0cm} \hspace{-0.6cm}
\includegraphics[height=.24\textheight, angle =0]{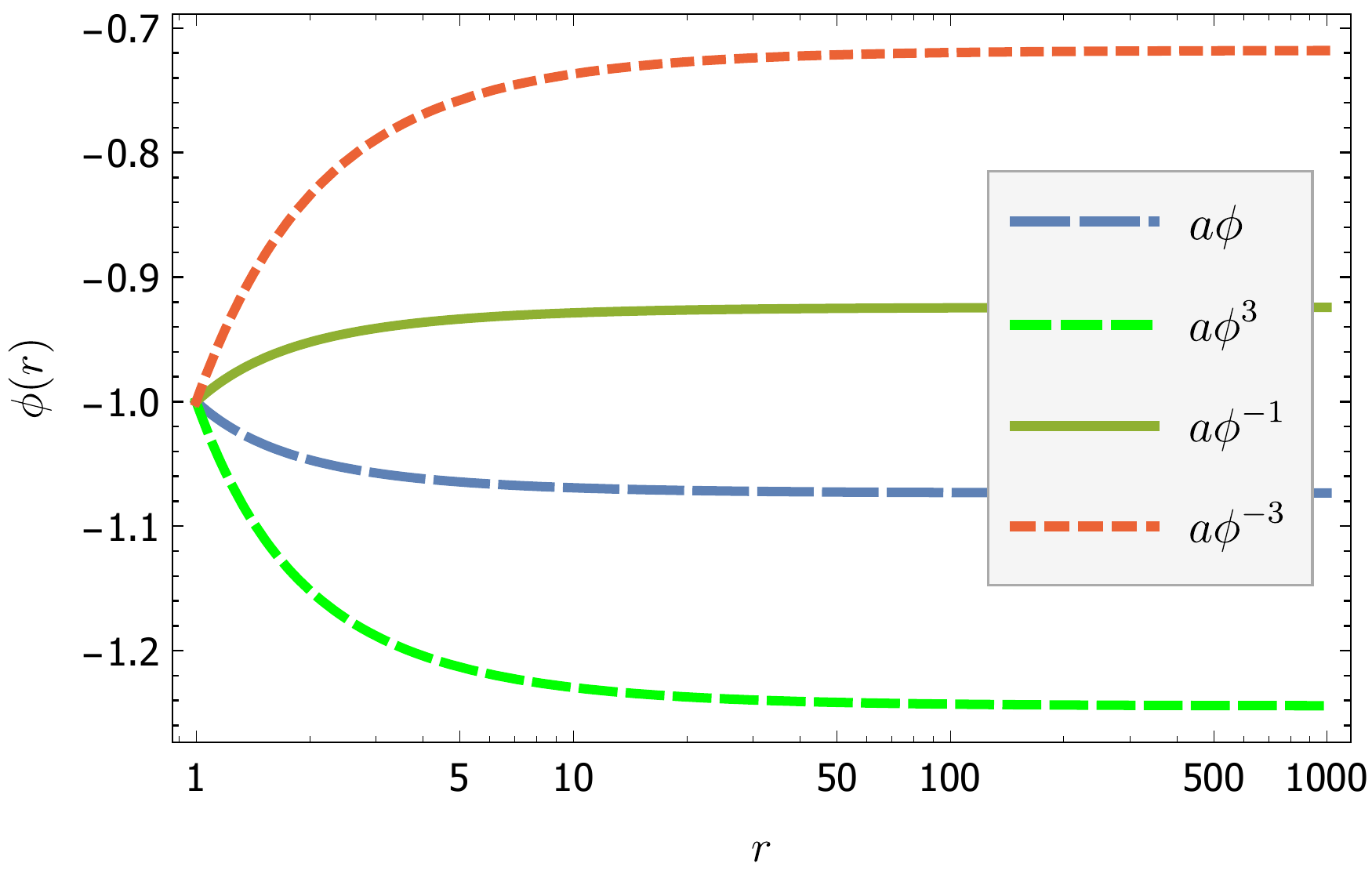}
\hspace{0.52cm} \hspace{-0.6cm}
\includegraphics[height=.24\textheight, angle =0]{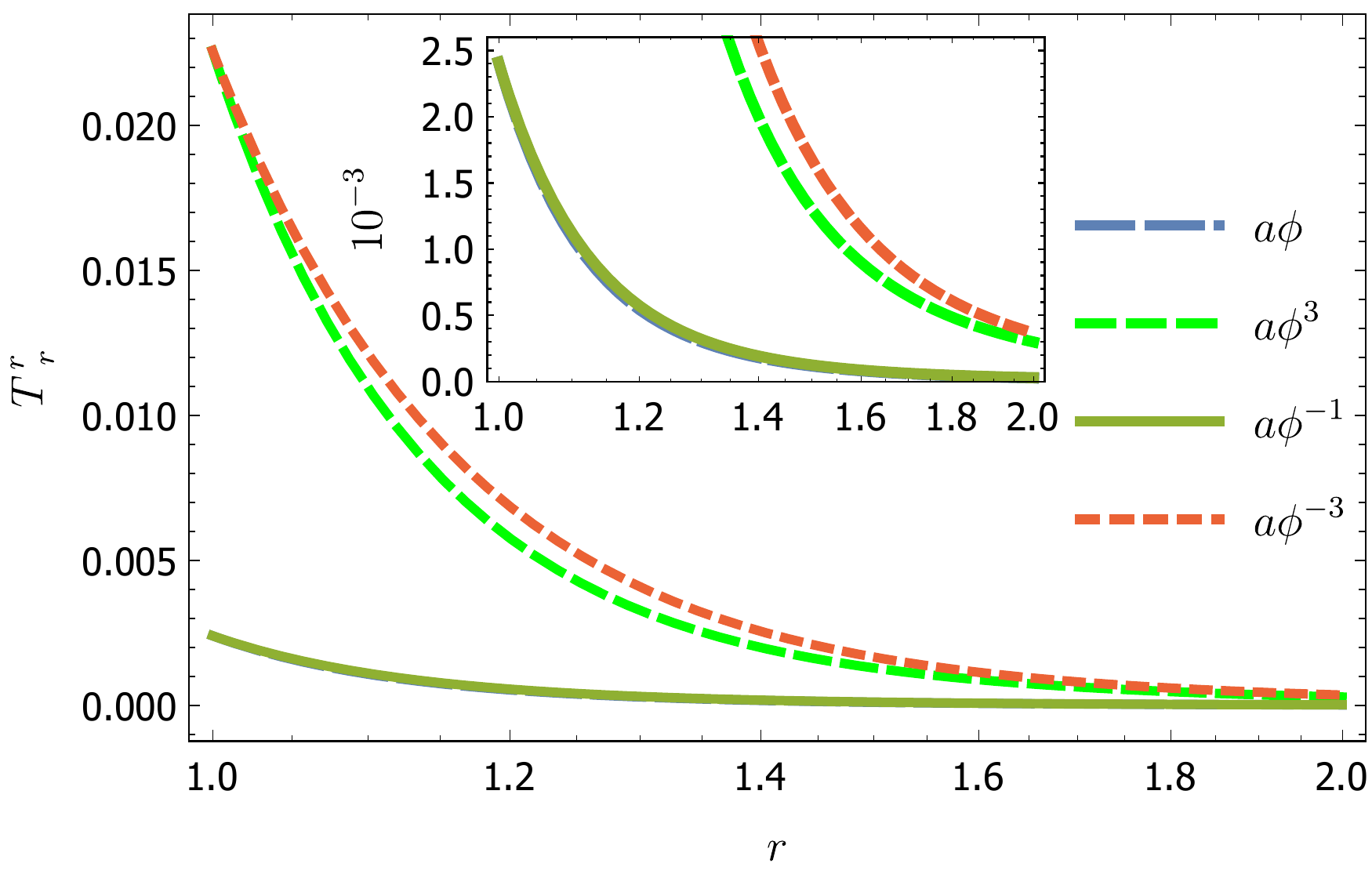}
\\
\hspace*{0.7cm} {(a)} \hspace*{7.5cm} {(b)}  \vspace*{-0.5cm}
\end{center}
\caption{(a) Regular solutions for the scalar field  with $f(\f)<0$ ``forbidden" by  the old No-Hair theorem but not by the novel No-Hair theorem and (b) the corresponding $T_r^r$ component of the energy-momentum tensor.}
\label{nesc}
\end{figure} 

The two plots in Fig. \ref{Phi_phi3}, (a) and (b), depict the
same quantities but for the case $f(\phi)=\alpha \phi^3$. Their profile
agrees with that expected for a regular, black-hole solution with a scalar
hair. The alerted reader may notice that, here, we have chosen to present
solutions with $\phi_h<0$; for an odd polynomial coupling function, these
should have been prohibited under the constraint $f(\phi)>0$, that follows from
the old no-hair theorem \cite{NH-scalar1,NH-scalar2, ABK1}. Nevertheless, regular
black-hole solutions with a
non-trivial scalar field do emerge, that do not seem to satisfy $f(\phi)>0$.
A set of such solutions are shown in   Fig. \ref{Phi_phi3}(a) (we
refrain from showing the set of solutions with $\phi_h>0$ as these have
similar characteristics). As we observe, all of them
obey the $\phi'_h<0$ constraint imposed by the evasion of the novel
no-hair theorem, and lead to the expected behaviour of $T_{\mu\nu}$;
the latter may be clearly seen in  Fig. \ref{Phi_phi3}(b) where
such a ``prohibited'' solution is plotted. The behaviour of the metric
components and the GB term continue to be given by plots similar to
the ones in Fig. \ref{Metric_GB}.

As shown in Fig. \ref{nesc}, the emergence of solutions that violate the constraint $f(\phi)>0$ is in
fact a general feature of our analysis and not an isolated finding in the
case of odd polynomial coupling function. Apparently, the presence of the
coupling of the scalar field to the GB term not only opens the way for 
black-hole solutions to emerge but renders the old no-hair theorem
incapable of dictating when this may happen. Looking more carefully at
the argument on which the old no-hair theorem was based on
\cite{NH-scalar1,NH-scalar2, ABK1}, one readily
realizes that this involves the integral of the scalar equation over the
entire exterior regime, and thus the global solution of the field equations
whose characteristics cannot be predicted beforehand. In contrast, the
evasion of the novel no-hair theorem is based on local quantities, such
as the energy-momentum components and their derivatives at particular
radial regimes, which may be easily computed. In addition, it is indifferent
to the behaviour of the solution in the intermediate regime, which may
indeed exhibit an arbitrary profile as the one presented in the plots of
Fig. \ref{Phi_alpha_phi2}. In fact, all solutions found
in the context of our analysis, with no exception, satisfy the constraints 
that ensure the evasion of the novel no-hair theorem.

\begin{figure}[t!] 
\begin{center}
\hspace{0.0cm} \hspace{-0.6cm}
\includegraphics[height=.24\textheight, angle =0]{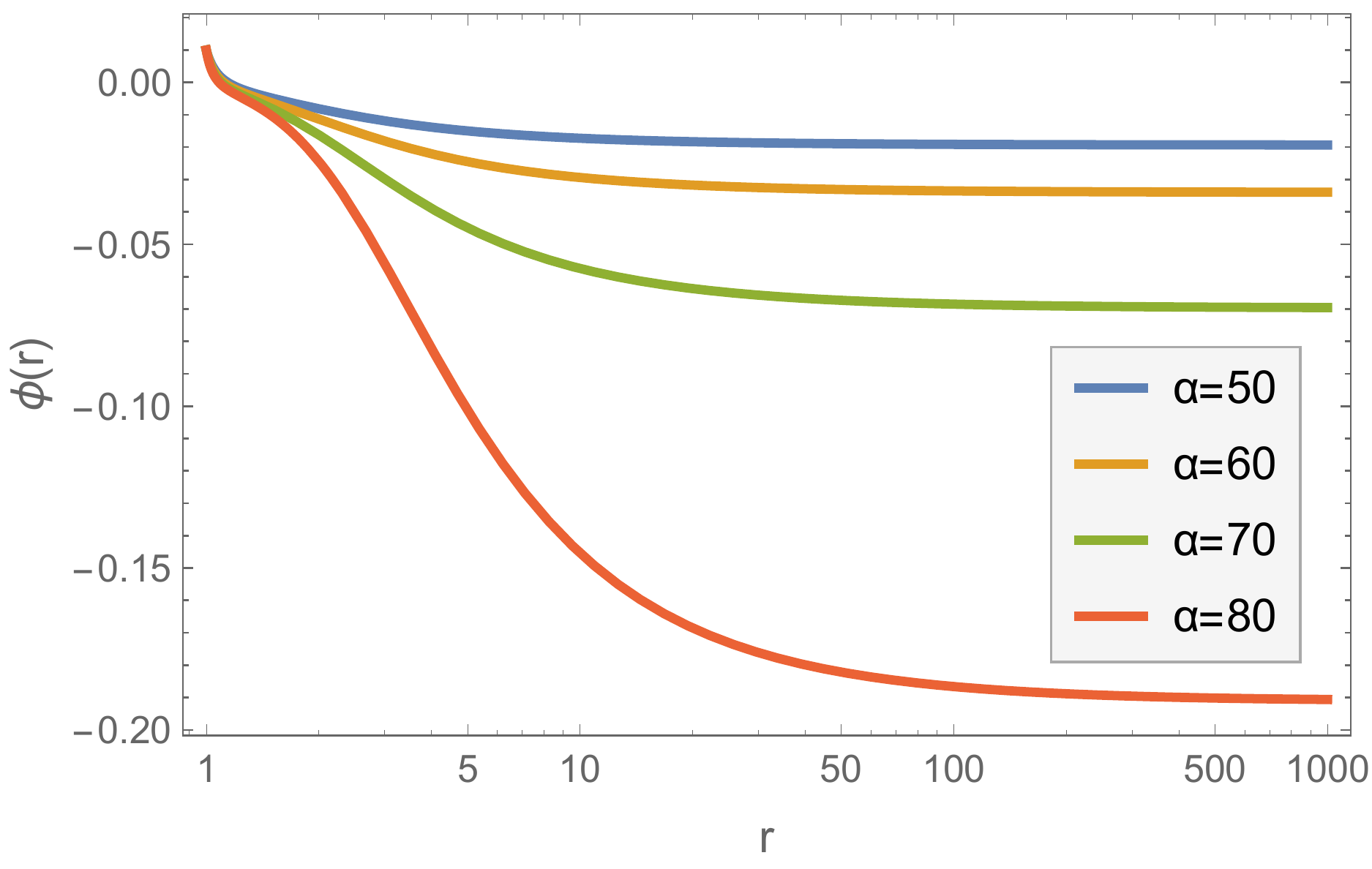}
\hspace{0.52cm} \hspace{-0.6cm}
\includegraphics[height=.24\textheight, angle =0]{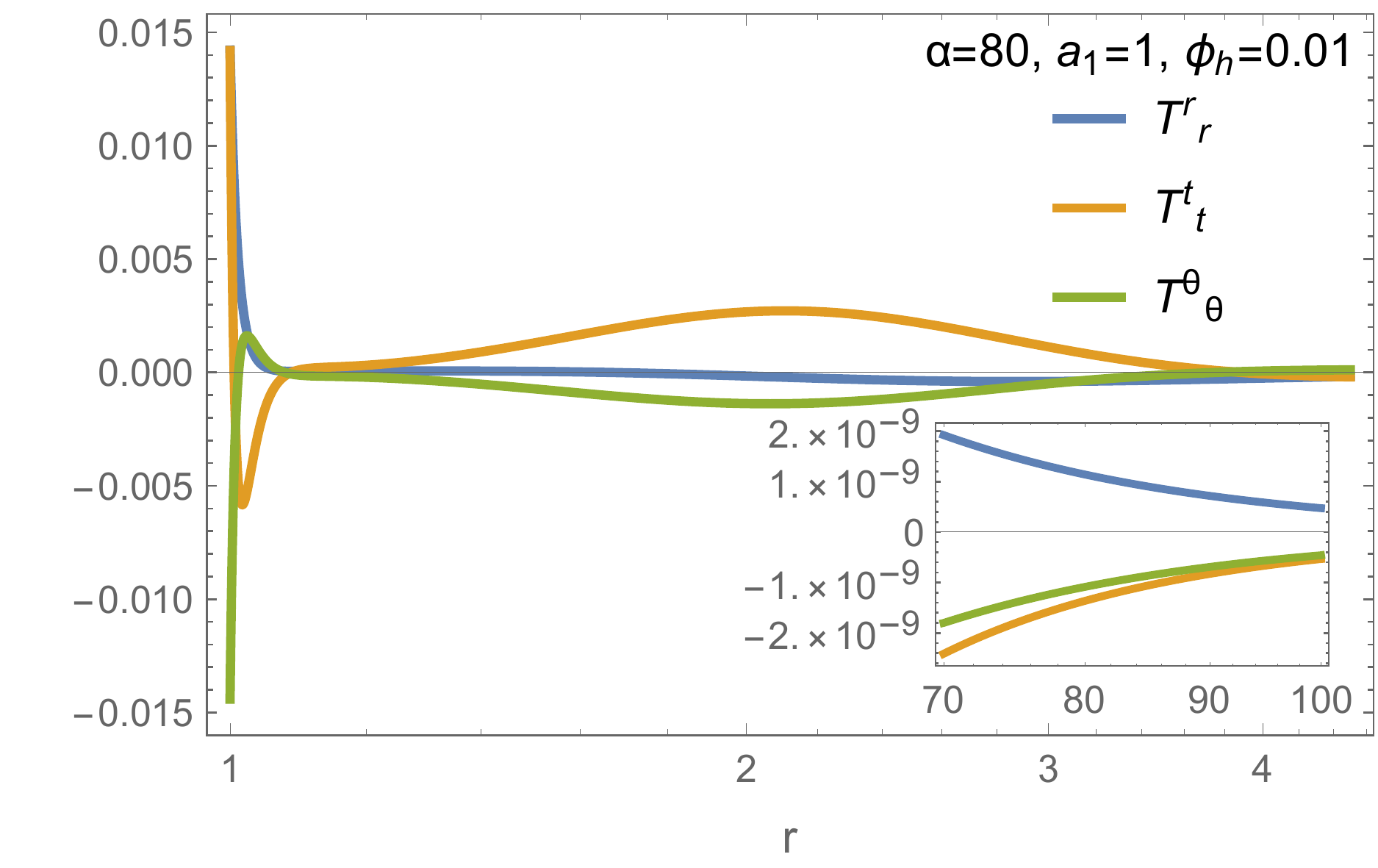}
\\
\hspace*{0.7cm} {(a)} \hspace*{7.5cm} {(b)}  \vspace*{-0.5cm}
\end{center}
\caption{(a) The scalar field $\phi$, and (b) the energy-momentum tensor
$T_{\mu\nu}$ 
 in terms of the radial coordinate $r$, for $f(\phi)=\alpha \phi^3$ and a variety of values of the coupling constant $\alpha$.}
\label{Phi_alpha_phi3}
\end{figure}

In this case, too, one may derive solutions for a variety of values of the
coupling constant $\alpha$, as long as these obey the constraint (\ref{con-f1}).
In Fig. \ref{Phi_alpha_phi3}(a), we depict a family of solutions with $\phi_h=0.01$
and a variety of values of $\alpha$. This family of solutions present a less
monotonic profile compared to the one exhibited by the solutions in Fig. \ref{Phi_phi3}.
The components of the energy-momentum tensor for one of these solutions is
depicted in  Fig. \ref{Phi_alpha_phi3}(b), and they present a more evolved
profile with the emergence of minima and maxima between the black-hole horizon and
radial infinity. We also notice that the solutions for the scalar field, although they start
from the positive-value regime ($\phi_h=0.01$), they cross to negative values for fairly
small values of the radial coordinate. This behavior causes the odd coupling function
to change sign along the radial regime, a feature that makes any application of the
old no-hair theorem even more challenging.

\begin{figure}[t!] 
\begin{center}
\hspace{0.0cm} \hspace{-0.6cm}
\includegraphics[height=.24\textheight, angle =0]{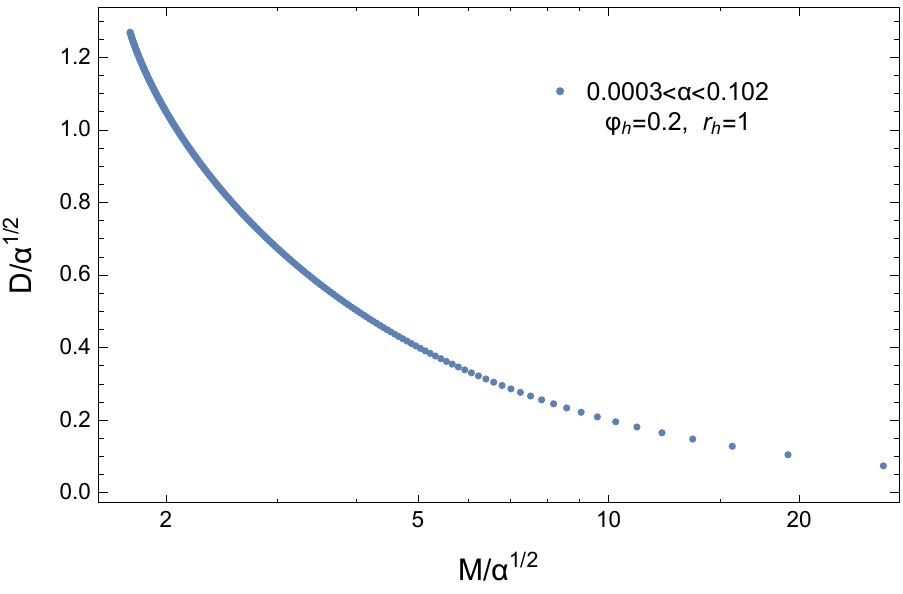}
\hspace{0.52cm} \hspace{-0.6cm}
\includegraphics[height=.24\textheight, angle =0]{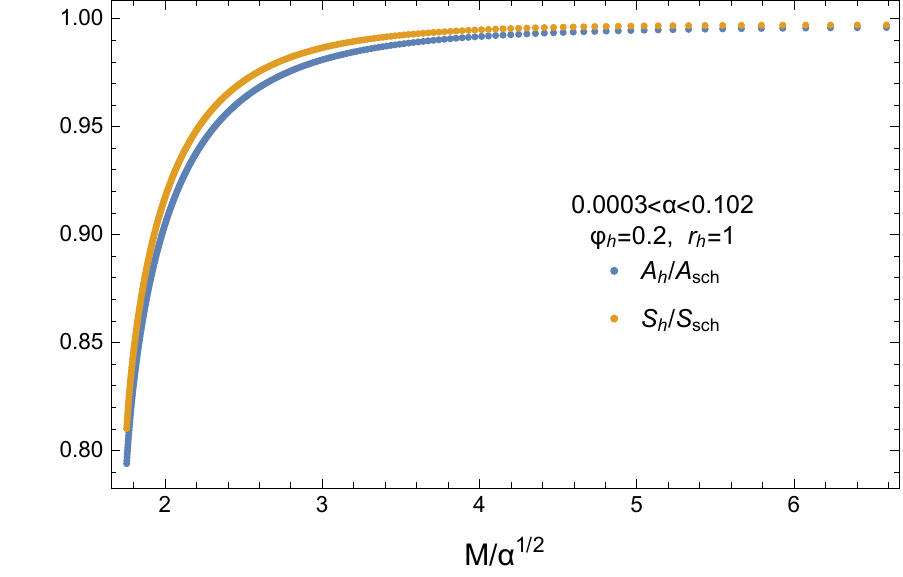}
\\
\hspace*{0.7cm} {(a)} \hspace*{7.5cm} {(b)}  \vspace*{-0.5cm}
\end{center}
\caption{(a) The scalar charge $D$, and (b) the ratios $A_h/A_{Sch}$ and $S_h/S_{Sch}$ 
 in terms of the mass  $M$, for $f(\phi)=\alpha \phi$.}
\label{D-AS-linear}
\end{figure} 
\begin{figure}[b!] 
\begin{center}
\hspace{0.0cm} \hspace{-0.6cm}
\includegraphics[height=.24\textheight, angle =0]{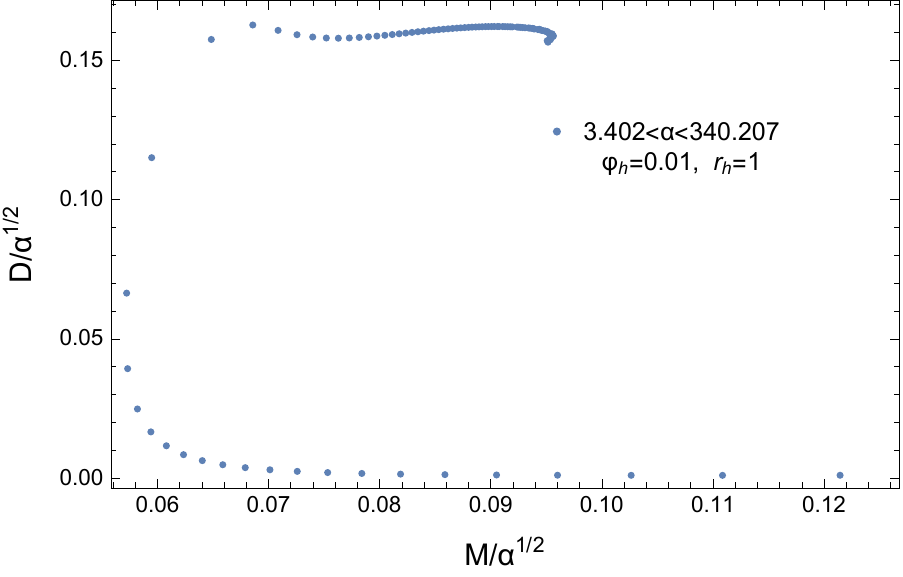}
\hspace{0.52cm} \hspace{-0.6cm}
\includegraphics[height=.24\textheight, angle =0]{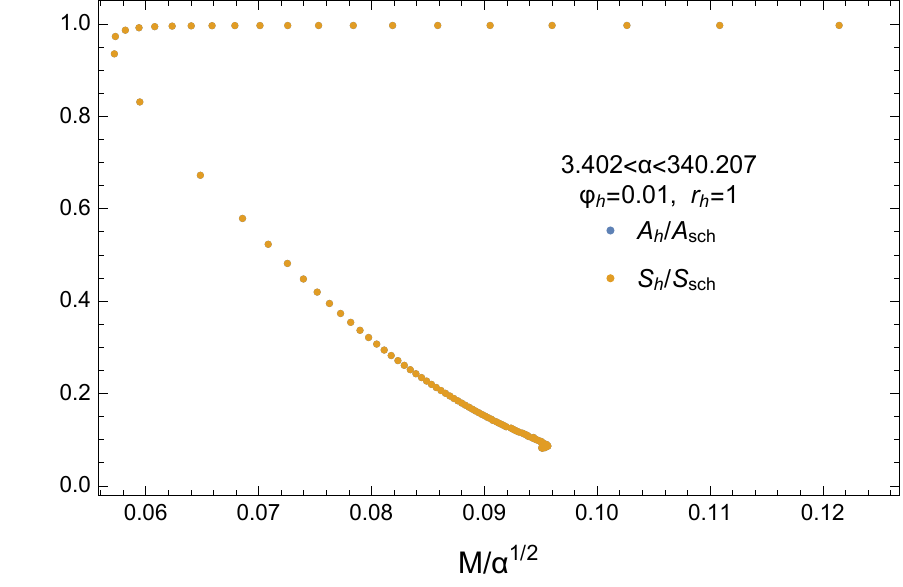}
\\
\hspace*{0.7cm} {(a)} \hspace*{7.5cm} {(b)}  \vspace*{-0.5cm}
\end{center}
\caption{(a) The scalar charge $D$, and (b) the ratios $A_h/A_{Sch}$ and $S_h/S_{Sch}$ 
 in terms of the mass  $M$, for $f(\phi)=\alpha \phi^3$.}
\label{D-AS-qubic}
\end{figure} 

Turning again to the characteristics of the `odd polynomial' black-hole solutions, in
Fig. \ref{D-AS-linear}(a), we depict the scalar charge $D$ in terms of the
mass $M$ of the black hole, for the linear coupling function $f(\phi)=\alpha \phi$:
here, the function $D(M)$ is monotonic and approaches a vanishing asymptotic
value as $M$ increases. The scalar charge $D$ has no dependence on the initial
scalar-field value $\phi_h$ since it is the first derivative $f'(\phi)$ that appears
in the scalar equation (\ref{phi-eq22}) and, for a linear function, this is merely a
constant. The horizon area $A_h$ and entropy $S_h$ exhibit again an increasing profile
in terms of $M$ similar to
that of Fig. \ref{AS-exp}(a) for the exponential case. The more informative
ratios $A_h/A_{Sch}$ and $S_h/S_{Sch}$ are given in   Fig. \ref{D-AS-linear}(b):
as in the case of quadratic and quartic coupling functions, both quantities remain
smaller than unity and interpolate between a lowest value corresponding to the
black-hole solution with the lowest mass and the Schwarzschild limit acquired at
the large mass limit.


We now address separately the characteristics of the black-hole solutions arising
in the case of the cubic coupling function $f(\phi)=\alpha \phi^3$ since here we
find a distinctly different behaviour. As mentioned above, also in this case, as
the coupling constant $\alpha$ increases, from zero to its maximum value (for
$r_h$ and $\phi_h$ fixed), solutions with no monotonic profile in terms of the
radial coordinate arise (Fig. \ref{Phi_alpha_phi3}(a)). We depict
the scalar charge $D$ in terms of the mass $M$ of the black hole, for the whole
$\alpha$-regime, in    Fig. \ref{D-AS-qubic}(a): we may easily observe
the emergence of two different branches of solutions (with a third, short one 
appearing at the end of the upper branch) corresponding to the same mass $M$. 
These branches appear at the small-mass limit of the solutions whereas for
large masses only one branch survives with a very small scalar charge. In
   Fig. \ref{D-AS-qubic}(b), we show the ratio $S_h/S_{Sch}$ in terms
of the mass $M$: this quantity also displays the existence of three branches 
with the one that is smoothly connected to the Schwarzschild limit having the
higher entropy. The two additional branches with the larger values of scalar
charge, compared to the one of the `Schwarzschild' branch, have a lower entropy
and they are probably less thermodynamically stable. This behavior was not
observed in the case of the quadratic coupling function where more evolved
solutions for the scalar field also appeared (see Fig. \ref{Phi_phi2}(a)):
there, the function $D(M)$ was not monotonic but was always single-valued; that
created short, disconnected  `branches' of solutions with slightly different
values of entropy ratio $S_h/S_{Sch}$ but all lying below unity. Let us finally
note that the horizon area ratio $A_h/A_{Sch}$, not shown here for brevity,
has the same profile as the one displayed in Fig. \ref{D-AS-linear} for the
linear function while the $D(\phi_h)$ function shows the anticipated increasing
profile as $\phi_h$, and thus the GB coupling, increases.


\subsection{Inverse Polynomial Function}

The next case to consider is the one where $f(\phi)=\alpha \phi^{-k}$, where
$k>0$, and $\alpha$ is also assumed to be positive, for simplicity. Let us 
consider directly some indicative cases:

\begin{figure}[t!] 
\begin{center}
\hspace{0.0cm} \hspace{-0.6cm}
\includegraphics[height=.26\textheight, angle =0]{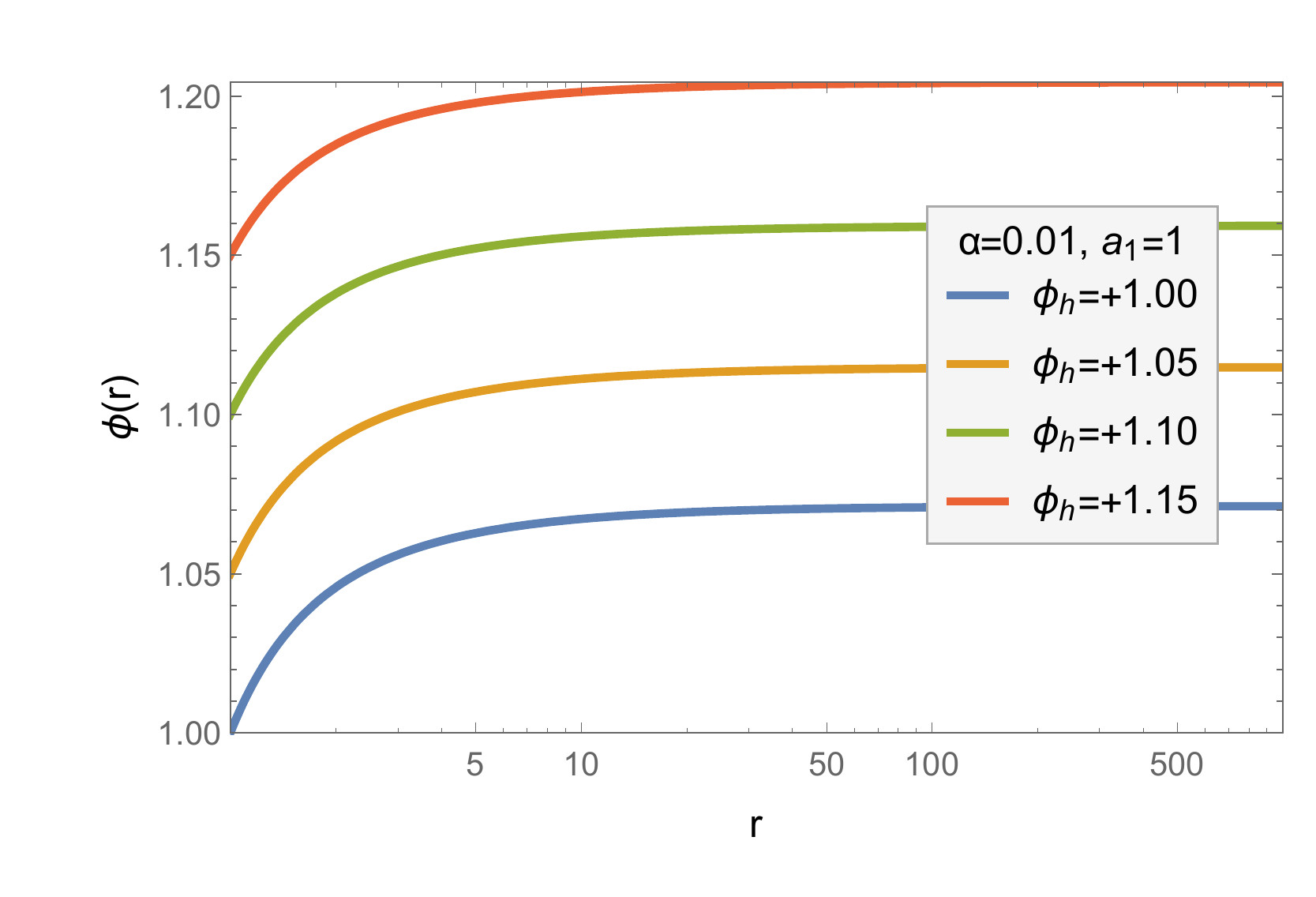}
\hspace{0.52cm} \hspace{-0.6cm}
\includegraphics[height=.26\textheight, angle =0]{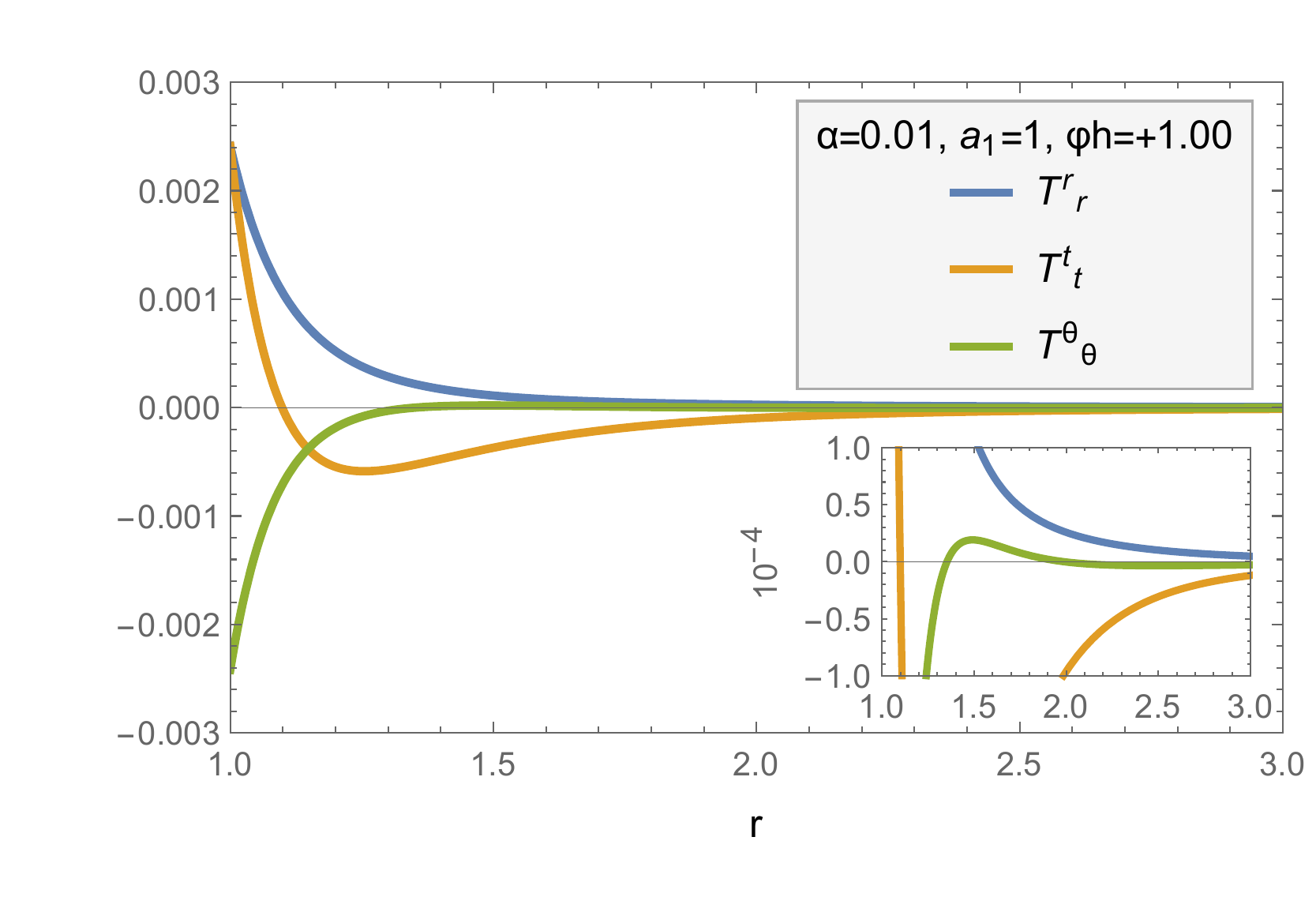}
\\
\hspace*{0.7cm} {(a)} \hspace*{7.5cm} {(b)}  \vspace*{-0.5cm}
\end{center}
\caption{(a) The scalar field $\phi$, and (b) the energy-momentum tensor
$T_{\mu\nu}$ 
 in terms of the radial coordinate $r$, for $f(\phi)=\alpha/\phi$.}
\label{Phi_neg1}
\end{figure} 
\begin{figure}[t!] 
\begin{center}
\hspace{0.0cm} \hspace{-0.6cm}
\includegraphics[height=.26\textheight, angle =0]{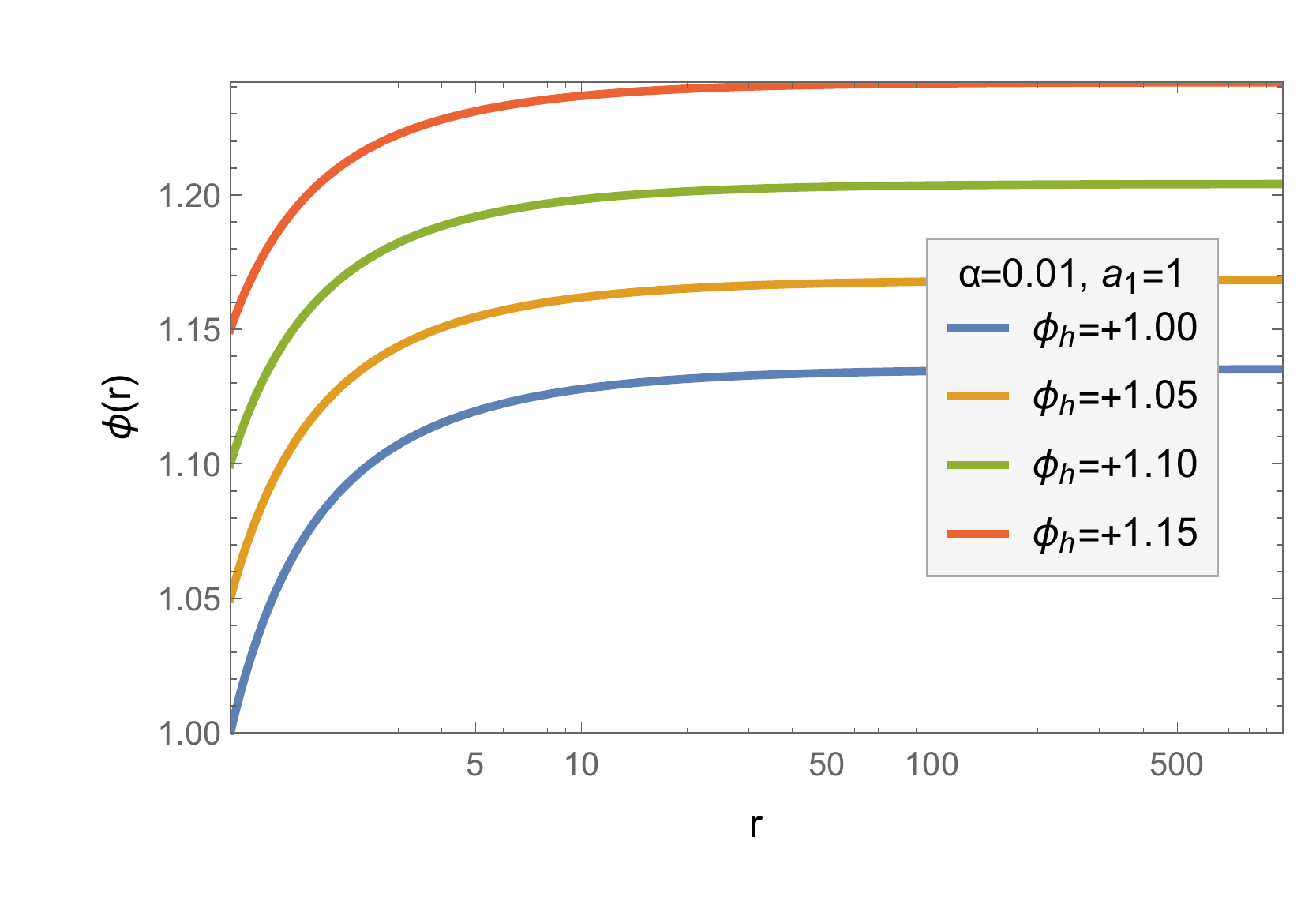}
\hspace{0.52cm} \hspace{-0.6cm}
\includegraphics[height=.26\textheight, angle =0]{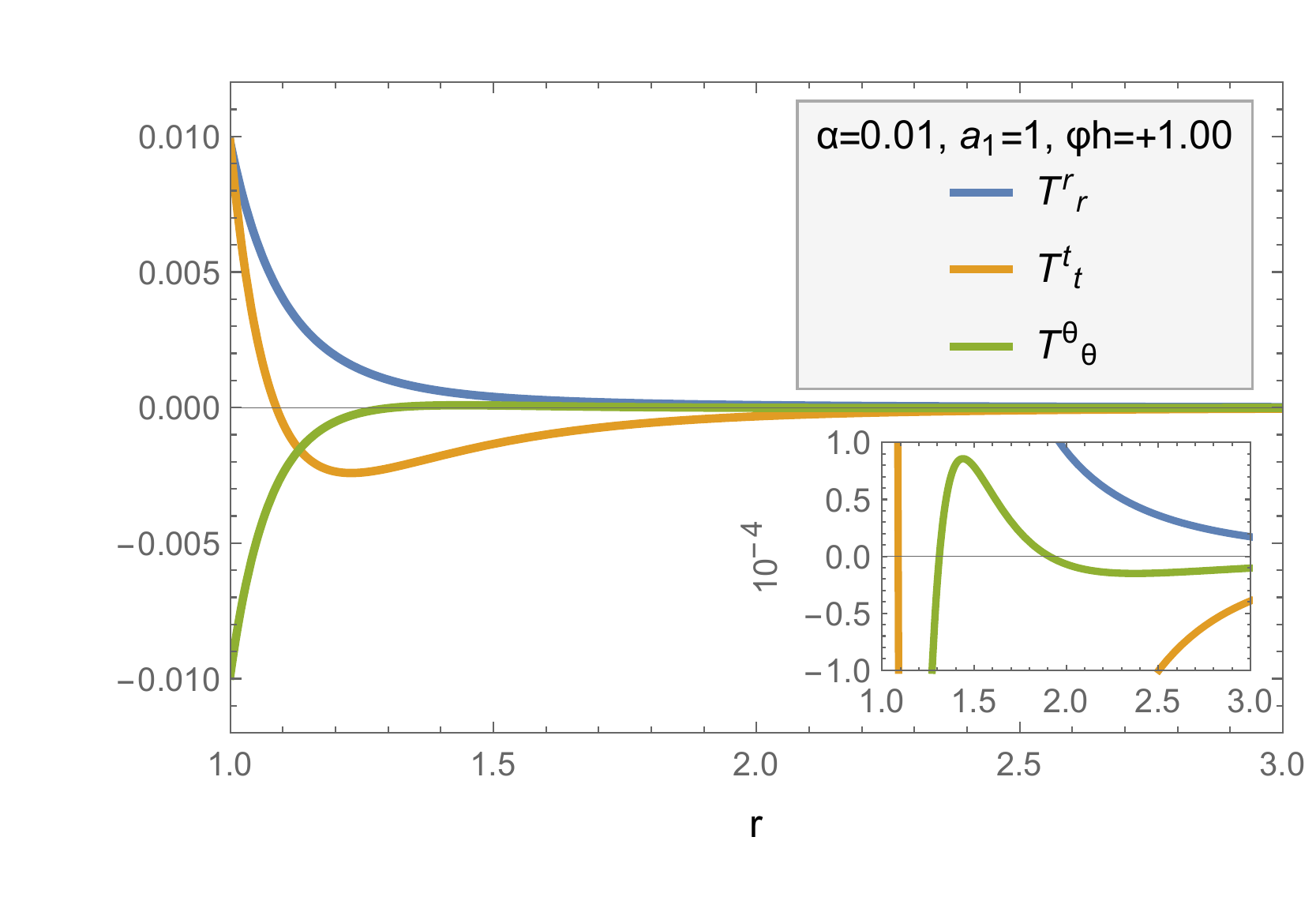}
\\
\hspace*{0.7cm} {(a)} \hspace*{7.5cm} {(b)}  \vspace*{-0.5cm}
\end{center}
\caption{(a) The scalar field $\phi$, and (b) the energy-momentum tensor
$T_{\mu\nu}$ 
 in terms of the radial coordinate $r$, for $f(\phi)=\alpha/\phi^2$.}
\label{Phi_neg2}
\end{figure} 

\begin{itemize}
\item $k=1$: In this case, the constraint for the evasion of the novel no-hair
theorem becomes: $\dot f \phi'=-2\alpha \phi'/\phi^2<0$ which demands $\phi'_h>0$
for all solutions. In 
Fig. \ref{Phi_neg1}(a), we present a family of solutions for the scalar field emerging
for this coupling function. All solutions are increasing away from the black-hole
horizon in accordance to the above comment. The components of the energy-momentum
tensor are also well behaved, as may be seen from   Fig. \ref{Phi_neg1}(b).
As in the case of the odd polynomial function, an additional set of solutions arises
with $\phi_h<0$ with similar characteristics. 

\item $k=2$: In this case, the constraint becomes: $\dot f \phi'=-\alpha \phi'/\phi^3<0$
which demands $\phi_h \phi'_h>0$. A family of solutions for the scalar field emerging
for this coupling function, with $\phi_h>0$ and increasing with $r$, are presented in Fig. \ref{Phi_neg2}(a) -- a complementary family of solutions with $\phi_h<0$
and decreasing away from the black-hole horizon were also found. The components of the
energy-momentum tensor for an indicative solution are depicted in  
Fig. \ref{Phi_neg2}(b), and clearly remain finite over the whole exterior regime.

\end{itemize}

\begin{figure}[t!] 
\begin{center}
\hspace{0.0cm} \hspace{-0.6cm}
\includegraphics[height=.24\textheight, angle =0]{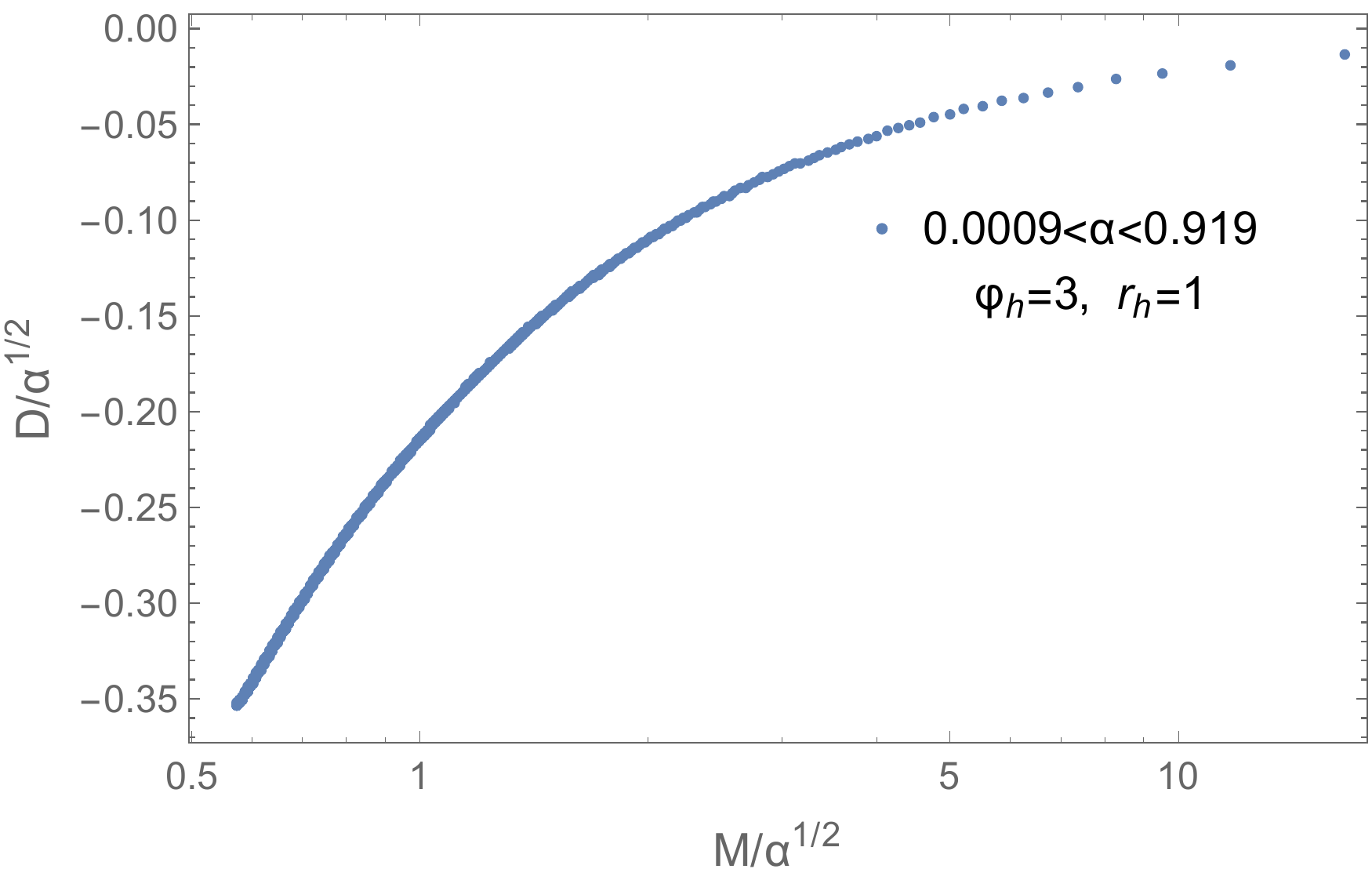}
\hspace{0.52cm} \hspace{-0.6cm}
\includegraphics[height=.24\textheight, angle =0]{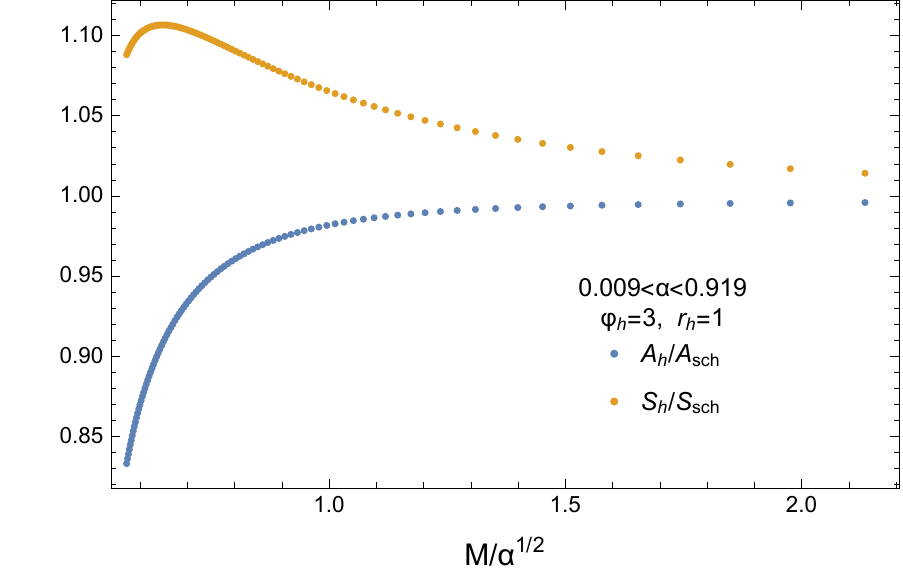}
\\
\hspace*{0.7cm} {(a)} \hspace*{7.5cm} {(b)}  \vspace*{-0.5cm}
\end{center}
\caption{(a) The scalar charge $D$, and (b) the ratios $A_h/A_{Sch}$ and $S_h/S_{Sch}$ 
(right plot) in terms of the mass  $M$, for $f(\phi)=\alpha/\phi$.}
\label{D-AS-inv_lin}
\end{figure} 

Concerning the characteristics of the `inverse polynomial' GB black holes, we find
again an interesting behaviour -- in Fig. \ref{D-AS-inv_lin}, we depict the inverse
linear case, as an indicative one. The scalar charge $D$ exhibits a monotonic
decreasing behaviour in terms of the mass $M$, as one may see in the left plot
of the figure, approaching zero at the large-mass limit.
In terms of the input parameter $\phi_h$, the scalar charge presents the
anticipated behaviour (and thus is not shown here): for an inverse coupling function,
$D$ increases as the value of $\phi_h$, and thus of the GB coupling, decreases. 
The quantities $A_h$ and $S_h$ increase once again quickly with the mass $M$,
as in Fig. \ref{AS-exp}. The ratio $A_h/A_{Sch}$ is shown in  
Fig. \ref{D-AS-inv_lin}(b), and reveals again the constantly smaller size of the GB
black holes compared to the asymptotic Schwarzschild solution, as well as the
existence of a lowest-mass solution. The ratio $S_h/S_{Sch}$, depicted also
in Fig. \ref{D-AS-inv_lin}(b), reveals that the entire class of these black-hole
solutions  -- independently of their mass -- have a higher entropy compared to
the asymptotic Schwarzschild solution. This feature is in fact unique for the
inverse linear function; a similar analysis for the inverse quadratic coupling
function has produced similar results for the quantities $D(M)$, $D(\phi_h)$,
$A_h/A_{Sch}$ and $S_h/S_{Sch}$ with the only difference being that the very-low
mass regime of the `inverse quadratic' GB black holes have a lower entropy than
the Schwarzschild solution, i.e. the situation resembles more the one depicted in
Fig. \ref{AS-exp}.


\subsection{Logarithmic Function}

\begin{figure}[b!] 
\begin{center}
\hspace{0.0cm} \hspace{-0.6cm}
\includegraphics[height=.26\textheight, angle =0]{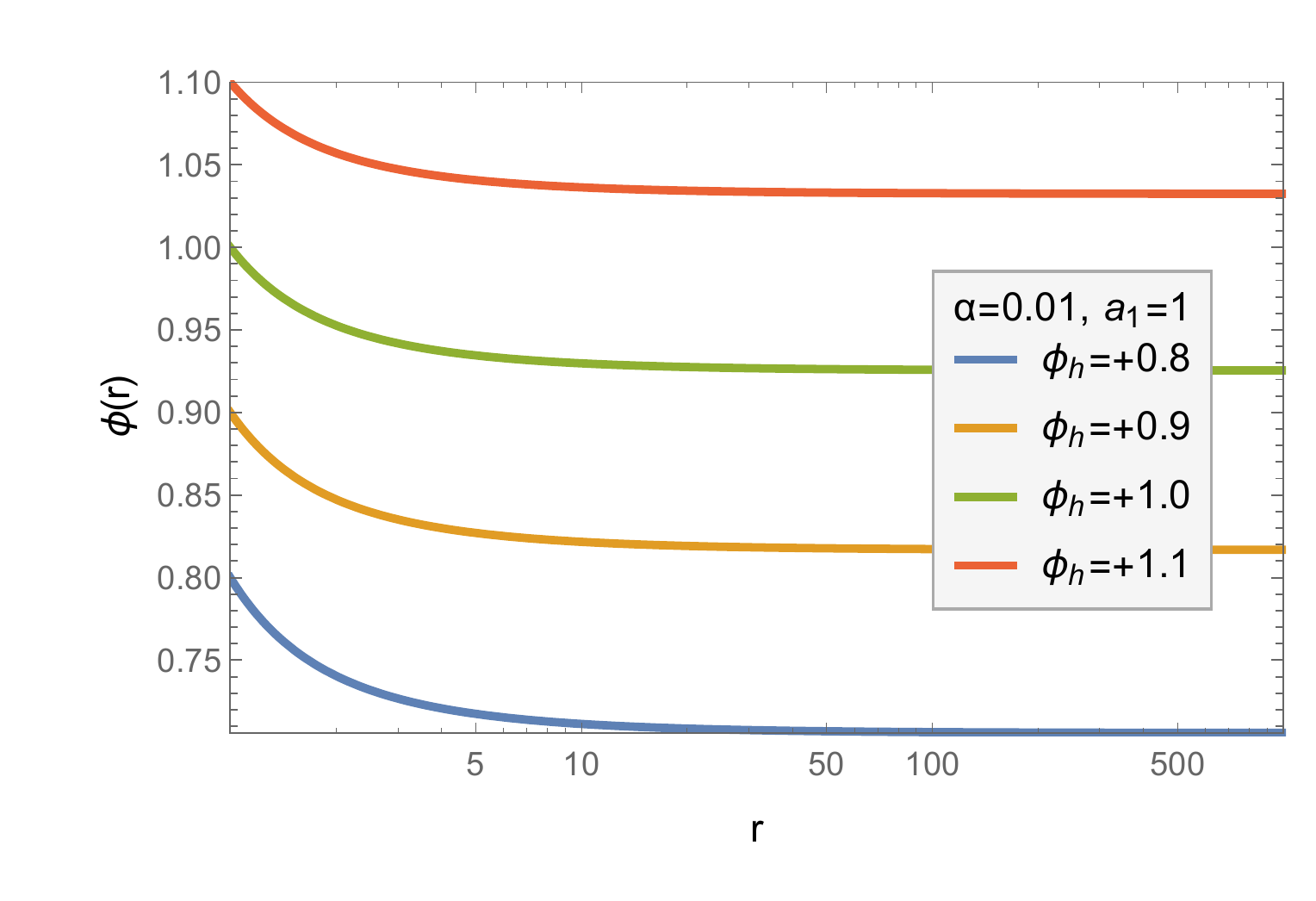}
\hspace{0.52cm} \hspace{-0.6cm}
\includegraphics[height=.26\textheight, angle =0]{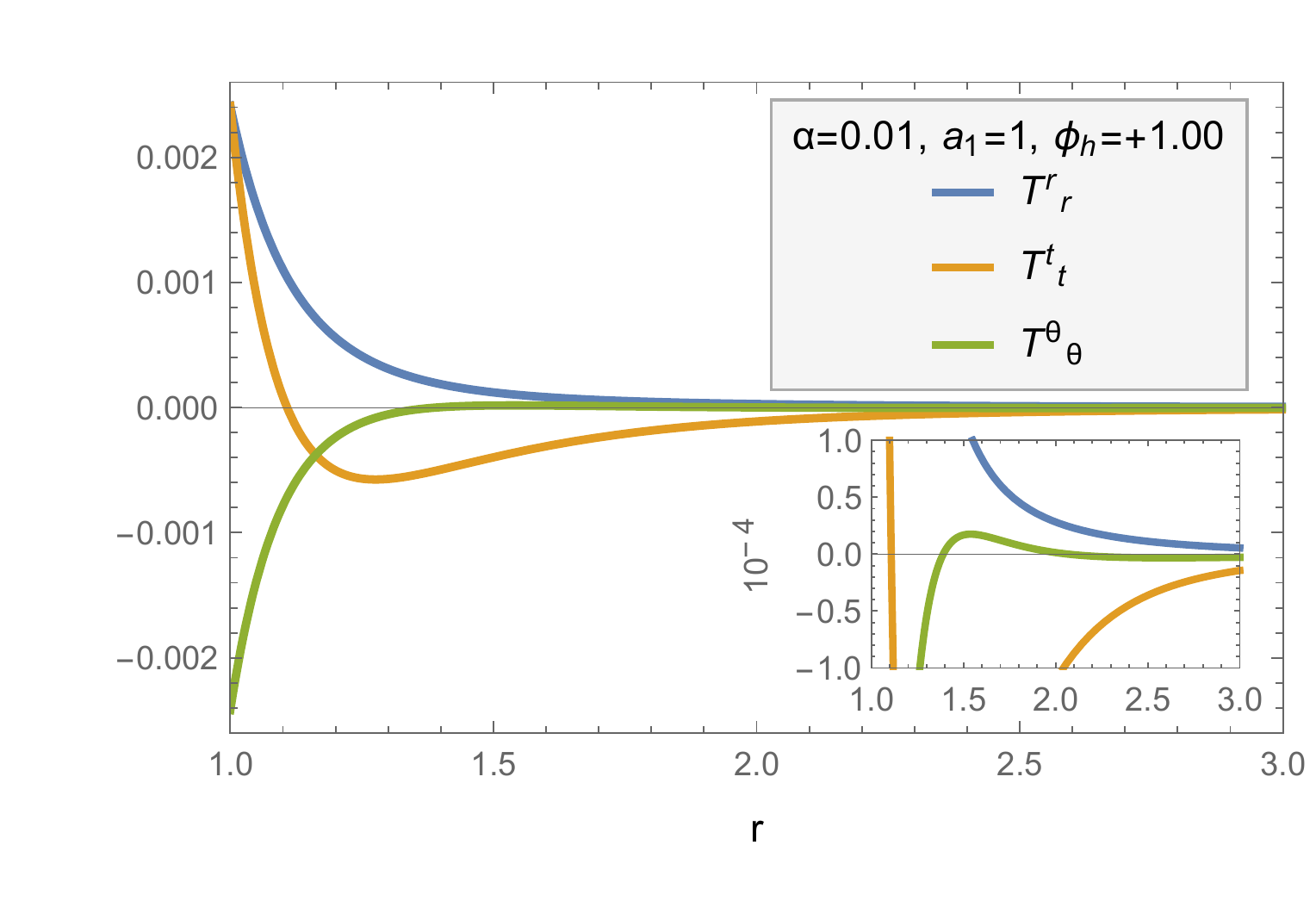}
\\
\hspace*{0.7cm} {(a)} \hspace*{7.5cm} {(b)}  \vspace*{-0.5cm}
\end{center}
\caption{(a) The scalar field $\phi$, and (b) the energy-momentum tensor
$T_{\mu\nu}$ 
  in terms of the radial coordinate $r$, for $f(\phi)=\alpha\,\ln\phi$.}
\label{Phi_log}
\end{figure} 

We finally address the case where $f(\phi)=\alpha \ln(\phi)$. Here, black-hole solutions
emerge for $\dot f \phi' =\alpha \phi'/\phi<0$, near the black-hole horizon; for $\alpha>0$,
this translates to $\phi'_h<0$ (since the argument of the logarithm must be a positive number,
i.e. $\phi>0$). As a result, the solutions for the scalar field are restricted to have a decreasing
behaviour as we move away from the black-hole horizon - this is indeed the behaviour
observed in the class of solutions depicted in Fig. \ref{Phi_log}(a). One may
also observe that the plot includes solutions with either $\phi_h>1$ or $\phi_h<1$, 
or equivalently with $f>0$ or $f<0$. Once again, the old no-hair theorem is proven to
be inadequate to exclude the presence of regular black holes with scalar hair even
in subclasses of the theory (\ref{act2}). In contrast, the derived solutions continue to
satisfy the constraints for the evasion of the novel no-hair theorem.  

The components of the energy-momentum tensor are presented in  Fig. \ref{Phi_log}(b): they exhibit the same characteristics as in the cases
presented in the previous subsections with the most important being the monotonic,
decreasing profile of the $T^r_{\,\,r}$ component. The coupling constant $\alpha$
can also take a variety of values as long as it satisfies Eq. (\ref{con-f1}); in
this case, the monotonic behaviour of $\phi$ over the whole exterior space of
the black hole is preserved independently of the value of $\alpha$. The
energy-momentum tensor also assumes the same form as in Fig. \ref{Phi_log},
and thus we refrain from presenting any new plots.

 \begin{figure}[t!] 
\begin{center}
\hspace{0.0cm} \hspace{-0.6cm}
\includegraphics[height=.24\textheight, angle =0]{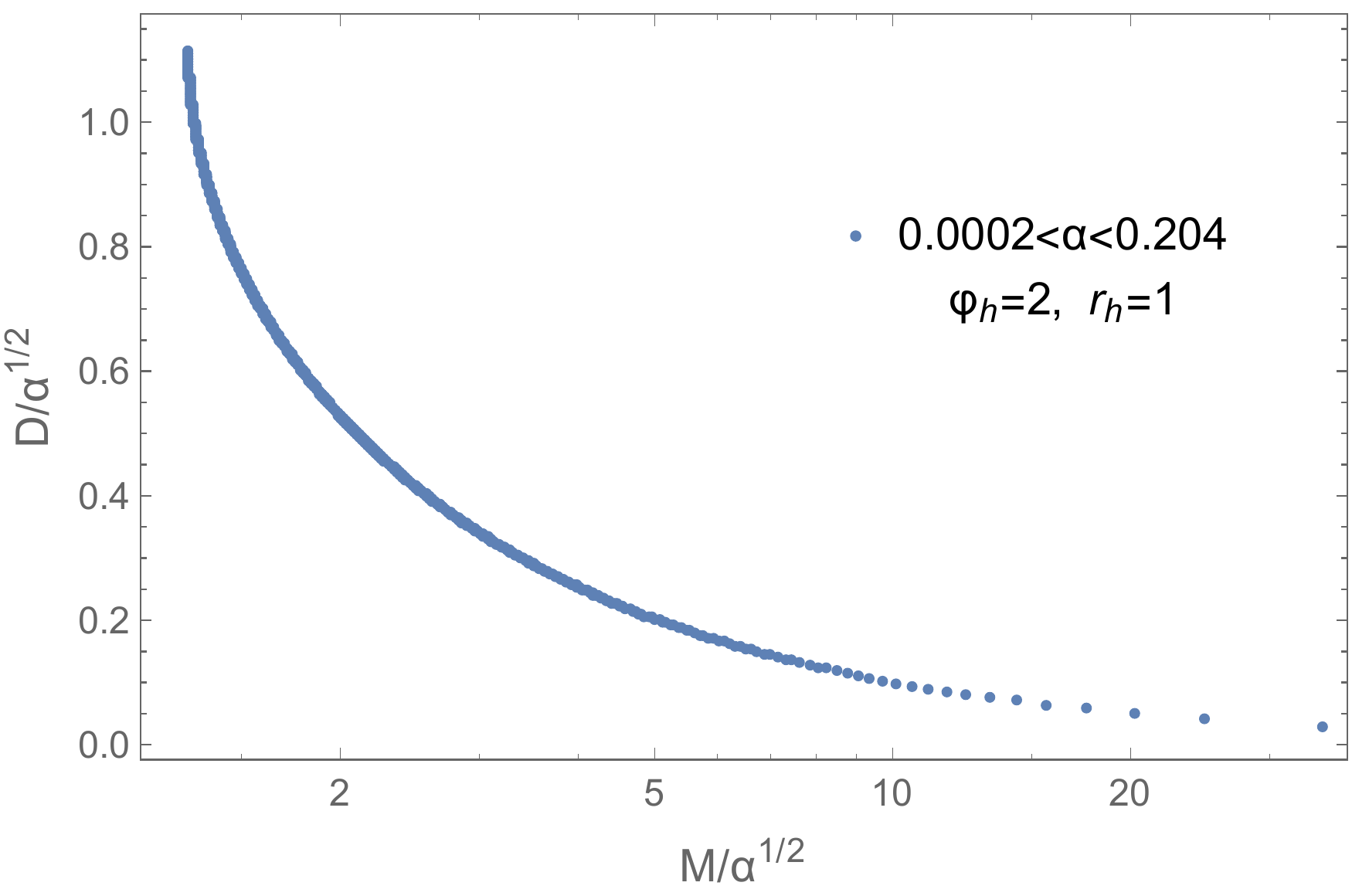}
\hspace{0.52cm} \hspace{-0.6cm}
\includegraphics[height=.24\textheight, angle =0]{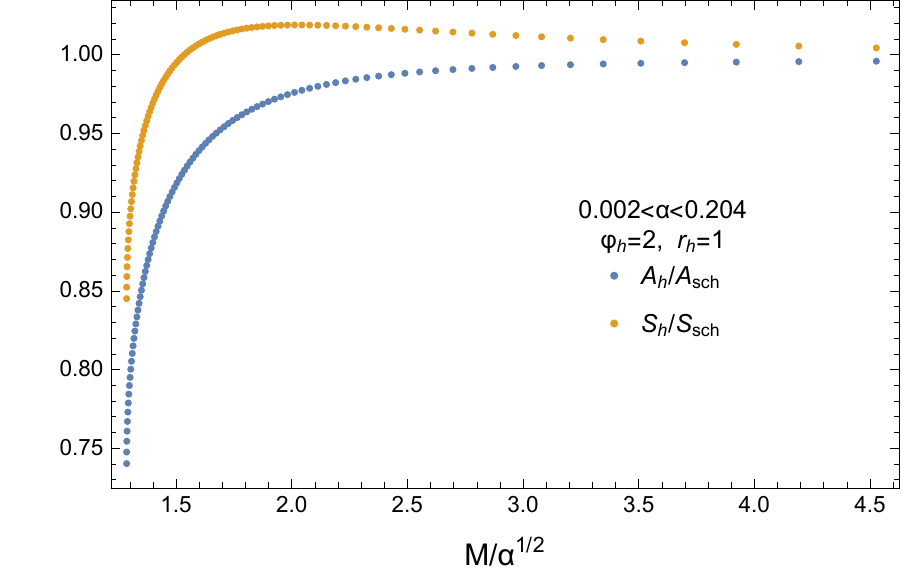}
\\
\hspace*{0.7cm} {(a)} \hspace*{7.5cm} {(b)}  \vspace*{-0.5cm}
\end{center}
\caption{(a) The scalar charge $D$, and (b) the ratios $A_h/A_{Sch}$ and $S_h/S_{Sch}$ 
  in terms of the mass  $M$, for $f(\phi)=\alpha\,\ln\phi$.}
\label{D-AS-log}
\end{figure} 

The scalar charge $D$ of the `logarithmic' GB black holes, in terms of the mass
$M$ is shown in  Fig. \ref{D-AS-log}(a). Once again, we observe that,
as the mass of the black-hole solution increases, $D$ decreases towards a
vanishing  value -- our previous analysis has shown that this feature
is often connected with the thermodynamical stability of the solutions. Indeed,
as it may be seen from
 Fig. \ref{D-AS-log}(b), the ratio $S_h/S_{Sch}$ is above unity for
a very large part of the mass-regime; it is again the small-mass regime that
is excluded from the thermodynamical stable solutions. The scalar charge $D$
has again the anticipated behaviour in terms of the parameter $\phi_h$: since
it is the $f'(\phi)=\alpha/\phi$ that enters the scalar equation (\ref{phi-eq22}),
the scalar charge increases as $\phi_h$ decreases. The ratio $A_h/A_{Sch}$ is
also shown in Fig. \ref{D-AS-log}(b): as in the previous cases,
it corresponds to a class of smaller black-hole solutions, compared to the
Schwarzschild solution of the same mass, with a minimum-allowed mass that
approaches the asymptotic Schwarzschild solution as $M$ reaches large values.


\section{Discussion}\label{dis3}

In this chapter, we have extended the analysis of the previous chapter \cite{ABK1} by considering
several subclasses of the general theory that contained the Ricci scalar, a scalar
field and the GB term. We have studied a large number of choices for the
coupling function $f(\phi)$ between the scalar field and the GB term:
exponential, polynomial (even and odd), inverse polynomial (even and odd)
and logarithmic. In each case, employing the constraint (\ref{con-phi'1})
for the value of $\phi'_h$, we constructed a large number of exact black-hole
solutions with scalar hair, and studied in detail their characteristics. 
Our solutions
were characterised by a universal behaviour of the components of the metric
tensor, having the expected behaviour near the black-hole horizon and
asymptotic flatness at radial infinity. All curvature invariant quantities
were examined, and found to have a similar universal profile, independently
of the form of the coupling function $f$, that ensured the finiteness of 
the space-time and thus the regularity of all solutions. 

The same regularity characterised the components of the energy-momentum
tensor over the whole radial regime. In fact, the first constraint on
the value of $\phi'_h$ -- necessary for the evasion of the novel no-hair
theorem -- simultaneously guarantees the regularity of the scalar field
at the black-hole horizon and therefore, the regularity of the solution.
The second constraint for the evasion of the no-hair theorem involves
both $\phi'_h$ and $\phi''_h$, and determines the behaviour of 
$T^r_{\,\,r}$ near the black-hole horizon; this constraint was automatically
satisfied by all the constructed solutions and demanded no further action
or fine-tuning of the parameters. It is worth noting that these constraints
were local as they applied to the black-hole radial regime, and were
therefore easy to check. On the other hand, the old
no-hair theorem, based on an integral constraint over the whole radial
regime, fails to lead to a unique constraint whose violation or not
would govern the existence of regular black-hole solutions. A special
form of it, i.e. $f(\phi)>0$, was found to lead indeed to novel solutions
when satisfied but it could not exclude their emergence in the opposite case. An analysis of the old no-hair theorem that bypasses the constraint on the sign of the coupling function, and allows solutions with $f(\f)<0$,  may be found here \cite{Lee:2018zym}. 

The profile for the scalar field $\phi$ was found in all of the cases considered,
and was indeed regular over the entire radial domain. The scalar field
increased or decreased away from the black-hole horizon, always in accordance
with the constraint (\ref{con-phi'1}), and approached a constant value at
asymptotic infinity. The scalar charge $D$ was determined in each case, and
its dependence on either $\phi_h$ or $M$ was studied. In terms of the first
parameter, whose value determines the magnitude of the GB coupling in the
theory, its behaviour was the anticipated one: for large values of that coupling,
$D$ assumed a large value, while for small couplings $D$ tended to zero. 
Its profile in terms of the mass $M$ of the black hole had a universal
behaviour in the large-$M$ limit, with $D$ assuming increasingly smaller values,
however, in the small-$M$ limit, each family of solutions presented a different
behaviour (either monotonic or not monotonic). In all cases, however, the
scalar charge is an $M$-dependent quantity, and therefore our solutions have
a non-trivial scalar field but with a ``secondary'' hair.

The horizon area $A_h$ and entropy $S_{h}$ of the solutions were also found
in each case. Both quantities quickly increase with the mass $M$, with the
former always dominating over the latter. The function $A(M)$ also revealed
a generic feature of all black-hole solutions found, namely the existence
of a lower value for the horizon radius $r_h$ and thus of the horizon
area $A_h$ of the black hole; this is due to the constraint (\ref{con-f1})
that, for fixed $\phi_h$ and parameter $\alpha$, does not allow for regular
solutions with horizon radius smaller than a minimum value, given by 
$(r_h^2)_{min} = 4\sqrt{6}\,|\dot f_h|$, to emerge. This, in turn, imposes
a lower-bound on the mass of the black hole solution, and therefore all
curves $A(M$) terminate at a specific point in the low-mass regime. 

The study of the ratios $A_h/A_{Sch}$ and $S_h/S_{Sch}$, with respect to the
corresponding quantities of the Schwarzschild solution with the same mass,
had even more information to offer. The first ratio remained below unity
for all classes of black-hole solutions found and for all mass regimes;
as a result, we may conclude that the presence of the additional, gravitational
GB term leads to the formation of more compact black holes compared to the
standard General Relativity (GR). In the large-mass limit, the horizon area
of all black-hole solutions approached the Schwarzschild value -- the same
was true for the entropy ratio $S_h/S_{Sch}$; these two features together
suggest that, for large masses, it will be extremely difficult to distinguish
between GB black holes and their GR analogues.

Do we really expect to detect any of these classes of GB black holes in the
universe? This depends firstly on their stability behaviour, a topic that
needs to be studied carefully and individually for each class of solutions
presented in this work. The curves $S_h/S_{Sch}(M)$, that we produced, may
provide hints for their stability: as mentioned above, the entropy of all
solutions found here approached, in the large-$M$ limit, the Schwarzschild value
thus it is quite likely that large GB black holes share the stability of 
the Schwarzchild solution. For smaller masses, where the GB black
holes are expected to differ from their GR analogue, different profiles
were observed: the `exponential', `inverse-quadratic' and `logarithmic' 
GB black holes had a ratio $S_h/S_{Sch}$ larger than unity for the entire
intermediate mass-regime but smaller than unity in the very-low-mass regime.
These results point towards the thermodynamical stability of solutions
with intermediate and large masses but to an instability for solutions
with small masses (although even in the latter case, an accretion of mass
from their environment could lead to an increase in their mass and to a
change in their (in)stability). On the other hand, the `quadratic', `quartic'
and `linear' GB black holes had their entropy ratio $S_h/S_{Sch}$ below
unity over the entire mass-regime, and may not lead to stable
configurations. Finally, two classes of solutions, the `inverse-linear'
and the first branch of the `cubic' GB black holes have their ratio
$S_h/S_{Sch}$ larger than unity for all values of the black-hole mass -
small, intermediate and large - and may hopefully lead to stable solutions
with a variety of masses. In all cases, a careful study of all the above
solutions under perturbations is necessary in order to verify or refute
the above expectations. The first class of GB black-hole solutions that 
have been studied under linear perturbations are the `exponential' ones
that were found to be indeed stable \cite{DBH1,DBH2} in accordance to the
above comments. In addition, the last years, stability analyses have been performed for the case of the spontaneously scalarized black holes \cite{Don-Kunz1,Silva:2018qhn}. These works indicate that the stability of the black-hole solutions seems to be dependent on the explicit form of the coupling function and therefore each case needs to be studied separately. 

Assuming therefore that one or more classes of the aforementioned black-hole
solutions are stable, we then need a number of signatures or observable
effects that would distinguish them from their GR analogues and convince
us of their existence. A generic feature of all GB black holes is their
minimum horizon radius: if, in the small-mass limit, certain families of
GB black holes are more favourable to emerge compared to the Schwarzschild
solution -- from the stability point-of-view, then the observed black holes
will not have an arbitrarily small mass. Also, in the small-mass limit,
observable effects may include deviations from standard GR in the calculation
of the bending angle of light, the precession observed in near-horizon
orbits and the spectrum from their accretion discs. Studies of this type
have been performed \cite{Chakra1,Chakra2} for black holes in the
Einstein-scalar-GB theory with a linear coupling \cite{SZ} -- a special
case of our analysis -- and shown that the near-horizon strong dynamics
may leave its imprint on all of these observables.

Our GB black-hole solutions are characterized also by a scalar charge. A
previous analysis of dilatonic (`exponential') GB black holes \cite{Kunz} has
revealed that scalar radiation is rather suppressed, especially for
non-spinning black holes, unless particular couplings are introduced in
the theory between the scalar field and ordinary matter. In addition, in
\cite{Yagi} it was demonstrated that the scalar charge of neutron stars,
emerging in the context of the same theory, is extremely small. We would
like to add to this that, according to our analysis, the more stable
configurations tend to correspond to black holes with small scalar charge.
Perhaps, future observations of gravitational waves from black-hole 
or neutron-star processes could lead to clear signatures (or impose
constraints) on the existence of GB compact objects provided that these
objects have a small mass and/or a large scalar charge. Finally, the
measurement of the characteristic frequencies of the quasi-normal modes
(especially the polar sector) will also help to distinguish these solutions
from their GR analogues \cite{Kunz}.

\clearpage
\thispagestyle{empty}

\chapter{Novel Black-Hole Solutions in Einstein-Scalar-Gauss-Bonnet Theories with a Cosmological Constant}\label{5}

\section{Introduction}

In this chapter, we will extend the  analysis of the   previous two chapters \cite{ABK1,ABK2}, that aimed at deriving
asymptotically-flat black-hole solutions, by introducing in our theory a cosmological
constant $\Lambda$, either positive or negative. In the context of this theory, we will
investigate whether the previous, successful synergy between the Ricci scalar, the scalar
field and the Gauss-Bonnet term survives in the presence of $\Lambda$. The question of
the existence of black-hole solutions in the context of a scalar-tensor theory, 
with scalar fields minimally-coupled or conformally-coupled to gravity, and a 
cosmological constant has been debated in the literature for decades
\cite{Narita, Winstanley-Nohair1,Winstanley-Nohair2, Bhatta2007, Bardoux, Kazanas}. 
In the case of a {\it positive} cosmological constant, the existing studies predominantly
excluded the presence of a regular, black-hole solution with an asymptotic de Sitter
behaviour - a counterexample of a black hole in the context of a theory with a
conformally-coupled scalar field \cite{Martinez-deSitter} was shown later to be unstable 
\cite{Harper}. On the other hand, in the case of a {\it negative} cosmological constant,
a substantial number of solutions with an asymptotically Anti-de-Sitter behaviour have been found
in the literature (for a non-exhaustive list, see \cite{Martinez1, Martinez2, Martinez3, Radu-Win, Anabalon, Hosler,
Kolyvaris1, Kolyvaris2, Ohta1, Ohta2, Ohta3, Ohta4, Saenz, Caldarelli, Gonzalez, Gaete1, Gaete2, Giribet, BenAchour}). 

Here, we perform a comprehensive study of the existence of black-hole solutions
with a non-trivial scalar hair and an asymptotically (Anti)-de Sitter behaviour in 
the context of a general class of theories containing the higher-derivative, quadratic
GB term. Prior to our work, the only similar study was the one performed in the special
case of the shift-symmetric Galileon theory \cite{Hartmann1, Hartmann2}, i.e. with a linear coupling
function between the scalar field and the GB term. In this chapter, we consider the most general
class of this theory by considering an arbitrary form of the coupling function $f(\phi)$,
and look for regular black-hole solutions with non-trivial scalar hair. Since the uniform
distribution of energy associated with the cosmological constant permeates the whole
spacetime, we expect $\Lambda$ to have an effect on both the near-horizon and
far-field asymptotic solutions. We will thus repeat our analytical calculations both in
the small and large-$r$ regimes to examine how the presence of $\Lambda$ affects
the asymptotic solutions both near and far away from the black-hole horizon. As we
will see, our set of field equations admits regular solutions near the black-hole horizon
with a non-trivial scalar hair for both signs of the cosmological constant. At the 
far-away regime, the analysis needs to be specialised since a positive or negative
sign of $\Lambda$ leads to either a cosmological horizon or an asymptotic
Schwarzschild-Anti-de Sitter-type
gravitational background, respectively. Our results show that the emergence of a black-hole
solution with a non-trivial scalar hair strongly depends on the type of asymptotic background
that is formed at large distances, and thus on the sign of $\Lambda$. 

In the   case of $\Lambda<0$, we present a large number of novel black-hole
solutions with a regular black-hole horizon, a non-trivial scalar field and a
Schwarzschild-Anti-de Sitter-type asymptotic behaviour at large distances, These solutions
correspond to a variety of forms of the coupling function $f(\phi)$: exponential, polynomial
(even or odd), inverse polynomial (even or odd) and logarithmic. Then, we proceed to study
their physical properties such as the temperature, entropy, and horizon area.  The case with $\L>0$ was also investigated. We construct asymptotically de Sitter solutions and we discuss their properties. Finally, we
investigate features of the asymptotic profile of the scalar field, namely its effective potential
and rate of change at large distances since this greatly differs from the asymptotically-flat case.

The outline of this chapter is as follows: in Section \ref{theo5}, we present our theoretical framework
and perform our analytic study of the near and far-way radial regimes as well as of their
thermodynamical properties. In Section \ref{num5}, we present our numerical results for   both    cases 
of $\Lambda<0$ and $\L>0$. In section \ref{uco} we  briefly present solutions for ultra-compact black holes using a potential instead of a cosmological constant.  We finish with our conclusions in Section \ref{con5}. The analysis of this chapter is based on \cite{ABK3}


\section{The Theoretical Framework}\label{theo5}

We consider the class of higher-curvature gravitational theories 
described by the following action functional:
\begin{equation}
S=\frac{1}{16\pi}\int{d^4x \sqrt{-g}\left[R-\frac{1}{2}\,\partial_{\mu}\phi\partial^{\mu}\phi+f(\phi)R^2_{GB}- 2\L\right]},
\label{action3}
\end{equation}
employed in the previous two chapters. But now, a
 cosmological constant $\Lambda$, that
may take either a positive or a negative value, is also present in the theory.
By varying the action (\ref{action3}) with respect to the metric tensor $g_{\mu\nu}$
and the scalar field $\phi$, we derive the gravitational field equations and the
equation for the scalar field, respectively. These are found to have again the form:
\begin{equation}
G_{\mu\nu}=T_{\mu\nu}\,, \4\4\text{and}\4\4
\nabla^2 \phi+\dot{f}(\phi)R^2_{GB}=0\,, \label{field-eqs}
\end{equation}
where  the effective energy-momentum tensor,
has the following form
\begin{equation}\label{Tmn}
T_{\mu\nu}=-\frac{1}{4}g_{\mu\nu}\partial_{\rho}\phi\partial^{\rho}\phi+\frac{1}{2}\partial_{\mu}\phi\partial_{\nu}\phi-\frac{1}{2}\left(g_{\rho\mu}g_{\lambda\nu}+g_{\lambda\mu}g_{\rho\nu}\right)
\eta^{\kappa\lambda\alpha\beta}\tilde{R}^{\rho\gamma}_{\quad\alpha\beta}
\nabla_{\gamma}\partial_{\kappa}f(\phi)- \L\,g_{\m\n}\,.
\end{equation}
 
Compared to the theory studied in the previous two chapters \cite{ABK1, ABK2}, where $\Lambda$ was zero, the changes
in Eqs. (\ref{field-eqs}) look minimal: the scalar-field equation
remains unaffected while the energy-momentum tensor $T^\mu_{\;\, \nu}$ receives a
constant contribution $-\Lambda \delta^{\mu}_{\;\, \nu}$.
However, as we will see, the presence of the cosmological constant affects both of the
asymptotic solutions, the properties of the derived black holes and even their existence.

In the context of this chapter, we will investigate the emergence of regular, static,
spherically-symmetric but non-asymptotically flat black-hole solutions with a non-trivial
scalar field. The line-element of space-time will accordingly take the form
\begin{equation}\label{metric3}
{ds}^2=-e^{A(r)}{dt}^2+e^{B(r)}{dr}^2+r^2({d\theta}^2+\sin^2\theta\,{d\varphi}^2)\,.
\end{equation}
The scalar field will also be assumed to be static and spherically-symmetric, $\phi=\phi(r)$.
The coupling function $f(\phi)$ will retain a general form during the first part
of our analysis, and will be chosen to have a particular form only at the stage of
the numerical derivation of specific solutions.

The non-vanishing components of the Einstein tensor $G^\mu_{\;\, \nu}$ may be easily
found by employing the line-element (\ref{metric3}), and they read
\begin{align}
 G^t_{\;\, t} &= \frac{e^{-B}}{r^2}(1-e^B-rB'),\label{Gtt}\\
 G^r_{\;\, r} &= \frac{e^{-B}}{r^2}(1-e^B+rA'),\label{Grr}\\
 G^\theta_{\;\,\theta} &=G^\phi_{\;\,\phi}=
 \frac{e^{-B}}{4r}\left[r{A'}^2-2B'+A'(2-rB')+2rA''\right].\label{Gthth}
\end{align}
As before, the prime denotes differentiation with respect to the radial
coordinate $r$. Using Eq. (\ref{Tmn}), the components of the energy-momentum tensor
$T^\mu_{\;\, \nu}$ take in turn the form 
\begin{align}
T^t_{\;\,t}=&-\frac{e^{-2B}}{4r^2}\left[\phi'^2\left(r^2e^B+16\ddot{f}(e^B-1)\right)-8\dot{f}\left(B'\phi'(e^B-3)-2\phi''(e^B-1)\right)\right]-\L, \label{Ttt3}\\[0mm]
T^r_{\;\,r}=&\frac{e^{-B}\phi'}{4}\left[\phi'-\frac{8e^{-B}\left(e^B-3\right)\dot{f}A'}{r^2}\right] -\L, \label{Trr3}\\[2mm]
T^{\theta}_{\;\,\theta}=&T^{\varphi}_{\;\,\varphi}=-\frac{e^{-2B}}{4 r}\left[\phi'^2\left(re^B-8\ddot{f}A'\right)-4\dot{f}\left(A'^2\phi'+2\phi'A''+A'(2\phi''-3B'\phi')\right)\right]
-\L. \label{Tthth3}
\end{align}
Matching the corresponding components of $G^\mu_{\;\, \nu}$ and $T^\mu_{\;\, \nu}$,
the explicit form of Einstein's field equations may be easily derived. These are supplemented
by the scalar-field equation (\ref{field-eqs}) whose explicit form remains unchanged and reads
\begin{equation}
2r\phi''+(4+rA'-rB')\,\phi'+\frac{4\dot{f}e^{-B}}{r}\left[(e^B-3)A'B'-(e^B-1)(2A''+A'^2)\right]=0\,. \label{phi-eq3}
\end{equation}

As in the previous analyses, only two of the three unknown functions $A(r)$, $B(r)$ and
$\phi(r)$ are independent. The metric function $B(r)$ may be found through the $(rr)$-component of field equations which takes again the form of
a second-order polynomial with respect to $e^B$, i.e. $\alpha e^{2B}+\beta e^{B}+\gamma=0$. This easily leads to the following solution 
\begin{equation}\label{Bfunction3}
e^B=\frac{-\beta\pm\sqrt{\beta^2-4\a\gamma}}{2\a},
\end{equation}
where
\beq
\a= 1-\L r^2, \qquad
\beta=\frac{r^2{\phi'}^2}{4}-(2\dot{f}\phi'+r) A'-1,\label{abc}\qquad
\gamma=6\dot{f}\phi'A'.
\eeq
Employing the above expression for $e^B$, the quantity $B'$ may be also found to have the form
\begin{equation}\label{B'3}
B'=-\frac{\gamma'+\beta' e^{B}+\a' e^{2B}}{2\a e^{2B}+\beta e^{B}}.
\end{equation}
Therefore, by using Eqs. (\ref{Bfunction3}) and (\ref{B'3}), the metric function $B(r)$ may be
completely eliminated from the field equations. The remaining three equations then form a
system of only two independent, ordinary differential equations of second order for
the functions $A(r)$ and $\phi(r)$:
\begin{align}
A''=&\frac{P}{S}\,,  \label{A-sys} \\
\phi''=&\frac{Q}{S}\,. \label{phi3}
\end{align} 
The expressions for the quantities $P$, $Q$ and $S$, in terms of $(r, \phi', A', \dot f, \ddot f)$,
are given for the interested reader in Appendix \ref{appb1} as they are quite complicated. 

\subsection{Asymptotic Solution at Black-Hole Horizon}\label{2221} 

As we are interested in deriving novel black-hole solutions, we will first investigate
whether an asymptotic solution describing a regular black-hole horizon is admitted
by the field equations. Instead of assuming the usual power-series
expression in terms of $(r-r_h)$, where $r_h$ is the horizon radius, we will construct
the solution as was also done in \cite{DBH1,DBH2, ABK1,ABK2}. To this end, we demand that,
near the horizon, the metric function $e^{A(r)}$ should vanish (and $e^{B(r)}$ should
diverge) whereas the scalar field must remain finite. The first demand is reflected in the
assumption that $A'(r)$ should diverge as $r \rightarrow r_h$ -- this will be
justified again {\it a posteriori} -- while $\phi'(r)$ and $\phi''(r)$ must be finite in the
same limit. 

Assuming the aforementioned behaviour near the black-hole horizon, Eq. (\ref{Bfunction3})
may be expanded in terms of $A'(r)$ as follows\footnote{Note, that only the (+)-sign
in the expression for $e^B$ in Eq. (\ref{Bfunction3}) leads to the desired black-hole
behaviour.}
\bea\label{expb}
e^B=\frac{(2\dot f \phi '+ r)}{1-\Lambda  r^2}\,A'  - 
\frac{2 \dot f \phi ' \left(r^2 \phi '^2-12 \Lambda  r^2+8\right)+r \left(r^2 \phi '^2-4\right)}
{4 (1-\Lambda r^2)\,(2 \dot f \phi '+r)} + \mathcal{O}\left(\frac{1}{A'}\right).
\eea
Then, substituting the above into the system (\ref{A-sys})-(\ref{phi3}), we obtain 
\begin{align}
A''=&\frac{W_1}{W_3}\,A'^2+\mathcal{O}\left(A'\right),\label{expa}\\[3mm]
\f''=&\frac{W_2}{W_3}\,(2 \dot f \phi '+r) A' + \mathcal{O}(1),\label{expf}
\end{align}
where
\bea\label{w1}
W_1&=& -(r^4+4 r^3 \dot f \phi' + 4 r^2 \dot f^2 \phi '^2-24 \dot f^2) + 
24 \Lambda ^2 r^4 \dot f^2 \nn\\[2mm]
   &&+   \Lambda \left[4 r^5 \dot f \phi'+4 r^2 \dot f^2 \left(r^2 \phi'^2-16\right)
   -64 r \dot f^3 \phi '-64 \dot f^4 \phi'^2+r^6\right], 
\eea
\bea\label{w2}
W_2&=& -r^3 \phi ' \left(1-\Lambda  r^2\right) -32 \Lambda  \dot f^3 \phi '^2+
       16 \Lambda  r \dot f ^2 \phi ' \left(\Lambda  r^2-3\right)\nn\\[2mm]
  && -2 \dot f \left[6+ r^2\phi '^2 + 2 \Lambda ^2 r^4-\Lambda  r^2 \left(r^2 \phi'^2+4\right)\right],
\eea
and
\beq
W_3=\left(1-\Lambda  r^2 \right)
   \left[r^4 + 2 r^3 \dot f \phi '-16 \dot f^2 \left(3-2 \Lambda  r^2\right)-
    32 \Lambda  r \dot f^3 \phi '\right].\label{w3}
\eeq
%
From Eq. (\ref{expb}), we conclude that the combination $(2\dot f \phi '+ r)$ near the horizon
must be non-zero and positive for the metric function $e^B$ to have the correct behaviour,
that is to diverge as $r \rightarrow r_h$ while being positive-definite. Then, Eq. (\ref{expf})
dictates that, if we want $\phi''$ to be finite, we must necessarily have 
\begin{equation}
W_2|_{r=r_h}=0\,.
\end{equation}
The above constraint may be written as a second-order polynomial with respect to $\phi'$,
which can then be solved to yield
\begin{align}\label{solf}
\f'_h=-\frac{r_h^3 (1- \L r_h^2)+16 \Lambda  r_h \dot f^2_h (3-\Lambda  r_h^2)
      \pm (1-\Lambda  r_h^2)\sqrt{C}}{4 \dot f \,\Bigl[r_h^2 -\Lambda (r_h^4-16
   \dot f^2_h)\Bigr]},
\end{align}
where all quantities have been evaluated at $r=r_h$. The quantity $C$ under the square root
stands for the following combination
\begin{equation}
C=256 \Lambda  \dot f^4_h \left(\Lambda  r_h^2-6\right) +
   32 r_h^2 \dot f^2_h \left(2\Lambda  r_h^2-3\right)+r_h^6 \geq 0\,,
\label{C-def}
\end{equation}
and must always be non-negative for $\phi'_h$ to be real. This combination may be written
as a second-order polynomial for $\dot f^2_h$ with roots
\beq 
\dot f^2_{\pm}=\frac{r_h^2 \left[3-2 \Lambda r_h^2 \pm \sqrt{3}\sqrt{3-2 \Lambda r_h^2 +
\Lambda^2 r_h^4}\right]}{16 \Lambda\,(-6 + \Lambda r_h^2)}\,.
\label{con-C}
\eeq
Then, the constraint on $C$ becomes
\beq
C=(\dot f^2_h -\dot f^2_{-})\,(\dot f^2_h -\dot f^2_{+}) \geq 0\,.
\label{C-newdef}
\eeq
Therefore, the allowed regime for the existence of regular, black-hole solutions with 
scalar hair is given by $\dot f^2_h \leq \dot f^2_{-}$ or $\dot f^2_h \geq \dot f^2_{+}$,
since $\dot f^2_{+}>\dot f^2_{-}$. To obtain some physical insight on these inequalities,
we take the limit of small cosmological constant; then, the allowed ranges are
\beq
\dot f^2_h \leq  \frac{r_h^4}{96} \left(1 + \frac{\Lambda r_h^2}{6} + ... \right),
\qquad {\rm or,} \qquad
\dot f^2_h \geq  \frac{r_h^4}{48} \left(1 - \frac{3}{\Lambda r_h^2} + ... \right),
\label{f-ineq}
\eeq
respectively. In the absence of $\Lambda$, Eq. (\ref{C-def}) results into the
simple constraint $\dot f^2_h \leq r_h^4/96$, and defines a sole branch of solutions with a
minimum allowed value for the horizon radius (and mass) of the black hole \cite{ABK1,ABK2}.
In the presence of a cosmological constant, this constraint is now replaced by 
$\dot f^2_h \leq \dot f^2_{-}$, or by the first inequality presented in Eq. (\ref{f-ineq})
in the small-$\Lambda$ limit. This inequality leads again to a branch of solutions that --
for chosen $f(\phi)$, $\phi_h$ and $\Lambda$ -- terminates at a 
black-hole solution with a minimum horizon radius $r_{h}^{min}$. We observe that, at least for
small values of $\Lambda$, the presence of a positive cosmological constant relaxes the constraint
allowing for smaller black-hole solutions, while a negative cosmological constant pushes the
minimum horizon radius towards larger values. The second inequality in Eq. (\ref{f-ineq})
describes a new branch of black-hole solutions that does not exist when $\Lambda=0$;
this was also noted in \cite{Hartmann1, Hartmann2} in the case of the linear coupling function. This branch
of solutions describes a class of very small GB black holes, and terminates instead at a
black hole with a maximum horizon radius $r_{h}^{max}$. 

Returning now to Eq.  (\ref{A-sys}) and employing the constraint (\ref{solf}), the former takes the form
\beq
A''=-A'^2+\mathcal{O}\left(A'\right).\label{expa2}
\eeq
Integrating the above, we find that $A'(r) \sim 1/(r-r_h)$, a result that justifies the diverging
behaviour of this quantity near the horizon that we assumed earlier. A second integration
yields  $A(r) \sim \ln (r-r_h)$, which then uniquely determines the expression of the metric
function $e^A$ in the near-horizon regime. Employing Eq. (\ref{expb}), the metric
function $B$ is also determined in the same regime. Therefore, the asymptotic solution
of Eqs. (\ref{Bfunction3}), (\ref{A-sys}) and (\ref{phi3}), that describes a regular, black-hole
horizon in the limit $r \rightarrow r_h$, is given by the following expressions
\bea
&&e^{A}=a_1 (r-r_h) + ... \,, \label{A-rh3}\\[1mm]
&&e^{-B}=b_1 (r-r_h) + ... \,, \label{B-rh3}\\[1mm]
&&\phi =\phi_h + \phi_h'(r-r_h)+ \phi_h'' (r-r_h)^2+ ... \,, \label{phi-rh3}
\eea
where $a_1, b_1$ and $\phi_h$ are integration constants.
We observe that the above asymptotic solution constructed for the case of a non-zero
cosmological constant has exactly the same functional form as the one constructed
in \cite{ABK1,ABK2} for the case of vanishing $\Lambda$. The presence of the cosmological
constant modifies though the exact expressions of the basic constraint (\ref{solf})
for $\phi'_h$ and of the quantity $C$ given in (\ref{C-def}), the validity of which 
ensures the existence of a regular black-hole horizon. As in \cite{ABK1,ABK2}, the
exact form of the coupling function $f(\phi)$ does not affect the existence of the
asymptotic solution, therefore regular black-hole solutions may emerge for a wide
class of theories of the form (\ref{action3}). 

The regularity of the asymptotic black-hole solution is also reflected in the non-diverging
behaviour of the components of the energy-momentum tensor and of the scale-invariant
Gauss-Bonnet term. The components of the former quantity in this regime assume the form
\begin{align}
T^t_{\;\,t}&=\frac{2e^{-B}}{r^2}\,B'\f' \dot f - \L +\mathcal{O}(r-r_h),\label{Ttt_rh}\\[3mm]
T^r_{\;\,r}&=-\frac{2e^{-B}}{r^2}\,A'\f' \dot f - \L+\mathcal{O}(r-r_h),\label{Trr_rh}\\[3mm]
T^\theta_{\;\,\theta}&=\frac{e^{-2B}}{r}\,(2A'' +A'^2-3A'B')\,\f' \dot f - \L+\mathcal{O}(r-r_h).
\label{Tthth_rh}
\end{align}
Employing the asymptotic expansions (\ref{A-rh3})-(\ref{phi-rh3}), one may see that all
components remain indeed finite in the vicinity of the black-hole horizon.
For future use, we note that the cosmological constant adds a positive contribution to 
all components of the energy-momentum tensor $T^\mu_{\;\,\nu}$ for $\Lambda<0$, while it
subtracts a positive contribution for $\Lambda>0$. Also, all scalar curvature quantities,
the explicit form of which may be found in Appendix \ref{apa2}, independently exhibit a regular
behaviour near the black-hole horizon -- when these are combined, the GB term, in
the same regime, takes the form
\beq
R^2_{GB} = +\frac{12  e^{-2 B}}{r^2}A'^2 +\mathcal{O}(r-r_h)\,, \label{GB-rh3}
\eeq
exhibiting, too, a regular behaviour as expected.


\subsection{Asymptotic Solutions at Large Distances}

The form of the asymptotic solution of the field equations at large distances from the
black-hole horizon depends strongly on the sign of the cosmological constant. Therefore,
in what follows, we study separately the cases of positive and negative $\Lambda$. 


\subsubsection{Positive Cosmological Constant}

In the presence of a positive cosmological constant, a second horizon, the cosmological one, 
is expected to emerge at a radial distance $r=r_c>r_h$. We demand that this horizon is
also regular, that is that the scalar field $\phi$ and its derivatives remain finite in its
vicinity. We may in fact follow a method identical to the one followed in section \ref{2221} near
the black-hole horizon: we again demand that, at the cosmological horizon, $g_{tt} \rightarrow 0$
while $g_{rr} \rightarrow \infty$; then, using that $A'$ diverges there, the regularity of $\phi''$
from Eq. (\ref{phi3}) eventually leads to the constraint 
\begin{align}\label{solfc}
\f'_c=-\frac{r_c^3 (1- \L r_c^2)+16 \Lambda  r_c \dot f^2_c (3-\Lambda  r_c^2)
      \pm (1-\Lambda  r_c^2)\sqrt{\tilde C}}{4 \dot f \,\Bigl[r_c^2 -\Lambda (r_c^4-16
   \dot f^2_c)\Bigr]},
\end{align}
with $\tilde C$ now being given by the non-negative expression
\begin{equation}\label{tildeC-def}
\tilde C=256 \Lambda  \dot f^4_c \left(\Lambda  r_c^2-6\right) +
   32 r_c^2 \dot f^2_c \left(2\Lambda  r_c^2-3\right)+r_c^6 \geq0 \,.
\end{equation} 
Employing Eq. (\ref{solfc}) in Eq. (\ref{A-sys}), the solution for the metric function $A$ may
be again constructed. Overall, the asymptotic solution of the field equations near a regular,
cosmological horizon will have the form
\begin{align}
e^A&=a_2\,(r_c-r)+... , \label{A-rc}\\[3mm]
e^{-B}&=b_2\,(r_c-r)+... , \label{B-rc}\\[3mm]
\f&=\f_c+\f_c'(r_c-r)+\f_c''(r_c-r)^2+... , \label{phi-rc}
\end{align}
where care has been taken for the fact that $r \leq r_c$. One may see again that
the above asymptotic expressions lead to finite values for the components of the energy-momentum
tensor and scalar invariant quantities. Once again, the explicit form of the coupling function
$f(\phi)$ is of minor importance for the existence of a regular, cosmological horizon.

\subsubsection{Negative Cosmological Constant}\label{2222}

For a negative cosmological constant, and at large distances from the black-hole horizon,
we expect the spacetime to assume a form close to that of the Schwarzschild-Anti-de Sitter
solution. Thus, we assume the following approximate forms for the metric functions
\bea
e^{A(r)}&=& \left(k-\frac{2M}{r}-\frac{\L_{eff}}{3}\,r^2+\frac{q_2}{r^2}\right)
\left(1+\frac{q_1}{r^2}\right)^2,\label{alfar1}\\[3mm]
e^{-B(r)}&=& k-\frac{2M}{r}-\frac{\L_{eff}}{3}\,r^2 +\frac{q_2}{r^2},\label{bfar1}
\eea
where $k$, $M$, $\Lambda_{eff}$ and $q_{1,2}$ are, at the moment, arbitrary constants. 
Substituting the above expressions into the scalar field equation (\ref{phi-eq3}), we obtain at first
order the constraint
\beq
\phi''(r)+\frac{4}{r}\,\phi'(r) - \frac{8 \Lambda_{eff} \dot f}{r^2}=0\,.
\label{phi-Anti-far}
\eeq
The gravitational equations, under the same assumptions, lead to two additional
constraints, namely
\beq
\Lambda - \Lambda_{eff} +\frac{\Lambda_{eff}\,r^2  \phi'}{12}\left(\phi'-
\frac{16 \Lambda_{eff} \dot f}{r}\right)=0, \label{con-far-1}
\eeq
\beq
\Lambda - \Lambda_{eff} -\frac{4}{9}\,\dot f \Lambda_{eff}^2 r^2 \left(\phi'' +
\frac{3 \phi'}{r}\right) -\frac{\Lambda_{eff}\,r^2}{12}\,\phi'^2 \left(1+
\frac{16 \Lambda_{eff} \ddot f}{3}\right)=0\,.
\label{con-far-2}
\eeq

Contrary to what happens close to the horizons (either black-hole or cosmological ones),
the form of the coupling function $f(\phi)$ now affects the asymptotic form of the scalar
field at large distances. The easiest case is that of a linear coupling function,
$f(\phi)=\alpha \phi$ - that case was first studied in \cite{Hartmann1, Hartmann2}, however, we
review it again in the context of our analysis as it will prove to play a more general
role.  The scalar field, at large distances, may be shown to have the approximate form
\beq
\phi(r) = \phi_\infty + d_1 \ln r +\frac{d_2}{r^2} +  \frac{d_3}{r^3}+ ...\,, \label{phi-far-Anti}
\eeq
where again $(\phi_\infty, d_1, d_2, d_3)$ are arbitrary constant coefficients. The coefficients
$d_1$ and $\Lambda_{eff}$ may be determined through the first-order constraints 
(\ref{phi-Anti-far}) and (\ref{con-far-1}), respectively, and are given by
\beq
d_1=\frac{8}{3}\,\alpha \Lambda_{eff}\,, \qquad 
\Lambda_{eff} \left( 3 +\frac{80 \alpha^2 \Lambda_{eff}^2}{9}\right)=3 \Lambda\,.
\label{d1-Leff}
\eeq
The third first-order constraint, Eq. (\ref{con-far-2}), is then trivially satisfied. 
In order to determine the values of the remaining coefficients, one needs to derive
higher-order constraints. For example, the coefficients $k$, $q_1$ and $d_2$ are found
at third-order approximation to have the forms
\beq
k=\frac{81 + 864\,\alpha^2 \Lambda_{eff}^2 + 1024\,\alpha^4 \Lambda_{eff}^4}
{81 + 1008\,\alpha^2 \Lambda_{eff}^2 + 2560\,\alpha^4 \Lambda_{eff}^4}\,,
\qquad 
q_1=\frac{24\,\alpha^2 \Lambda_{eff}\,(9+ 64\,\alpha^2 \Lambda_{eff}^2)}
{(9 + 32\,\alpha^2 \Lambda_{eff}^2)\,(9 + 80\,\alpha^2 \Lambda_{eff}^2)}\,,
\nonumber
\eeq
\beq
d_2=-\frac{12 \alpha\,(27 + 288\,\alpha^2 \Lambda_{eff}^2 + 512\,\alpha^4 \Lambda_{eff}^4)}
{81 + 1008\,\alpha^2 \Lambda_{eff}^2 + 2560\,\alpha^2 \Lambda_{eff}^2}\,,
\label{k-q1}
\eeq
while for $q_2$ or $d_3$ one needs to go even higher. In contrast, the coefficient $M$ remains
arbitrary and may be interpreted as the gravitational mass of the solution.

In the perturbative limit (i.e. for small values of the coupling constant $\alpha$
of the GB term), one may show that the above asymptotic solution is valid for
all forms of the coupling function $f(\phi)$. Indeed, if we write
\beq
\phi(r)= \phi_0 +\sum_{n=1}^\infty \alpha^n\,\phi_n(r)\,,
\label{phi-pert}
\eeq
and define $f(\phi)=\alpha \tilde f(\phi)$, then, at first order, 
$\dot f \simeq \alpha \,\dot{\tilde f}(\phi_0)$.
Therefore, independently of the form of $f(\phi)$, at first order in the perturbative limit, 
$\dot f$ is a constant, as in the case of a linear coupling function. Then, a solution
of the form of Eqs. (\ref{alfar1})-(\ref{bfar1}) and (\ref{phi-far-Anti}) is easily 
derived~\footnote{In the perturbative limit, at first order, one finds $d_1=8 \Lambda \dot f(\phi_0)/3$, $\Lambda_{eff}=\Lambda$, $k=1$, $q_1=0$, and $d_2=-4 \dot f(\phi_0)$.} with $\alpha$ in Eqs. (\ref{d1-Leff}) and (\ref{k-q1}) being
now replaced by $\dot f(\phi_0)$.

For arbitrary values of the coupling constant $\alpha$, though, or for a non-linear
coupling function $f(\phi)$, the approximate solution described by Eqs. (\ref{alfar1}),
(\ref{bfar1}) and (\ref{phi-far-Anti}) will not, in principle, be valid any more. Unfortunately,
no analytic form of the solution at large distances may be derived in these cases. However,
as we will see in section \ref{num5}, numerical solutions do emerge with a non-trivial scalar field
and an asymptotic Anti-de Sitter-type behaviour at large distances. These solutions are also
characterised by a finite GB term and finite, constant components of the
energy-momentum tensor at the far asymptotic regime.


\subsection{Thermodynamical Analysis}

In this subsection, we calculate the thermodynamical properties of the sought-for 
black-hole solutions, namely their temperature and entropy. The first quantity may be
easily derived by using the following definition \cite{York, GK}
\beq 
T=\frac{k_h}{2\pi}=\frac{1}{4\pi}\,\left(\frac{1}{\sqrt{|g_{tt} g_{rr}|}}\,
\left|\frac{dg_{tt}}{dr}\right|\right)_{r_h}=\frac{\sqrt{a_1 b_1}}{4\pi}\,,
\label{Temp-def3}
\eeq
that relates the black-hole temperature $T$ to its surface gravity $k_h$. Note that the above formula
has exactly the same form with the corresponding equation for the case of the asymptotically flat solutions, Eq. (\ref{Temp-def}), since our solutions are spherically-symmetric. In the previous chapter we calculated the horizon entropy by using the Euclidean approach Eq. (\ref{entropy-def}). However, in the case of a non-asymptotically-flat
behaviour, the above method needs to be modified: in the case of a de-Sitter-type
asymptotic solution, the Euclidean action needs to be integrated only over the causal spacetime
$r_h \leq r \leq r_c$ whereas, for an Anti-de Sitter-type asymptotic solution, the Euclidean action
needs to be regularised \cite{HP, Dutta}, by subtracting the diverging, `pure' AdS-spacetime contribution.

Alternatively, one may employ the Noether current approach developed in \cite{Wald} to calculate
the entropy of a black hole. In this, the Noether current of the theory under diffeomorphisms
is determined, with the Noether charge on the horizon being identified with the entropy of the
black hole. In \cite{Iyer}, the following formula was finally derived for the entropy 
\begin{equation}
S=-2\pi \oint{d^2x \sqrt{h_{(2)}}\left(\frac{\partial \mathcal{L}}{\partial R_{abcd}}\right)_\mathcal{H}\hat{\epsilon}_{ab}\,\hat{\epsilon}_{cd}}\,,
\label{entropy_AdS}
\end{equation}
where $\mathcal{L}$ is the Lagrangian of the theory, $\hat{\epsilon}_{ab}$ the binormal to
the horizon surface $\mathcal{H}$, and $h_{(2)}$ the 2-dimensional projected metric on 
$\mathcal{H}$. The equivalence of the two approaches has been demonstrated in \cite{Dutta},
in particular in the context of theories that contain higher-derivative terms such as the GB term. 
Here, we will use the Noether current approach to calculate the entropy of the black holes
as it leads faster to the desired result. 

To this end, we need to calculate the derivatives of the scalar gravitational quantities, appearing
in the Lagrangian of our theory (\ref{action3}), with respect to the Riemann tensor. In Appendix \ref{appb2}, we present a simple way to derive those derivatives. Then, substituting in
Eq. (\ref{entropy_AdS}), we obtain
\begin{align}
S=&-\frac{1}{8}\oint{d^2x \sqrt{h_{(2)}}\bigg\{\frac{1}{2}\left(g^{ac}g^{bd}-g^{bc}g^{ad}\right)+f(\phi)\Big[2R^{abcd}+}\nonumber\\
&-2\left(g^{ac}R^{bd}-g^{bc}R^{ad}-g^{ad}R^{bc}+g^{bd}R^{ac}\right)+R\left(g^{ac}g^{bd}-g^{bc}g^{ad}\right)\Big]\bigg\}_\mathcal{H}\hat{\epsilon}_{ab}\,\hat{\epsilon}_{cd}\,.
\label{entropy_1}
\end{align}
The first term inside the curly brackets of the above expression comes from the variation
of the Einstein-Hilbert term and leads to:
\begin{equation}
S_1=-\frac{1}{16}\oint{d^2x\sqrt{h_{(2)}}\left(\hat{\epsilon}_{ab}\,
\hat{\epsilon}^{\,ab}-\hat{\epsilon}_{ab}\,\hat{\epsilon}^{\,ba}\right)}.
\end{equation}
We recall that $\hat{\epsilon}_{ab}$ is antisymmetric, and, in addition, satisfies 
$\hat{\epsilon}_{ab}\,\hat \epsilon^{\,ab}=-2$. Therefore, we easily obtain the result
\begin{equation}
S_1=\frac{A_{\mathcal{H}}}{4}. \label{S1}
\end{equation}
where $A_{\mathcal{H}}=4\pi r_h^2$ is the horizon surface. The remaining terms in
Eq. (\ref{entropy_1}) are all proportional to the coupling function $f(\phi)$ and follow
from the variation of the GB term. To facilitate the calculation, we notice that, on the
horizon surface, the binormal vector is written as:
$\hat{\epsilon}_{ab}=\sqrt{-g_{00}\,g_{11}}\big|_{\mathcal{H}} \left(\delta^0_a\delta^1_b-\delta^1_a\delta^0_b\right)$.
This means that we may alternatively write:
\beq
\left(\frac{\partial \mathcal{L}}{\partial R_{abcd}}\right)_\mathcal{H}\hat{\epsilon}_{ab}\,\hat{\epsilon}_{cd}=4g_{00}\,g_{11}\big|_{\mathcal{H}}\left(\frac{\partial \mathcal{L}}{\partial R_{0101}}\right)_\mathcal{H}.
\eeq
Therefore, the terms proportional to $f(\phi)$ may be written as
\begin{align}
S_2=&-\frac{1}{2}\,f(\phi)\,g_{00}\,g_{11}\big|_{\mathcal{H}}\,\oint d^2x \sqrt{h_{(2)}}\,\bigg[2R^{0101} \nonumber \\[2mm] 
& \hspace*{3cm}-2\left(g^{00}R^{11}-g^{10}R^{01}-g^{01}R^{10}+g^{11}R^{00}\right)+g^{00}g^{11}R\bigg]_{\mathcal{H}}\,.
\label{S2}
\end{align}
To evaluate the above integral, we will employ the near-horizon asymptotic solution
(\ref{A-rh3})-(\ref{phi-rh3}) for the metric functions and scalar field. The asymptotic values of
all quantities appearing inside the square brackets above are given in Appendix \ref{appb2}.
Substituting in Eq. (\ref{S2}), we straightforwardly find 
\begin{equation}
S_2=\frac{f(\phi_h)A_{\mathcal{H}}}{r_h^2}= 4\pi f(\phi_h). \label{S2-final}
\end{equation}
Combining the expressions (\ref{S1}) and (\ref{S2-final}), we finally derive the result
\beq
S_h=\frac{A_h}{4} +4 \pi f(\phi_h)\,.
\label{entropy3}
\eeq
The above describes the entropy of a GB black hole arising in the context of the theory
(\ref{action3}), with a general coupling function $f(\phi)$ between the scalar field
and the GB term, and a cosmological constant term. We observe that the above expression
matches the one derived in the previous chapter Eq. (\ref{entropy}) \cite{ABK1,ABK2} in the context of the theory (\ref{action3}) but in
the absence of the cosmological constant. This was, in fact, expected on the basis of
the more transparent Noether approach used here: the $\Lambda$ term does not change
the overall topology of the black-hole horizon and it does not depend on the Riemann
tensor; therefore, no modifications are introduced to the functional form of the entropy of
the black hole due to the cosmological constant. However, the presence of $\Lambda$
modifies in a quantitative way the properties of the black hole and therefore the value
of the entropy, and temperature, of the found solutions.


\section{Numerical Solutions}\label{num5}

\subsection{ Anti-de Sitter Gauss-Bonnet Black Holes}
In order to construct the complete black-hole solutions in the context of the theory
(\ref{action3}), i.e. in the presence of both the GB and the cosmological constant terms,
we need to numerically integrate the system of Eqs. (\ref{A-sys})-(\ref{phi3}). The integration
starts at a distance very close to the horizon of the black hole, i.e. at
$r\approx r_h+\mathcal{O}(10^{-5})$ (for simplicity, we set $r_h=1$). The 
metric function $A$ and scalar field $\phi$ in that regime are described by the asymptotic
solutions (\ref{A-rh3}) and (\ref{phi-rh3}). The input parameter $\phi'_h$
is uniquely determined through Eq. (\ref{solf}) once the coupling function 
$f(\phi) =\alpha \tilde f(\phi)$ is selected and the values of the remaining parameters
of the model near the horizon are chosen. These parameters appear to be $\alpha$,
$\phi_h$ and $\Lambda$. However, the  first two are not independent: since it is their
combination $\alpha \tilde f(\phi_h)$ that determines the strength of the coupling between
the GB term and the scalar field, a change in the value of one of them may be absorbed in 
a corresponding change to the value of the other; as a result, we may fix $\alpha$
and vary only $\phi_h$. The values of $\phi_h$ and $\Lambda$ also cannot be
totally uncorrelated as they both appear in the expression of $C$, Eq. (\ref{C-def}),
that must always be positive; therefore, once the value of the first is chosen,
there is an allowed range of values for the second one for which black-hole solutions
arise. This range of values are determined by the inequalities $\dot f_h^2 \leq \dot f_-^2$
and $\dot f_h^2 \geq \dot f_+^2$ according to Eq. (\ref{C-newdef}), and lead in
principle to two distinct branches of solutions. In fact, removing the square, four
branches emerge depending on the sign of $\dot f_h$. However, in what follows
we will assume that $\dot f_h>0$, and thus study the two regimes $\dot f_h \leq \dot f_-$
and $\dot f_h \geq \dot f_+$; similar results emerge if one assumes instead that
$\dot f_h <0$. 

Before starting our quest for black holes with an (Anti)-de Sitter asymptotic behaviour
at large distances, we first consider the case with $\Lambda=0$ where upon we
successfully reproduce the families of asymptotically-flat back holes derived in 
\cite{ABK1, ABK2}. Then, for given values of the coupling constant $\alpha$ we select all the values of $\Lambda$ that validate the constraint (\ref{C-def}) and look for
novel black-hole solutions.
In this section we consider only solutions with negative cosmological constant $\L<0$ and thus we expect them to have an anti-de-Sitter behavior at infinity. 

\subsubsection{Exponential coupling function}

As mentioned above, the integration starts from the near-horizon regime with the
asymptotic solutions (\ref{A-rh3}) and (\ref{phi-rh3}), and it proceeds towards large
values of the radial coordinate until the form of the derived solution for the metric
resembles, for $\Lambda<0$, the asymptotic solution (\ref{alfar1})-(\ref{bfar1})
describing an Anti-de Sitter-type background. The arbitrary coefficient $a_1$,
that does not appear in the field equations, may be fixed by demanding that, at
very large distances, the metric functions satisfy the constraint $e^A \simeq e^{-B}$. 
We have considered a large number of
forms for the coupling function $f(\phi)$, and, as we will now demonstrate, we have
managed to produce a family of regular black-hole solutions with an Anti-de Sitter
asymptotic behaviour, for every choice of $f(\phi)$.

\begin{figure}[t!] 
\begin{center}
\hspace{0.0cm} \hspace{-0.6cm}
\includegraphics[height=.26\textheight, angle =0]{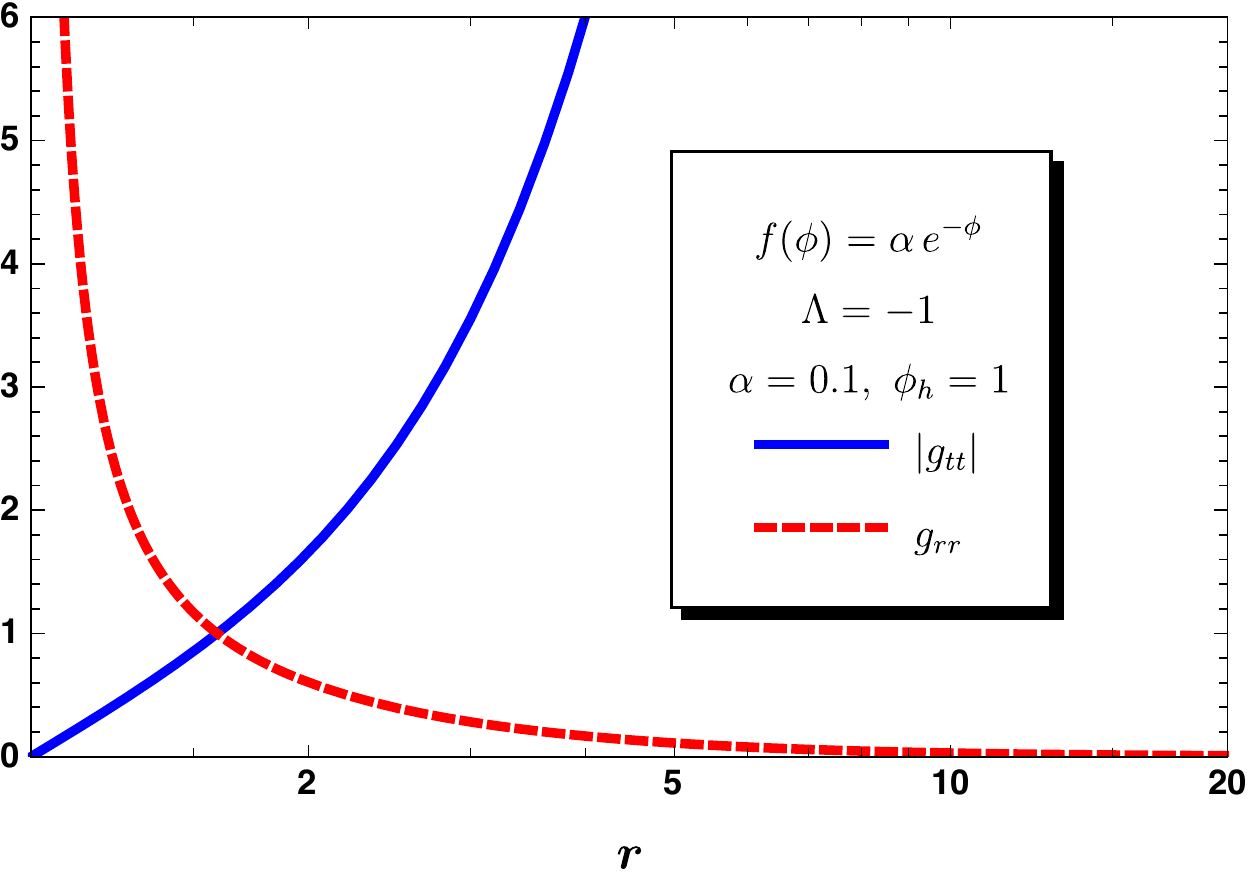}
\hspace{0.52cm} \hspace{-0.6cm}
\includegraphics[height=.26\textheight, angle =0]{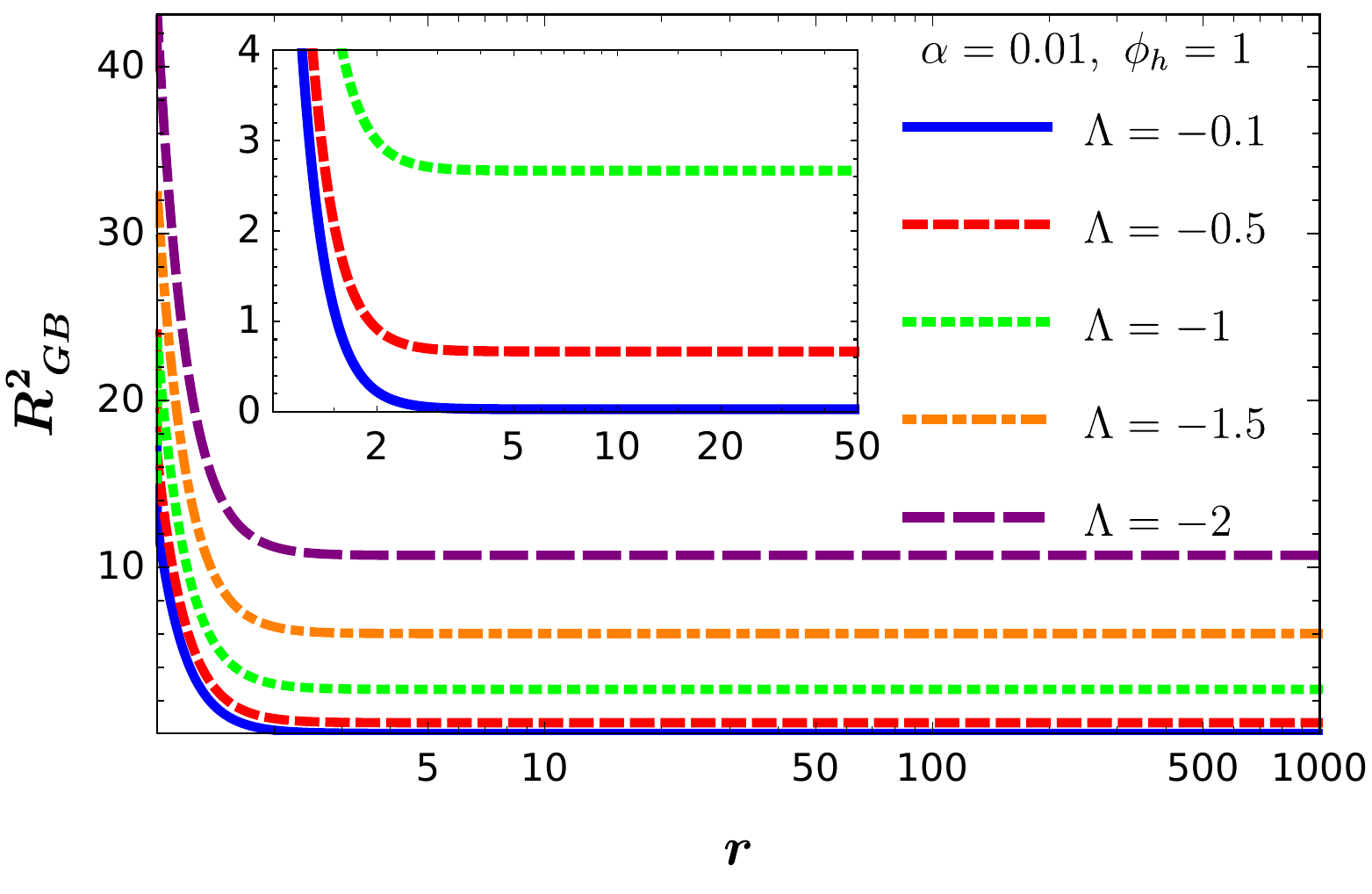}
\\
\hspace*{0.7cm} {(a)} \hspace*{7.5cm} {(b)}  \vspace*{-0.5cm}
\end{center}
\caption{(a) The metric components $|g_{tt}|$ and $g_{rr}$, and (b) the Gauss-Bonnet
term $R^2_{GB}$ in terms of the radial coordinate $r$, for $f(\phi)=\alpha e^{-\phi}$.}
   \label{Metric_GB3}
\end{figure} 

We will first discuss the case of an exponential coupling function, $f(\phi)=\alpha e^{-\phi}$. The
solutions for the metric functions $e^{A(r)}$ and $e^{B(r)}$ are depicted in  
Fig. \ref{Metric_GB3}(a). We may easily see that the near-horizon behaviour, with $e^{A(r)}$
vanishing and $e^{B(r)}$ diverging, is eventually replaced by an Anti-de Sitter regime with
the exactly opposite behaviour of the metric functions at large distances. The solution presented
corresponds to the particular values $\Lambda=-1$ (in units of $r_h^{-2}$), $\alpha=0.1$
and $\phi_h=1$, however,
we obtain the same qualitative behaviour for every other set of parameters satisfying the
constraint~\footnote{Here, we do not present black-hole solutions that satisfy the alternative
choice $\dot f_h \geq \dot f_+$ since this leads to solutions plagued by numerical instabilities,
that prevent us from deducing their physical properties in a robust way. The same ill-defined
behaviour of this second branch of solutions with very small horizon radii was also found in
\cite{Hartmann1, Hartmann2}.} $\dot f_h \leq \dot f_-$, that follows from Eq. (\ref{C-def}). The
spacetime is regular in the whole radial regime, and this is reflected in the form of
the scalar-invariant Gauss-Bonnet term: this is presented in  
Fig. \ref{Metric_GB3}(b), for $\alpha=0.01$, $\phi_h=1$ and for a variety of values
of the cosmological constant. We observe that the GB term acquires its maximum
value near the horizon regime, where the curvature of spacetime is larger, and
reduces to a smaller, constant asymptotic value in the far-field regime. This asymptotic
value is, as expected, proportional to the cosmological constant as this quantity determines
the curvature of spacetime at large distances.  

\begin{figure}[t!] 
\begin{center}
\hspace{0.0cm} \hspace{-0.6cm}
\includegraphics[height=.25\textheight, angle =0]{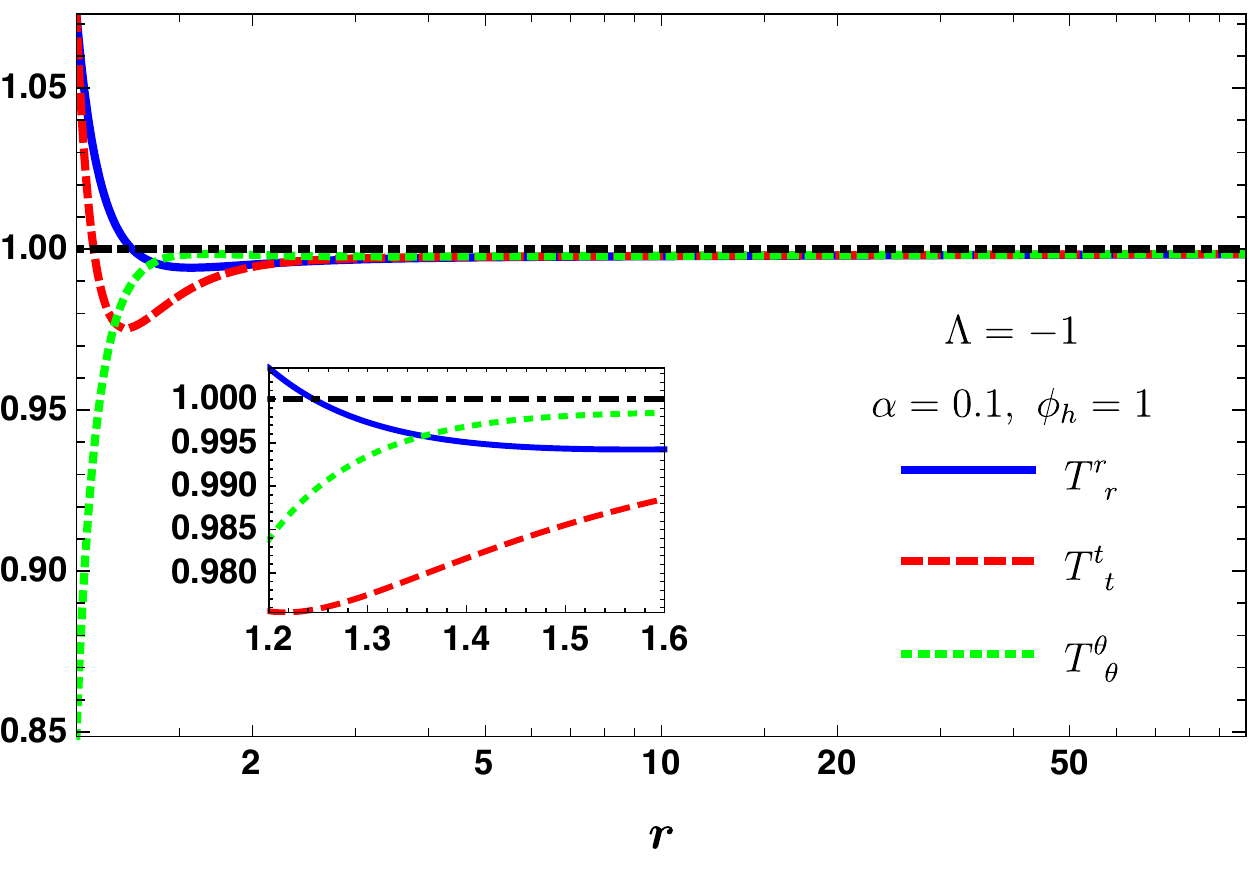}
\hspace{0.52cm} \hspace{-0.6cm}
\includegraphics[height=.25\textheight, angle =0]{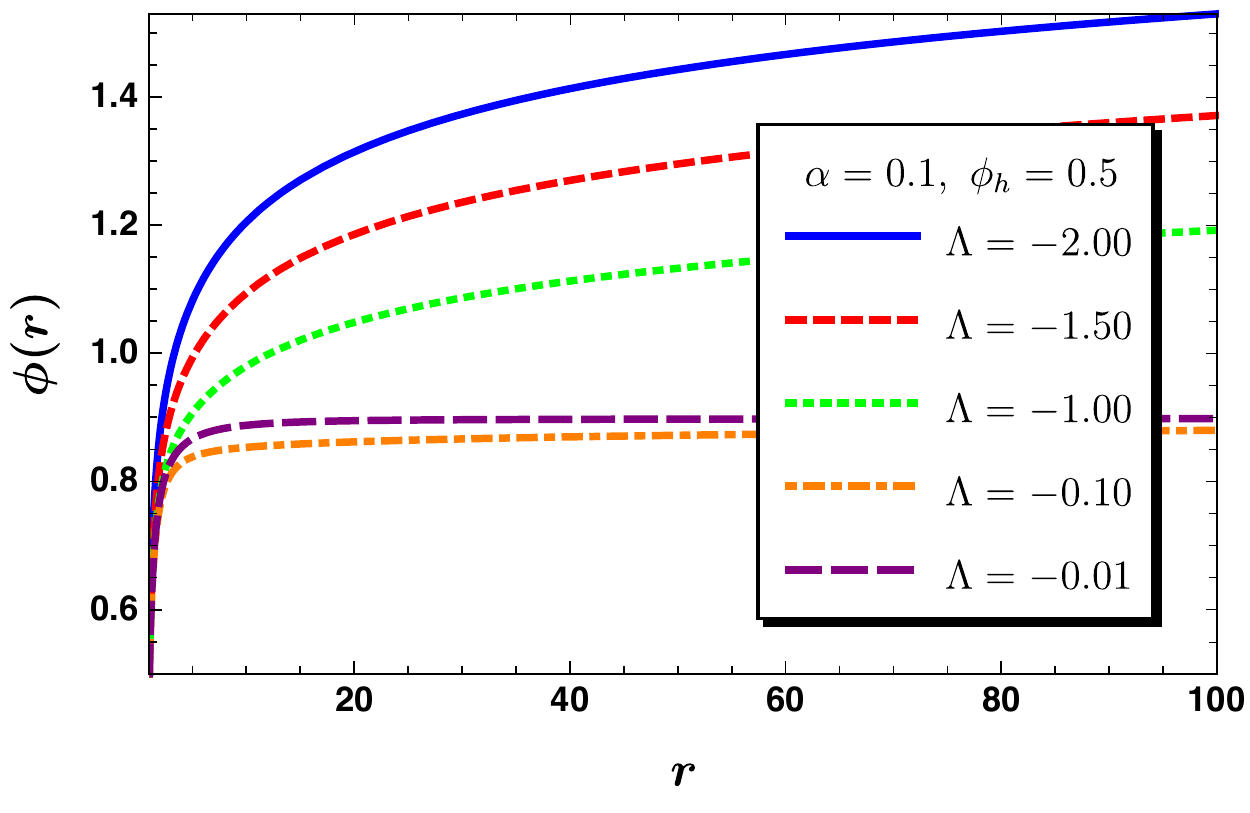}
\\
\hspace*{0.7cm} {(a)} \hspace*{7.5cm} {(b)}  \vspace*{-0.5cm}
\end{center}
\caption{(a) The energy-momentum tensor $T_{\mu\nu}$, and (b)  the scalar field $\phi$
  in terms of the radial coordinate $r$, for $f(\phi)=\alpha e^{-\phi}$.}
\label{exp-phi-Tmn}
\end{figure}

Although in Section \ref{2222}, we could not find the analytic form of the scalar field at large
distances from the black-hole horizon for different forms of the coupling function $f(\phi)$,
our numerical results ensure that its behaviour is such that the effect of the scalar field
at the far-field regime is negligible, and it is only the cosmological term that determines
the components of the energy-momentum tensor. In   Fig. \ref{exp-phi-Tmn}(a),
we display all three components of $T^{\mu}_{\;\,\nu}$ over the whole radial regime, for
the indicative solution $\Lambda=-1$, $\alpha=0.1$ and $\phi_h=1$.
Far away from the black-hole horizon, all components reduce to $-\Lambda$, in
accordance with Eqs. (\ref{Ttt3})-(\ref{Tthth3}), with the effect of both the scalar field and
the GB term being there negligible. Near the horizon, and according to the asymptotic behaviour
given by Eqs. (\ref{Ttt_rh})-(\ref{Tthth_rh}), we always have $T^r_{\;\,r}\approx T^t_{\;\,t}$,
since, at $r \simeq r_h$, $A' \simeq -B'$; also, the $T^\theta_{\;\,\theta}$ component
always has the opposite sign to that of $T^r_{\;\,r}$ since $A'' \simeq -A'^2$. This
qualitative behaviour of $T^{\mu}_{\;\,\nu}$ remains the same for all forms of the coupling
function we have studied and for all solutions found, therefore we refrain from giving
additional plots of this quantity for the other classes of solutions found.

From the results depicted in  Fig. \ref{exp-phi-Tmn}(a), we see that, near the
black-hole horizon, we always have $T^r_{\;\,r}\approx T^t_{\;\,t}>0$. Comparing this
behaviour with the asymptotic forms (\ref{Ttt_rh})-(\ref{Tthth_rh}), we deduce that, close
to the black-hole horizon where $A'>0$, we must have $(\phi' \dot f)_h <0$. 
In the case of vanishing cosmological constant, the negative value of this quantity was of
paramount importance for the evasion of the no-hair theorem \cite{Bekenstein} and the
emergence of novel, asymptotically-flat black-hole solutions \cite{ABK1,ABK2}. We observe that also in
the context of the present analysis with $\Lambda \neq 0$, this quantity turns out to be again
negative, and to lead once again to novel black-hole solutions. Coming back to our assumption
of a decreasing exponential coupling function and upon choosing to consider $\alpha>0$,
the constraint
$(\phi' \dot f)_h <0$ means that $\phi'_h>0$ independently of the value of $\phi_h$. In
 Fig. \ref{exp-phi-Tmn}(b), we display the solution for the scalar field in
terms of the radial coordinate, for the indicative values of $\alpha=0.1$, $\phi_h=0.5$ and
for different values of the cosmological constant. The scalar field satisfies indeed the 
constraint $\phi'_h>0$ and increases away from the black-hole horizon\footnote{A
complementary family of  solutions arises if we choose $\alpha<0$, with the scalar
profile now satisfying the constraint $\phi'_h<0$ and decreasing away from the black-hole
horizon.}. At large distances, we observe that, for small values of the cosmological
constant, $\phi(r)$ assumes a constant value; this is the behaviour found for asymptotically-flat
solutions \cite{ABK1,ABK2} that the solutions with small $\Lambda$ are bound to match.
For increasingly larger values of $\Lambda$ though, the profile of the scalar field deviates
significantly from the series expansion in powers of $(1/r)$ thus allowing for a $r$-dependent
$\phi$ even at infinity -- in the perturbative limit, as we showed in the previous section,
this dependence is given by the form $\phi(r) \simeq d_1 \ln r$.


\subsubsection{Polynomial  coupling functions}

\begin{figure}[t!] 
\begin{center}
\hspace{0.0cm} \hspace{-0.6cm}
\includegraphics[height=.23\textheight, angle =0]{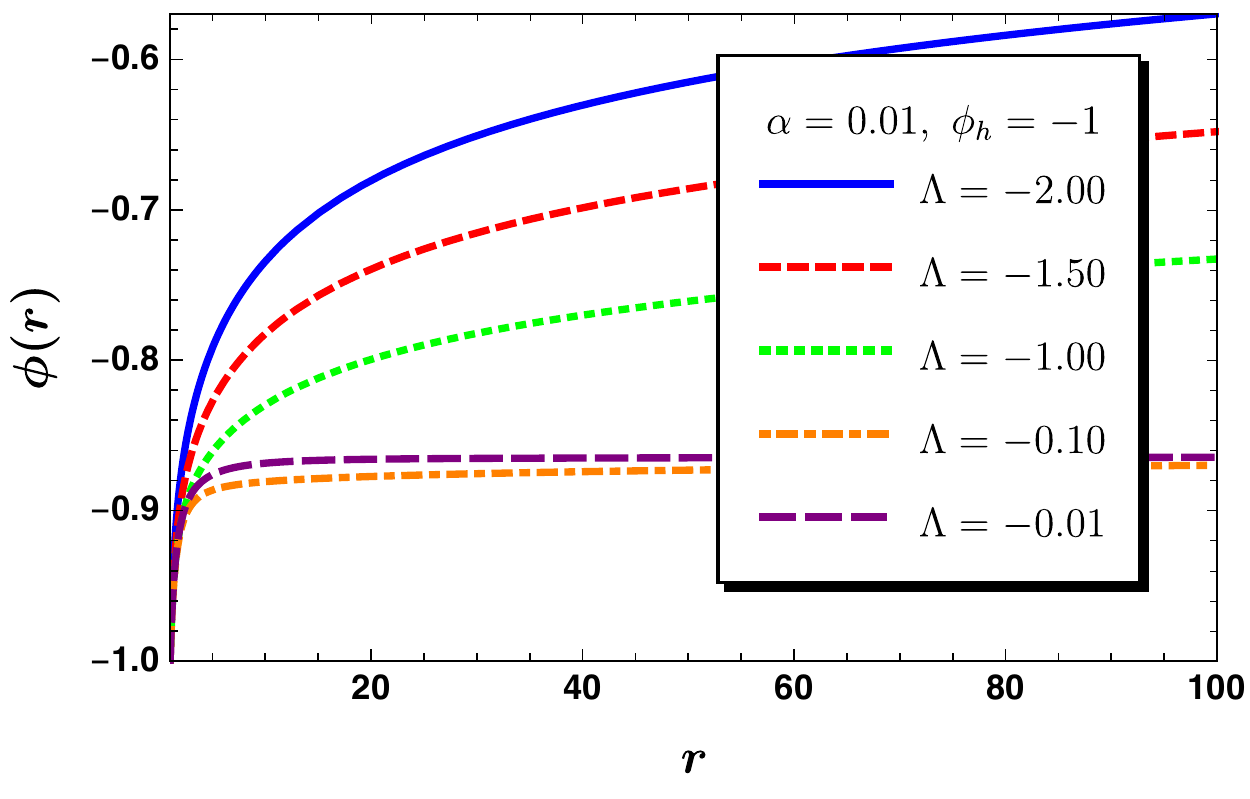}
\hspace{0.52cm} \hspace{-0.6cm}
\includegraphics[height=.23\textheight, angle =0]{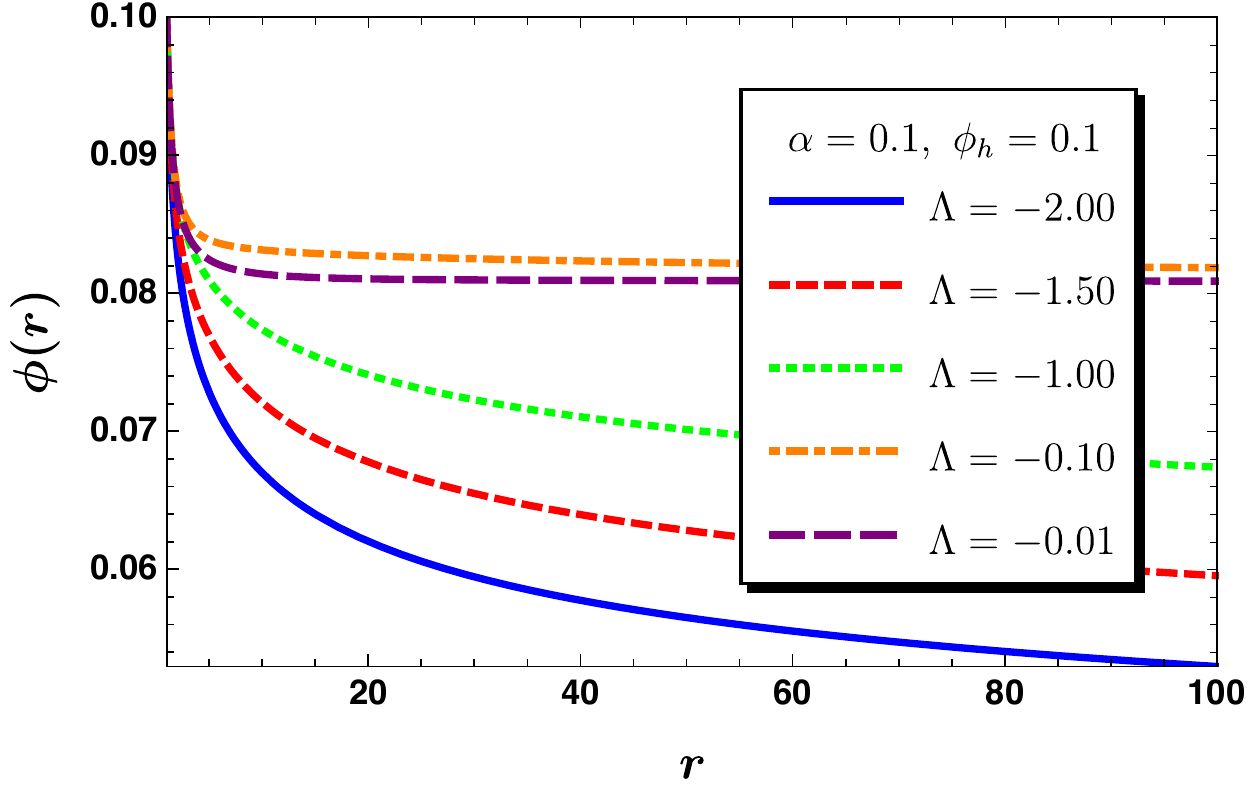}
\\
\hspace*{0.7cm} {(a)} \hspace*{7.5cm} {(b)}  \vspace*{-0.5cm}
\end{center}
\caption{ The scalar field $\phi$  in terms of the radial coordinate $r$, (a) for
$f(\phi)=\alpha\phi^2$, and (b) for  $f(\phi)=\alpha\phi^3$.}
\label{fig_quad_qub}
\end{figure}

We will now consider the case of an even polynomial coupling function of the form
$f(\phi)=\alpha \phi^{2n}$ with $n \geq 1$. The behaviour of the solution for the metric
functions matches the
one depicted\,\footnote{Let us mention at this point that, for extremely large values of either
the coupling constant $\alpha$ or the cosmological constant $\Lambda$, that are nevertheless
allowed by the constraint (\ref{C-def}), solutions that have their metric behaviour deviating
from the AdS-type form (\ref{alfar1})-(\ref{bfar1}) were found; according to the obtained
behaviour, both metric functions seem to depend logarithmically on the radial coordinate
instead of polynomially. As the physical interpretation of these solutions is not yet clear,
we omit these solutions from the remaining of our analysis.} in  Fig.
\ref{Metric_GB3}(a). The same is true for the behaviour of the GB term and the energy-momentum tensor,
whose profiles are similar to the ones displayed in Figs. \ref{Metric_GB3}(b) and
\ref{exp-phi-Tmn}(a), respectively. The positive-definite value of $T^r_{\,\,r}$
near the black-hole horizon implies again that, there, we should have $(\dot{f}\phi')_h<0$,
or equivalently $\phi_h\phi'_h<0$,  for $\alpha>0$. Indeed, two classes of solutions arise
in this case: for positive values of $\phi_h$, we obtain solutions for the scalar field that
decrease away from the black-hole horizon, while for $\phi_h<0$, solutions that increase with
the radial coordinate are found. In Fig. \ref{fig_quad_qub}(a), we present a family of
solutions for the case of the quadratic coupling function (i.e. $n=1$), for $\phi_h=-1$ and
$\alpha=0.01$, arising for different values of $\Lambda$ -- since $\phi_h<0$, the scalar
field exhibits an increasing behavior as expected. 

\begin{figure}[b!] 
\begin{center}
\hspace{0.0cm} \hspace{-0.6cm}
\includegraphics[height=.24\textheight, angle =0]{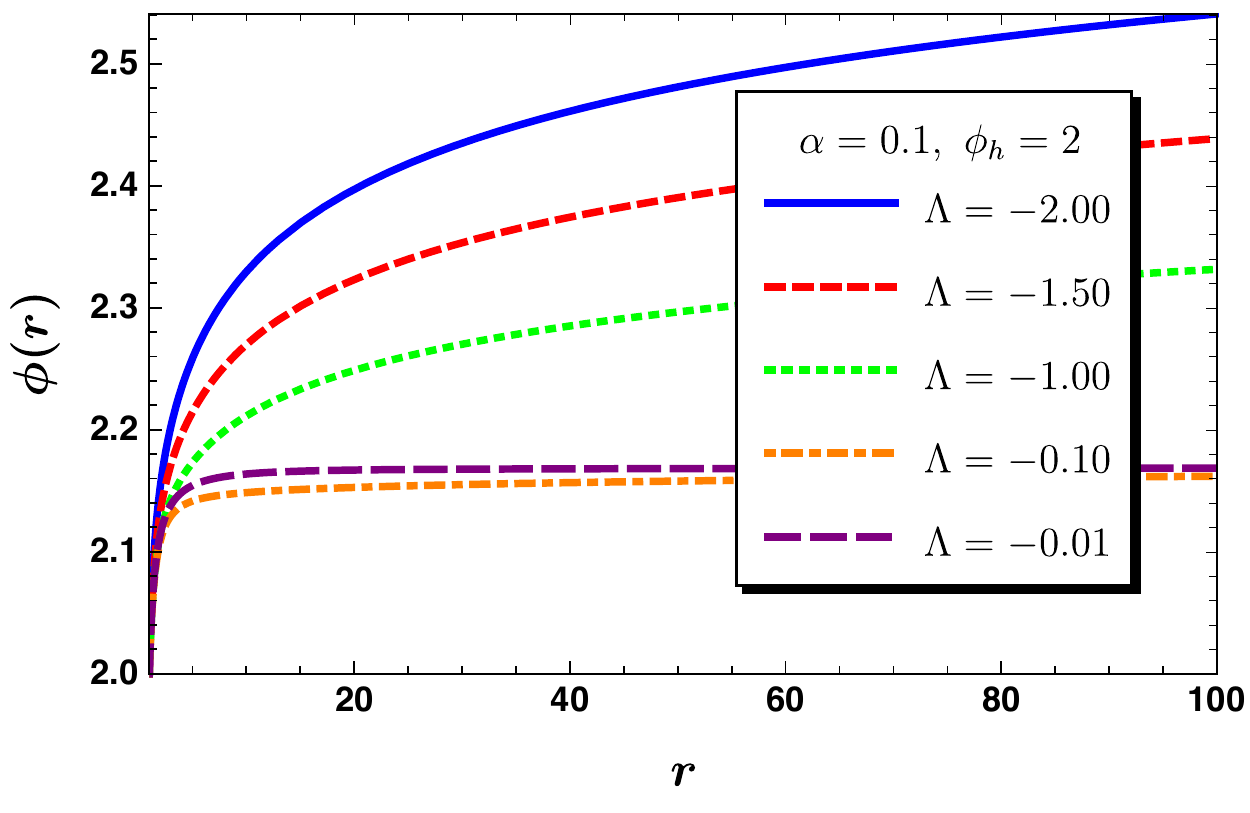}
\hspace{0.52cm} \hspace{-0.6cm}
\includegraphics[height=.24\textheight, angle =0]{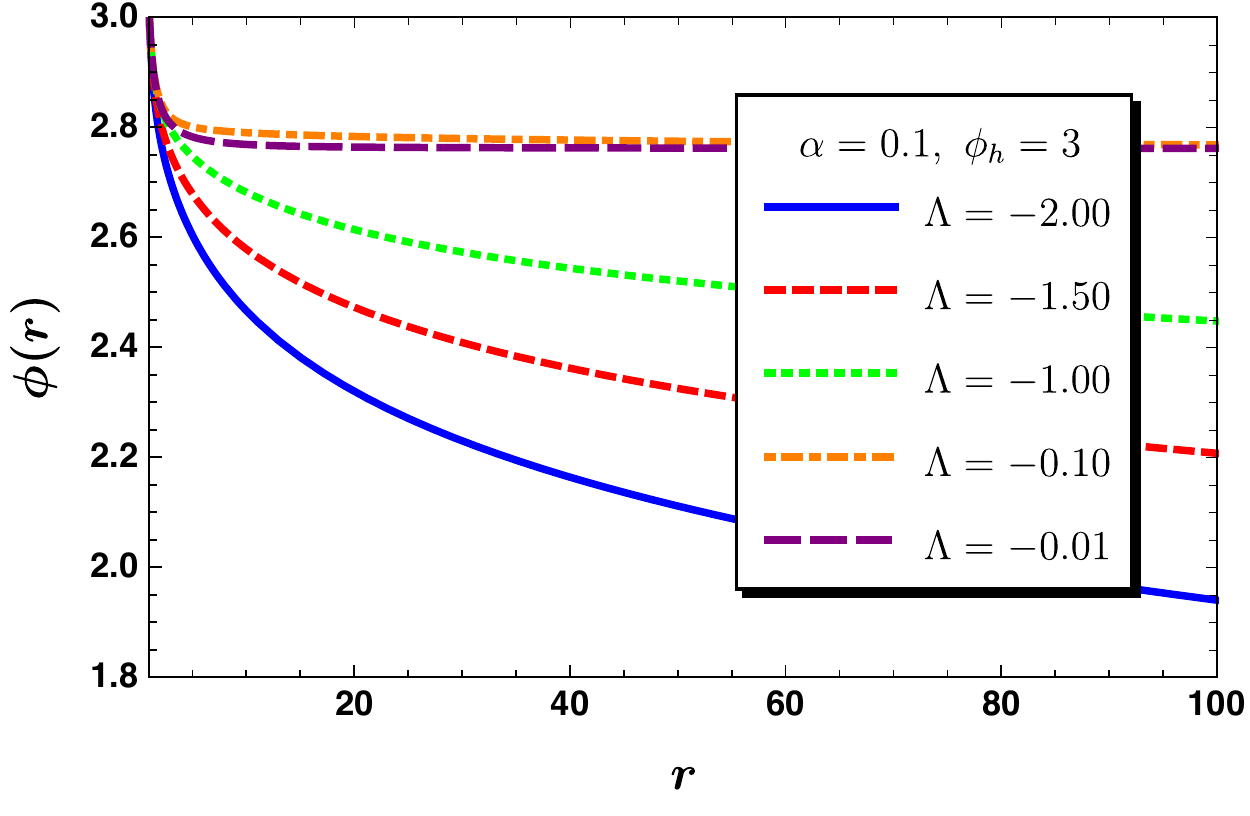}
\\
\hspace*{0.7cm} {(a)} \hspace*{7.5cm} {(b)}  \vspace*{-0.5cm}
\end{center}
\caption{ The scalar field $\phi$  in terms of the radial coordinate $r$, (a) for
$f(\phi)=\alpha/\phi$, and (b) for  $f(\phi)=\alpha\ln\phi $.}
\label{fig_inv_log}
\end{figure} 


Let us examine next the case of an odd polynomial coupling function, $f(\phi)=\alpha\phi^{2n+1}$
with $n \geq 0$. The behaviour of the metric functions, GB term and energy-momentum tensor have
the expected behaviour for an asymptotically AdS background, as in the previous cases. The solutions
for the scalar field near the black-hole horizon are found to satisfy the constraint
$\alpha (\phi^{2n}\phi')_h<0$ or simply $\phi_h'<0$, when $\alpha>0$. As this holds independently of
the value of $\phi_h$, all solutions for the scalar field are expected to decrease away from the
black-hole horizon. Indeed, this is the profile depicted in  Fig. \ref{fig_quad_qub}(b)
where a family of solutions for the indicative case of a qubic coupling function (i.e. $n=1$) is
presented for $\alpha=0.1$, $\phi_h=0.1$ and various values of $\Lambda$. 


The case of an inverse polynomial coupling function, $f(\phi)=\alpha\phi^{-k}$, with $k$ either an
even or odd positive integer, was also considered. For odd $k$, i.e. $k=2n+1$, the positivity
of $T^r_{\,\,r}$ near the black-hole horizon demands again that $(\dot{f}\phi')_h<0$, or that $-\alpha/\phi^{2n+2}\phi'<0$. For $\alpha>0$, the solution for the scalar field should thus
always satisfy $\phi'_h>0$, regardless of our choices for $\phi_h$ or $\Lambda$. As an indicative
example, in Fig. \ref{fig_inv_log}(a), we present the case of $f(\phi)=\alpha/\phi$
with a family of solutions arising for $\alpha=0.1$ and $\phi_h=2$. The solutions for the
scalar field clearly satisfy the expected behaviour by decreasing away from the black-hole horizon.
On the other hand, for even $k$, i.e. $k=2n$, the aforementioned constraint now demands that
$\phi_h\,\phi'_h<0$. As in the case of the odd polynomial coupling function, two subclasses of solutions
arise: for $\phi_h>0$, solutions emerge with $\phi'_h<0$ whereas, for $\phi_h<0$, we find solutions 
with $\phi'_h>0$. The profiles of the solutions in this case are similar to the ones found before,
with $\phi$ approaching, at large distances, an almost constant value for small $\Lambda$ 
but adopting a more dynamical behaviour as the cosmological constant gradually takes on
larger values.

\subsubsection{Logarithmic  coupling function}


As a final example of another form of the coupling function between the scalar field and the
GB term, let us consider the case of a logarithmic coupling function, $f(\phi)=\alpha \ln\phi$.
Here, the condition near the horizon of the black hole gives $\alpha \phi'/\phi<0$, therefore,
for $\alpha>0$, we must have $\phi'_h\phi_h<0$; for $\phi_h>0$, this translates to a
decreasing profile for the scalar field near the black-hole horizon. In  Fig.
\ref{fig_inv_log}(b), we present a family of solutions arising for a logarithmic coupling function
for fixed $\alpha=0.01$ and $\phi_h=1$, while varying the cosmological constant $\Lambda$.
The profiles of the scalar field agree once again with the one dictated by the near-horizon
constraint, and they all decrease in that regime. As in the previous cases, the metric
functions approach asymptotically an Anti-de Sitter background, the scalar-invariant GB term
remains everywhere regular, and the same is true for all components of the energy-momentum
tensor that asymptotically approach the value $-\Lambda$.

\subsubsection{Effective potential and physical characteristics}

\begin{figure}[t!] 
\begin{center}
\hspace{0.0cm} \hspace{-0.6cm}
\includegraphics[height=.24\textheight, angle =0]{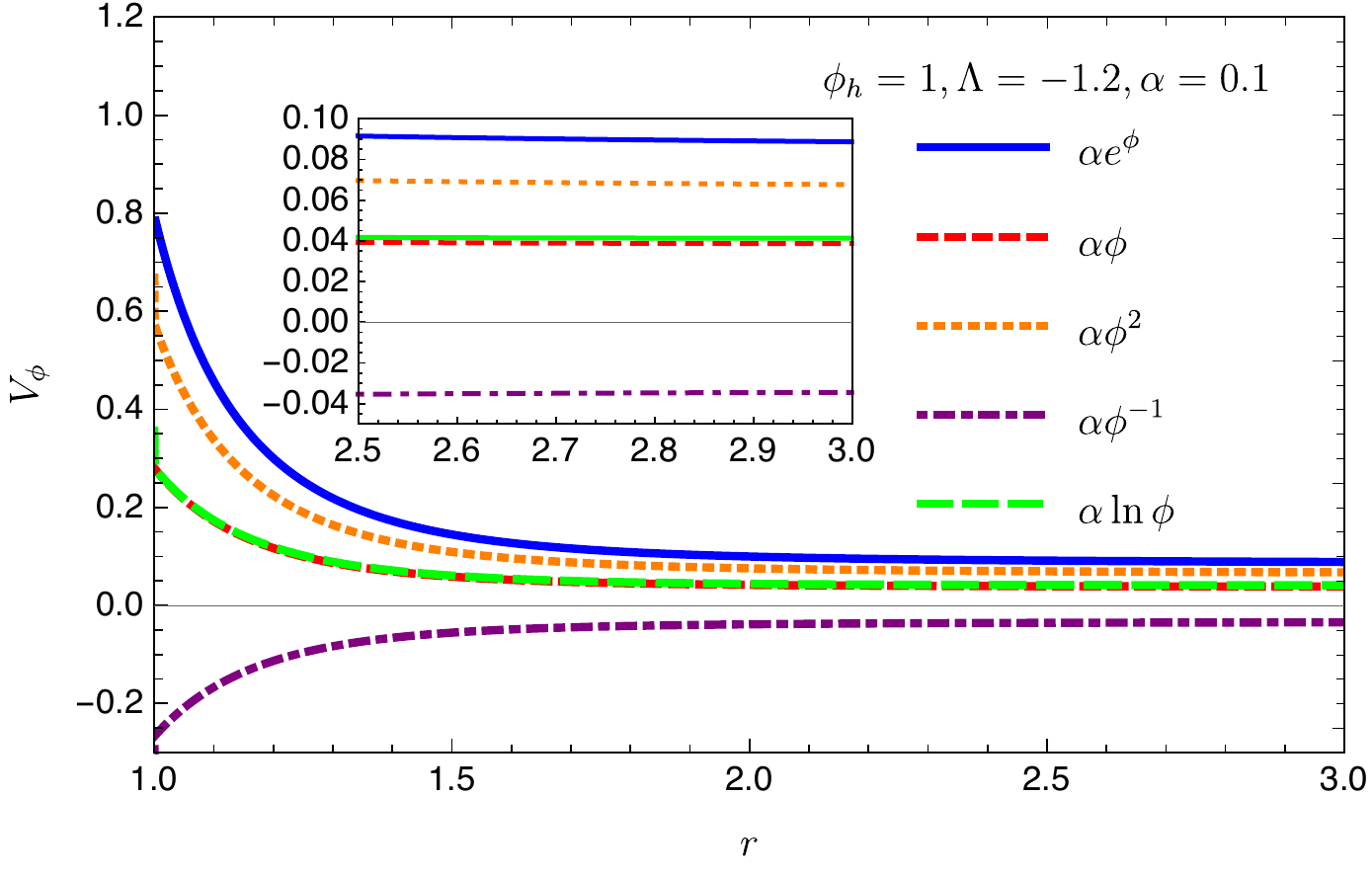}
\hspace{0.52cm} \hspace{-0.6cm}
\includegraphics[height=.24\textheight, angle =0]{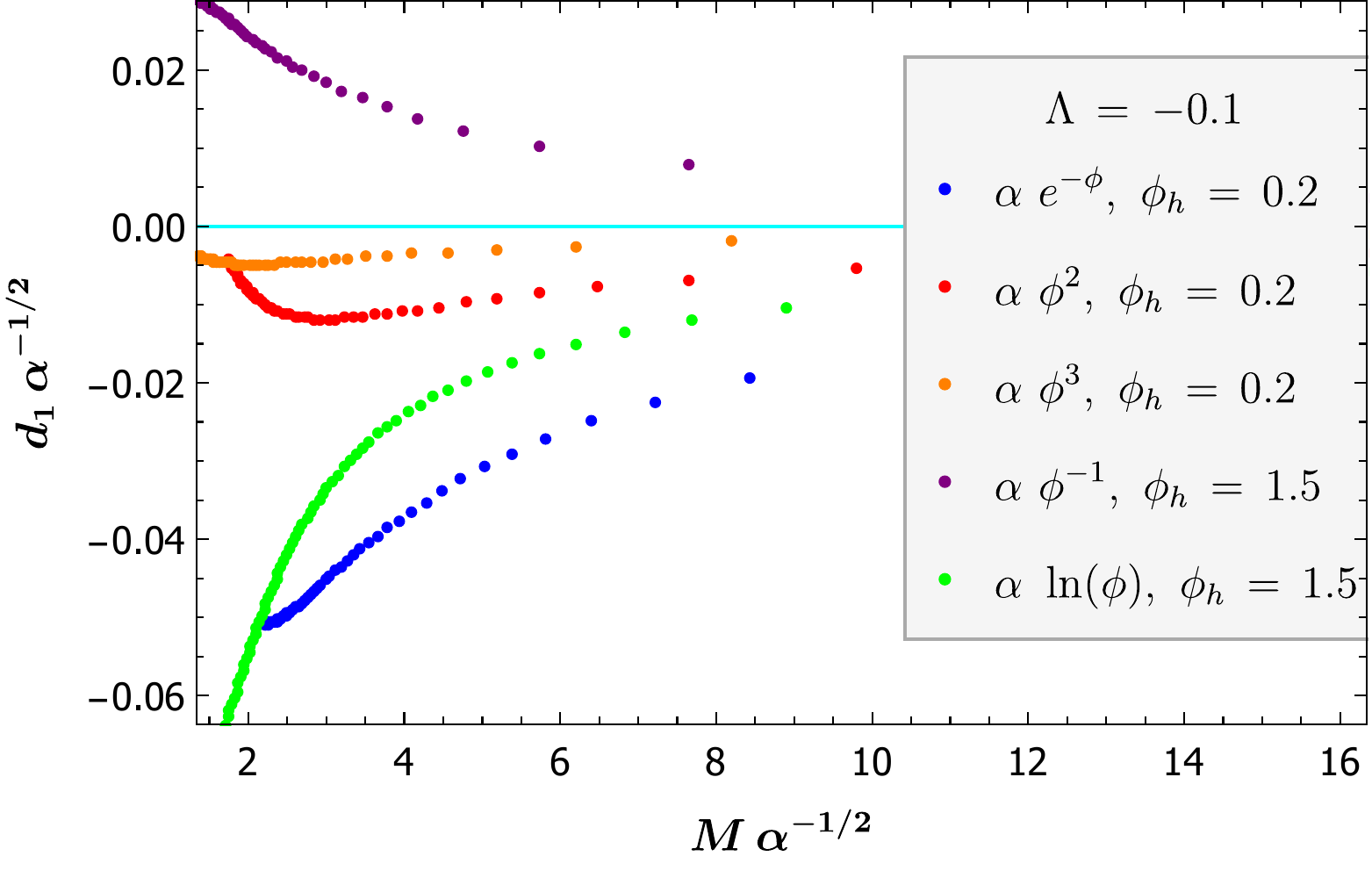}
\\
\hspace*{0.7cm} {(a)} \hspace*{7.5cm} {(b)}  \vspace*{-0.5cm}
\end{center}
\caption{(a) The effective potential $V_\phi$ of the scalar field, in terms of the
radial coordinate, and (b) the coefficient $d_1$ 
in terms of the mass $M$, for various forms of $f(\phi)$.}
   \label{V-d1}
\end{figure}

It is of particular interest to study also the behaviour of the effective potential of the scalar
field, a role that in our theory is played by the GB term together with the coupling function,
i.e. $|V_\phi| \equiv \dot f(\phi)\,R^2_{GB}$. In  Fig. \ref{V-d1}(a), we present a combined graph
that displays its profile in terms of the radial coordinate, for a variety of forms of the
coupling function $f(\phi)$. As expected, the potential $V_\phi$ takes on its maximum
value always near the horizon of the black hole, where the GB term is also maximized and 
thus sources the non-trivial form of the scalar field. On the other hand, as we move
towards larger distances, $V_\phi$ reduces to an asymptotic constant value. Although
this asymptotic value clearly depends on the choice of the coupling function, its common
behaviour allows us to comment on the asymptotic behaviour of the scalar field at
large distances. Substituting a constant value $V_\infty$ in the place of $V_\phi$ in
the scalar-field equation (\ref{phi-eq3}), we arrive at the intermediate result
\beq
\partial_r \left[e^{(A-B)/2} r^2 \phi' \right] = -e^{(A+B)/2} r^2\,V_\infty\,.
\eeq
Then, employing the asymptotic forms of the metric functions at large distances
(\ref{alfar1})-(\ref{bfar1}), the above may be easily integrated with respect to the radial
coordinate to yield a form for the scalar field identical to the one given in 
Eq. (\ref{phi-far-Anti}). We may thus conclude that the logarithmic
form of the scalar field may adequately describe its far-field behaviour even beyond
the perturbative limit of very small $\alpha$. 

We now proceed to discuss the physical characteristics of the derived solutions. Due to
the large number of solutions found, we will present, as for $V_\phi$, combined graphs
for different forms of the coupling function $f(\phi)$. Starting with the scalar field, we
notice that no conserved quantity, such as a scalar charge, may be associated with
the solution at large distances in the case of asymptotically Anti-de Sitter black holes:
the absence of an ${\cal O}(1/r)$ term in the far-field expression (\ref{phi-far-Anti})
of the scalar field, that would signify the existence of a long-range interaction term,
excludes the emergence of such a quantity, even of secondary nature. One could attempt
instead to plot the dependence of the coefficient $d_1$, as a quantity that predominantly
determines the rate of change of the scalar field at the far field, in terms of the mass
of the black hole. This is displayed in  Fig. \ref{V-d1}(b) for the indicative
value $\Lambda=-0.1$ of the cosmological constant.
We see that, for small values of the mass $M$, this coefficient takes in general a non-zero
value, which amounts to having a non-constant value of the scalar field at the far-field
regime. As the mass of the black hole increases though, this coefficient asymptotically
approaches a zero value. Therefore, the rate of change of the scalar field at infinity
for massive GB black holes becomes negligible and the scalar field tends to a constant.
This is the `Schwarzschild-AdS regime', where the GB term decouples from the theory and
the scalar-hair disappears - the same behaviour was observed also in the case of the
asymptotically-flat GB black holes studied in the previous two chapters\cite{ABK1,ABK2} where, in the limit of large mass, all
of our solutions merged with the Schwarzschild ones.

\begin{figure}[t!] 
\begin{center}
\hspace{0.0cm} \hspace{-0.6cm}
\includegraphics[height=.24\textheight, angle =0]{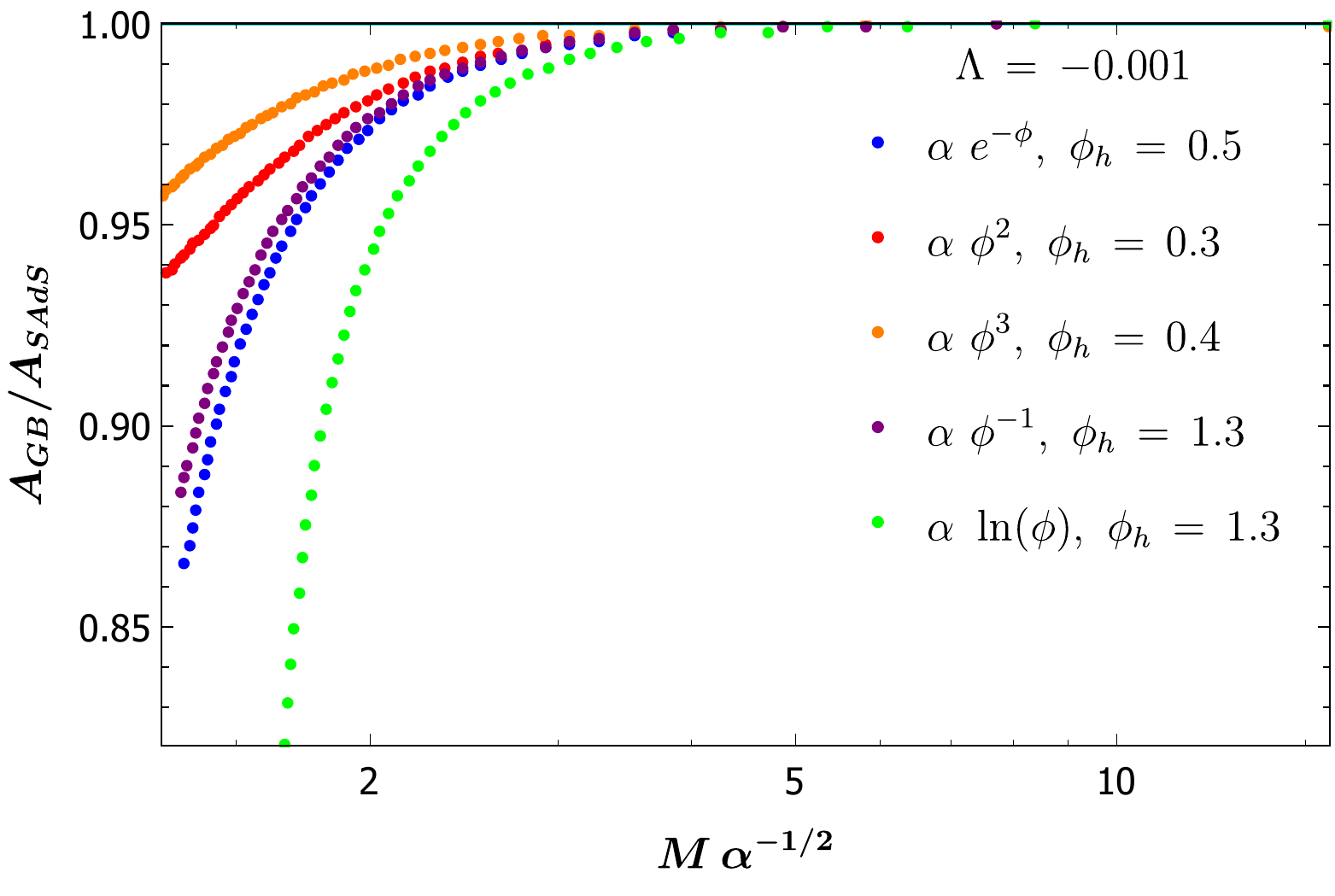}
\hspace{0.52cm} \hspace{-0.6cm}
\includegraphics[height=.24\textheight, angle =0]{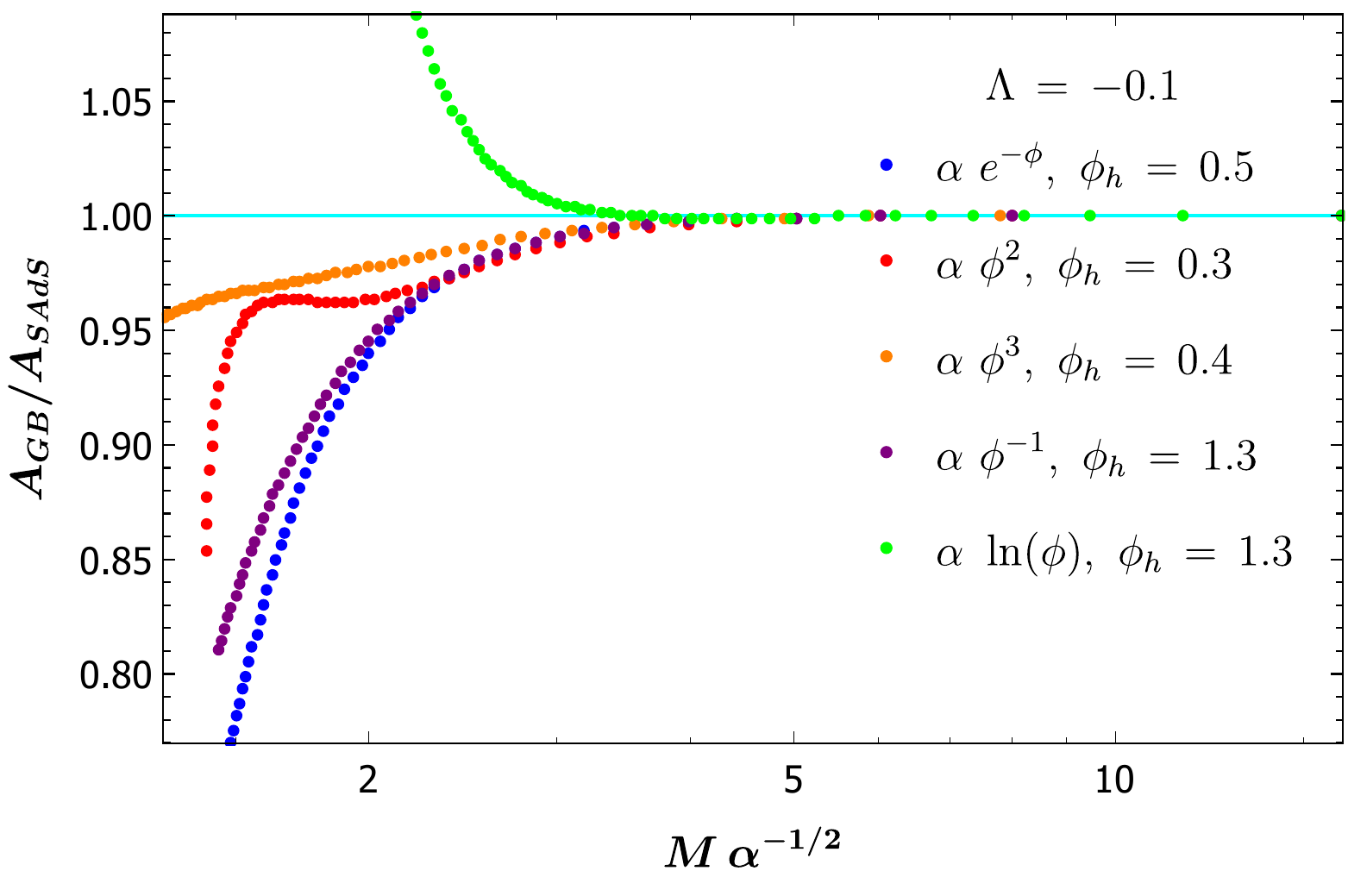}
\\
\hspace*{0.7cm} {(a)} \hspace*{7.5cm} {(b)}  \vspace*{-0.5cm}
\end{center}
\caption{The area ratio $A_{GB}/A_{SAdS}$ of our solutions as a function of the
mass $M$ of the black hole, for various forms of $f(\phi)$,  (a) for $\Lambda=-0.001$
 and (b) $\Lambda=-0.1$.}
   \label{Area}
\end{figure} 

We present next the ratio of the horizon area of our solutions compared to the
horizon area of the SAdS one with the same mass, for the indicative values of the
negative cosmological constant $\Lambda=-0.001$ and $\Lambda=-0.1$ in the
two plots of Fig. \ref{Area}. These plots provide further evidence for the merging of our
GB black-hole solutions with the SAdS solution in the limit of large mass. The left
plot of Fig. \ref{Area} reveals that, for small cosmological constant, all our GB solutions
remain smaller than the scalar-hair-free SAdS solution independently of the choice
for the coupling function $f(\phi)$ - this is in complete agreement with the profile
found in the asymptotically-flat case \cite{ABK1,ABK2}. This behaviour persists for even
larger values of the negative cosmological constant for all classes of solutions
apart from the one emerging for the logarithmic function whose horizon area
is significantly increased in the small-mass regime, as may be seen from the
right plot of Fig. \ref{Area}. These plots verify also the
termination of all branches of solutions at the point of a minimum horizon, or
minimum mass, that all our GB solutions exhibit as a consequence of the
inequality (\ref{C-def}). We also observe that, as hinted by the small-$\Lambda$
approximation given in Eq. (\ref{C-newdef}), an increase in the value of the 
negative cosmological constant pushes upwards the lowest allowed value of the
horizon radius of our solutions.

\begin{figure}[t!] 
\begin{center}
\hspace{0.0cm} \hspace{-0.6cm}
\includegraphics[height=.24\textheight, angle =0]{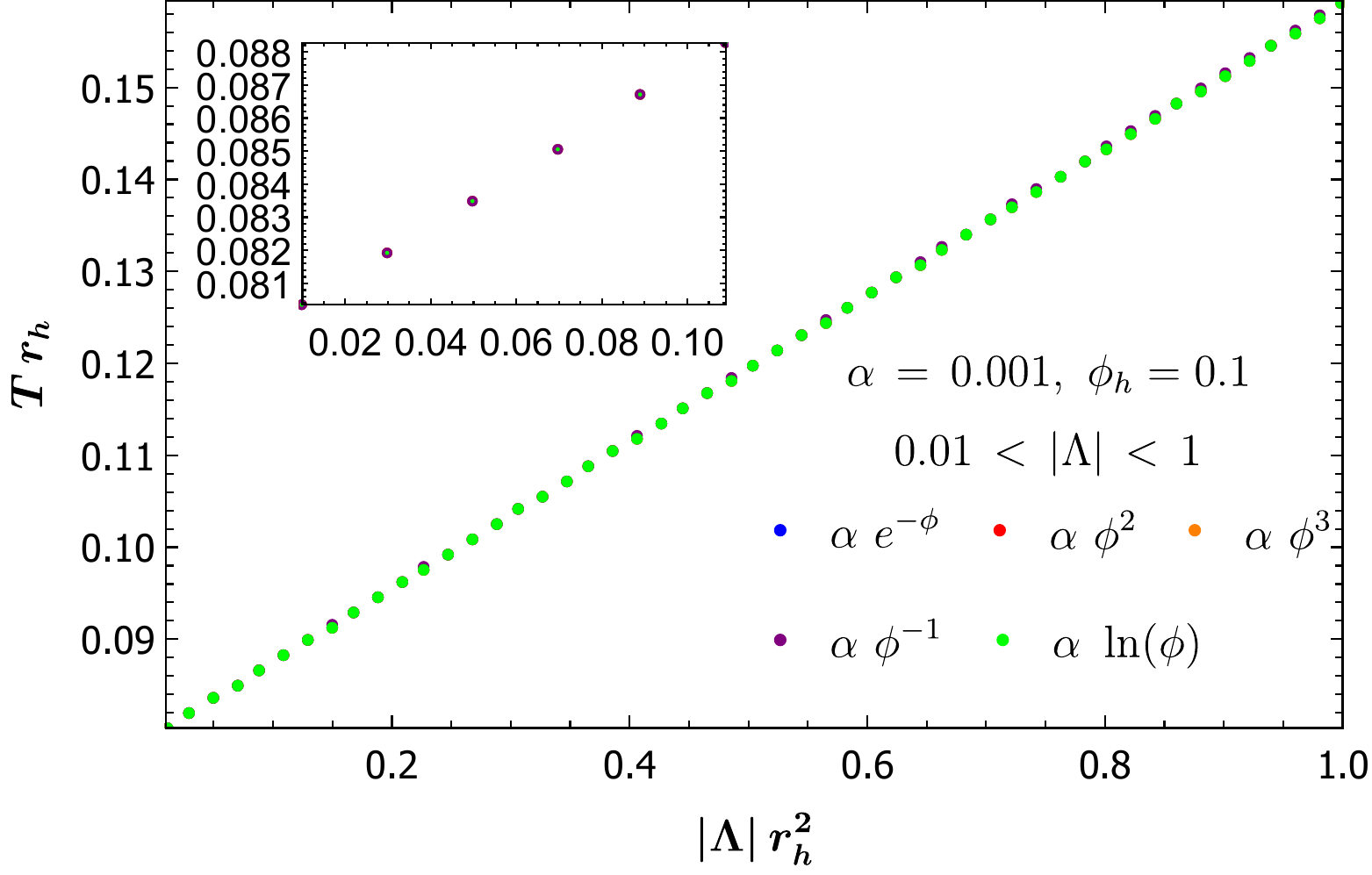}
\hspace{0.52cm} \hspace{-0.6cm}
\includegraphics[height=.24\textheight, angle =0]{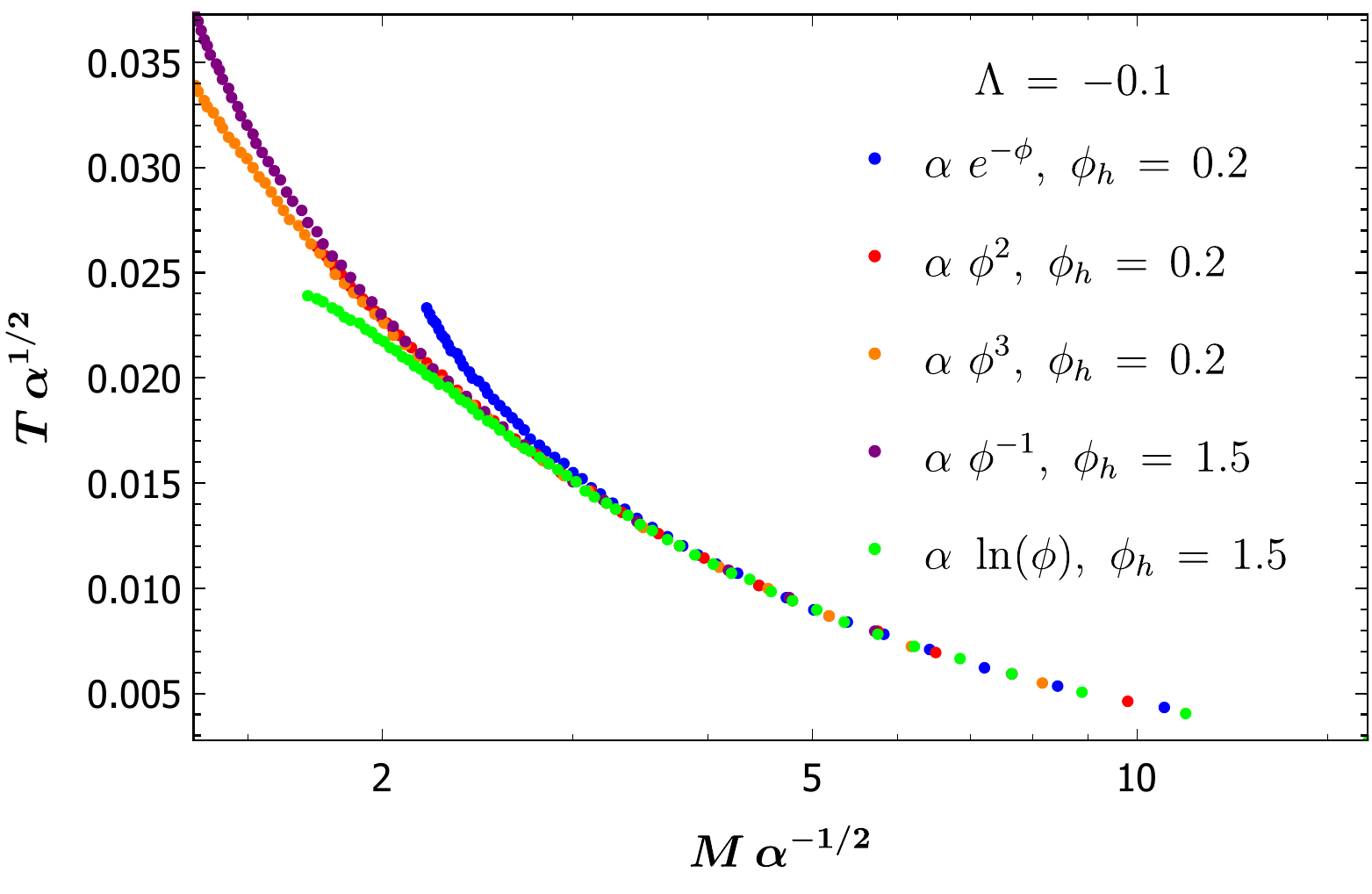}
\\
\hspace*{0.7cm} {(a)} \hspace*{7.5cm} {(b)}  \vspace*{-0.5cm}
\end{center}
\caption{(a) The temperature  $T$ of the black hole as a function of the cosmological
constant $\Lambda$  and (b) the mass $M$ of the black hole, for
various forms of $f(\phi)$.}
   \label{Temp-plot}
\end{figure}

We now move to the thermodynamical quantities of our black-hole solutions. We start
with their temperature $T$ given by Eq. (\ref{Temp-def3}) in terms of the near-horizon
coefficients $(a_1,b_1)$.  In   Fig. \ref{Temp-plot}(a), we display its
dependence in terms of the cosmological constant $\Lambda$, for several forms
of the coupling function. We observe that $T$ increases, too, with $|\Lambda|$;
we thus conclude that the more negatively-curved the spacetime is, the hotter
the black hole, that is formed, is. Note that the form of the coupling function
plays almost no role in this relation with the latter thus acquiring a universal 
character for all GB black-hole solutions. The dependence of the temperature
of the black hole on its mass, as displayed in  
Fig. \ref{Temp-plot}(b), exhibits a decreasing profile, with the obtained
solution being colder the larger its mass is. For small black-hole solutions,
the exact dependence of $T$ on $M$ depends on the particular form of the
coupling function but for solutions with a large mass its role becomes unimportant
as a common `Schwarzschild-AdS regime' is again approached.

\begin{figure}[t!] 
\begin{center}
\hspace{0.0cm} \hspace{-0.6cm}
\includegraphics[height=.24\textheight, angle =0]{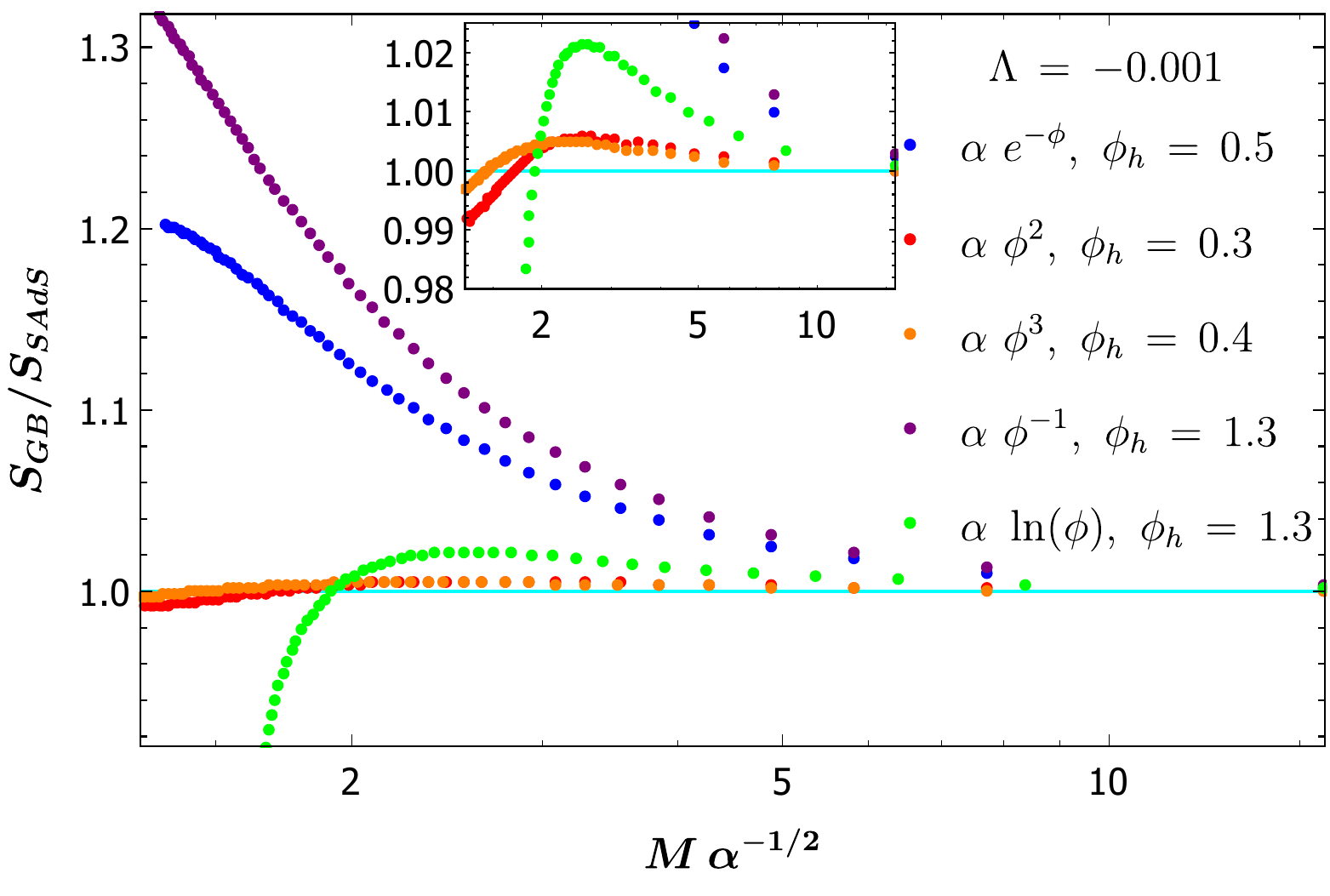}
\hspace{0.52cm} \hspace{-0.6cm}
\includegraphics[height=.24\textheight, angle =0]{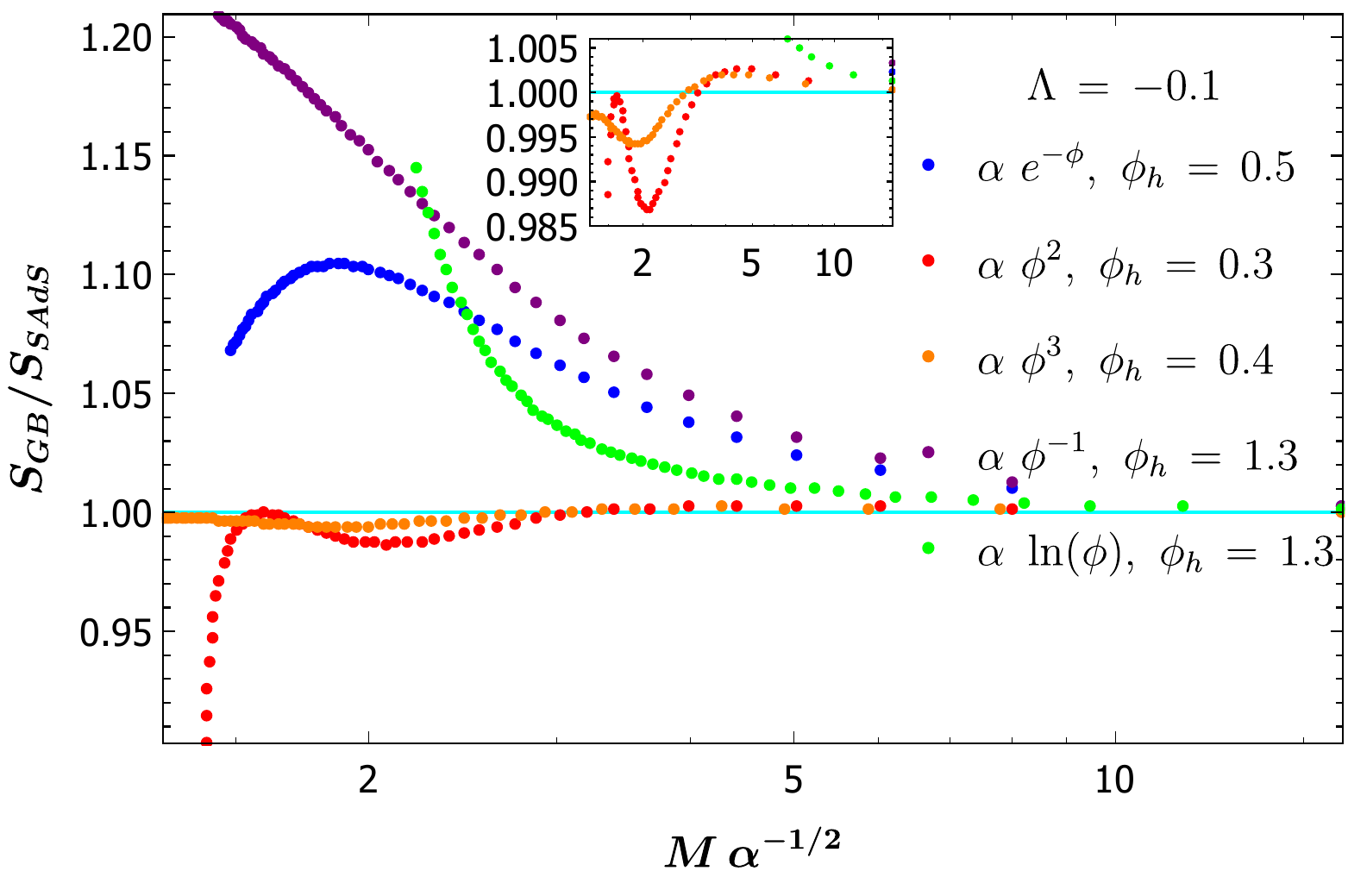}
\\
\hspace*{0.7cm} {(a)} \hspace*{7.5cm} {(b)}  \vspace*{-0.5cm}
\end{center}
\caption{The entropy ratio $S_{GB}/S_{SAdS}$ of our solutions as a function of the
mass $M$ of the black hole, for various forms of $f(\phi)$, (a) for $\Lambda=-0.001$
 and (b) $\Lambda=-0.1$.}
   \label{St3}
\end{figure}

Let us finally study the entropy of the derived black-hole solutions. In Fig. \ref{St3}, we
display the ratio of the entropy of our GB solutions over the entropy of the corresponding
Schwarzschild-Anti-de Sitter solution with the same mass, for the same indicative values
of the negative cosmological constant as for the horizon area. i.e. for
$\Lambda=-0.001$ (left plot) and $\Lambda=-0.1$ (right plot).
We observe that the profile of this quantity depends strongly on the choice of the coupling
function $f(\phi)$, for solutions with small masses, whereas in the limit of large mass,
where our solutions reduce to the SAdS ones, this ratio approaches unity as expected.
For small values of $\Lambda$, the left plot of Fig. \ref{St3} depicts a behaviour similar
to the one found in the asymptotically-flat case \cite{ABK1,ABK2}: solutions emerging for the
linear and the quadratic coupling functions exhibit smaller entropy compared to the
SAdS one, while solutions for the exponential, logarithmic and inverse-linear coupling
functions lead to GB black holes with a larger entropy over the whole mass range or
for particular mass regimes. As we increase the value of the cosmological constant
(see right plot of Fig. \ref{St3}), the entropy ratio is suppressed for all families of GB
black holes apart from the one emerging for the logarithmic coupling function, which
exhibits a substantial increase in this quantity over the whole mass regime. Together
with the solutions for the exponential and inverse-linear coupling functions, they have
an entropy ratio larger than unity while this ratio is now significantly lower than
unity for all the other polynomial coupling functions.
Although the question of the stability of the derived solutions
is an important one and must be independently studied for each family of solutions
found, the entropy profiles presented above may provide some hints regarding the
thermodynamical stability of our solutions compared to the Schwarzschild-Anti-de Sitter ones. 


\subsection{de Sitter Gauss-Bonnet Black Holes}


We now address the case of a positive cosmological constant, $\Lambda>0$. In order to find solutions with a positive cosmological constant we use a slightly different form for the metric
\begin{equation}\label{metric31}
{ds}^2=-e^{-2\d(r)}N(r){dt}^2+\frac{1}{N(r)}{dr}^2+r^2({d\theta}^2+\sin^2\theta\,{d\varphi}^2)\,,
\end{equation}
with 
\begin{equation}
N(r)=1-\frac{2m(r)}{r}-\frac{\L}{3}r^2.
\end{equation}
Using the expansions near the horizon (\ref{A-rh3})-(\ref{phi-rh3}) it is easy to find the corresponding asymptotic expressions for the new metric functions
\begin{align}
m(r)=\frac{r_h}{2}-\frac{r_h^3\L}{6}+m_1(r-r_h)+\mathcal{O}\left(\left(r-r_h\right)^2\right),\4\4\d(r)=\d_0+\mathcal{O}(r-r_h),\label{ho11}
\end{align}
where,
\begin{align}
m_1=-\frac{\left(  -1+r_h^2\L \right)\f'_h \dot{f}(\f_h)  }{r_h+2\f'_h \dot{f}(\f_h)}\label{ho12}.
\end{align}
The expansion for the scalar field Eq. (\ref{phi-rh3}) and the regularity constraint Eq. (\ref{solf}) remain unchanged.
We start our
integration process at a distance close to the black-hole horizon, using the asymptotic
solutions (\ref{phi-rh3}) and  (\ref{ho11})-(\ref{ho12}) and choosing $\phi_h$ to satisfy again the regularity
constraint (\ref{solf}). The coupling function $f(\phi)$ is assumed to take on a variety
of forms -- namely exponential, even and odd polynomial, inverse even and odd
polynomial, and logarithmic forms -- as in the case of the negative cosmological constant.
The numerical integration then proceeds outwards to meet the corresponding asymptotic
solution  near the cosmological horizon.

\begin{figure}[t!] 
\begin{center}
\hspace{0.0cm} \hspace{-0.6cm}
\includegraphics[height=.24\textheight, angle =0]{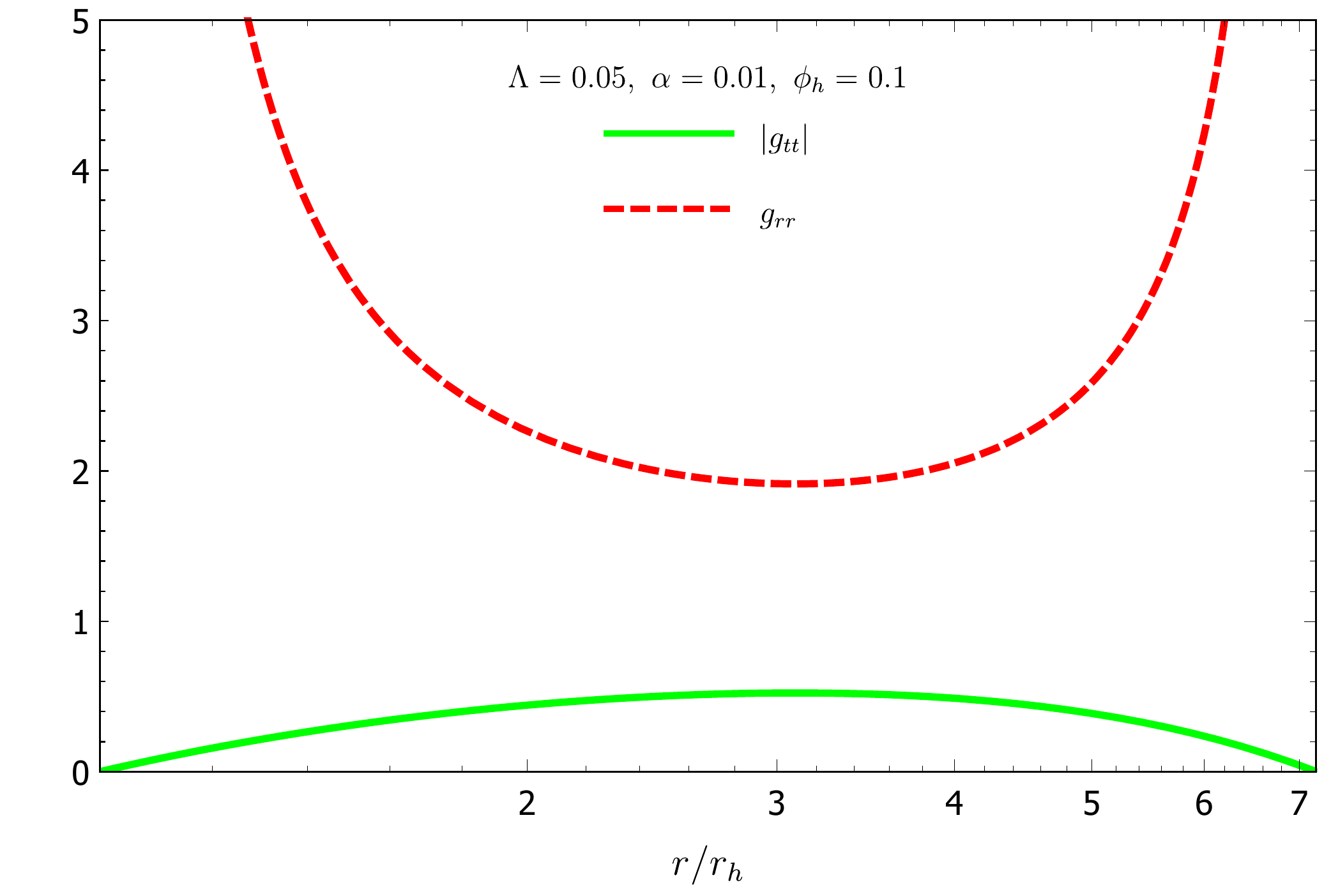}
\hspace{0.52cm} \hspace{-0.6cm}
\includegraphics[height=.24\textheight, angle =0]{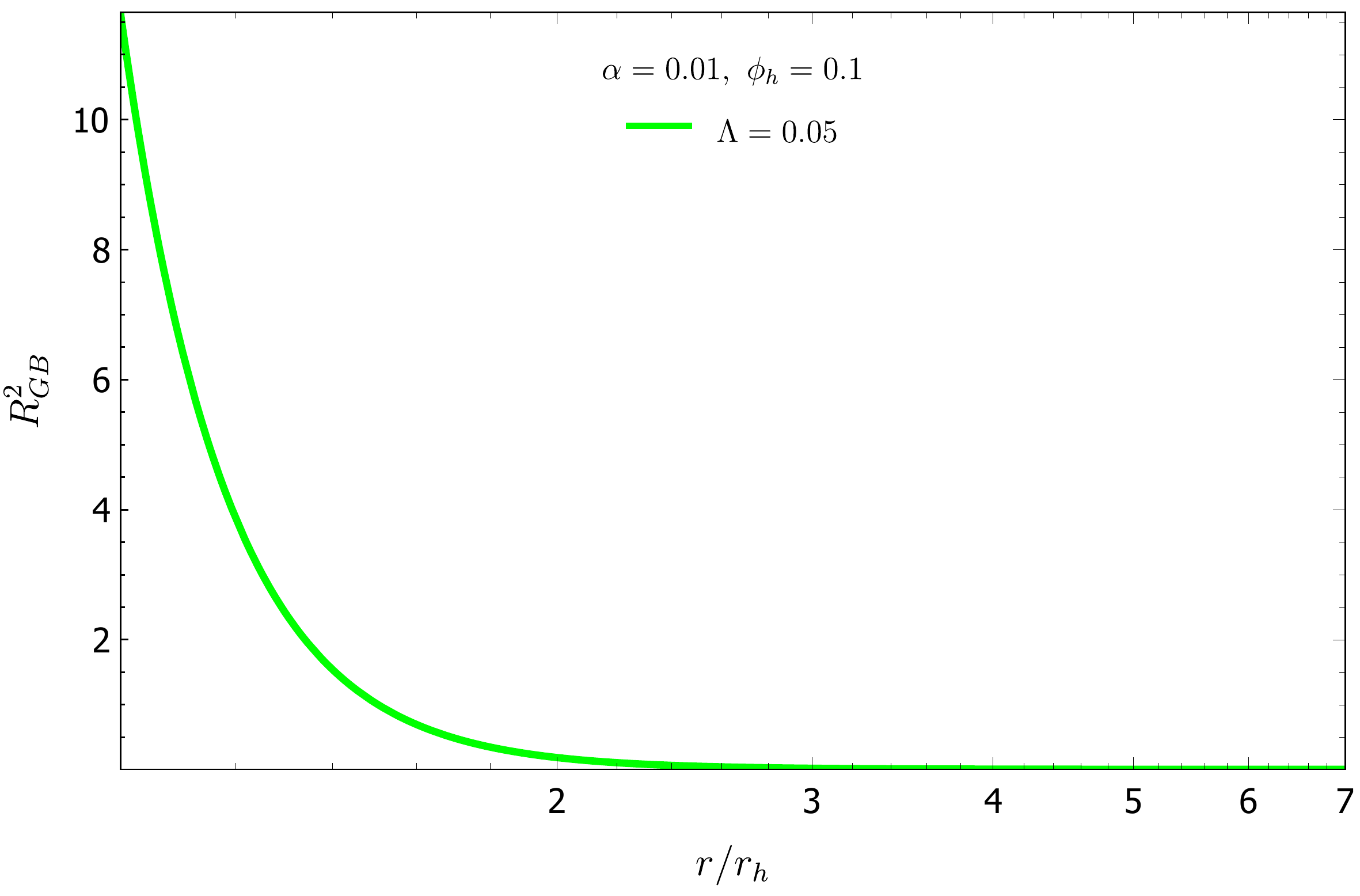}
\\
\hspace*{0.7cm} {(a)} \hspace*{7.5cm} {(b)}  \vspace*{-0.5cm}
\end{center}
\caption{(a) The metric functions  $|g_{tt}|$ and $g_{rr}$ of the spacetime 
 and (b) the Gauss-Bonnet term $R^2_{GB}$ in terms of the radial coordinate $r$, for
a positive cosmological constant and coupling function $f(\phi)=\alpha e^{-\phi}$.}
   \label{pos11}
\end{figure} 

In Fig. \ref{pos11}(a) we depict the metric functions of a Gauss-Bonnet  black-hole solution with a positive cosmological constant for the exponential coupling function. We note that both the metric functions have the anticipated behavior at the two asymptotic regions. The spacetime is asymptotically de Sitter and the location of the cosmological horizon for $\L=0.05$ is calculated numerically at $r_c/r_h=7.1967$. The regularity of the spacetime  is confirmed by the finiteness of the Gauss-Bonnet term  at both horizons and in the intermediate radial regime as depicted in Fig. \ref{pos11}(b). The de Sitter spacetime has a constant curvature everywhere given by Eq. (\ref{rimcu}) while the Gauss-Bonnet term near the cosmological horizon has the form $R^2_{GB}\approx 8\L^2/3$.  The plateau in the form of the Gauss-Bonnet term near the cosmological horizon in Fig. \ref{pos11}(b) also verifies the de Sitter behavior of the spacetime there.  Finally, the validity of the expression $R^2_{GB}\approx 8\L^2/3$  has been also verified numerically for our solutions. 

 \begin{figure}[t!] 
\begin{center}
\hspace{0.0cm} \hspace{-0.6cm}
\includegraphics[height=.24\textheight, angle =0]{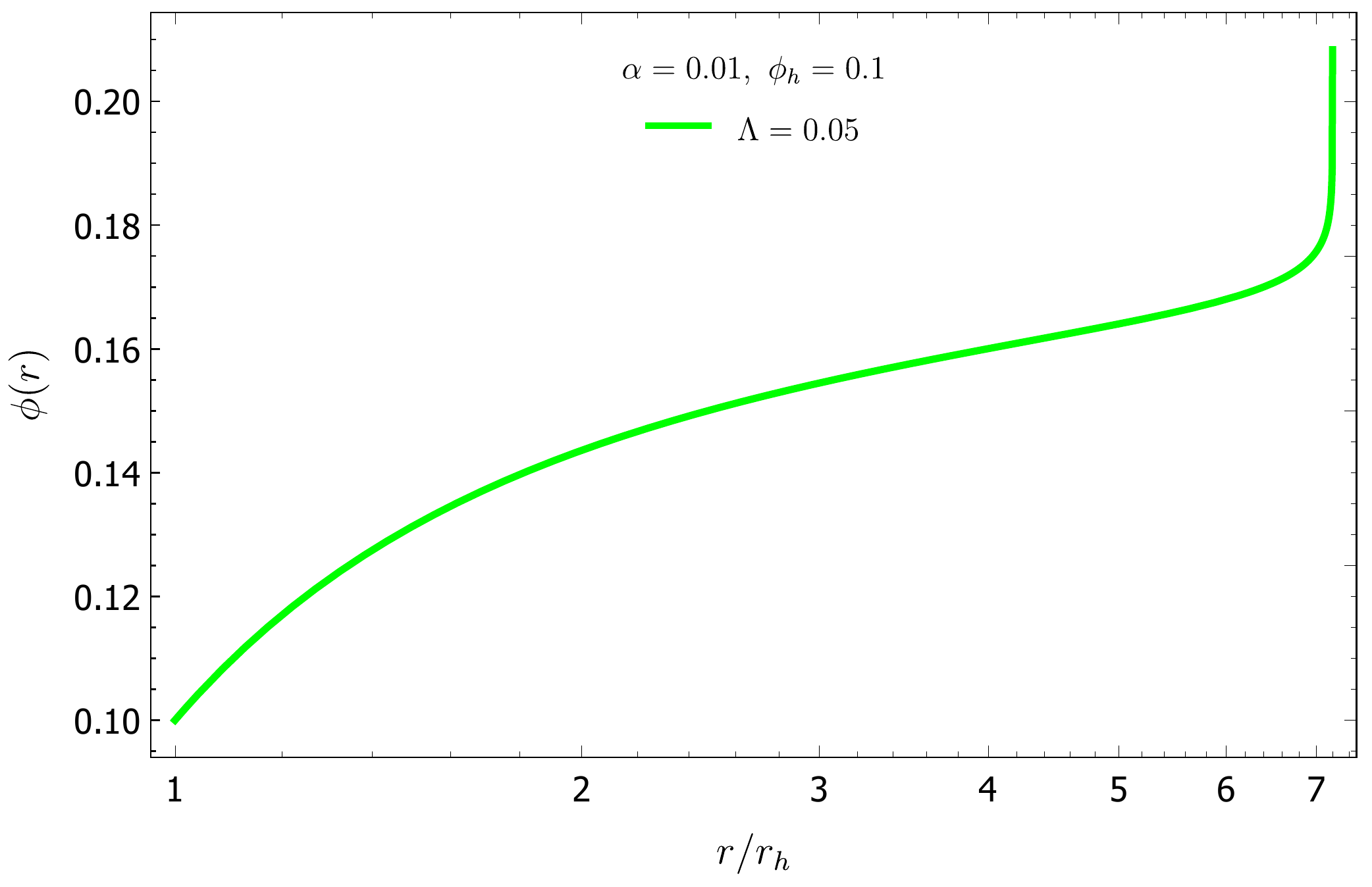}
\hspace{0.52cm} \hspace{-0.6cm}
\includegraphics[height=.24\textheight, angle =0]{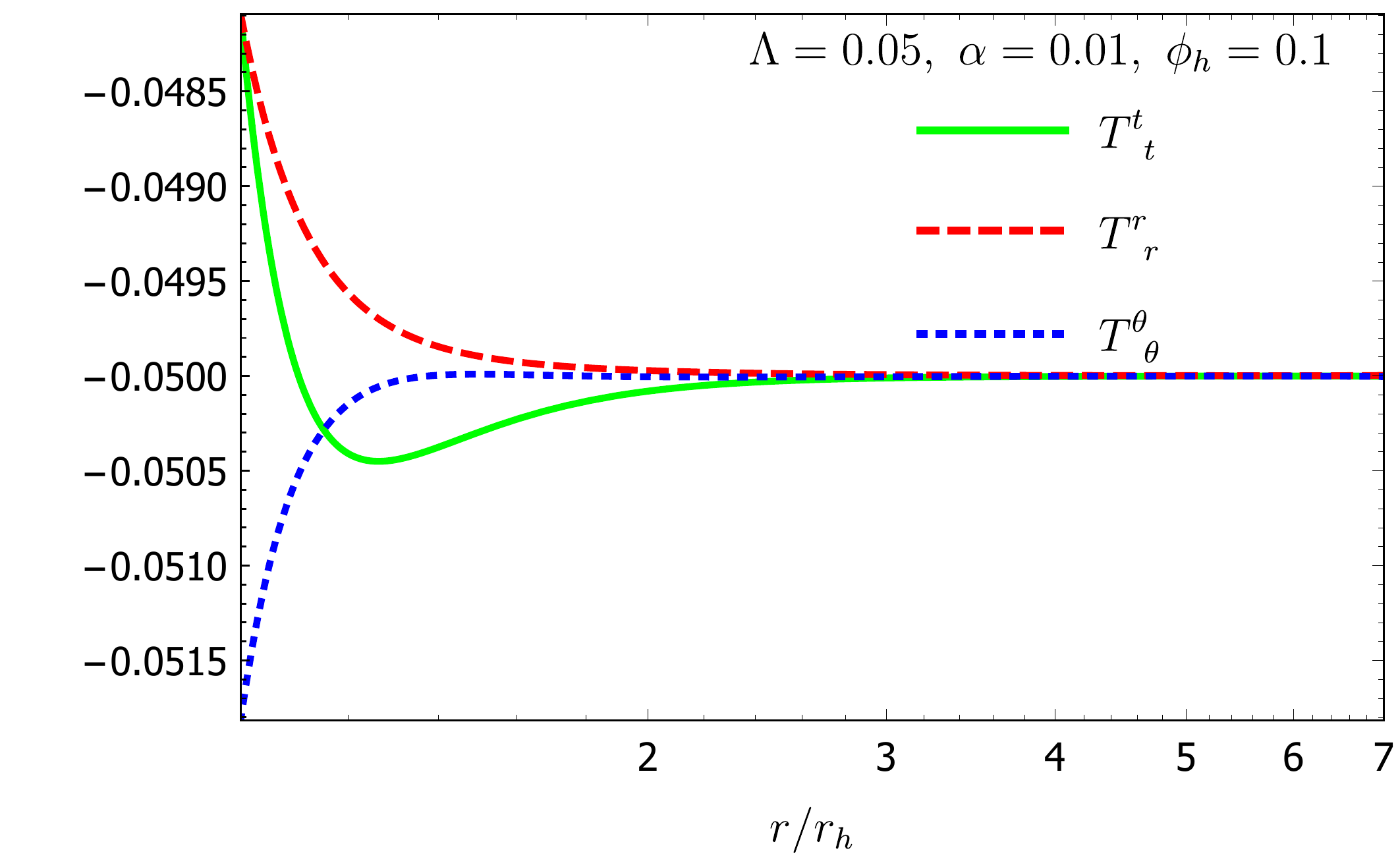}
\\
\hspace*{0.7cm} {(a)} \hspace*{7.5cm} {(b)}  \vspace*{-0.5cm}
\end{center}
\caption{(a) The scalar field $\f$ and (b) the components of the energy momentum tensor 
  in terms of the radial coordinate $r$, for
a positive cosmological constant and coupling function $f(\phi)=\alpha e^{-\phi}$.}
   \label{pos12}
\end{figure} 

The form of the scalar field is presented in Fig. \ref{pos12}(a). We observe that while the scalar field is regular at the horizon of the black hole, it diverges at the cosmological horizon.  Nevertheless,     the energy-momentum tensor shown in Fig. \ref{pos12}(b) is everywhere  regular as is also the scalar invariant Gauss-Bonnet term. 
   Using the small-$\a$ (large mass) approximation, we may investigate the divergence of the scalar field analytically. If the coupling constant is small $(\a/r_h^2\ll1)$  we may find perturbative solutions
around the Schwarzschild-de Sitter solution of the form
\begin{align}
g_{tt}&=-\left(1-\frac{2M}{r}-\frac{\L}{3}r^2\right)\left(1+\sum_{i=1}^\infty \a^i A_i(r)\right),\label{gttt}\\[3mm]
g_{rr}&=\left(1-\frac{2M}{r}-\frac{\L}{3}r^2\right)^{-1}\left(1+\sum_{i=1}^\infty \a^i B_i(r)\right),\label{grrr}\\[3mm]
\f(r)&=\f_0+\sum_{i=1}^\infty \a^i \f_i(r).\label{phiii}
\end{align}
By replacing the above expressions into the field equations and by performing the expansion we find that the field equations are satisfied with the first corrections of the metric functions being $A_1=0$ and $B_1=0$. The first correction of the scalar field $\f_1$ has a complicated form, however its first derivative is simpler and is given by
\begin{align}
\phi_1(r)=\frac{3 C a r^3 + 8 \dot f(\f_0) \left( \L^2  r^6 - 18 M^2\right)}{3 r^4\left(\L  r^3 +6M - 3r \right)},\label{dphiii}
\end{align} 
where $C$ is an integration constant. We observe that for a positive cosmological constant, the above equation diverges at both horizons. By fixing the value of the integration constant $C$ we may remove the divergence from the black-hole horizon but we cannot remove it simultaneously from the cosmological horizon. A physical divergence in the scalar field would result to a divergence in the trace of the energy momentum tensor making our spacetime  singular. However we already know from Fig. \ref{pos12}(b) that the energy-momentum tensor is everywhere regular. We can easily interpret this behavior by looking at the Einstein's field equations $G_{\m\n}=T_{\m\n}$: for our solutions Eqs. (\ref{gttt})-(\ref{grrr}), the left part of the Einstein's field equations $G_{\m\n}$ is everywhere regular (between the two horizons). Since Eq. (\ref{dphiii}) is a solution  of the Einstein's field equations and $G_{\m\n}$ is everywhere regular, the energy-momentum tensor $T_{\m\n}$ should be also regular. Similar results have been found for all the other forms of the coupling function $f(\f)$ that we have considered. The case of the positive cosmological constant was also investigated in \cite{Brihaye:2019gla} where solutions with an asymptotic de Sitter behavior were found exhibiting also a scalar-field singularity at a point beyond the cosmological horizon. Also, in the context of the EsGB theory,  particle-like solutions with a diverging scalar field at the origin and a regular energy-momentum tensor were found in \cite{Kleihaus:2019rbg, Kleihaus:2020qwo}.



\section{Ultra-compact solutions with a Self-interacting Scalar Field}\label{uco}

In a follow-up work \cite{bakoppk} we considered the Einstein-scalar-Gauss-Bonnet theory, and studied the case where the
negative cosmological constant --usually introduced
in an ad hoc way in the theory-- is replaced by a more realistic, negative scalar-field
potential $V(\f)$:
\begin{equation}
S=\frac{1}{16\pi}\int{d^4x \sqrt{-g}\left[R-\frac{1}{2}\,\partial_{\mu}\phi\partial^{\mu}\phi+f(\phi)R^2_{GB}
- 2\L V(\phi)\right]}.
\label{action22}
\end{equation}
The quantity $\L$ now has the role of a coupling constant. However, by setting $V(\f)=1$ in the above action we recover the action (\ref{action3}) in which $\L$ is treated as the cosmological constant. By varying the above action with respect to the metric tensor we obtain the Einstein's field equations that have the same form as the left expression in Eq. (\ref{field-eqs}). The effective energy-momentum tensor has the form of Eq. (\ref{Tmn}) with the change $\L\rightarrow \L V(\f)$. Finally the variation with respect to the scalar field leads to the following scalar field equation
\begin{equation}
\nabla^2 \phi+\dot{f}(\phi)R^2_{GB}-2\Lambda \dot V(\phi)=0\,. \label{phi-eq_011}
\end{equation}

We are interested in deriving regular, static, spherically-symmetric black-hole solutions thus we use the line element of Eq. (\ref{metric3}). By repeating the calculations of sections \ref{theo5} and \ref{2221} we find that the solutions near the horizon of the black hole have the same form as Eqs. (\ref{A-rh3}-\ref{phi-rh3}). However, in this case the regularity constraint for the scalar field demands that 
\begin{align}\label{solphi'22}
\f'_h=-\frac{r_h^3(1-\Lambda  V r_h^2) +16 \Lambda V r_h \dot f^2(3-\Lambda V r_h^2)
-8 \Lambda  \dot V r_h^3 \dot f \pm (1-\Lambda  V r_h^2)\sqrt{\tilde C}}
{4 \dot f \Bigl[r_h^2\,(1-4 \Lambda  \dot V \dot f)-\Lambda  V (r_h^4-16
   \dot f^2)\Bigr]},
\end{align}
where all quantities have been evaluated at $r=r_h$. The quantity $\tilde C$ is given by the expression
\begin{equation}\label{C-def22}
\tilde C=384 \Lambda  r_h^2 \dot f^3 \dot V+32 r_h^2 \dot f^2 \left(2
   \Lambda V r_h^2 -3\right)+256 \Lambda  V \dot f^4
   \left(\Lambda V r_h^2-6\right)+r_h^6\,.
\end{equation}
%
\begin{figure}[t!] 
\begin{center}
\hspace{0.0cm} \hspace{-0.6cm}
\includegraphics[height=.24\textheight, angle =0]{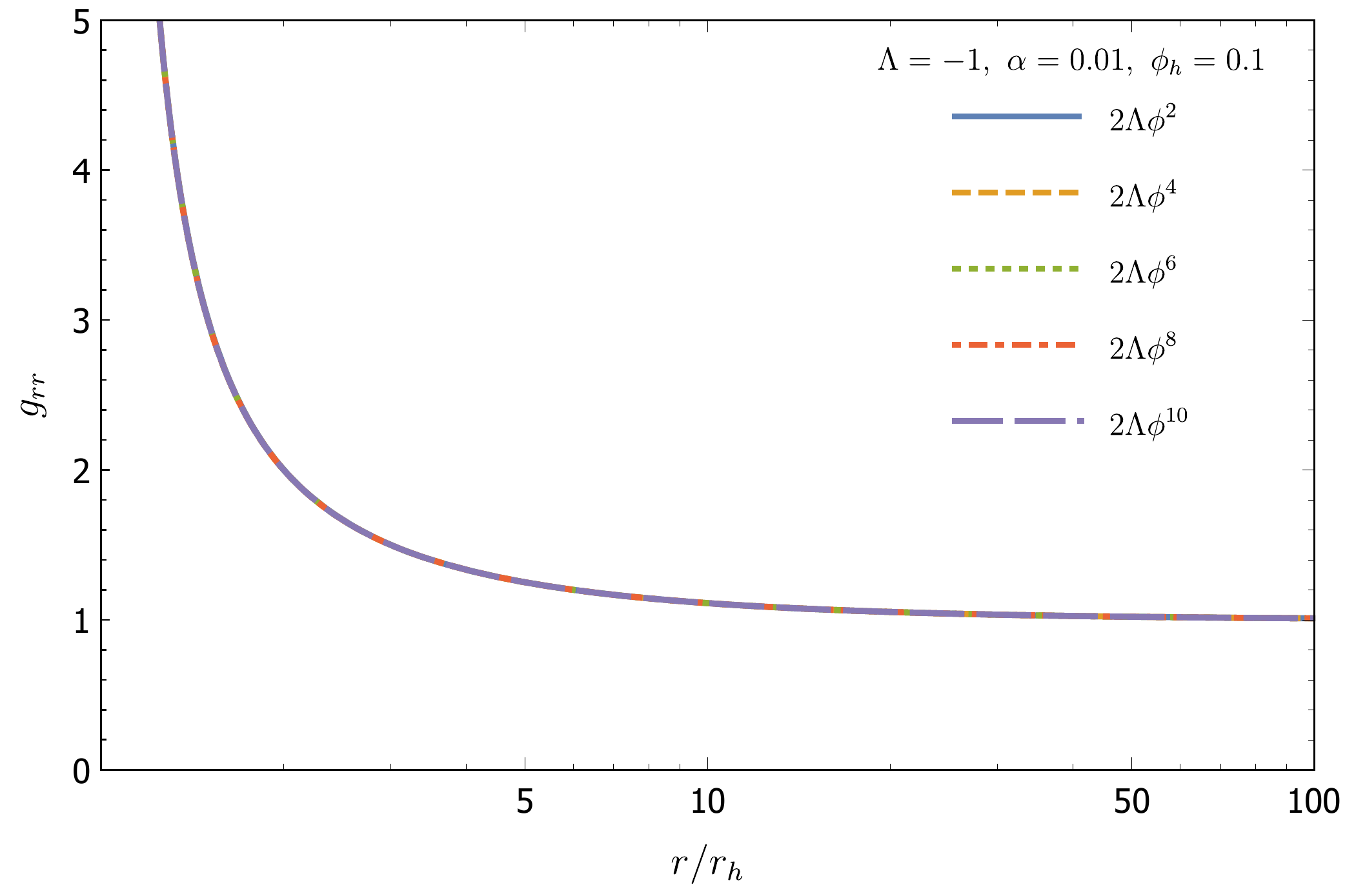}
\hspace{0.52cm} \hspace{-0.6cm}
\includegraphics[height=.24\textheight, angle =0]{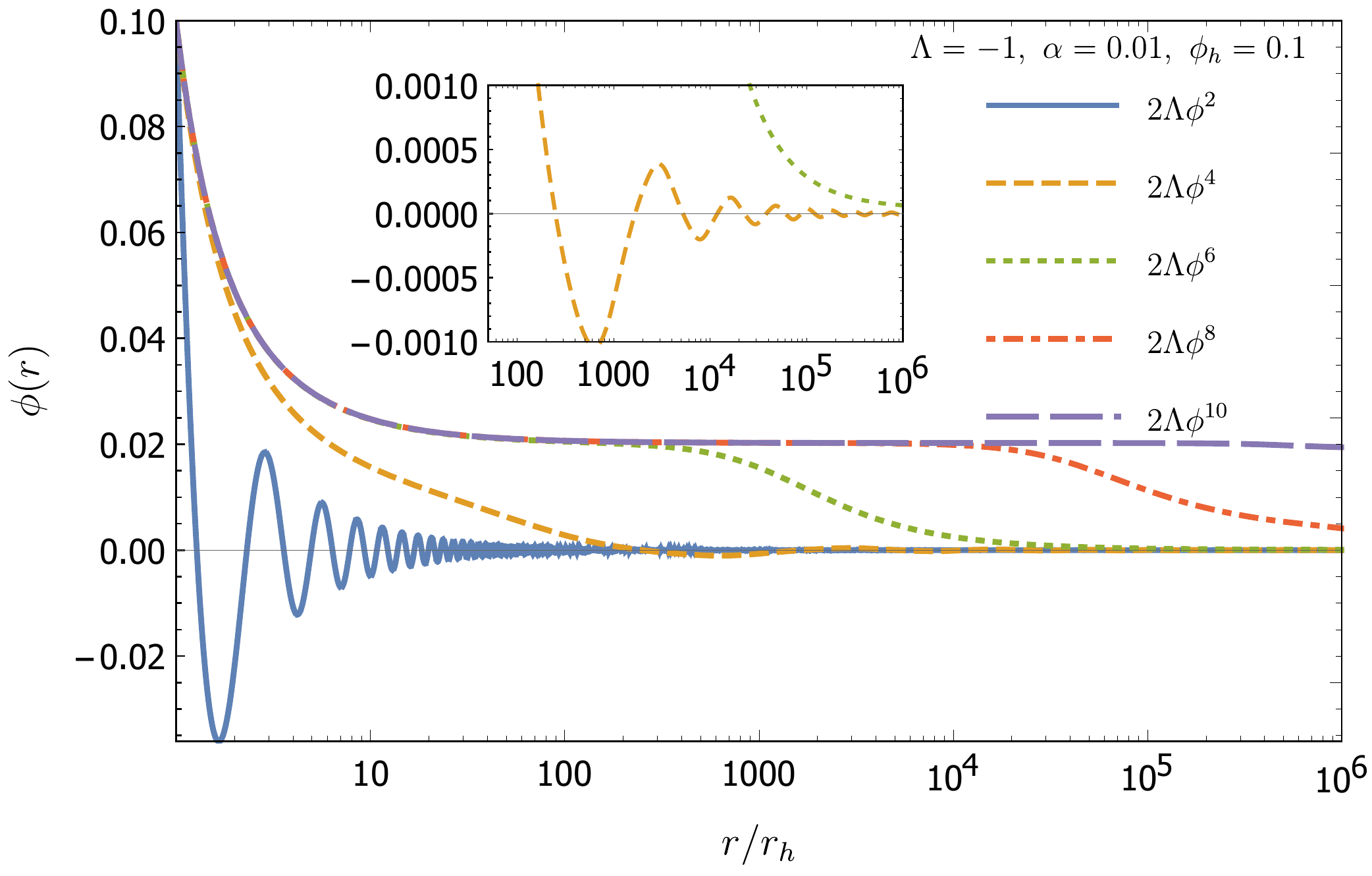}
\\
\hspace*{0.7cm} {(a)} \hspace*{7.5cm} {(b)}  \vspace*{-0.5cm}
\end{center}
\caption{(a) The $g_{rr}$ metric component and (b) the scalar field $\f$ in terms of
the radial coordinate $r$, for $f(\phi)=\alpha e^{\phi}$ and a variety of potentials $V(\phi)$
(with $\Lambda=-1$).}
\label{ulco1}
\end{figure} 
In the other asymptotic region, the form of the solutions depend on the behavior of the effective potential $V_{eff}= -f(\phi)\,R^2_{GB} + 2\Lambda\,V(\phi)$. In principle, the effective potential could approach  a constant value at infinity which would serve  as an effective cosmological constant $\L_{eff}=\L V^\infty_{eff}$. In this case, the metric would be asymptotically (anti-)de Sitter. However, in \cite{bakoppk} we studied the case  of power law potential for the scalar field and the effective potential vanishes at infinity $V^\infty_{eff}=0$ resulting to asymptotically flat black-hole solutions. Substituting a Schwarzschild-type metric form into the scalar equation (\ref{phi-eq_011}) we find the differential equation obeyed by the scalar field at infinity
\begin{equation}
r \phi'' + 2 \phi'-2 r \Lambda \frac{dV}{d\phi}=0\,.\label{phifty}
\end{equation}
The first form of the potential that one should use is a mass term i.e. $V(\phi)=\phi^2$ with $\Lambda=m^2/2$. This case was investigated in \cite{Doneva:2019vuh}. However, in \cite{bakoppk} we considered even polynomial potentials $V(\f)=\phi^{2n}$ with $\L<0$. For the case with $V(\phi)=\phi^2$ and $\Lambda=-m^2/2$, the solution of Eq. (\ref{phifty}) takes the form 
\begin{equation}
\phi(r) \simeq \frac{1}{r}\,[C_1 \cos (m r) + C_2\,\sin (mr)]\,.
\label{phi-neg-mass}
\end{equation}
Unfortunately, Eq. (\ref{phifty})
cannot be analytically solved for any of the other forms of the scalar potential $V(\phi)$ that we studied in \cite{bakoppk}. Nevertheless, the behavior found considerably resembles the one for the negative quadratic one. 

In order to construct the black-hole solutions in the context of the theory (\ref{action22}) we follow a similar procedure with the one presented in section \ref{num5}. In Fig. \ref{ulco1} (a) we depict solutions for the $g_{rr}$ component of the metric for the exponential coupling function and many different choices for the potential. We observe that the $g_{rr}$ component has the anticipated behavior, diverging near the black-hole horizon   while   approaching  the Minkowski limit at large distances. The metric function $|g_{tt}|$ --which is not presented there-- has also the anticipated behavior, vanishing in the near horizon region and approaching the Minkowski spacetime at infinity. The behavior of the metric functions both at small and large distances from the black-hole horizon is in fact independent of the form of the scalar-field potential $V(\phi)$. The profile of the scalar field is depicted in  Fig. \ref{ulco1} (b), for a variety of forms of its potential $V(\phi)$. The scalar field is regular near the black-hole horizon independently of the form of its potential. However, the subsequent evolution does strongly depend on the particular form of $V(\phi)$. We observe that, despite the negative sign of $\Lambda$, the profile of the scalar field remains finite in the entire radial regime. For all of the forms of $V(\phi)$ employed, the scalar field decreases at the near-horizon regime and reduces to a  vanishing value at asymptotic infinity. The  solution for the quadratic potential justifies the oscillatory behavior given in Eq. (\ref{phi-neg-mass}) as is clearly observed in the lower curve of  Fig. \ref{ulco1} (b), which corresponds to this case. Finally Fig. \ref{ulco1} (b) implies that the scalar field decreases asymptotically to zero in an oscillatory way also for the cases with $n>1$; however, as $n$ increases, the frequency of the oscillations is strongly damped and the asymptotic vanishing value is reached at an increasingly larger distance. Similar results are found for  alternative forms of the coupling function $f(\f)$.

\begin{figure}[t!] 
\begin{center}
\hspace{0.0cm} \hspace{-0.6cm}
\includegraphics[height=.24\textheight, angle =0]{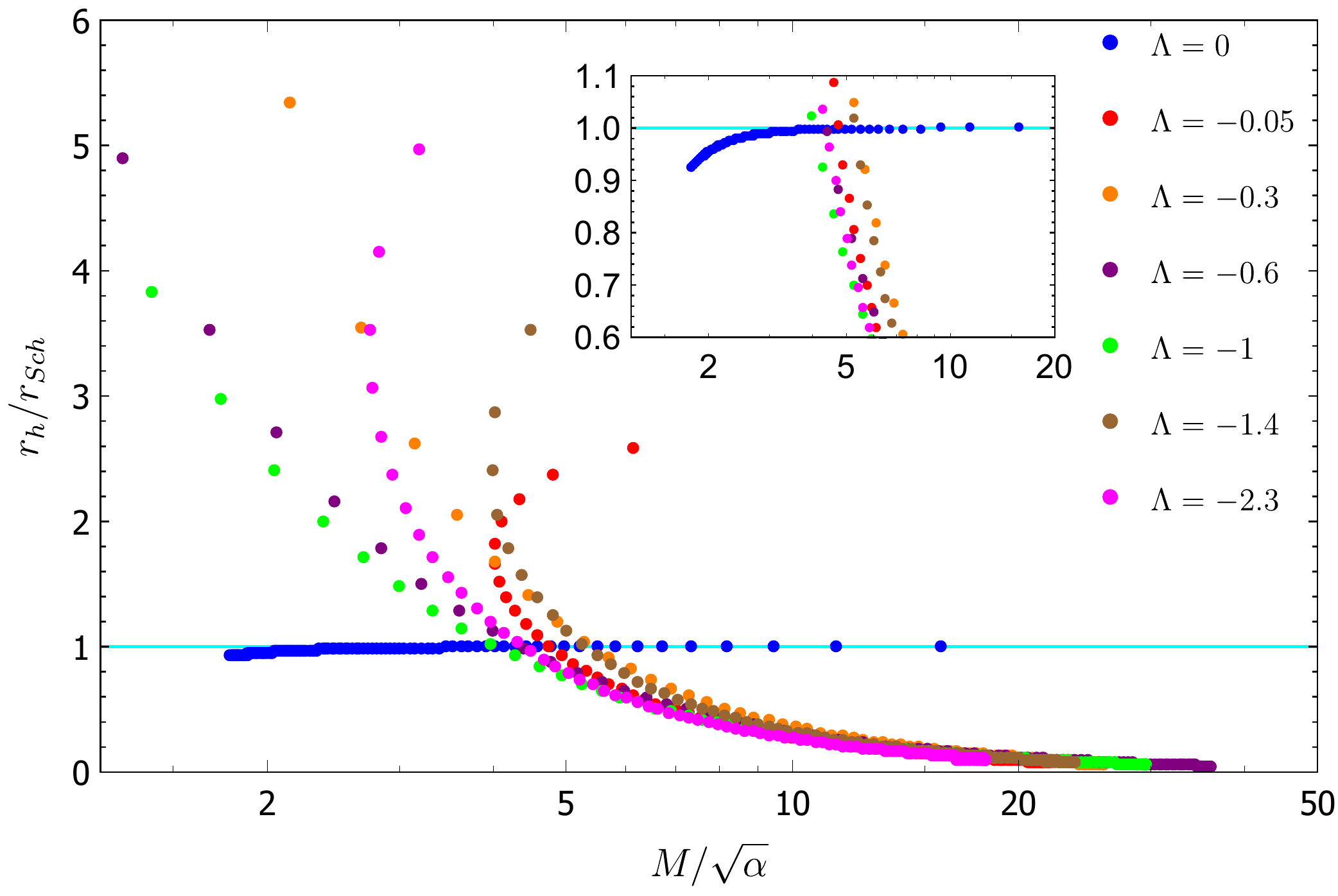}
\hspace{0.52cm} \hspace{-0.6cm}
\includegraphics[height=.24\textheight, angle =0]{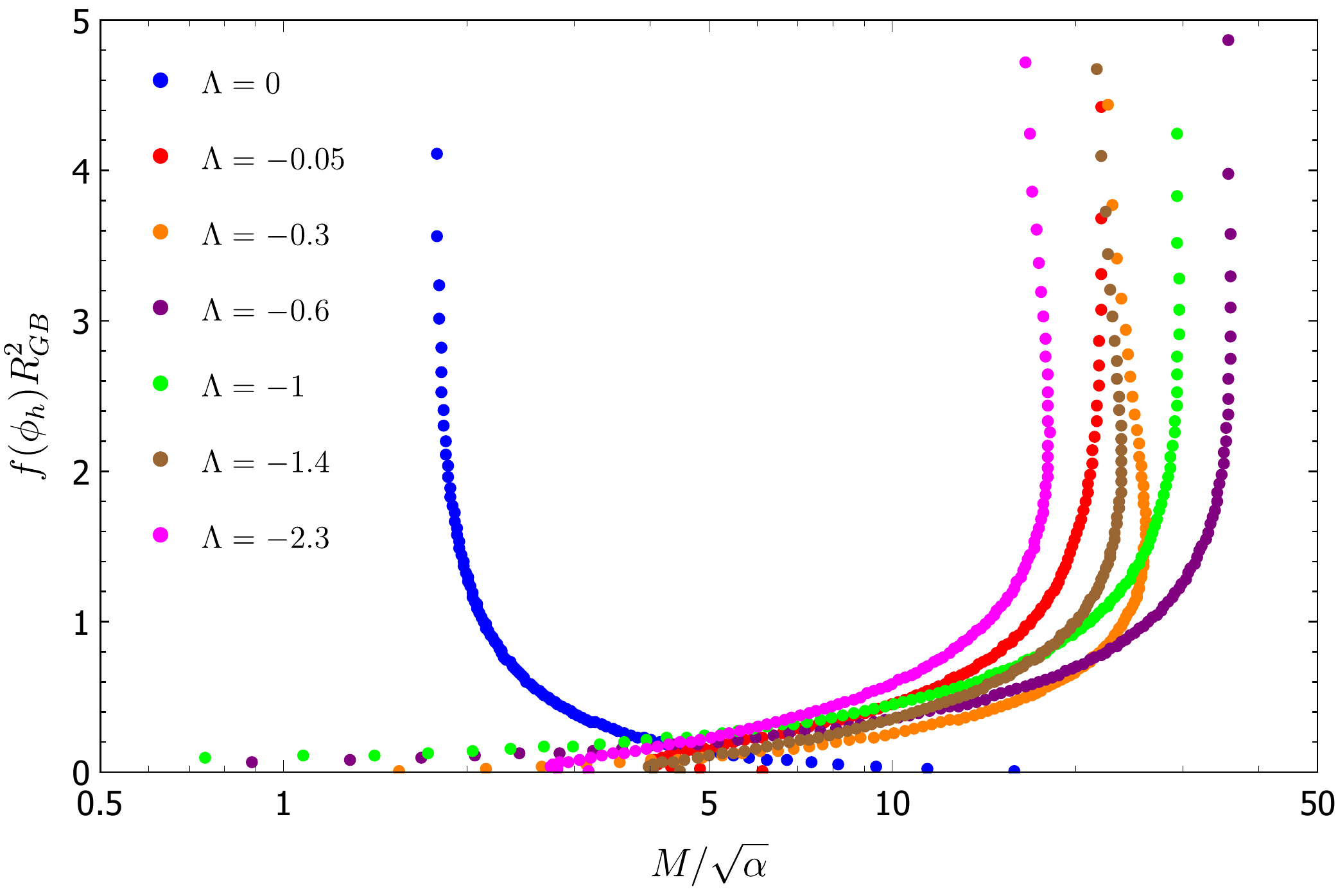}
\\
\hspace*{0.7cm} {(a)} \hspace*{7.5cm} {(b)}  \vspace*{-0.5cm}
\end{center}
\caption{(a) The horizon radius $r_h$  normalised
with respect to the corresponding Schwarzschild values and (b) the combination $f(\phi)\,R^2_{GB}$ at the horizon radius in terms of the mass $M$, for $f(\phi)=\alpha e^{\phi}$, 
$V(\phi)=\phi^2$ and various values of $\Lambda<0$. The blue dots correspond
to the dilatonic black holes with no potential ($\Lambda=0$).}
\label{ulco2}
\end{figure} 

In Fig. \ref{ulco2} (a) , we plot the horizon radius $r_h$  normalised with respect to the corresponding Schwarzschild values, for solutions obtained for $f(\phi)=\alpha e^{\phi}$, $V(\phi)=\phi^2$ and various negative values of the parameter $\Lambda$. The case with $\Lambda=0$ corresponding to the asymptotically-flat dilatonic GB black holes with no potential, presented in chapters \ref{3} and \ref{4}, \cite{DBH1, ABK1, ABK2} is also shown for comparison. From  Fig. \ref{ulco2} (a), we observe that whereas the dilatonic black holes are always smaller than the corresponding Schwarzschild black holes, the black-hole solutions with a negative quadratic potential may be either smaller, equal or larger than the Schwarzschild solution with respect to their horizon values. 
 The most distinctive feature of the case with a negative quadratic potential is the behavior  of the black-hole solutions in the limit of large mass. We observe that, as the mass of the black-hole solution increases, the horizon radius decreases reaching eventually a very small value. Thus, in the limit of large mass, a branch of massive,  ultra-compact black holes seems to emerge.
 
The branch of the massive ultra-compact GB black holes is also present for every non-vanishing, negative value of  the coupling parameter $\Lambda$. In fact, the exact value of $\Lambda$ determines the value of $M$ where this branch terminates. As the mass $M$ increases, the horizon radius gets smaller and the question emerges of whether the end point is indeed a black hole with a small but non-vanishing horizon radius or perhaps a naked singularity. To investigate this, in Fig. \ref{ulco2} (b)   we plot the values of the GB term at the location of the horizon (multiplied by $f(\phi_h)$ for scaling purposes) in terms of the black-hole mass $M$ of the solutions. The case of the dilatonic black holes with $\Lambda=0$ is again shown for comparison.  We observe that, in the presence of a negative potential, a reversal takes place in the behaviour of the GB term: for $\Lambda=0$, it is the low-mass GB black-hole solutions that are the most compact, and therefore create the most curved background around them; for $\Lambda <0$, on the other hand, it is the black-hole solutions near the end point of the branch of ultra-compact objects, i.e. the black holes with the smallest horizon radius and the largest mass, that naturally create the most curved background. Although the GB term reaches a large value, it never becomes infinite -- the solutions at the end of the branch have a GB term at their horizon which is comparable to the one for a dilatonic black hole with no potential. This signifies the fact that the horizon radius, although  very small, remains in fact non-vanishing and therefore these massive, ultra-compact objects are extremely small-sized black holes. For all the other forms of the coupling function $f(\f)$ we considered and for a negative quadratic potential, we also found ultra-compact solutions. However the most compact solutions appear for the linear and the exponential coupling functions.   Finally, for all other forms of the scalar-field potential $V(\f)$, this branch of solutions is absent since the horizon radius approaches, in the limit of large mass, values which are very close to the Schwarzschild ones.

\section{Discussion}\label{con5}

In this chapter, we have extended the analysis of the  previous two chapters \cite{ABK1, ABK2}, on the emergence of
novel, regular black-hole solutions in the context of the Einstein-scalar-Gauss-Bonnet theory, to include
the presence of a positive or negative cosmological constant. Since the uniform distribution
of energy associated with the cosmological constant permeates the whole spacetime, we
expected $\Lambda$ to have an effect on both the near-horizon and far-field asymptotic
solutions. Indeed, our analytical calculations in the small-$r$ regime revealed that the 
cosmological constant modifies the constraint that determines the value of $\phi'_h$ for
which a regular, black-hole horizon forms. In addition, it was demonstrated that
such a horizon is indeed formed, for either positive or negative $\Lambda$ and for all
choices of the coupling function $f(\phi)$. 

In contrast, the behaviour of the solution in the far-field regime depended strongly on
the sign of the cosmological constant. For $\Lambda>0$, a second horizon, the cosmological
one, was expected to form at a distance $r_c>r_h$, whereas for $\Lambda<0$, an Anti-de Sitter
type of solution was sought for at asymptotic infinity. Both types of solutions were analytically
shown to be admitted by the set of our field equations at the limit of large distances, thus
opening the way for the construction of complete black-hole solutions with an 
(Anti)-de Sitter asymptotic behaviour. 

The complexity of the field equations prevented us from constructing such a solution
analytically, therefore we turned to numerical analysis. Using our near-horizon analytic
solution as a starting point, we integrated the set of field equations from the black-hole
horizon and outwards. For a negative cosmological constant ($\Lambda<0$), we 
demonstrated that regular black-hole solutions with an Anti-de Sitter-type asymptotic
behaviour arise with the same easiness that their asymptotically-flat counterparts emerge. 
We have produced solutions for an exponential, polynomial (even or odd), inverse
polynomial (even or odd) and logarithmic coupling function between the scalar field and
the GB term. In each and every case, once $f(\phi)$ was chosen, selecting the input
parameter $\phi_h'$ to satisfy the regularity constraint (\ref{solf}) and the second
input parameter $\phi_h$ to satisfy the inequality (\ref{C-def}) a regular black hole
solution always emerged. The metric components exhibited the expected behaviour near
the black-hole and asymptotic infinity with the scalar invariant GB term being everywhere
regular. All solutions possessed non-trivial scalar hair, with the scalar field having a
non-trivial profile both close to and far away from the black-hole horizon. For small
negative values of $\Lambda$, we recovered the power-law fall-off of the scalar field
at infinity, found in the asymptotically-flat case \cite{ABK1, ABK2} whereas for large
negative values of $\Lambda$ the profile of $\phi$ was dominated by a logarithmic
dependence on the radial coordinate. This behaviour was analytically shown to emerge
both in the linear coupling-function case and in the perturbative limit, in terms of the
coupling parameter $\alpha$, but it was numerically found to accurately describe all
of our solutions at large distances.  

The absence of a $(1/r)$-term in the expression of the scalar field at large
distances excludes the presence of a scalar charge, even a secondary one. The 
coefficient $d_1$ in front of the logarithmic term in the expression of $\phi$ can
give us information on how much the large-distance behaviour of the scalar
field deviates from the power-law one valid in the asymptotically-flat case. We have
found that this deviation is stronger for GB black holes with a small mass 
whereas the more massive ones have a $d_1$ coefficient that tends to zero.
The temperature of the black holes was found to increase with the cosmological
constant independently of the form of the coupling function. The latter plays a
more important role in the relation of $T$ with the black-hole mass: while 
the temperature decreases with $M$ for all classes of solutions found, the
lighter ones exhibit a stronger dependence on $f(\phi)$. The same dependence
on the form of the coupling function is observed in the entropy and horizon area
of our solutions. For small masses, the entropy of each class of solutions has 
a different behaviour, with the ones for the exponential, inverse-linear polynomial and
logarithmic coupling functions exhibiting a ratio $S_{GB}/S_{SAdS}$ (over the
entropy of the Schwarzschild-Anti-de Sitter black hole with the same mass)
larger than unity for the entire mass range, for large values of $\Lambda$.
This feature hints towards the enhanced thermodynamical stability of our
solutions compared to their General Relativity (GR) analogues.
In the limit of large mass, the entropy of all classes of our solutions tend to the one of 
the Schwarzschild-Anti-de Sitter black hole with the same mass. The same holds
for the horizon area: while for small masses, each class has its own pattern
with $M$, will all solutions being smaller in size than the corresponding SAdS one
apart from the logarithmic case, for large masses all black-hole solutions match
the horizon area of the SAdS solution. 

Based on the above, we conclude that our GB black-hole solutions with a negative
cosmological constant smoothly merge with the SAdS ones, in the large mass limit. 
As in the asymptotically-flat case, it is the small-mass range that provides
the characteristic features for the GB solutions. These solutions have a modified
dependence of both their temperature and horizon area on their mass compared to
the SAdS solution. Another characteristic is also the minimum horizon, or minimum
mass, that all our GB solutions possess due to the inequality (\ref{C-def}). Finally, we have demonstrated that the general classes of theories that contain
the GB term and lead to novel black-hole solutions, continue to do so even in the
presence of a negative cosmological constant in the theory.  A further investigation is clearly necessary since the relevance of the
GB solutions with an Anti-de Sitter-type asymptotic solution in the context of the AdS-CFT
correspondence should be inquired.

For the case of the positive cosmological constant $\L>0$ we also found numerically asymptotically de Sitter solutions for black holes.  While these solutions are characterized by   regular   metric functions at both horizons, the solution for the scalar field diverges at the cosmological horizon. However, the components of the energy momentum tensor are everywhere regular. In order to investigate further the behavior of the scalar field we used the small$-\a$ approximation in order to construct analytic solutions. We found that the regularity of the metric ensures the regularity of the energy momentum tensor. 
The observable quantities of our solutions are the components of the energy momentum tensor and not the   value of the scalar field. Therefore, since the divergence of the scalar field does not affect the observable quantities of the spacetime  we conclude that this behavior is not problematic.  

Finally, we have also investigated the case of the EsGB theory where the scalar field possesses also a self-interacting potential \cite{bakoppk}. This potential was considered to be negative and thus to have the opposite sign in the action compared to a traditional self-interacting term. This choice was motivated by our desire to investigate whether this theory, which yielded a plethora of novel black-hole solutions with a non-trivial scalar field in the presence of a negative cosmological constant \cite{ABK3}, could still support similar types of solutions in the more realistic case where the constant negative distribution of energy is replaced by a negative field potential. However by considering even polynomial forms for the potential we found that asymptotically flat black-holes may also exist with very interesting and new properties. We have found
a distinctly different behaviour of the parameters of the black holes obtained for either $V(\f)=\f^2$ or for $V(\f)=\f^{2n}$, with $n > 1$. In the latter case, branches of solutions, either smaller or larger than the Schwarzschild solution depending on their mass and value of $n$, were found; these solutions were characterised by a minimum mass and approached the properties of the Schwarzschild solution in the limit of large mass, as all previously found GB black holes \cite{DBH1, ABK1, ABK2}. In the former case, however, of a negative quadratic potential, the solutions are divided in two subgroups: the first subgroup comprises the
small-mass GB black holes which all have a larger horizon radius  compared to the Schwarzschild solution; the second subgroup includes the more massive black holes the horizon radius of which gets increasingly smaller
as their mass increases. Whereas the GB black holes obtained
in the case of a positive quadratic potential approach the Schwarzschild solution in the
limit of large mass \cite{Doneva:2019vuh}, the GB solutions supported by a negative quadratic potential
exhibit a regime of very massive but ultra-compact subgroup of black holes.

 
\clearpage
\thispagestyle{empty}

\chapter{ Novel Einstein–scalar-Gauss-Bonnet wormholes without exotic matter }\label{6}

\section{Introduction}

A particularly interesting property emerging in the EdGB solutions is the
presence of regions with negative {\sl effective} energy density  -- this
is due to the presence of the higher-curvature GB term and is therefore
of purely gravitational nature \cite{DBH1,DBH2,Kanti:2011jz}.
Consequently, the EdGB theory allows for Lorentzian, traversable wormhole
solutions without the need for exotic matter \cite{Kanti:2011jz,Kanti:2011yv}.
It is tempting to conjecture that the more general EsGB theories should also
allow for traversable wormhole solutions. 
Indeed, traversable wormholes require violation of the energy conditions
\cite{Morris:1988cz,Visser:1995cc}.
But whereas in General Relativity this violation is typically 
achieved by a phantom field
\cite{Ellis:1973yv,Ellis:1979bh,Bronnikov:1973fh,Kodama:1978dw,Kleihaus:2014dla,Chew:2016epf},
in EdGB theories it is the effective stress-energy tensor 
that allows for this violation \cite{Kanti:2011jz,Kanti:2011yv}.

Thus, in this chapter, we consider
a general class of EsGB theories with an arbitrary coupling function for the
scalar field. We first readdress the case of the exponential coupling function,
and show that the EdGB theory is even richer than previously thought, since
it features also wormhole solutions with a double throat and an equator in
between. Then, we consider alternative forms of the scalar coupling function,
and demonstrate that the EsGB theories 
always allow for traversable wormhole solutions,
featuring both single and double throats. The scalar field may vanish or be
finite at infinity, and it may have nodes. 
We also map the domain of existence (DOE)
of these wormholes in various exemplifications, evaluate their global charges
and throat areas and demonstrate that the throat remains open without the
need for any exotic matter.

The outline of this chapter is as follows: in section \ref{se2} we briefly recall EsGB theory and discuss the throat geometry
for single and double throat configurations. We here present the asymptotic
expansions near the throat/equator and in the two asymptotically flat regions,
and we also derive the formulae necessary to study the violation 
of the energy conditions.
We present our numerical solutions in section \ref{se3}, and discuss some of
their properties, their domains of existence and the energy conditions.
In order to impose symmetry of the solutions under reflection of the
spatial radial coordinate, $\eta \rightarrow -\eta$, we study 
in section \ref{se4} the junction conditions at the throat/equator, 
and show that we can solve these by including only a shell of ordinary matter. In section \ref{se5} we present embedding diagrams of the wormhole solutions,
and finally we conclude in section \ref{se6}. The analysis of this chapter is based on \cite{ABK4}

\section{The Einstein-scalar-Gauss-Bonnet Theory}\label{se2}

We consider the same quadratic scalar-tensor theory employed in chapters \ref{3} and \ref{4} and described by the following effective action
%
\begin{eqnarray}  
S=\frac{1}{16 \pi}\int d^4x \sqrt{-g} \left[R - \frac{1}{2}\,
 \partial_\mu \phi \,\partial^\mu \phi
 + F(\phi) R^2_{\rm GB}   \right].
\label{action4}
\end{eqnarray} 
The GB term, a topological invariant in four dimensions, is coupled to the scalar
field through a coupling function $F(\phi)$, that as before will be left arbitrary. Note that, for notational convenience, we use the capital letter $F$ to denote the coupling function. 

The Einstein and scalar field equations are obtained by variation of the action with
respect to the metric, respectively the scalar field, and have the form of Eqs. (\ref{Einstein-eqs})-(\ref{f}). Again,
 we consider only static, spherically-symmetric solutions of the field equations. 
To this end, we employ the following line-element
\begin{equation}
ds^2 
= -e^{f_0(\eta)}dt^2+e^{f_1(\eta)}\left\{d\eta^2
+\left(\eta^2+\eta_0^2\right)\left(d\theta^2+\sin^2\theta d\varphi^2 \right)\right\}\,.
\label{metric4} 
\end{equation}
The substitution of the above metric in the Einstein and scalar field equations  leads to three second-order and one 
first-order ordinary differential equations (ODEs) that are displayed in Appendix
\ref{appC}. Due to the Bianchi identity, only three of these four equations
are independent. In our analysis, we choose to solve the three second-order equations
while the first-order one will serve as a constraint on the unknown quantities
$f_0$, $f_1$ and $\phi$. 

\subsection{Single and Double-Throat Geometry}

In a previous analysis by Kanti, Kleihaus and Kunz \cite{Kanti:2011jz, Kanti:2011yv}, traversable wormhole solutions were determined 
using the following line-element
\begin{equation}
ds^2 
= -e^{f_0(l)}dt^2+p(l)\,dl^2
+\left(l^2+r_0^2\right)\left(d\theta^2+\sin^2\theta d\varphi^2 \right).
\label{metricSold} 
\end{equation}
The wormhole geometry is characterised by the circumferential radius $R_c$ defined as
\beq
R_c = \frac{1}{2\pi}\,\int_0^{2 \pi} \sqrt{g_{\varphi \varphi}}|_{\theta=\pi/2}\, d\varphi\,.
\label{circum-radius}
\eeq
A general property of a wormhole is the existence of a throat. A minimum of $R_c$ corresponds to a wormhole throat, whereas a local maximum corresponds to
an equator. For the line-element (\ref{metricSold}), the circumferential radius is
$R_c(l) = \sqrt{l^2+r_0^2}$.  This expression clearly possesses a minimum at $l=0$,
corresponding to a throat of radius $r_0$, and it does not allow for a local
maximum, i.e. an equator. Consequently, wormholes with only a throat and no equator
were presented in \cite{Kanti:2011jz,Kanti:2011yv}.

For the set of coordinates defined in Eq. (\ref{metric4}), the circumferential radius
is expressed as $R_c(\eta) = e^{f_1/2}\sqrt{\eta^2+\eta_0^2}$. As we will see, this expression
allows for the existence of one or two local minima (i.e. throats) and of a local maximum
(i.e. equator). We introduce the distance variable in a coordinate-independent way as
\beq
\xi = \int_0^\eta \sqrt{g_{\eta \eta}}\,d\tilde \eta = \int_0^\eta e^{f_1(\tilde\eta)/2}\,d\tilde\eta \,.
\eeq
The conditions for a throat, respectively equator, at $\eta=0$ then read
\beq
\frac{dR_c}{d\xi}\Biggr|_{\eta=0}=0\,, \qquad \frac{d^2R_c}{d\xi^2}\Biggr|_{\eta = 0} \gtrless 0\,,
\label{cond-1}
\eeq
where the greater sign ($>$) refers to a throat and the smaller sign ($<$) to an equator. Using
the metric (5), these conditions yield
\beq
f_1'(0)=0\,, \qquad \eta_0^2 \,f_1''(0) + 2 \gtrless 0 \,. \label{cond-2}
\eeq
where the prime denotes derivative with respect to $\eta$. In the degenerate case, when the
throat and equator coincide, the inequalities in Eqs. (\ref{cond-1}) and (\ref{cond-2})
become equalities. If a throat/equator is located at  $\eta=0$, then its area is given by 
$A_{t,e} = 4\pi R^2_c(0)=4\pi \eta_0^2 e^{f_1(0)}$, while for a double-throat wormhole
with the throat located at $\eta_{\rm t}$, 
$A_{t} = 4\pi R^2_c(\eta_{\rm t})=4\pi (\eta_{\rm t}^2 + \eta_0^2 )e^{f_1(\eta_{\rm t})}$.


\subsection{Asymptotic Expansions}
\subsubsection{Expansion near the throat/equator}

A traversable wormhole solution is characterised by the absence of horizons or singularities.
In order to ensure that this is the case for our solutions, we consider the following
regular expansions for the metric functions and scalar field, near the throat/equator at
$\eta=0$,
\bea
    e^{f_0}&=&a_0\,(1+a_1 \eta +a_2 \eta^2+...)\,,\label{appf0}\\[1mm]
    e^{f_1}&=&b_0\,(1+b_1 \eta +b_2 \eta^2+...)\,,\label{appf1}\\[1mm]
    \phi&=&\phi_0+\phi_1 \eta +\phi_2 \eta^2+...\,.\label{appff}
\eea
where $a_0=e^{f_0}|_{\eta=0}$, $b_0=e^{f_1}|_{\eta=0}$ are the values of the $|g_{tt}|$ and $g_{\eta\eta}$ metric components respectively at the $\eta=0$ value of the radial coordinate. The Lorentzian signature of spacetime demands that both parameters $a_0$ and $b_0$ must be positive;
in addition, they should be finite and non-vanishing. As discussed in the previous subsection, 
the emergence of an extremum in the circumferential radius $R_c$ dictates that $f_1'(0)=0$;
this leads to the result $b_1=0$. The $(\eta\eta)$ component of the Einstein equations, given in
Eq. (\ref{hheq}), yields near the throat/equator a constraint equation:
\begin{equation}\label{eq10}
\left[\left(\eta_0^2\phi'^2 + 4\right) e^{f_1} - 8 f_0'\phi' \dot{F}\right]_{\eta=0}= 0\,.
\end{equation}
The remaining three equations may then be solved to express the second-order coefficients
$(a_2, b_2, \phi_2)$ in terms of the zero and first-order coefficients in the $\eta$-expansions
(\ref{appf0})-(\ref{appff}). These are found to have the form:
\bea
    a_2&=&\frac{b_0 \left[4 b_0 \dot{F}_0^2 \phi _1^2 \left(\eta _0^2 \phi _1^2+4\right){}^2+b_0^3 \eta _0^2
   \left(\eta _0^2 \phi _1^2+4\right){}^2-128 \dot{F}_0^2 \eta _0^2 \phi _1^6 \ddot{F}_0\right]}{256
   \dot{F}_0^2 \phi _1^2 \left(b_0^2 \eta _0^2+4 \dot{F}_0^2 \phi _1^2\right)},\\
    b_2&=&-\frac{2 b_0 \phi _1^2 \ddot{F}_0}{b_0^2 \eta _0^2+4 \dot{F}_0^2 \phi _1^2},\\[1mm]
   \phi_2&=&-\frac{4 b_0 \dot{F}_0^2 \phi _1^2 \left(\eta _0^2 \phi _1^2+4\right)+b_0^3 \eta _0^2 \left(\eta _0^2
   \phi _1^2+4\right)+64 \dot{F}_0^2 \phi _1^4 \ddot{F}_0}{32 \left(b_0^2 \dot{F}_0 \eta _0^2+4
   \dot{F}_0^3 \phi _1^2\right)}.
\eea
From the above expressions, it seems that there are six free parameters in our theory: the coefficients
$(\eta_0, \;\phi_0,\; \phi_1,\;a_0,\; b_0)$ and the coupling constant $\alpha$ which is defined through the
relation  $F(\phi)=\alpha\tilde{F}(\phi)$, where $\tilde{F}$ is a dimensionless quantity. However, the
actual number of free parameters is much smaller. First of all, we notice that the field equations
(\ref{tteq})-(\ref{sceq}) are invariant under the simultaneous scaling of the coordinate $\eta$, the
constant $\eta_0$, and the scalar-field coupling constant $\alpha$,
\beq
\eta \rightarrow \lambda \eta \,, \qquad \eta_0 \rightarrow \lambda \eta_0\,, 
\qquad \alpha \rightarrow \lambda^2 \alpha\,,
\eeq
where $\lambda$ is an arbitrary constant. Therefore, we may fix $\eta_0$, which determines the scale
of the wormhole's equator/throat, to a specific value, or equivalently introduce a dimensionless
coupling parameter $\alpha/\eta_0^2$. We can also fix three of the remaining four parameters by
applying appropriate boundary conditions at infinity. Thus, by demanding asymptotic flatness, expressed
by the conditions
\beq
\lim_{\eta\to\infty}|g_{tt}|=1\\, \qquad \lim_{\eta\to\infty}g_{\eta\eta}=1\,,
\label{asym-flat}
\eeq
we may fix the $a_0$ and $b_0$ parameters, while the condition $\lim_{\eta\to\infty}\phi=\phi_\infty$
allows us to fix $\phi_1$.  Concluding, the only free parameters in the near-equator/throat area are the
dimensionless coupling constant $\alpha/\eta_0^2$ and the value $\phi_0$ of the scalar field at $\eta=0$.

Let us also examine the components of the effective stress-energy-momentum tensor Eq.(\ref{f}) near the
throat/equator. We find:
\bea
    &&T^t_{\;t}=\frac{1}{b_0 \eta _0^2}-\frac{4 \phi _1^2 \ddot{F}_0}{b_0^2 \eta _0^2+4 \dot{F}_0^2 \phi
   _1^2}+O\left(\eta\right),\label{Ttt-thr}\\
   &&T^\eta_{\;\eta}=-\frac{1}{b_0 \eta _0^2}+O\left(\eta \right),\label{eq15}\\
   &&T^\theta_{\;\theta}=\frac{-b_0 \eta _0^2 \phi _1^2 \ddot{F}_0 \left(\eta _0^2 \phi _1^2+4\right)+2 b_0^2 \eta
   _0^2+8 \dot{F}_0^2 \phi _1^2}{8 b_0 \dot{F}_0^2 \eta _0^2 \phi _1^2+2 b_0^3 \eta
   _0^4}+O\left(\eta\right). \label{Thh-thr}
\eea
We observe that, as desired, all components of the stress-energy tensor are finite at $\eta=0$, i.e. at the location
of the throat or equator of the solution.

\subsubsection{Expansion at large distances}
As the radial coordinate approaches infinity we expect that the space-time will be flat and the scalar field constant. Our solutions are expanded in a power series form in terms of $1/\eta^n$:
\bea
    e^{f_0}&=&1+\sum_{n=1}^\infty {\frac{p_n}{\eta^n}},\\
    e^{f_1}&=&1+\sum_{n=1}^\infty {\frac{q_n}{\eta^n}},\\
    \phi&=&\phi_\infty+\sum_{n=1}^\infty {\frac{d_n}{\eta^n}}\,.
\eea
 Substituting the above expansions into the field
equations (\ref{tteq})-(\ref{sceq}), we may determine the unknown coefficients 
$(p_n, q_n, d_n)$ in terms of only two coefficients that remain arbitrary: $d_1=-D$, where $D$
is the scalar charge of the wormhole, and $p_1=-2M$, where $M$ is the Arnowitt-Deser-Misner
(ADM) mass of the wormhole. Thus, the number of free parameters at infinity is also two,
similarly to the near-throat/equator regime. We have calculated the remaining coefficients up
to order $\mathcal{O}(1/r^5)$, and the asymptotic solutions have the following form:
\begin{align}
    e^{f_0}&=1-\frac{2M}{\eta}+\frac{2M^2}{\eta^2}-\frac{M(D^2+36M^2-12\eta_0^2)}{24\eta^3}\nn\\[1mm]
    &+\frac{D^2M^2+12(M^4-M^2\eta_0^2-4D M\dot{F}_\infty)}{12\eta^4}+\mathcal{O}\left(\frac{1}{\eta^5}\right),\label{inff0}\\[1mm]
    e^{f_1}&=1+\frac{2M}{\eta}+\frac{12M^2-D^2-4\eta_0^2}{8\eta^2}+\frac{M\left[12(M^2-3\eta_0^2)-5D^2\right]}{24\eta^3}\nonumber\\[1mm]
    &+\frac{3D^4+D^2(96\eta_0^2-104M^2)+48(M^4-24M^2\eta_0^2+7\eta_0^4)+1536D M\dot{F}_\infty}{768\eta^4}\nonumber\\[1mm]
    &+\mathcal{O}\left(\frac{1}{\eta^5}\right),\label{inff1}\\
    \phi&=\phi_\infty-\frac{D}{\eta}-\frac{D^3+4D(M^2-3\eta_0^2)}{48\eta^3}-\frac{4M^2\dot{F}_\infty}{\eta^4}+\mathcal{O}\left(\frac{1}{\eta^5}\right).\label{infff}
\end{align}
We observe that the above solutions have exactly the same form as the corresponding solutions
which describe asymptotically-flat black holes found in chapters \ref{3} and \ref{4} \cite{ABK1, ABK2}. Apparently, the emergence
of an asymptotically-flat limit does not depend on the choice of the boundary condition at the
other asymptotic regime i.e. the horizon of a black hole or the throat/equator of a wormhole.
The main difference is that, in the case of black holes, the mass $M$ and the scalar charge $D$
are related parameters -- which makes black holes a one-parameter family of solutions -- while,
in the case of wormholes, these two parameters are independent.  Also, the aforementioned asymptotic
solutions at infinity are almost independent of the functional form of the coupling function
$F(\phi)$, since the latter does not enter in the expansions earlier than in the fourth order. 

Finally, if we make use of the expansions above, we may calculate again the effective stress-energy tensor
components Eq. (\ref{f}) at large distances. These are found to be 
\begin{equation}
    T^t_{\;t}=-T^\eta_{\;\eta}=T^\theta_{\;\theta}=T^\varphi_{\;\varphi}\approx -\phi'^2/4 \approx -D^2/4\eta^4+\mathcal{O}\left(1/\eta^5\right). \label{Tmn-far4}
\end{equation}
As we expect, the above expressions have exactly the same form as the corresponding ones for the asymptotically
flat black holes. We observe that, at large distances where the curvature of spacetime is small, the stress-energy
tensor is dominated by the kinetic term of the scalar field which is itself decaying fast.


\subsection{Violation of Energy Conditions}
In the previous works by Kanti, Kleihaus and Kunz \cite{Kanti:2011jz,Kanti:2011yv}, it was shown that for any single-throat wormhole
the null energy condition is violated at least in some region near the throat. Here, we
review that analysis, and show that the violation of the null energy condition also
holds for double-throat wormholes.

The null energy condition (NEC) is expressed as $ T_{\mu\nu} n^\mu n^\nu \geq 0$, where
$n^\mu$ is any null vector satisfying the condition $n^\mu n_\mu=0$. We may define the
null vector as $n^\mu=\left(1,\sqrt{-g_{tt}/g_{\eta\eta}},0,0\right)$ with its contravariant
form being $n_\mu=\left(g_{tt},\sqrt{-g_{tt}\,g_{\eta\eta}},0,0\right)$. For a
spherically-symmetric spacetime, the NEC takes the form:
\beq
T_{\mu\nu}n^\mu n^\nu=T^t_t n^t n_t + T^\eta_\eta n^\eta n_\eta
=-g_{tt}\,(-T^t_t +T^\eta_\eta)\,.
\eeq
Then, the NEC holds if $ -T_t^t + T_\eta^\eta \geq 0$. Alternatively, we may choose
$n^\mu=\big(1,0,$ $\sqrt{-g_{tt}/g_{\theta \theta}},0\big)$, and a similar analysis
leads to the condition $-T_t^t + T_\theta^\theta \geq 0 $. 

For a wormhole solution to emerge, it is essential that these two conditions
are violated \cite{Morris:1988cz}. Indeed, using the expansion of the wormhole solution at the
throat/equator, we find 
\beq
\left[ -T_t^t + T_\eta^\eta \right]_{\eta_{t,e}} 
  = -2 \left[e^{-f_1} R_c''/R_c \right]_{\eta_{t,e}}\,.
  \eeq
Consequently, the NEC is always violated at the throat(s), 
since $R_c$ possesses a minimum there, implying $R_c''(\eta_{t}) > 0$, while
no violation occurs at the equator, where $R_c''(\eta_{e}) < 0$. For example,
for a single-throat solution with the throat at $\eta=0$, we obtain the
explicit expressions
\begin{align}
\left[ -T_t^t + T_\eta^\eta \right]_{\eta=0} &  =  
\left[ -\frac{2 e^{-f_1}}{\eta_0^2}
      +\frac{4 \ddot{F}\phi'^2}{e^{2f_1}\eta_0^2 +4 \dot{F}^2\phi'^2}\right]_{\eta=0}=
      -\frac{2}{b_0 \eta_0^2} + \frac{4 \phi _1^2 \ddot{F}_0}{b_0^2 \eta _0^2+4 \dot{F}_0^2 \phi _1^2}\,, 
\label{nec1} \\[3mm]
\left[ -T_t^t + T_\theta^\theta \right]_{\eta=0} &  =  
\left[\frac{\ddot{F}\phi'^2\left(4-\eta_0^2\phi'^2\right)}
       {2\left(e^{2f_1}\eta_0^2 +4 \dot{F}^2\phi'^2\right)}\right]_{\eta=0} 
       =\frac{\phi _1^2 \ddot{F}_0 \left(4-\eta _0^2 \phi _1^2\right)}
       {2 b_0^2 \eta _0^2+8 \dot{F}_0^2 \phi_1^2}\,,
\label{nec2}   
\end{align}
where we have used the approximate expressions Eqs. (\ref{Ttt-thr})-(\ref{Thh-thr}) near the wormhole
throat. We note that the desired violation of the NEC follows not from the presence of an
exotic form of matter but from the synergy between the scalar field and the quadratic GB term.

In the far-asymptotic regime, we may use the expansions at infinity Eqs. (\ref{inff0})-(\ref{infff})
to find that the two Null Energy Conditions take the form:
\bea
    -T^t_{\;t}+T^\eta_{\;\eta}&=&\,\frac{D^2}{2\eta^4}+\mathcal{O}\left(\frac{1}{\eta^5}\right),\\[3mm]
    -T^t_{\;t}+T^\theta_{\;\theta}&=&-\frac{40DM\dot{F}_\infty}{\eta^6}+\mathcal{O}\left(\frac{1}{\eta^7}\right).
\eea
We observe that if $D\dot{F}_\infty>0$, the second Null Energy Condition is also violated at spatial infinity.

Let us also examine the Weak Energy Condition (WEC), which suggests that the energy density measured by
any observer has to be greater than or equal to zero. This is expressed through the inequality:
$T_{\mu\nu}V^\mu V^\nu \ge 0$,
where $V^\mu$ is any timelike vector. If we choose $V^\mu=(1/\sqrt{-g_{tt}},0,0,0)$, 
and impose the condition $V_\mu V^\mu = -1$, then $V_\mu=(-\sqrt{-g_{tt}},0,0,0)$, and the WEC is simply
$T^t_{\;t}\le 0$. Near the throat/equator, we found that $T^t_{\;t}$ is given by Eq. (\ref{Ttt-thr});
this expression is not sign-definite, therefore the WEC may also be violated in the small $\eta$-regime. 
On the other hand, at asymptotic infinity, where $T_{\mu\nu}$ is dominated by the kinetic term of the
scalar field, the $T^t_{\;t}$ component is given by Eq. (\ref{Tmn-far4}) and clearly obeys the WEC.


\subsection{Smarr Relation}
Up to this point we have managed to construct approximate analytical solutions in both the asymptotic regions.  We have also found that our wormholes are two-parameter solutions. We may relate the near $\eta=0$ parameters with the parameters at spatial infinity using the Smarr relation. 
In order to derive the Smarr-like mass relation for our wormhole solutions we start with the definition of the Komar mass formula \cite{Kanti:2011jz,Kanti:2011yv}
\begin{equation}
    M=M_{\text{th}}+\frac{1}{4\pi}\int_\Sigma R_{\mu\nu}\xi^\mu n^\nu dV=M_{\text{th}}-\frac{1}{4\pi}\int \sqrt{-g}R^0_0d^3x,\label{komar}
\end{equation}
where $\xi^\mu$ is the timelike Killing vector, $\Sigma$ is a space-like hypersurface, $n^\nu$ is a normal vector on the surface $\Sigma$ and $dV$ is the natural volume element on the surface $\Sigma$. Here the mass term $M_{\text{th}}$ denotes the contribution of the throat. 
\begin{equation}
  M_{\text{th}}=\frac{1}{2}A_{\text{th}} \frac{\kappa_\text{th}}{2\pi},
\end{equation}
in which with $A_{\text{th}}$ we denote the area of the throat and $\kappa$ is the surface gravity at the throat,
\begin{equation}
    \kappa_\text{th}^2=-\frac{1}{2}\left( \nabla_\mu\xi_\nu  \right)\left( \nabla^\mu\xi^\nu   \right)=\frac{1}{2}\left(\frac{1}{\sqrt{|g_{tt}g_{\eta\eta}|}}  \left| \frac{dg_{tt}}{d\eta}\right|  \right)_{\eta=\eta_\text{th}}.\label{scur}
\end{equation}
Now we need to express  the $R_0^0\sqrt{-g}$ as a total derivative in order to calculate the integral in Eq. (\ref{komar}). To do that we first write the $R_0^0$ in terms of the effective energy-momentum tensor Eq. (\ref{f})
\begin{align}
R_0^0\sqrt{-g}=\left(T^0_0-\frac{1}{2}T^\mu_\mu+ \beta\left[ \nabla^2 \phi+\dot{F}(\phi)R^2_{GB}  \right]  \right)\sqrt{-g},\label{77}
\end{align}
where we have added a zero in the form of the scalar-field equation multiplied by a constant parameter $\b$. By replacing the explicit form of the field equations and performing some straightforward algebraic manipulations we find
\begin{align}
R_0^0\sqrt{-g}=\left(\dot F H-V F\right)'+\beta \,\partial_\mu\left(\sqrt{-g}\,\partial^\mu \phi \right)+\frac{1}{2}\sqrt{-g} R^2_{GB} \left( F + \beta\,\dot F \right).\label{10}
\end{align}
For simplicity, in the above equation, we define the functions $H(r)$ and $V(r)$ as
\begin{align}
H&=\frac{V \phi'}{A'}-e^{\frac{1}{2} \left(f_0-f_1\right)} f_0' \phi ' \left(\left(\eta ^2+\eta
   _0^2\right) f_1'+2 \eta \right),\\[4mm]
V&=\frac{e^{\frac{1}{2} \left(f_0-f_1\right)} f_0' \left(\eta ^3 f_1' \left(\eta 
   f_1'+4\right)+2 \eta _0^2 \left(\eta ^2 f_1'^2+2 \eta 
   f_1'-2\right)+\eta _0^4 f_1'^2\right)}{2 \left(\eta ^2+\eta
   _0^2\right)}.
\end{align}
Equation (\ref{10}) may be expressed  as a total derivative only if  $\left( F + \beta\,\dot F \right)=0$, or equivalently $F=a\,e^{-\phi/\beta }.$

From the above analysis it is clear that we may construct an analytic form of the Smarr Formula only for the case of the exponential coupling function. Here --following a similar notation with the works in \cite{Kanti:2011jz,Kanti:2011yv}-- we choose a slightly different form of the exponential coupling  $F=\alpha e^{-\gamma \phi}$, where $\gamma$ is a constant. The Smarr formula then has the following form
\begin{equation}
    M=2S_{\text{th}}\frac{\kappa_{\text{th}}}{2\pi}-\frac{D}{2\gamma}+\frac{D_{\text{th}}}{2\gamma}\label{smarrth}
\end{equation}
where the quantities $D_{\text{th}}$ and $S_{\text{th}}$ are defined at the throat as
\begin{align}
    S_{\text{th}}\,&=\,\frac{1}{4}\int\sqrt{h_{\text{th}}}\left(1+2\alpha e^{-\gamma\phi_{\text{th}}}\tilde{R}_{\text{th}}\right)d^2x,\label{sthr}\\[2mm]
    D_{\text{th}}\,&=\,\frac{1}{4\pi}\int \sqrt{h_{\text{th}}}e^{\frac{f_{0\text{th}}}{2}}n^\mu_{\text{th}}\partial_\mu\phi_{\text{th}}\left(1+2\alpha e^{-\gamma\phi_{\text{th}}}\tilde{R}_{\text{th}}\right)d^2x.\label{dthr}
\end{align}
In the above expressions with $\phi_{\text{th}}$ and $f_{0{\text{th}}}$ we denote the function values on the throat. $h_\text{th}$ is the induced spatial metric on the throat, $\tilde{R}_\text{th}$ the scalar curvature on $h_\text{th}$, and $n^\mu_\text{th}$ is the normal vector on the throat. More information on the derivation of the Smarr relation for the exponential coupling function may be found here \cite{Kanti:2011jz,Kanti:2011yv}. In case of a double throat wormhole a very similar Smarr relation can be derived by replacing the throat with
the equator,
\begin{equation}
    M=2S_{\text{eq}}\frac{\kappa_{\text{eq}}}{2\pi}-\frac{D}{2\gamma}+\frac{D_{\text{eq}}}{2\gamma},\label{smarreq}
\end{equation}
where $\kappa_{\text{eq}}$, $D_{\text{eq}}$ and $S_{\text{eq}}$ are defined analogously to Eqs. (\ref{scur}), (\ref{sthr})-(\ref{dthr}), but evaluated at the equator instead of the
throat. The difference of Eqs. (\ref{smarrth})  and (\ref{smarreq}) yields a relation of quantities defined at the throat and quantities
defined at the equator.
\begin{equation}
    2S_{\text{eq}}\frac{\kappa_{\text{eq}}}{2\pi}+\frac{D_{\text{eq}}}{2\gamma}= 2S_{\text{th}}\frac{\kappa_{\text{th}}}{2\pi}+\frac{D_{\text{th}}}{2\gamma}.\label{smarrdif}
\end{equation}

\section{Numerical solutions}\label{se3}
We now turn to the derivation of the wormhole solutions by numerically integrating
the three second-order, ordinary differential equations (\ref{tteq}), (\ref{ffeq})
and (\ref{sceq}). In order to find asymptotically-flat, regular wormhole solutions,
we have to impose appropriate boundary conditions at asymptotic infinity and at the
throat/equator, as discussed in the previous section. For completeness, we list
here the full set of these boundary conditions:
\begin{eqnarray}
& & f_0(\infty) = f_1(\infty) = 0 \ , \ \ \phi(\infty) = \phi_\infty \ , 
\label{bcinfty}\\
& & f_1'(0) = 0\ , \ \ 
\left[\left(\eta_0^2\phi'^2 + 4\right) e^{f_1} - 8 f_0'\phi' \dot{F}\right]_{\eta=0}= 0 \ . 
\label{bcs} 
\end{eqnarray}
For the numerical integration, we use the compactified coordinate $x=\eta/(\eta+\eta_0)$
to cover the range $0\leq \eta < \infty$. 
We choose $\eta_0=1$ for all our numerical solutions.
In order to
solve the three, second-order ODEs with the aforementioned boundary conditions, we constructed two independent codes: one using the software package COLSYS, and the other using the Mathematica software.

In our analysis, we have found wormhole solutions with either vanishing or non-vanishing
asymptotic values of the scalar field, namely for $\phi_\infty=0$ and $\phi_\infty=1$.
We have also considered several forms of $F(\phi)$, including exponential $F=\alpha e^{-\gamma\phi}$,
$F=\alpha e^{-\gamma\phi^2}$, power-law $F=\alpha \phi^n$ with $n\neq 0$, inverse power-law
$F=\alpha \phi^{-n}$, and logarithmic  $F=\alpha \ln(\phi)$ functions. We have found wormhole
solutions in every single case studied. Due to the qualitative similarity of the obtained
behaviour for the metric functions and scalar field, in this work we will mainly focus on the
presentation of results for the cases with coupling functions $F=\alpha e^{-\phi}$ and 
$F=\alpha \phi^2$, and present combined graphs for different forms of $F(\phi)$ whenever
possible. During our quest for regular, physically-acceptable wormhole solutions,
  scalarized black holes also emerged in multitude thus confirming the
results of \cite{ABK1,ABK2}.

%

\begin{figure}[t!] 
\begin{center}
\hspace{0.0cm} \hspace{-0.6cm}
\includegraphics[height=.25\textheight, angle =0]{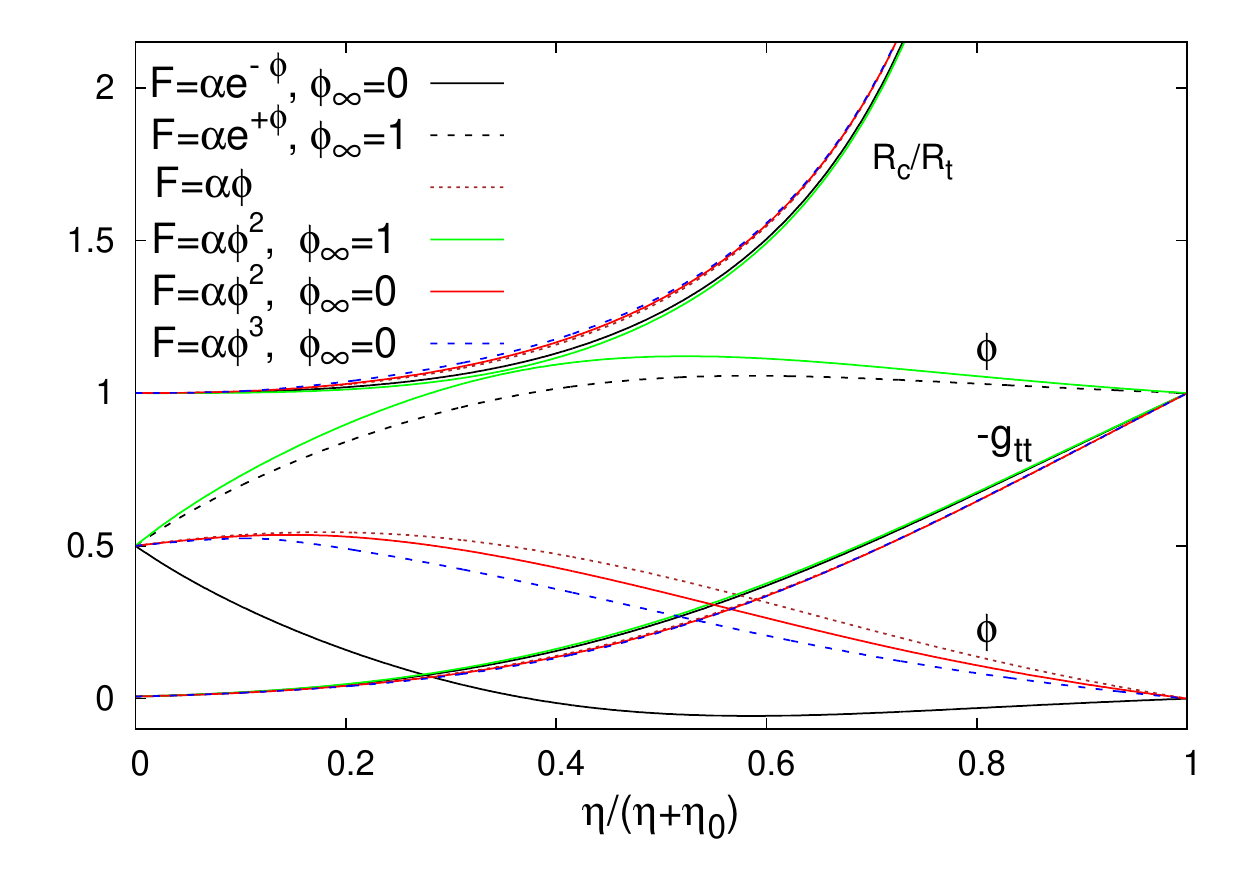}
\hspace{0.52cm} \hspace{-0.6cm}
\includegraphics[height=.25\textheight, angle =0]{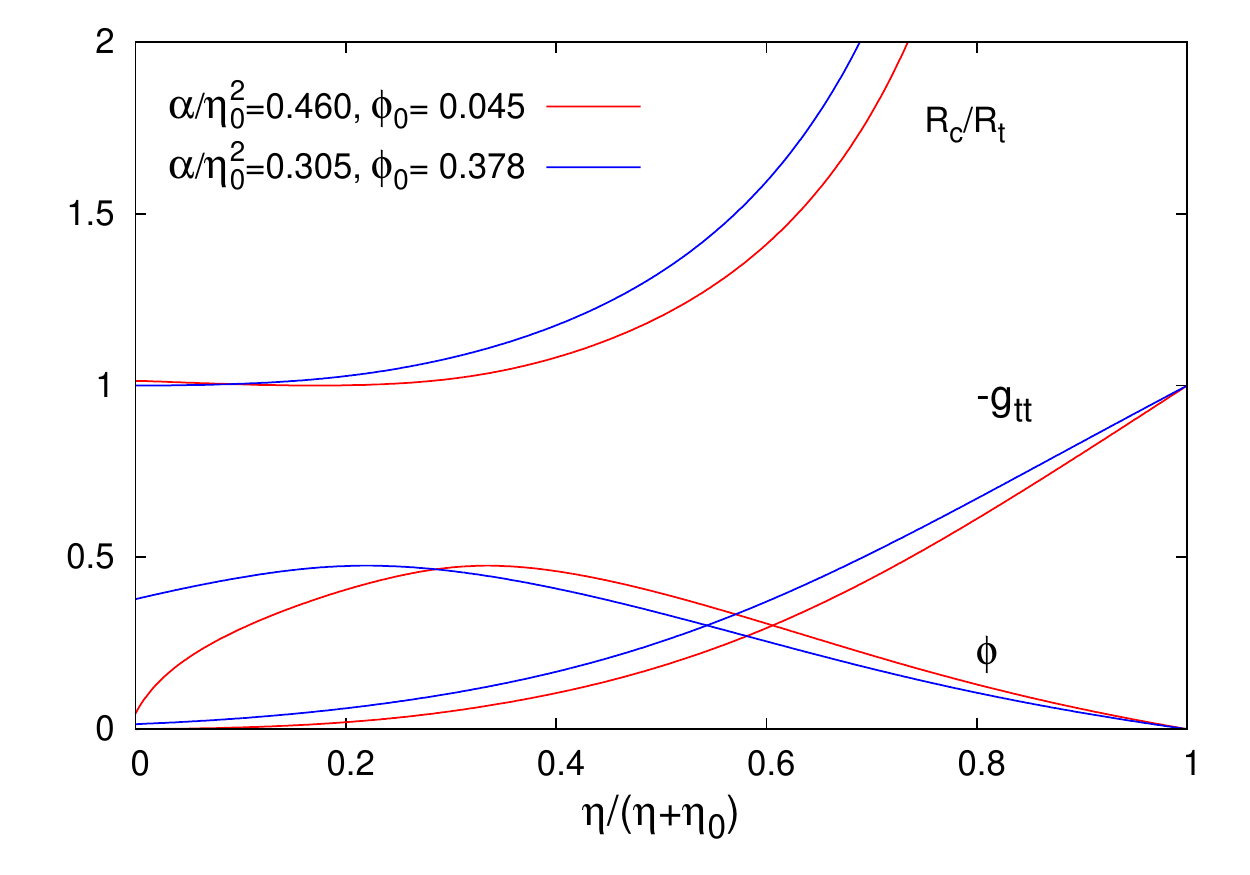}
\\
\hspace*{0.7cm} {(a)} \hspace*{7.5cm} {(b)}  \vspace*{-0.5cm}
\end{center}
\caption{Solutions:
(a) The metric component $-g_{tt}$, the scalar field $\phi$ and the scaled circumferential radius
$R_c/R_t$ are shown as functions of the compactified coordinate $\eta/(1+\eta)$ for different
coupling functions. All solutions are characterized by the same values of $f_0(0)$ and $\phi(0)$.
(b) The metric component $-g_{tt}$, the scalar field $\phi$ and the scaled circumferential radius
$R_c/R_t$ are shown as function of the compactified coordinate $\eta/(1+\eta)$ for
a single throat wormhole (blue) and a double throat wormhole (red) for the same values 
of the scaled scalar charge and the scaled throat area.
\label{fig1}}
\end{figure}


In Fig.~\ref{fig1}(a), we depict the metric component $-g_{tt}=e^{f_0}$, the scalar field $\phi$ 
and the scaled circumferential radius $R_c/R_t$ for several coupling functions $F(\phi)$.
All solutions are characterised by the same boundary values $f_0(0)=-5$ and
$\phi(0)=0.5$ but we have allowed for two different asymptotic values for $\phi$, namely
$\phi_\infty=0$ and $\phi_\infty=1$. We observe that the behaviour of the metric component
$-g_{tt}$ and the circumferential radius $R_c/R_t$ depends rather mildly on the form
of the coupling function or the asymptotic value $\phi_\infty$. 
On the other hand, both of these factors considerably affect the profile of the scalar field
as may be clearly seen from the plot.

We have found both single and double-throat wormhole solutions for every form of the
coupling function $F(\phi)$. In  Fig.~\ref{fig1}(b), we compare single and double-throat
wormholes for the same values of the scaled scalar charge $D/M$ and scaled throat area
$A_t/16\pi M^2$. Once again, it is the scalar field that is mostly affected by the
different geometry near the throat or equator. We note for future reference that the
derivatives of the $-g_{tt}$ and $\phi$ do not vanish at $\eta=0$, i.e. at the throat,
for single-throat wormholes, or at the equator, for double-throat ones. This feature
will lead to the introduction of a distribution of matter, albeit a physically-acceptable
one, at $\eta=0$ when we attempt to symmetrically continue our wormhole solutions to the
negative regime of the $\eta$ coordinate. This process and the implications of the
associated junction conditions will be studied in Section IV. 


\begin{figure}[t!] 
\begin{center}
\hspace{0.0cm} \hspace{-0.6cm}
\includegraphics[height=.25\textheight, angle =0]{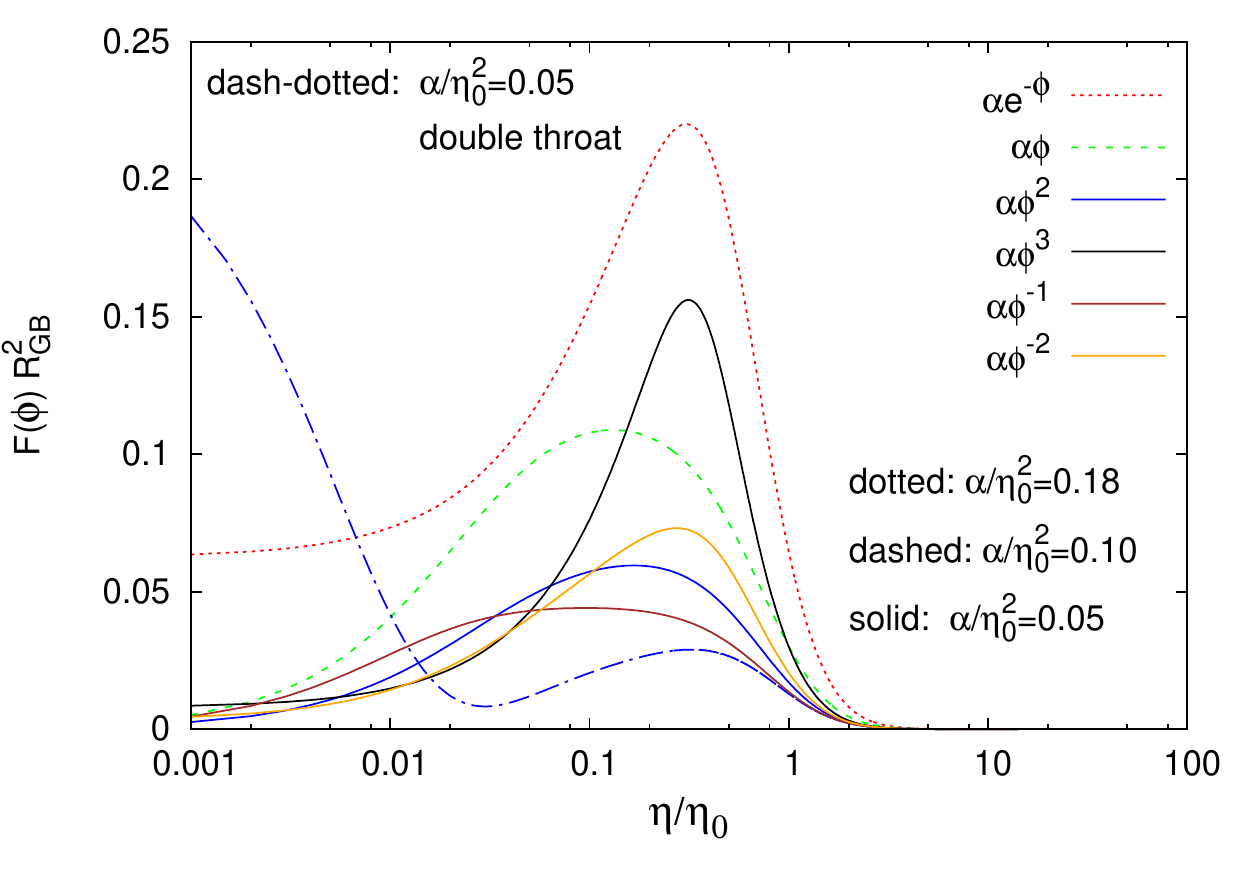}
\hspace{0.52cm} \hspace{-0.6cm}
\includegraphics[height=.25\textheight, angle =0]{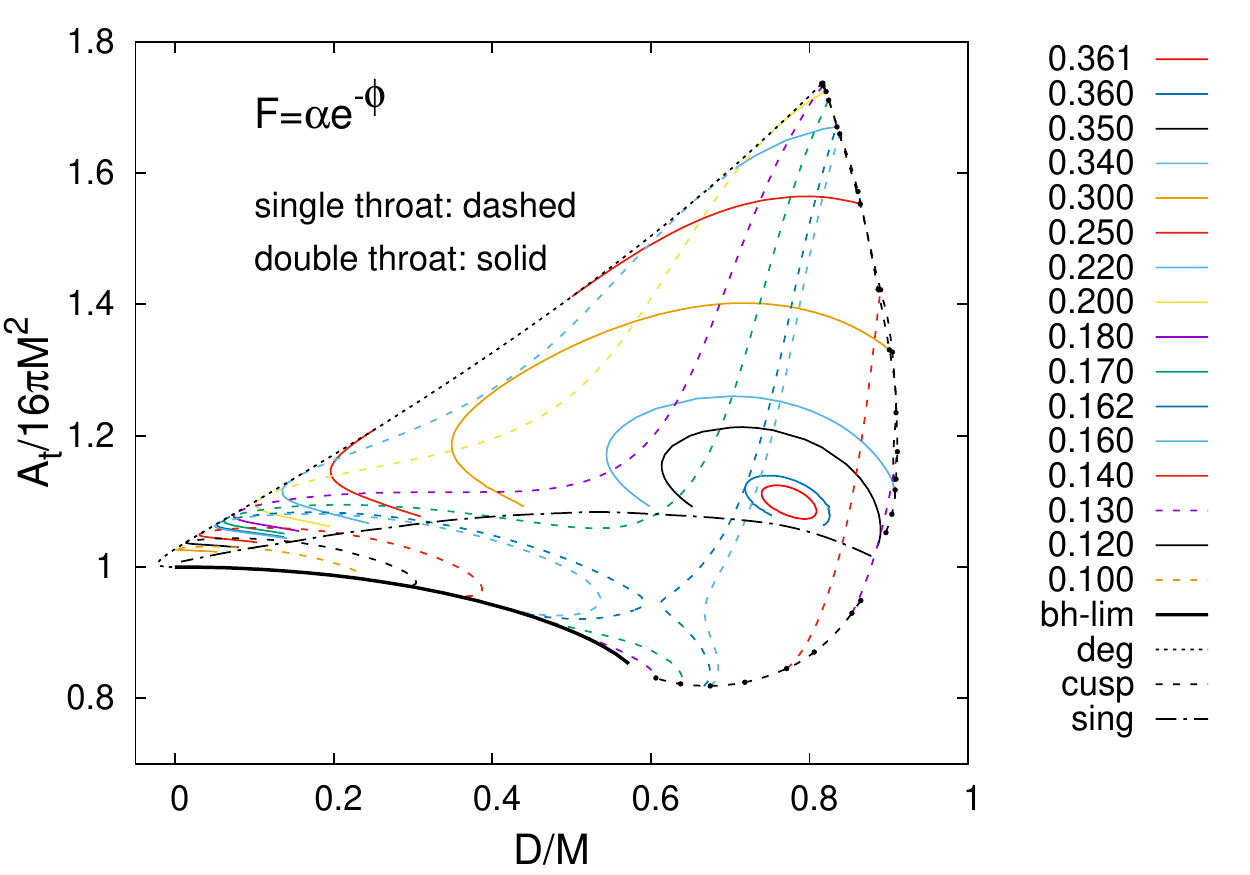}
\\
\hspace*{0.7cm} {(a)} \hspace*{7.5cm} {(b)}  \vspace*{-0.5cm}
\end{center}
\caption{(a) The quantity $F(\phi) R^2_{GB}$ as function of $\eta$ for several forms of the coupling
function $F(\phi)$.  
(b) Domain of existence: The scaled throat area of single and double throat wormholes
is shown as function of the scaled scalar charge for the coupling function
$F=\alpha e^{-\phi}$ for several values of $\alpha/\eta_0^2$. 
\label{fig2}}
\end{figure} 


The spacetime around our wormhole solutions is finite for all values of the radial
coordinate $\eta \in [\,0, \infty)$. All curvature invariant quantities remain 
everywhere finite, as expected. In Fig. \ref{fig2}(a), we depict the profile of the
quantity $F(\phi) R^2_{GB}$, for a variety of forms of the coupling function $F(\phi)$ and
for the same set of values of the free parameters for easy comparison. We observe
that the combination $F(\phi) R^2_{GB}$ is indeed finite, vanishes at asymptotic infinity
as anticipated while its profile in the small $\eta$-regime depends on the form of $F(\phi)$.
We also note that the double-throat solution presents a different profile from the 
single-throat ones; this is due to the fact that the value of the scalar field at the
equator is different from its value at the throat.


\begin{figure}[t!] 
\begin{center}
\hspace{0.0cm} \hspace{-0.6cm}
\includegraphics[height=.25\textheight, angle =0]{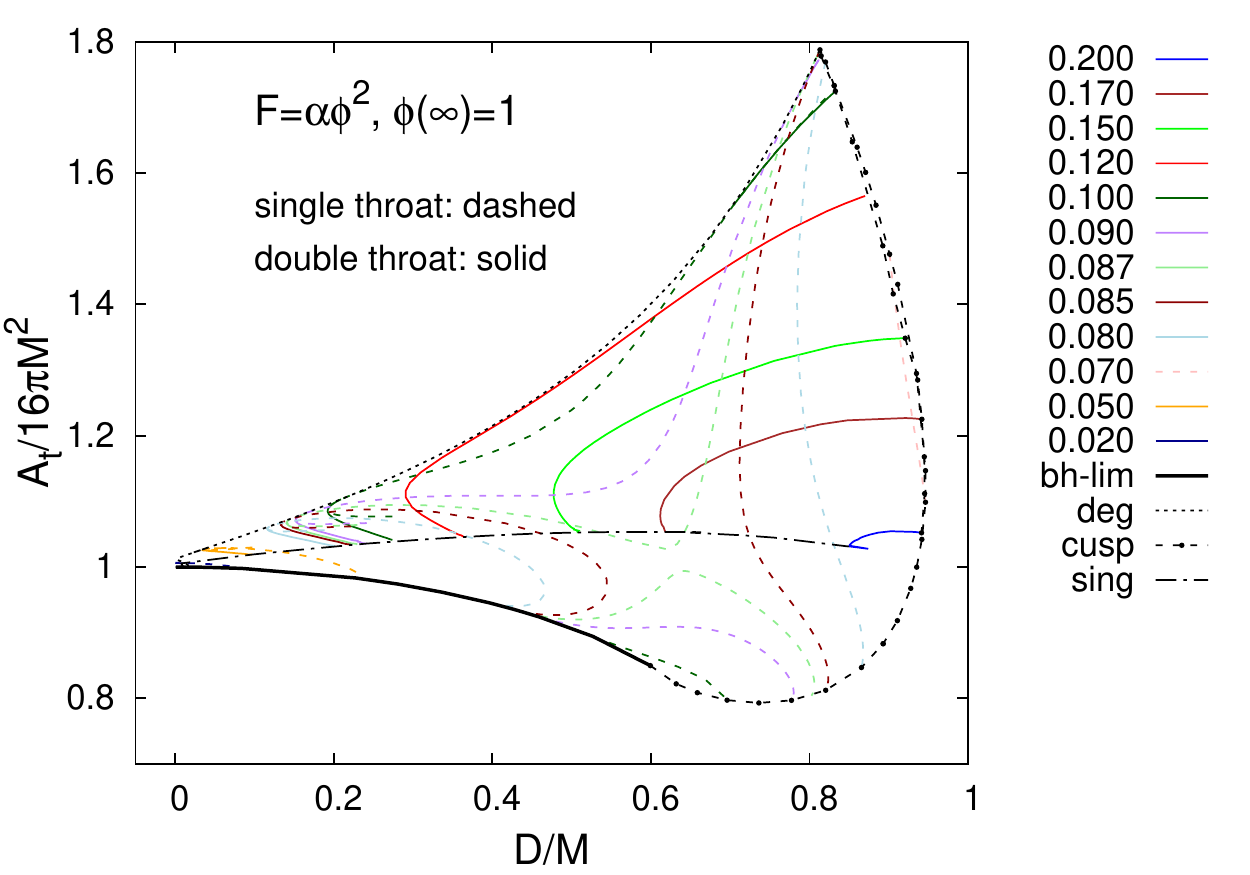}
\hspace{0.52cm} \hspace{-0.6cm}
\includegraphics[height=.25\textheight, angle =0]{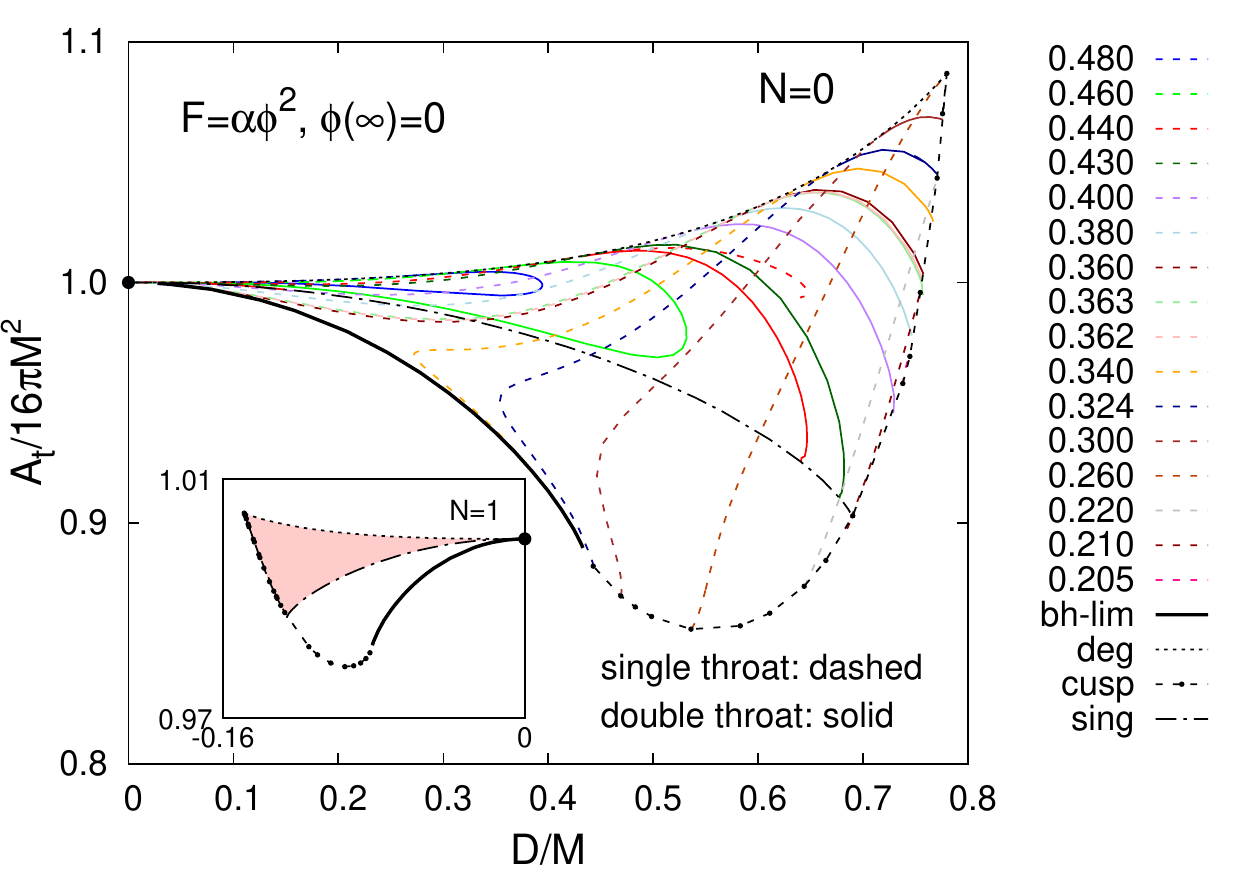}
\\
\hspace*{0.7cm} {(a)} \hspace*{7.5cm} {(b)}  \vspace*{-0.5cm}
\end{center}
\caption{Domain of existence for the coupling function $F=\alpha \phi^2$ for several
values of $\alpha/\eta_0^2$ : 
The scaled throat area of single and double throat wormholes is shown as function of
the scaled scalar charge for (a) $\phi_\infty =1$, and (b) $\phi_\infty =0$.
The dot indicates the Schwarzschild black hole. The inlet shows the domain of existence
for wormholes with one node of the scalar field -- the red area indicates the domain
where single and double-throat wormholes co-exist.
\label{fig3}}
\end{figure}


Next, we discuss the domain of existence (DOE) of the wormhole solutions,
in terms of the scaled scalar charge and the scaled throat area, and restrict our
discussion to the indicative cases of the exponential and quadratic coupling functions.
In Fig.~\ref{fig2}(b), we show the DOE for the exponential case,
$F=\alpha e^{-\phi}$. The different curves correspond to families of wormholes
for a fixed value of $\alpha$ with single throat (dashed) and double throat (solid). 
Solutions emerge for arbitrarily small values of $\alpha$
up to some maximal value - here, we depict a variety of solutions arising up to
the value $\alpha/\eta_0^2=0.361$.
The boundary of the DOE is formed by the black hole solution with
scalar hair (solid black), the wormhole solutions with a degenerate throat (dotted black),
configurations with cusp singularities outside the throat (dashed black) and configurations
with singularities at the equator (dashed-dotted black). We note that the part of the
DOE above the dashed-dotted curve comprises both single-throat and
double-throat wormholes. The single-throat wormholes of this area can in fact be
obtained from the double-throat ones -- we will return to this point in Section IV.
The region of the domain of existence below the dashed-dotted curve contains only
single-throat wormholes which are not related to double-throat solutions.

We now turn to the case of the quadratic coupling function, $F=\alpha \phi^2$. 
Contrary to what happens in the case of the exponential coupling function, in this case,
the DOE depends on the asymptotic value of the scalar field. For $\phi(\infty)=1$,
the quantity $\dot{F}(\phi)$ assumes a non-zero asymptotic value, as in the exponential case,
therefore the DOE, depicted in Fig.~\ref{fig3}(a), is similar to the one displayed in Fig. \ref{fig2}(b).
In contrast, if $\phi(\infty)=0$, then $\dot{F}$ vanishes asymptotically and the range
of $\alpha$, for which wormholes arise, is also limited from below. The DOE
in this case is shown in Fig.~\ref{fig3}(b) -- now, wormholes emerge only
if $0.205 < \alpha/\eta_0^2 <0.480$. The Schwarzschild black holes are now part of
the boundary of the DOE, as indicated by the dot in Fig.~\ref{fig3}(b),
since the constant configuration $\phi \equiv \phi_\infty=0$ solves the scalar field equation trivially.
Moreover, wormhole solutions exist for which the scalar field may possess $N$ nodes.
The boundary of the DOE for $N=1$ is shown in the inlet in Fig.~\ref{fig3}(b).
Note that the range of $\alpha$ in this case is approximately $1.85\leq \alpha/\eta_0^2 \leq 2.75$, 
i.~e.~considerably larger than for $N=0$.

\begin{figure}[t!] 
\begin{center}
\hspace{0.0cm} \hspace{-0.6cm}
\includegraphics[height=.25\textheight, angle =0]{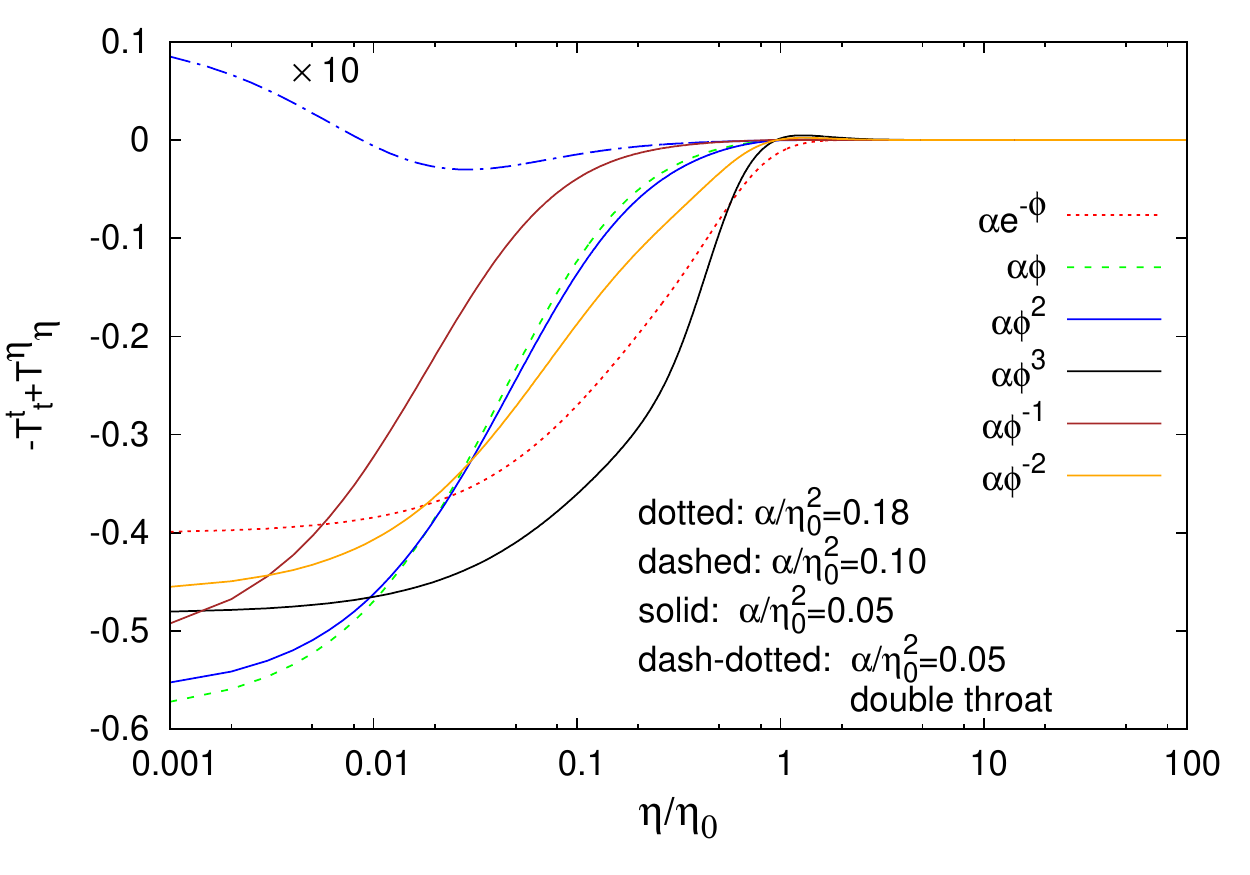}
\hspace{0.52cm} \hspace{-0.6cm}
\includegraphics[height=.25\textheight, angle =0]{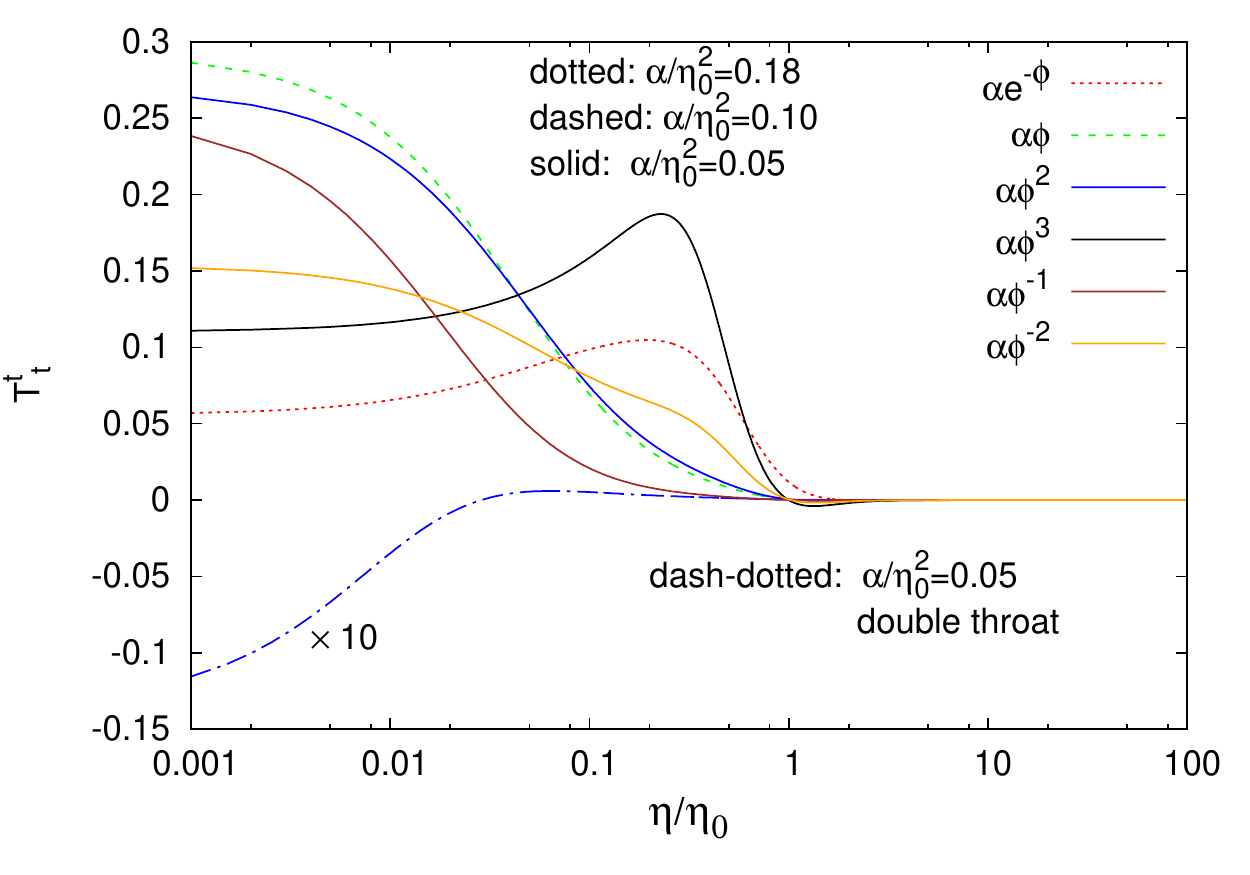}
\\
\hspace*{0.7cm} {(a)} \hspace*{7.5cm} {(b)}  \vspace*{-0.5cm}
\end{center}
\caption{(a) The Null Energy Condition and (b) the Weak Energy Condition for
a variety of forms of the coupling function $F(\phi)$.
\label{fig4}}
\end{figure}

Let us finally address the issue of the violation of the Null and Weak Energy Conditions. In 
Fig.~\ref{fig4}(a), we display the quantity $-T_t^t + T_\eta^\eta$ for a number of wormhole
solutions arising for different forms of the coupling function $F(\phi)$.  It is evident that
the NEC is always violated near the throat of each solution by an amount which depends
on the form of the coupling function $F(\phi)$ since the latter determines the weight of the
GB term in the theory. On the other hand, the NEC is obeyed at asymptotic infinity. We note
that, in the case of the double-throat solution, the NEC is violated at the throat while it is
obeyed at the equator, according to the analysis of the previous section. A similar behaviour
is exhibited by the $T_t^t$ component depicted in Fig. ~\ref{fig4}(b): the WEC is again violated
at the small $\eta$-regime, by an amount determined by $F(\phi)$, while it is obeyed at
asymptotic infinity where the GB term becomes negligible. The double-throat solution 
again respects the WEC at the equator while it violates it near the throat.


\section{Junction conditions}\label{se4}

\begin{figure}[t!] 
\begin{center}
\hspace{0.0cm} \hspace{-0.6cm}
\includegraphics[height=.35\textheight, angle =0]{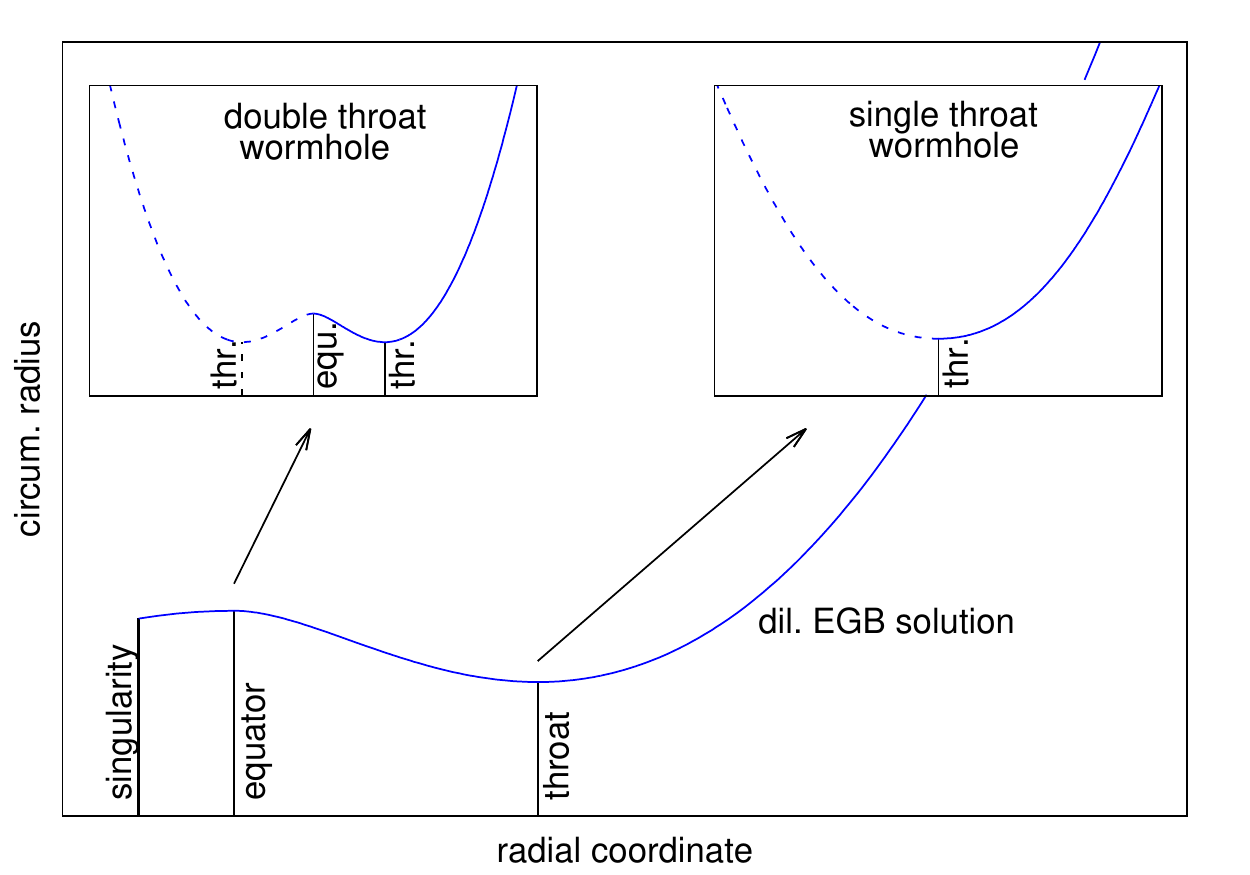}
\end{center}
\caption{Schematic picture for the construction of double-throat and single-throat
wormhole solutions.
\label{fig5}}
\end{figure}

Wormhole solutions may be either symmetric
or asymmetric under the change $\eta \rightarrow - \eta$. In the context of the EsGB theory
with an exponential coupling function \cite{Kanti:2011jz,Kanti:2011yv}, asymmetric wormholes were found
but they were plagued by curvature singularities lurking behind the throat. A regular
wormhole solution may then be constructed by imposing a symmetry under the change
$\eta \rightarrow - \eta$. The obtained solution then consists of two parts: the first
coincides with the part of the asymmetric solution which extends from the asymptotic
region at infinity to the location of the throat; the second part of the wormhole
solution is obtained by the symmetric continuation of the first part in the negative
$\eta$-regime.

A similar construction was performed in the context of the present analysis, in
the case of solutions with a single throat -- these solutions are the ones depicted in
Figs. \ref{fig2}(b) and \ref{fig3}(a,b) under the dashed-dotted curves. In the case
of singular wormhole solutions with a throat and an equator, a similar process may
give rise to double-throat wormholes and to single-throat wormholes since
now there are two options, as Fig. 5 depicts. The first option is to 
construct a regular wormhole by cutting at the throat and symmetrically continuing
to the left, as described above; in that case, the equator is removed from the
spacetime geometry and a single-throat wormhole is constructed. The second option
is to cut the singular solution at the equator, keep the regular part from the
asymptotic infinity to the equator and continue symmetrically to the left; in
this way, a double-throat wormhole solution, with an equator located exactly
between the throats, is constructed. Both wormholes possess the same mass and
scalar charge, since these quantities are extracted from the asymptotic region
that is common in both solutions. Hence, for any double-throat wormhole there
exist a single-throat wormhole with the same mass and scalar charge -- these are
the solutions depicted in Figs. \ref{fig2}(b) and \ref{fig3}(a,b) above the
dashed-dotted curves.

Let us now discuss in more detail the construction of symmetric, regular, and
thus traversable, wormholes
\footnote{As the coupling function may acquire different forms in the context of our analysis,
constructing also asymmetric wormholes, see e.g. \cite{Bronnikov:2017kvq}, may be indeed a possibility
that we plan to pursue in a future work.}.  
From Fig. \ref{fig1}, we observe that the derivatives of the $-g_{tt}$ and $\phi$ do not vanish
in general at $\eta=0$. Therefore, imposing a symmetry under
$\eta \rightarrow - \eta$ creates a ``cusp'' in the profile of the aforementioned quantities.
This feature may be attributed to the presence of a distribution of matter at $\eta=0$,
i.e. around the throat or the equator, for single or double-throat solutions, respectively. 
The embedding of this thin-shell matter distribution in the context of the complete
solution is determined through the junction conditions \cite{Israel:1966rt,Davis:2002gn}, that follow by
considering the jumps in the Einstein and scalar field equations (\ref{Einstein-eqs})-(\ref{scalar-eq}) as
$\eta \rightarrow -\eta$. These are found to have the form
\begin{equation}
\langle G^\mu_{\phantom{a}\nu} -T^\mu_{\phantom{a}\nu}\rangle = s^\mu_{\phantom{a}\nu} \ , \ \ \ 
\langle \nabla^2 \phi + \dot{F} R^2_{\rm GB}\rangle = s_{\rm scal} \ ,
\label{jumps}
\end{equation}
where $s^\mu_\nu$ denotes the stress-energy tensor of the matter
at the throat, resp. equator, and $s_{\rm scal}$ a source term for the scalar field.
For a physically-acceptable solution, this matter distribution should not be exotic.
We thus assume a perfect fluid with pressure $p$ and energy density $\rho$, and
a scalar charge $\rho_{\rm scal}$ at the throat, resp. equator, together with the 
gravitational source \cite{Kanti:2011jz,Kanti:2011yv}
\begin{equation}
S_\Sigma = \int \left[\lambda_1 + 2 \lambda_0 F(\phi) \bar{R}\right]\sqrt{-\bar{h}} d^3 x
\label{act_th}
\end{equation}
where $\lambda_1,\lambda_0$ are constants, $\bar{h}_{ab}$ is the three-dimensional
induced metric at the throat, resp. equator, and $\bar{R}$ is the corresponding Ricci
scalar. Substitution of the metric then yields the junction conditions
\begin{eqnarray}
8 \dot{F} \phi' e^{-\frac{3 f_1}{2}}
& = &
\lambda_1\eta_0^2 + 4\lambda_0 F  e^{-f_1}- \rho\eta_0^2 \ , 
\label{j_00}\\
e^{-\frac{f_1}{2}} f_0' 
& = &
\lambda_1 + p  \ , 
\label{j_tt}\\
e^{-f_1}\phi' - 4  \frac{\dot{F}}{\eta_0^2} f_0' e^{-2 f_1}
& = &
-4\lambda_0\frac{\dot{F}}{\eta_0^2} e^{-\frac{3 f_1}{2}} +\frac{\rho_{\rm scal}}{2} \ ,
\label{j_ss}
\end{eqnarray}
%
%
%
where all quantities are taken at $\eta=0$. 
The above junction conditions determine $\rho$, $p$ and $\rho_{scal}$
in terms of the arbitrary constants $\lambda_0$ and $\lambda_1$ and the form of the scalar
field and metric functions close to the boundary. 


\begin{figure}[t!] 
\begin{center}
\hspace{0.0cm} \hspace{-0.6cm}
\includegraphics[height=.35\textheight, angle =0]{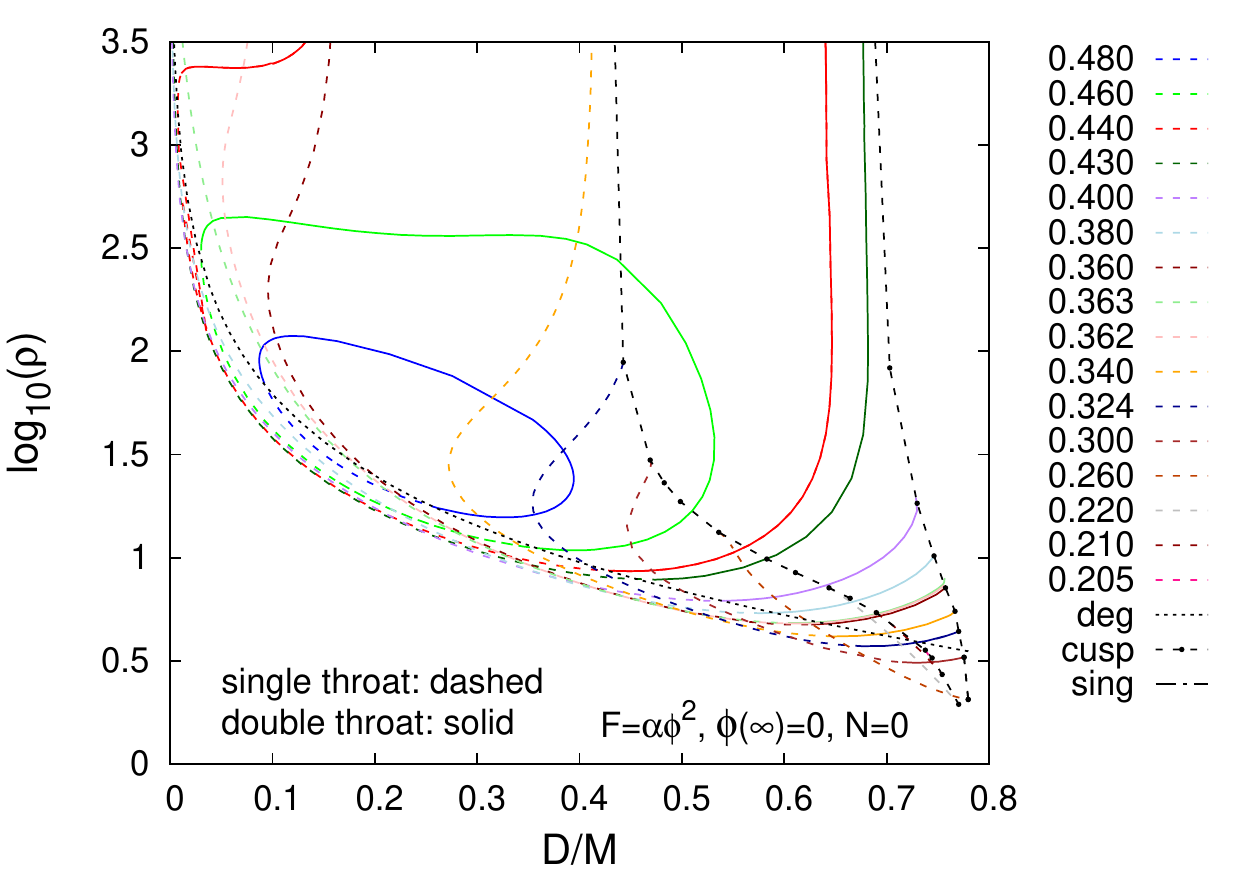}
\end{center}
\caption{The energy density $\rho$ at $\eta=0$, for $F(\phi)=\alpha \phi^2$ for several values of
$\alpha/\eta_0^2$, $p=0$ (dust) and specific values of $(\lambda_0, \lambda_1)$.
\label{fig6}}
\end{figure}


For every form of the coupling function $F(\phi)$, we may find an extensive
($\lambda_0$, $\lambda_1$)-parameter regime over which $\rho$ is always positive, and the
necessity of the exotic matter is thus avoided. An interesting special case is when the
matter distribution around the throat has a vanishing pressure, i.e. $p=0$, and therefore
its equation of state is the one of dust. In this case, Eq. (\ref{j_tt}) gives
$\lambda_1=e^{-f_1/2}f_0'$. If we choose $\lambda_0=\lambda_1$, Eqs. (\ref{j_00}) and
(\ref{j_ss}) easily yield
\beq
    \rho=\frac{e^{-\frac{3 f_1}{2}}}{\eta_0}\left[  \left(  4 F + \eta_0^2 e^{f_1} \right)f_0'
    -8\dot F \phi'     \right], \qquad 
    \rho_{\rm sc}=2 e^{-f_1} \phi',\label{rhomat} 
\eeq
respectively, where again all quantities are evaluated at $\eta=0$.
In Fig. \ref{fig6}, we depict the energy density $\rho$ at the throat, resp. equator,
as a function of the scaled scalar charge $D/M$, for a variety of wormhole
solutions arising for $F(\phi)=\alpha \phi^2$ and for
the aforementioned values of $p$, $\lambda_0$ and $\lambda_1$. We note that in this
example the energy density $\rho$ is positive for all wormhole solutions.
As in the construction of the solution, where the synergy of an ordinary distribution
of matter with a gravitational source kept the throat, resp. equator, open, here
a similar synergy creates a symmetric wormhole free of singularities.


\section{Embedding Diagram}\label{se5}

A useful way to visualize the geometry of a given manifold is the construction of the
corresponding embedding diagram. In this case, we consider the isometric embedding of
the equatorial plane of our wormhole solutions, defined as the line-element (\ref{metric4})
for $t=const.$ and $\theta=\pi/2$. The isometric embedding follows by equating the
line-element of the two-dimensional equatorial plane with a hypersurface in the
three-dimensional, Euclidean space, namely 
\begin{equation}
e^{f_1}\,[d\eta^2 +(\eta^2+\eta_0^2)\,d\varphi^2]=
dz^2 +dw^2 +w^2 d\varphi^2\,, \label{emb}
\end{equation}
where ($z$, $w$, $\varphi$) is a set of cylindrical coordinates on the hypersurface.
Considering $z$ and $w$ as functions of $\eta$, we find
\bea
    w&=&e^{f_1/2}\sqrt{\eta^2+\eta_0^2}, \label{wdef}\\[2mm]
    \left(\frac{dw}{d\eta}\right)^2+\left(\frac{dz}{d\eta}\right)^2&=&e^{f_1}\,.\label{zzde}
\eea
Then, combining the above equations, we find
\beq
   z(\eta)=\pm \int_0^\eta \sqrt{e^{f_1(\tilde \eta)} -\left( \frac{d}{d\tilde \eta}
   \left[  e^{f_1(\tilde \eta)/2}\sqrt{\tilde \eta^2+\eta_0^2} \right]  \right)^2}d\tilde \eta. \label{zeq}
\eeq 
Therefore, $\{w(\eta), z(\eta)\}$ is a parametric representation of a slice of the embedded
$\theta = \pi/2$-plane for a fixed value of the $\varphi$ coordinate, while the corresponding
surface of revolution is the three-dimensional representation of the wormhole's geometry. 


\begin{figure}[t!] 
\begin{center}
\hspace{0.0cm} \hspace{-0.6cm}
\includegraphics[height=.25\textheight, angle =0]{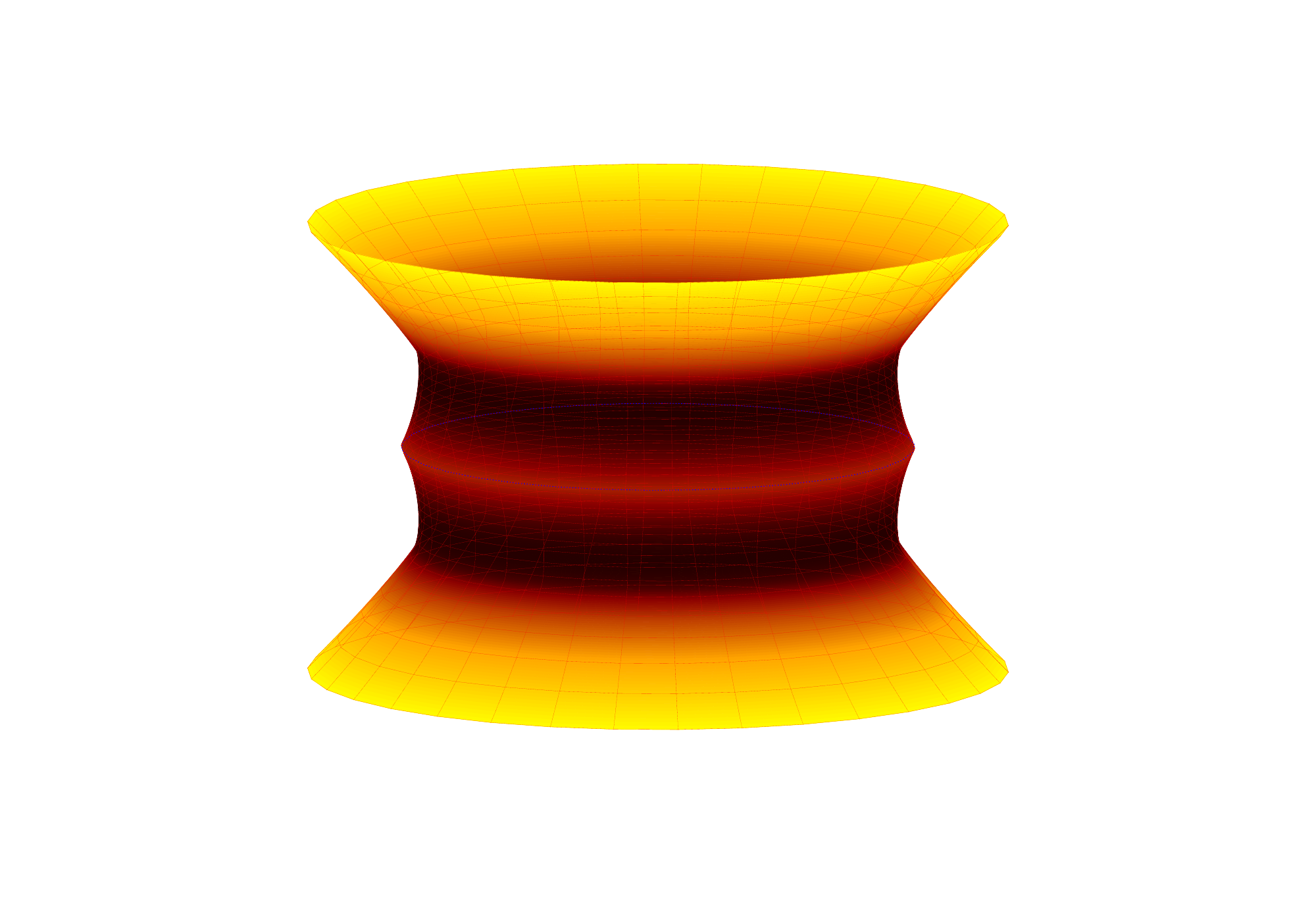}
\hspace{0.52cm} \hspace{-0.6cm}
\includegraphics[height=.25\textheight, angle =0]{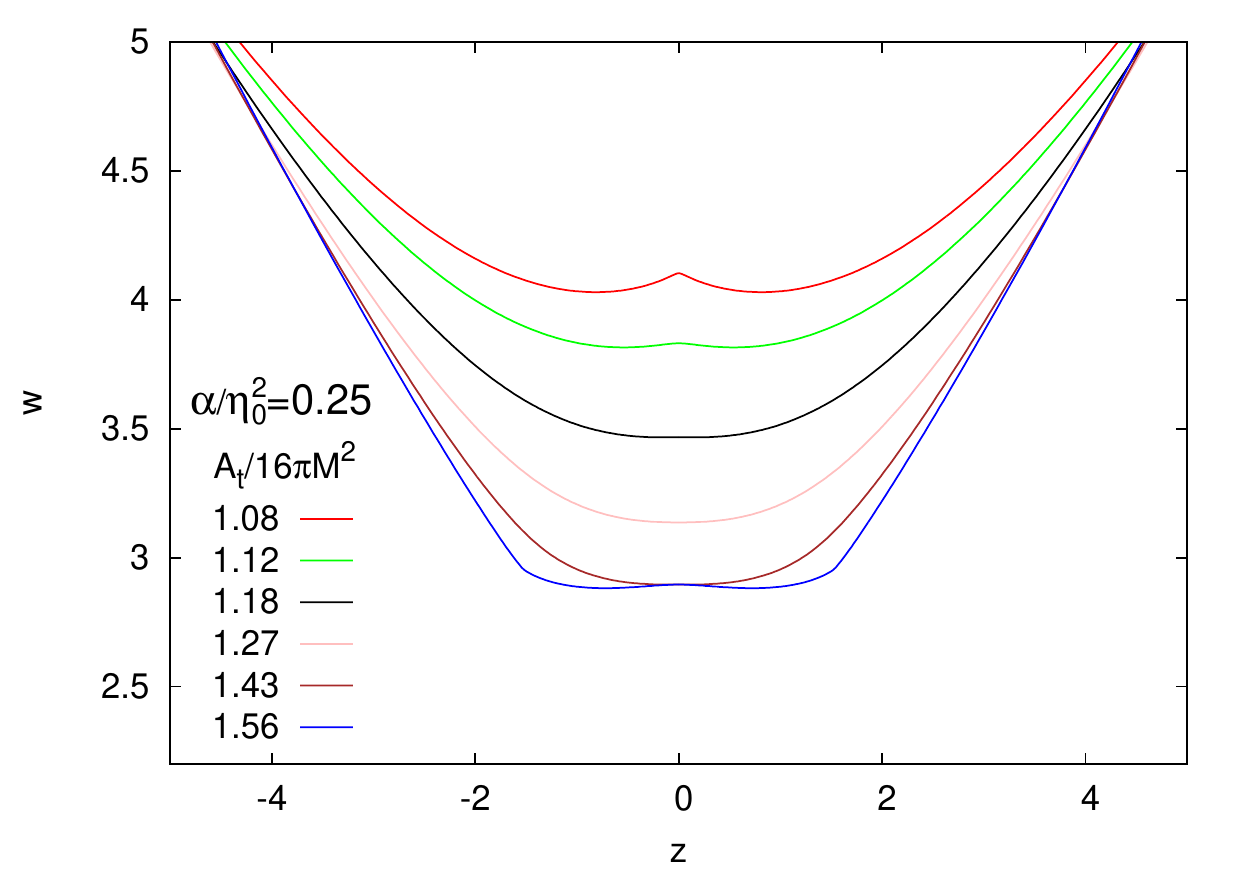}
\\
\hspace*{0.7cm} {(a)} \hspace*{7.5cm} {(b)}  \vspace*{-0.5cm}
\end{center}
\caption{\label{fig7}
(a) The embedded equatorial plane is shown for the double-throat wormhole with 
$\alpha/\eta_0^2=0.35$ and $D/M=0.886$.
(b) The profiles of the isometric embedding
are shown for a sequence of solutions for the coupling function $F=0.25 e^{-\phi}$.}
\end{figure}


In Fig. \ref{fig7}(a), we depict the isometric embedding of the geometry of a
symmetric, traversable, double-throat wormhole solution. 
The three-dimensional view of the surface follows from the parametric plot
$(w(\eta)\cos\varphi,w(\eta)\sin\varphi,z(\eta))$ as described above.  
The 
diagram clearly features an equator and two throats smoothly connected to two
asymptotic regimes. In Fig.~\ref{fig7}(b), we also show the geometry transition
between single and double-throat wormholes, by plotting $w$ vs. $z$, for a sequence
of solutions for fixed $\alpha/\eta_0^2=0.25$. We observe that, with increasing scaled
throat area, the double-throat wormholes develop a degenerate throat and turn into
single-throat ones. If the scaled throat area is increased further, a second transition
takes place where the single-throat wormholes turn again to double-throat ones.

\section{Bounds on the EsGB theory}

Let us address at this point the issue of the existing bounds on the GB coupling constant. The parameters of any modified gravitational theory, including the EsGB theory, may be constrained by processes and observations in strong gravitational regimes. We have already mentioned in section \ref{esgbi} that, in the cosmological Friedmann–Lemaître–Robertson –Walker (FLRW) background  the EsGB theory predicts a propagation speed for the gravitational waves different from the speed of light \cite{DeFelice}. Therefore, after the detection of the GW170817 gravitational wave event and its GRB170817A gamma-ray burst counterpart \cite{gw2, gw3, gw4} the Einstein-scalar-Gauss-Bonnet theory was ruled out as cosmological dark energy model. Although the theory is ruled out as a late time cosmological model, we can still use it as a model for early time cosmological solutions \cite{Odintsov:2019clh}  or even for local solutions like black-holes or wormholes\cite{Nair:2019iur,gwww1,gwww2}. 

Actually, a Gauss-Bonnet black hole or wormhole may affect the speed of a gravitational wave but only locally. For the Gauss-Bonnet solutions the gravitational background at first approximation reduces very fast to that of the corresponding GR solution at distances a few times the characteristic scale of the solution e.g. the horizon radius of the black hole or the throat radius of the wormhole, for example see Figs. \ref{Metric_GB}(b) and \ref{fig2}(a). What is clear is that the derivation of the dispersion relation for cosmological backgrounds may be very different from that in a black-hole background. Indeed, in \cite{Ayzen}, the authors performed a linear stability analysis of the dynamical quadratic gravity, and working in the geometric optics limit found that there are corrections to the dispersion relation but these decay very fast compared to the GR terms.

Shortly afterwards, a more rigorous analysis was performed \cite{Koba1,Koba2} where perturbations
were considered around a static, spherically-symmetric BH solution arising in the context
of quadratic gravity. They found that, indeed, the speed of gravitational waves is modified,
and a formula was produced. Using our EsGB theory and the behaviour of the scalar and
metric functions at asymptotic infinity, we have found that
\begin{equation}
c^2_{gw}=\frac{\mathcal{G}}{\mathcal{F}}=1+\frac{8D\dot{F}(\f_\infty)}{r^3}+\frac{4D\ddot{F}(\f_\infty)+8MD\dot{F}(\f_\infty)}{r^4}+\cdots ,
\end{equation}
where the expressions for the functions $\mathcal{F}$ and $\mathcal{G}$ may be found in appendix \ref{apa3}, and for their calculation we employ the expansions at infinity Eqs. (\ref{inff0})-(\ref{infff}). Therefore, if $\dot{F}(\f_\infty)\neq 0$ and $\ddot{F}(\f_\infty)\neq 0$, in principle, the gravitational-wave speed may be modified but the effect is proportional to the coupling which is assumed to be small. Also, $D$ is the scalar charge that, as we know, neutron stars do not have any. Therefore, this particular event GW170817 cannot impose any constraint on our solutions. Even if we use the other gravitational wave detections that involve black-holes instead of neutron stars, the correction terms scale at least as $1/r^3$; these turn out to be very small given the fact that most of the mergers were detected at distances larger than 300 Mpc. 

The last years, many researchers have  found bounds for the EsGB theory using astrophysical or cosmological data. Using data for the time delay produced by the gravitational field of the Sun for light to travel from Earth to the Cassini spacecraft and back to Earth while passing close to the Sun \cite{Bertotti:2003rm}, the following bound was derived on the value of the GB coupling constant \cite{Amendola}: 
$
\saa\leq 1.25\times10^{12} \,\,\rm{cm}.
$
In the same work, another weaker bound was derived based on the uncertainty in the values of the semi-major axes of the planetary orbits around the Sun (particularly Mercury’s):
$
\saa\leq1.9\times10^{13} \,\,\rm{cm}.
$
In \cite{DBH1}, a theoretical upper bound was imposed on the GB coupling constant so that black-hole solutions exist in the EdGB theory \cite{Pani:2009wy} (see also Eq. (\ref{con-f})): 
\begin{equation}
\saa\leq\left(\frac{M}{M_\odot}\right)\times10^5 \,\,\rm{cm}.
\end{equation}
 This bound imposes a minimum mass for black-hole solutions and thus must be checked against the minimum black-hole masses that have ever been detected. Using $M\sim 5M_\odot$ we found
$
\saa\leq 6.2\times 10^5 \,\,\rm{cm}.
$

 The existence of compact stars, i.e. neutron stars, in the EdGB theory was investigated in \cite{Pani1}. It was found that for a given central density of matter, there is a maximum allowed value of the coupling constant for which neutron stars may be formed. That bound was found to be
$
\saa\leq7.2\times10^5 \,\,\rm{cm}.
$
 A lot of work has been done on the modifications that quadratic gravity terms bring to the orbital decay rate of binary systems \cite{Yunes2011, Yagi, Yunes11, Yagi2}. The most stringent constraint on this direction 
$
\saa\leq1.9\times10^5 \,\,\rm{cm},
$
 was found in   \cite{Yagi3} using a low-mass X-ray binary. Finally, in \cite{Kunz}, the stability analysis of both axial and polar sectors of the static black-hole solutions in EdGB was performed. Apart from the linear stability of the solutions that was demonstrated, the authors studied the Quasi-Normal-Modes (QNMs), which will be relevant for the late-time behaviour of a black-hole merger. The measurement of two of the dominant modes may help us to place a bound on the coupling constant of the GB. According to the authors, this bound has the form:
\begin{equation}
\saa\leq 10\left( \frac{50}{\r}\right)\left(\frac{M}{10M_\odot}\right)\times 10^5 \,\,\rm{cm},
\end{equation}
  where $\r$ is the signal-to-noise ratio  of the measurement instrument. All LIGO-Virgo detections have a signal-to-noise ratio $\r$ around $10$ while for the Einstein Telescope, it will hold that $\r=100$. A list of all known bounds in EdGB theory may be found in Table \ref{R-GB}.

\begin{table}[t!]
\begin{center}
$\begin{array}{|l|l|} \hline\hline &  \\[-0.35cm] 
\hspace*{2cm}  \text{Method} \hspace*{1cm}&  \hspace*{0.6cm}
(\sqrt{\a})_{\rm max}  \hspace{2cm} \\[0.05cm] \hline \hline
& \\[-0.35cm]
\text{Solar System} & 1.25 \times 10^{12}\,\rm{cm} \\[0.2mm]
\text{QNMs from LIGO-Virgo} & \,\,\,\,\,25 \times 10^{5}\,\rm{cm} \\[0.2mm] 
\text{Binary Black-Hole Signals} &  10.1 \times 10^{5}\,\rm{cm}\\[0.2mm]
\text{Existence of Neutron Stars} & \,\,\, 7.2 \times 10^{5}\,\rm{cm} \\[0.2mm]
\text{Existence of Black Holes} & \,\,\, 6.2 \times 10^{5}\,\rm{cm} \\ 
\text{QNMs from Einstein Telescope} & \,\,\, 2.5 \times 10^{5}\,\rm{cm} \\[0.2mm]
\text{Orbital Decay Rate in Binaries} &  \,\,\,1.9 \times 10^{5}\,\rm{cm} \\[0.05cm] 
%
\hline \end{array}$
\end{center}
\caption{A list of the known bounds in EdGB theory.
\label{R-GB}}
\end{table}

The most recent bound on the GB coupling parameter $\alpha$ was set in \cite{Nair:2019iur} where
the effect of the scalar dipole radiation on the phase evolution of the gravitational
waveform was taken into account -- this radiation was emitted during the merging process
of two binary systems in which one of the constituents could be a scalarised black hole
(GW151226 and GW170608 as detected by LIGO). This bound was set on the value
\begin{equation}
\sqrt{\alpha} < 10.1\times10^5 \,\,\rm{cm},
\end{equation}
taking into account the different definitions of
$\alpha$; in dimensionless units, this translates to $\alpha/M^2<1.72$, where $M$
is the characteristic mass scale of the system, i.e the black-hole mass. In the 
absence of a direct bound on wormholes, since no such object has been detected so far,
and demanding that the EsGB theory should allow for both black-hole solutions and
wormholes to emerge, we apply the aforementioned bound by LIGO on our wormhole
solutions, too. For an exponential coupling function, all of our solutions satisfy
the bound $\alpha/M^2 < 0.91$ while for a quadratic coupling function we obtain
$\alpha/M^2 < 0.605$ (for solutions with no nodes for the scalar field), respectively
$\alpha/M^2 < 2.9$ (for solutions with one node). Thus the observational bound
leaves unaffected the aforementioned DOEs: all solutions in Fig. \ref{fig2}(b) and
Figs. \ref{fig3}(a,b) (with no nodes) fall entirely within the allowed range\footnote{Note that these plots include both the wormhole and the black hole solutions.}.


\section{Discussion}\label{se6}

In this chapter, we have considered a general class of EsGB theories with an arbitrary
coupling function between the scalar field and the quadratic Gauss-Bonnet term.
By employing a novel coordinate system, we have allowed for wormhole solutions 
with either single-throat or double-throat geometries to emerge. We have determined 
the asymptotic form of the metric functions and scalar field in the small and
large radial-coordinate regimes, and demonstrated that the Null and Weak Energy
Conditions may be violated, especially in the inner regime where the effect of
the GB term is dominant. 

We have then numerically integrated our set of field equations in order to determine
the complete wormhole solutions that interpolate between the derived asymptotic
solutions. We have found wormholes, with either a single throat or a double
throat and an equator, for every form of the coupling function we have tried. 
The spacetime is regular over the entire positive range of the radial coordinate,
as also is the non-trivial scalar field that characterizes every wormhole solution.
Our solutions are therefore characterized by two independent parameters, their mass
and scalar charge. The domain of existence has been studied in detail in each case,
and here we have presented the ones for the exponential and quadratic coupling functions
in order to discuss the qualitative differences as the form of the coupling function
and the value of the scalar field at asymptotic infinity varies. 

An important result of our analysis is that the EsGB theories always feature wormhole
solutions without the need for exotic matter, since the higher-curvature terms
allow for gravitational {\sl effective} negative energy densities. This has been
demonstrated by examining the Null and Weak Energy Conditions for our solutions and
showing that indeed the coupling between the scalar field and the GB term results in
a negative energy density near the throat/equator. The Null Energy Condition is
also violated since it is associated with the appearance of a throat that every
wormhole solution must possess. 

In order to construct traversable wormhole solutions with no spacetime singularities
beyond the throat or equator, our regular solution over the positive range of the
radial coordinate was extended in the negative range in a symmetric way. This
construction demands the introduction of a distribution of matter around the throat
or equator that nevertheless may be shown to consist of physically-acceptable particles.
We have provided an indicative example where this distribution of matter is described
by the equation of state of dust with a vanishing isotropic pressure and a positive
energy density.

Our next step will be to study the physical characteristics of our solutions in greater
detail and to generalise them to admit also rotation \cite{Kleihaus:2014dla}. In addition,
a linear stability analysis of these EsGB wormholes \cite{Kanti:2011jz,Kanti:2011yv, Evseev:2017jek, Cuyubamba:2018jdl} 
will be performed and their radial and quasi-normal modes, which could be observable
signatures of their existence, will be determined.

\clearpage
\thispagestyle{empty}

\chapter{Conclusions}\label{7}

Einstein’s General theory of Relativity  is the theory that describes gravity. It is a geometrical theory in which the gravitational interactions are related with the geometry of spacetime. Einstein’s theory not only includes (as a limit) the traditional Newtonian gravitational theory, which describes weak gravitational fields but also extends it providing a framework for the description of strong gravitational interactions. Although General Relativity is well tested, it is known that it is not a perfect theory. The Standard Model for Cosmology has many open problems, like the nature of the dark energy and matter, the accurate model for inflation or the initial singularity problem. All the above, together with theoretical reasons such as the non-renormalizability of General Relativity, motivate the need for modified gravitational theories.

The last decades many theories have been proposed as alternatives to General Relativity. However, the most simple and most studied modified theories of gravity are the scalar-tensor theories. In these, the degrees of freedom of the gravitational field are increased with the addition of one or more scalar fields. The last twenty years a scalar-tensor theory which is known as  the Einstein-scalar-Gauss-Bonnet theory has attracted the interest of many scientists. In this theory, the scalar field $\f$ has a  non-minimal coupling $f(\f)$ with the gravitational field through  the quadratic gravitational Gauss-Bonnet term. The most attractive feature of this theory is that it belongs to the family of Horndeski theories, thus the field equations are of second order and as a result, it avoids Ostrogradsky instabilities or problematic ghost states. In this work we focused on the derivation of novel local solutions in the framework of the Einstein-scalar-Gauss-Bonnet theory. 

One of the most interesting properties of the black-hole solutions in General Relativity is their uniqueness and simplicity: black holes are uniquely determined only by three physical quantities (mass, electromagnetic charge and angular momentum). No-Hair theorems, which forbid the association of the GR black holes with any other type of conserved ``charges''  were formulated quite early on. Even in the case of the scalar-tensor theories, No-scalar Hair theorems were proposed which made difficult the derivation of new black-hole solutions in these theories. However, in some scalar-tensor theories the No-Hair theorems are evaded and new solutions have been found.

 In chapter \ref{3} we discussed the evasion of the No-scalar hair theorem in the framework of the Einstein-scalar-Gauss-Bonnet theory. In order to kept our analysis as general as possible we keep arbitrary the form of the coupling function. At first we examined the evasion of   Bekenstein's old No-scalar-hair theorem. This theorem is based on the integration of the scalar field equation over the exterior regime of the black hole. The old No-Hair theorem may be evaded only for positive coupling functions $f(\f)>0.$ Then, we examined the evasion of novel Bekenstein's No-scalar-Hair theorem. The novel theorem, which is more general, relies its existence on the sign of the $T^r_r$ component of the energy momentum tensor in the two asymptotic regions of the black hole. Actually, we found that for the novel No-Hair theorem to evaded we only need to impose a constraint on $\f'_h$ and the following constraint $r_h^4>96\dot{f}_h^2$ on the horizon of the black hole. These constraints impose  bounds on the parameters of the theory and not on the explicit form of the coupling function. Thus, for every form of the coupling  function --as long as the parameters of the theory validate the constraints-- we may evade the No-Hair theorem. Also, the above constraint denotes the existence of a lower bound on the horizon radius $r_h$ and thus the mass of the black hole. 

Subsequently, we investigated the spontaneous scalarisation effect in the framework of a general coupling function. The spontaneous scalarisation effect for black holes was discovered first for the special cases of the exponential and quadratic coupling functions. This effect appears for coupling functions which allow the Schwarzschild black hole (with an everywhere constant scalar field $\f(r)=\f_0$) to be a solution of the Einstein-scalar-Gauss-Bonnet theory.  In these theories, which are identified by the constraints $\dot{f}(\f_0)=0$ and $\ddot{f}(\f_0)>0$, the Schwarzschild black hole with mass smaller than a critical mass $M_c$ becomes unstable and as a result the system assumes the more stable Gauss-Bonnet black hole as a solution. Although the spontaneous scalarisation effect may provide an explanation on how these solutions may arise it is not an existence theorem. The only way for a spontaneous scalarised solution to exist is to validate the constraint $r_h^4>96\dot{f}_h^2$ like every other solution in the framework of the Einstein-scalar-Gauss-Bonnet theory. Finally, we found that the restrictions of the No-scalar-Hair theorem apply in the case of the pure scalar-Gauss-Bonnet theory. In this theory we cannot construct a regular black hole horizon and therefore the synergy between the Ricci scalar and the Gauss-Bonnet term is important for the construction of novel, non-trivial black hole solutions. 

In chapter \ref{4} we extended the analysis of the previous chapter by considering several sub-classes of the Einstein-scalar-Gauss-Bonnet theory. We have studied a large number of choices for the coupling function $f(\phi)$ between the scalar field and the GB term: exponential, polynomial (even and odd), inverse polynomial (even and odd) and logarithmic. In each case, employing the appropriate boundary conditions, we numerically constructed  a large number of exact black-hole solutions with scalar hair, and studied in detail their characteristics. Our solutions are  characterized by a universal behavior of the components of the metric tensor, having the expected behavior near the black-hole horizon and asymptotic flatness at radial infinity. The solution for the scalar field is everywhere regular and is characterized by a finite value near the horizon, while it approaches a constant value at infinity. The behavior of the scalar field is indicated by the inequality $\dot f_h \f'_h<0$. For coupling functions with $\dot f_h>0$ the scalar field has a decreasing profile while for $\dot f_h<0$ has an increasing profile. All curvature invariant quantities were examined, and found to have a similar universal profile, independently of the form of the coupling function $f$, that ensured the finiteness and asymptotically flatness of  the space-time and thus the regularity of all solutions. Finally, for every solution, the components of the energy momentum tensor were also found to be regular over the entire radial domain. In accordance with the results of the previous chapter, that of the evasion of the No-Hair theorem, the $T^r_r$ component was found to be positive and decreasing in both asymptotic regimes and in all cases considered. 

The physical characteristics of our black holes were also extracted for every solution. In each case the scalar charge $D$ was determined and its dependence on the mass of the black hole $M$ was studied. In all cases the scalar charge is a mass-dependent quantity and therefore our black holes are characterized by a non-trivial scalar field with a secondary type of hair. For every solution the black-hole entropy $S$ and the horizon area $A$ were also calculated. However, instead of the entropy and the area themselves we mainly  studied their ratios $A_h/A_{Sch}$ and $S_h/S_{Sch}$, with respect to the
corresponding quantities of the Schwarzschild solution with the same mass, since they have even more information to offer.  In the large mass limit, for all cases considered, the scalar charge  $D$  vanishes while both   ratios  $A_h/A_{Sch}$ and $S_h/S_{Sch}$  approach  unity. Therefore large Gauss-Bonnet black holes have the same characteristics with the corresponding Schwarzschild black holes of the same mass. On the other hand, for small masses the Gauss-Bonnet black holes are quite different from the Schwarzschild ones. In all cases the area ratio is always below unity. The Gauss-Bonnet black holes are therefore always smaller than the corresponding Schwarzschild black holes. The Gauss-Bonnet term, as an extra gravitational term, causes the shrinking of the black holes. Contrariwise, for small masses, the entropy ratio may be higher or below unity. The behavior of the entropy ratio may provide indications for the stability of our solutions. Gauss-Bonnet black holes with entropy higher  than that of the Schwarzschild  solution tend to be more thermodynamically stable solutions. Finally, the behavior of the scalar charge is strongly dependent on the form of the coupling function. In the small-$M$ limit, each family of solutions presented a different behavior either monotonic or not monotonic.

In chapter \ref{5} we extended the analysis of the previous two chapters by adding a cosmological constant $\L$ in the theory. At first we investigated the existence of approximate solutions in the two boundaries. In the near horizon region we found solutions that have the same qualitative behavior with the solutions found in the previous chapters. For every form of the coupling function, we may construct regular black hole solutions. As in the previous chapters, the condition for the existence of the horizon implies the existence of a lower bound on the horizon radius and therefore the mass of the black hole.  In the other asymptotic region, that of large distances the solutions depend on the sign of the cosmological constant. For $\L<0$ we found that the spacetime assumes an anti-de Sitter form while the scalar field has a logarithmic form. For $\L>0$ we found that the spacetime assumes a  de Sitter form and the scalar field is characterized by a finite value at the cosmological horizon. In fact the approximate solutions at the cosmological horizon have a similar form with the solutions near the horizon of the black hole. However, from  the No-scalar-Hair theorem we know that the existence of regular solutions in the two boundaries does not guarantee the existence  of an intermediate solution that smoothly connects the two asymptotic solutions. While the addition of a cosmological constant does not alter significantly the form of the field equations, it makes the derivation of an evasion theorem extremely difficult. Therefore we investigated the existence of new solutions only numerically. 

In this chapter we focused on the derivation of numerical  asymptotically de Sitter and anti-de Sitter solutions. As in the previous chapter we used many different forms for the coupling function: exponential, polynomial, inverse-polynomial and logarithmic as well.  For $\L<0$ the spacetime solutions are characterized by a regular black hole horizon and by an asymptotically anti-de Sitter region at infinity in accordance with our  approximate analytic solutions. Also, the solution for the scalar field  is everywhere finite and is characterized by a logarithmic form at infinity $\f(r)= \f_0+d_1 \ln r+\mathcal{O}(r^{-2})$. This behavior along with the absence of an $1/r$ term indicates that we cannot associate a conserved scalar charge with these solutions. The coefficient $d_1$ in front of the logarithmic term  provides us information on how much the large-distance behavior of the scalar field deviates from the power-law one. We have found that this deviation is stronger for Gauss-Bonnet black holes with a small mass whereas for massive ones the $d_1$ coefficient vanishes. In addition, for every solution constructed, we calculated the ratios of the horizon area and entropy, $A_h/A_{SAdS}$ and $S_h/S_{SAdS}$, with respect to the corresponding quantities of the Schwarzschild Anti-de Sitter solution. As in the asymptotically flat case, in the majority of the anti-de Sitter solutions the area ratio is below unity. The anti-de Sitter Gauss-Bonnet black holes are also smaller than the corresponding solutions of the General Relativity. Only in one case of the logarithmic coupling function we found solutions with area ratio above unity. In addition, the entropy ratio provides hints on the stability of our new solutions. The behavior of the entropy ratio depends on both the explicit form of the coupling function and the value of the cosmological constant resulting to potentially stable or unstable solutions in every case considered. 

For $\L>0$ the spacetime solutions are characterized by a regular black hole horizon and by an asymptotically  de Sitter region at large distances with a regular cosmological horizon. While the scalar field is also regular at the black-hole horizon, at the cosmological horizon diverges. However, at the cosmological horizon the trace of the energy momentum tensor, which is a more fundamental quantity, remains finite. A similar behavior has been found in the case of the particle-like solutions in the framework of the EsGB theory. Finally in chapter \ref{5} we considered the more realistic case where the cosmological constant is replaced by a potential for the scalar field. We considered the case of negative polynomial potentials and we constructed  asymptotically flat black-holes. These solutions are everywhere regular and their most attractive feature is their size. Massive black holes tend to be ultra-compact with the most compact having approximately the $1/20$ of the horizon radius of the Schwarzschild solution with the same mass.

Finally, in chapter \ref{6} we constructed traversable wormhole solutions in the framework of the Einstein-scalar-Gauss-Bonnet theory. All the black hole solutions we have found, even the asymptotically de Sitter ones, are characterized by a negative effective energy density near the horizon of the black hole. This negative energy density leads to the violation of both the Weak and the Null Energy Conditions in the near horizon regime. This  violation, in the framework of the Einstein-scalar-Gauss-Bonnet theory, is attributed to the contribution from the Gauss-Bonnet term. The violation of the energy conditions leads to the emergence of   wormhole solutions since it is essential for the creation of the throat. However, since the violation is a purely gravitational  effect it could not be ascribed to the presence of exotic matter. In our case, the scalar field constitutes  ordinary matter and therefore our solutions are considered as realistic. As in the previous chapters we have found numerically   wormhole solutions using many different choices for the coupling function. 

The spacetime of our wormhole solutions  is regular over the entire radial regime and does not possess  horizons or singularities; thus our wormholes are traversable. Also, our wormholes may be  characterized by a  single or even a double throat and an equator in between whereas they are asymptotically flat.   The scalar field is everywhere regular and is characterized by a conserved scalar charge which,   contrary to the case of the black holes, is of primary hair type. In addition, in order to construct traversable wormhole solutions with no spacetime singularities beyond the throat or equator, our regular solution over the positive range of the radial coordinate was extended in the negative range in a symmetric way. This construction demands the introduction of a thin shell distribution of matter around the throat or equator that nevertheless may be shown to consist of physically-acceptable particles. In addition, for every solution we constructed the isometric embedding diagram and the domain of existence has been studied in detail as well. Finally, in this chapter we also discussed the bounds on the Einstein-scalar-Gauss-Bonnet theory and  we found that these bounds leave unaffected the majority of our solutions, both-black holes and wormholes.

Concluding, the Einstein-scalar-Gauss-Bonnet theory is a well-motivated gravitational theory which not only contains Einstein's General Relativity but also extends it with a quadratic curvature term  --as we would expect from a novel effective gravitational field describing gravity in the strong regime. The EsGB theory belongs to the family of the Horndeski theories and therefore it contains only one additional scalar field ant no more than second derivatives in the field equations. In addition, the EsGB theory, interdependently of the form of the coupling function $f(\f)$,     leads to a plethora of novel local gravitational  solutions like black holes, wormholes or even solitons with attractive properties. Finally, regarding the local solutions, the EsGB theory still survives from the imposed bounds of observational data.

\clearpage
\thispagestyle{empty}



\chapter*{   }
\addcontentsline{toc}{chapter}{Appendices}

\vspace{2cm}

\begin{center}
\Huge{ \textbf{Appendices}}
\end{center}

\clearpage
\thispagestyle{empty}

\appendix


\chapter{Black holes in Einstein-Scalar-Gauss-Bonnet theory}\label{appA}
\section{Set of Differential Equations}\label{apa1} 
Here, we display the explicit expressions of the coefficients $P$, $Q$ and $S$ that
appear in the system of differential equations (\ref{Aphi1})-(\ref{Aphi-system}) and (\ref{A}) and whose solution determines
the metric function $A$ and the scalar field $\phi$. They are:
\bea
P&=&+e^{4B}\Bigl(32A'\dot{f}-48r A'^2\dot{f}-8r^2\phi'-4r^3A'\phi'-64r\phi'^2\dot{f}\Bigr)
+e^{3B}\Bigl(-64A'\dot{f}+96rA'^2\dot{f}\nonumber\\
&&+48r^2A'^3\dot{f}+8r^2\phi'-4r^3A'\phi'-4r^4A'^2\phi'+128A'^2\phi'\dot{f}^2
+96rA'^3\phi'\dot{f}^2+64r\phi'^2\dot{f}\nonumber\\
&&+24r^2A'\phi'^2\dot{f}-20r^3A'^2\phi'^2\dot{f}-2r^4\phi'^3+96rA'\phi'^3\dot{f}^2
-16r^3\phi'^4\dot{f}+32r^2\phi'^3\ddot{f}\nonumber\\
&&+r^5A'\phi'^3 +16r^3A'\phi'^3\ddot{f}\Bigr)
+16e^{2B}\Bigl(8A'\dot{f}-12rA'^2\dot{f}-20r^2A'^3\dot{f}-64A'^2\phi'\dot{f}^2\nonumber\\
&&-112rA'^3\phi'\dot{f}^2 -14r^2A'\phi'^2\dot{f}+19r^3A'^2\phi'^2\dot{f}-
96A'^3\phi'^2\dot{f}^3-32rA'\phi'^3\dot{f}^2\nonumber\\
&&+36r^2A'^2\phi'^3\dot{f}^2 +8r^3\phi'^4\dot{f}-4r^4A'\phi'^4\dot{f}
-8r^2\phi'^3\ddot{f}+4r^3A'\phi'^3\ddot{f}-32r^2A'\phi'^4\dot{f}\ddot{f}\Bigr)\nonumber\\
&&+16e^B\Bigl(8A'^2\phi'\dot{f}^2+38rA'^3\phi'\dot{f}^2+64A'^3\phi'^2\dot{f}^3+
18rA'\phi'^3\dot{f}^2-17r^2A'^2\phi'^3\dot{f}^2\Bigr)\nonumber\\
&&-1152A'^3\phi'^2\dot{f}^3\,,\label{P}
\eea
\bea
Q&=&+32e^{5B}r-e^{4B}\left(64r+24r^2A'+160\phi'\dot{f}+48rA'\phi'\dot{f}+4r^3\phi'^2+128r\phi'^2\ddot{f}\right)\nonumber\\
&&+e^{3B}\left(32r+24r^2A'-8r^3A'^2+320\phi'\dot{f}+224rA'\phi'\dot{f}-32r^2A'^2\phi'\dot{f}-
12r^3\phi'^2\right.\nonumber\\
&&\left.+6r^4A'\phi'^2+256A'\phi'^2\dot{f}^2-32rA'^2\phi'^2\dot{f}^2-24r^2\phi'^3\dot{f}+12r^3A'\phi'^3\dot{f}
-32r\phi'^4\dot{f}^2\right.\nonumber\\
&&\left.-r^5\phi'^4+256r\phi'^2\ddot{f}+128r^2A'\phi'^2\ddot{f}+640\phi'^3\dot{f}\ddot{f}+256rA'\phi'^3\dot{f}\ddot{f}-16r^3\phi'^4\ddot{f}\right)\nonumber\\
&&+e^{2B}\left(128r^2A'^2\phi'\dot{f}-160\phi'\dot{f}
-176rA'\phi'\dot{f}-640A'\phi'^2\dot{f}^2+320rA'^2\phi'^2\dot{f}^2\right.\nonumber\\
&&\left.+152r^2\phi'^3\dot{f}-52r^3A'\phi'^3\dot{f}+128A'^2\phi'^3\dot{f}^3+256\phi'^4r\dot{f}^2-80r^2A'\phi'^4\dot{f}^2+4r^4\phi'^5\dot{f}\right.\nonumber\\
&&\left.-128r\phi'^2\ddot{f}-128r^2A'\phi'^2\ddot{f}-1280\phi'^3\dot{f}\ddot{f}-1280rA'\phi'^3\dot{f}\ddot{f}+16r^3\phi'^4\ddot{f}\right.\nonumber\\
&&\left.-1280A'\phi'^4\dot{f}^2\ddot{f}  +64r^2\phi'^5\dot{f}\ddot{f}\right)+e^B\left(384A'\phi'^2\dot{f}^2-672rA'^2\phi'^2\dot{f}^2-768A'^2\phi'^3\dot{f}^3\right.\nonumber\\
&&\left.-480r\phi'^4\dot{f}^2+144r^2A'\phi'^4\dot{f}^2+640\phi'^3\dot{f}\ddot{f}+
1024rA'\phi'^3\dot{f}\ddot{f}+3584A'\phi'^4\dot{f}^2\ddot{f}\right.\nonumber\\
&&\left.-64r^2\phi'^5\dot{f}\ddot{f}\right)+1152A'^2\phi'^3\dot{f}^3-
2304A'\phi'^4\dot{f}^2\ddot{f}\,,\label{Q}\nonumber
\eea
and
\bea
S &=& +128r\dot{f}e^{4B}+8e^{3B}\left(r^4\phi'-32r\dot{f}-16r^2A'\dot{f}-80\phi'\dot{f}^2
32rA'\phi'\dot{f}^2+4r^3\phi'^2\dot{f}\right)\nonumber\\
&&+32e^{2B}\left(4r\dot{f}+4r^2A'\dot{f}+40\phi'\dot{f}^2+40rA'\phi'\dot{f}^2-
3r^3\phi'^2\dot{f}+40A'\phi'^2\dot{f}^3-4r^2\phi'^3\dot{f}^2\right)\nonumber\\
&&+8e^B\left(32r^2\phi'^3\dot{f}^2-80\phi'\dot{f}^2-
128rA'\phi'\dot{f}^2-448A'\phi'^2\dot{f}^3\right)+2304A'\phi'^2\dot{f}^3\,.\label{S}
\eea
\section{Scalar quantities}\label{apa2}

By employing the metric components of the line-element (\ref{metric}), one may
compute the following scalar-invariant gravitational quantities:
\bea
R&=&+\frac{e^{-B}}{2r^2}\left(4e^B-4-r^2A'^2+4rB'-4rA'+r^2A'B'-2r^2A''\right),\label{A1}\\\nonumber\\
R_{\mu\nu}R^{\mu\nu}&=&+\frac{e^{-2B}}{16 r^4}\left[8(2-2e^B+rA'-rB')^2+r^2(rA'^2-4B'-rA'B'+2rA'')^2\right.\nonumber\\
&&\left.+r^2(rA'^2+A'(4-rB')+2rA'')^2\right],\\\nonumber\\
R_{\mu\nu\rho\sigma}R^{\mu\nu\rho\sigma}&=&+\frac{e^{-2B}}{4r^4}\left[r^4A'^4-2r^4A'^3B'-4r^4A'B'A''+r^2A'^2(8+r^2B'^2+4r^2A'')\right.\nonumber\\
&&\left.+16(e^B-1)^2+8r^2B'^2+4r^4A''^2\right],\\\nonumber\\
R_{GB}^2&=&+\frac{2e^{-2B}}{r^2}\left[(e^B-3)A'B'-(e^B-1)A'^2-2(e^B-1)A''\right].\label{A4}
   \eea

\section{Gravitational waves speed in Horndeski theory}\label{apa3}

The expression for the propagation speed of gravitational waves along the radial direction for a black hole in Horndeski theory was found at \cite{Koba1, Koba2} 
\begin{equation}
c^2_{gw}=\frac{\mathcal{G}}{\mathcal{F}}. 
\end{equation}
Using the metric (\ref{metric1}) we may express the functions $\mathcal{F}$ and $\mathcal{G}$ in terms of the Horndeski's $G_i$ functions 
\begin{align}
\mathcal{F}&=2\left( G_4+\frac{1}{2}e^B \f' X' G_{5X}-XG_{5\f}\right),\\[3mm]
\mathcal{G}&=2\left[ G_4-2XG_{4X}+X\left( \frac{A'}{2}e^B\f'G_{5X}+G_{5\f}    \right)\right] .  
\end{align}

\clearpage
\thispagestyle{empty}

\chapter{Black holes with a cosmological constant in EsGB theory}\label{appB}

\section{Set of Differential Equations}
\label{appb1}

Here, we display the explicit expressions of the coefficients $P$, $S$ and $Q$ that
appear in the system of differential equations (\ref{A-sys})-(\ref{phi3}) and whose solution determines
the metric function $A$ and the scalar field $\phi$. Note, that in these
expressions we have eliminated again, via Eq. (\ref{B'3}), $B'$, that involves
$A''$ and $\phi''$, but retained $e^B$ for notational simplicity. They are:
\begin{align}
P&=-128 e^{4 B} \Lambda ^2 r^3 \dot f \left(r A'+2 e^B-2\right)+16
   A'^3 \dot f \Bigl[     
   -2 e^B \left(-14 e^B+3 e^{2 B}+19\right)
   r \dot f \phi ' \nn \\
   &+8\left(-8 e^B+3 e^{2 B}+9\right) \dot f^2
   \phi '^2-e^{2 B} \left(3 e^B-5\right) r^2  \Bigr] 
   +4 e^B  A'^2 \bigg\{
   e^B r \dot f \Bigl[\left(5 e^B-19\right) r^2
    \phi'^2 \nn\\
    &+12 \left(e^B-1\right)^2\Bigr]
  -4 \dot f^2 \phi' \Big[\left(9 e^B-17\right) r^2
   \phi'^2+8 \left(e^B-1\right)^2\Big] +e^{2 B} r^4
   \phi '\bigg\}\nn\\  
  &+ 4 e^{2B} 2 \,\L \biggl\{-e^{2B} r^3 (-2+rA') \f' -16 A' \dot f^2 \f' \left[ 6(3-4e^B +e^{2B}) +(-5 +e^B)r A' \right] \nn\\
  &+ 4 e^B \dot f \Bigl[ -3 r^2 A'^2(1+e^B) +4 \Bigl(4(-1+e^B)^2-r^2\f'^2\Bigr)+2rA'(3-3e^B+r^2\f'^2)\Bigr]\biggr\} \nn\\
  &-2e^{2B} r \f' \biggl\{-8\dot f \f' \Big[4 e^B (-1+e^B) + r^2 \f'^2 (-2+e^B)\Big]
 -4re^B(-1+e^B) \nn\\
  &-r\f'^2\Bigl[ r^2e^B-16\ddot f(-1+e^B) \Bigr]\biggr\}
  - A' e^B \biggl\{ 32 r \dot f^2 \f^2 \f' (9-4e^B+3e^{2B}) \nn\\
  &-r^3\f' e^B\Big[4e^B(1+e^B)-\f'^2\Bigl(r^2 e^B+16\ddot f (1+e^B)\Big) \Big]\nn\\
  &+8 e^B \dot f \Big[ 4(-1+e^B)^2  +r^2 \f'^2(-7+3e^B)
  -2\f'^4(r^4 +8r^2\ddot f)\Big]\bigg\},
  \end{align}
\begin{align}
S&=2304 A'  \dot f^3 \phi'^2 + 8 e^B \Big[  -128 r A'  \dot f^2 \phi '-448 A' \dot f^3
   \phi'^2+32 r^2 \dot f^2 \phi'^3-80 \dot f^2 \phi' \Big] \nn\\
   &+8e^{2B} \Big[ 16 r^2 A' f'+160 r A' \dot f^2 \phi '+160 A'
   \dot f^3 \phi '^2-12 r^3 \dot f \phi'^2 -16 r^2 \dot f^2 \phi '^3\nn\\
   &-64 \Lambda  r^2 \dot f^2 \phi '+16 r \dot f+160
   \dot f^2 \phi ' \Big] + 8 e^{3B} \Big[ -16 r^2 A' \dot f
   -32 r A' \dot f^2 \phi '+4 r^3 \dot f \phi '^2 \nn\\
   &+16 \Lambda  r^3 \dot f+64 \Lambda  r^2 \dot f^2 \phi '-32 r \dot f
   -80 \dot f^2 \phi
   '+r^4 \phi ' \Big] +8 e^{4B} \Big[  16 r \dot f-16 \Lambda  r^3 \dot f \Big],
\end{align}
and
\begin{align}
Q&=2304 A'  \dot f^2 \ddot f  \phi'^4-1152
   A'^2 \dot f^3 \phi'^3 + e^B \Big[-144 r^2 A' \dot f^2 \phi'^4+672 r
   A'^2 \dot f^2 \phi '^2\nn\\
   &+768 A'^2 \dot f^3 \phi '^3-384 A'
   \dot f^2 \phi '^2-1024 r A' \dot f \ddot f
   \phi '^3-3584 A' \dot f^2 \ddot f \phi
   '^4\nn\\
   &+480 r \dot f^2 \phi '^4+64 r^2
   \dot f \ddot f \phi '^5-640 \dot f \ddot f \phi '^3 \Big] 
   + e^{2B} \Big[ 128 r^2 A' \ddot f \phi '^2+52 r^3 A' \dot f \phi
   '^3\nn\\
   &+80 r^2 A' \dot f^2 \phi '^4-128
   r^2 A'^2 \dot f \phi '-576 \Lambda  r^2 A'
   \dot f^2 \phi '^2-320 r A'^2
   \dot f^2 \phi '^2\nn\\
   &+176 r A' \dot f \phi '-128
   A'^2 \dot f^3 \phi '^3+640 A'
   \dot f^2 \phi '^2+1280 r A' \dot f \ddot f
   \phi '^3 +1280 A' \dot f^2 \ddot f \phi'^4\nn\\
   &-16 r^3 \ddot f \phi '^4+128 r \ddot f
   \phi '^2-4 r^4 \dot f \phi '^5-152 r^2
   \dot f \phi '^3 -256 r \dot f^2 \phi
   '^4+384 \Lambda  r \dot f^2 \phi'^2\nn\\
   &+160 \dot f \phi '-64 r^2 \dot f \ddot f \phi'^5
   -512 \Lambda  r^2 \dot f \ddot f \phi '^3+1280
   \dot f \ddot f \phi '^3\Big] + e^{3B} \Big[ -128 r^2 A' \ddot f \phi '^2\nn\\
   &+208 \Lambda  r^3 A' \dot f \phi '+32 r^2
    A'^2 \dot f \phi '+320 \Lambda  r^2 A'
   \dot f^2 \phi '^2+32 r A'^2
   \dot f'^2 \phi '^2-224 r A' \dot f \phi '\nn\\
   &-256
   A' \dot f^2 \phi '^2-256 r A' \dot f \ddot f
   \phi '^3-6 r^4 A' \phi '^2+8 r^3
   A'^2-24 r^2 A'+16 r^3 \ddot f \phi '^4\nn\\
   &+128 \Lambda  r^3 \ddot f \phi '^2-256 r \ddot f \phi
   '^2+16 \Lambda  r^4 \dot f \phi '^3+24 r^2 \dot f \phi
   '^3 -12 r^3 A' \dot f \phi'^3 +32 r \dot f^2 \phi '^4\nn\\
   &+224 \Lambda  r^2 \dot f \phi '-512 \Lambda  r \dot f^2
   \phi '^2-320 \dot f \phi '
   +512 \Lambda  r^2 \dot f \ddot f
   \phi '^3-640 \dot f \ddot f \phi '^3+r^5
   \phi '^4\nn\\
   &+12 r^3 \phi '^2-32 r \Big] +e^{4B} \Big[ -48 \Lambda  r^3 A' \dot f \phi '
   +48 r A' \dot f \phi '-24 \Lambda  r^4
   A'+24 r^2 A'\nn\\
   &-128 \Lambda  r^3 \ddot f \phi '^2+128 r
   \ddot f \phi '^2+128 \Lambda ^2 r^4 \dot f \phi'
   -224 \Lambda  r^2 \dot f
   \phi '+128 \Lambda  r \dot f^2 \phi
   '^2+160 \dot f \phi '\nn\\
   &-4 \Lambda  r^5 \phi'^2+4 r^3 \phi
   '^2-64 \Lambda  r^3+64 r \Big] +e^{5B} \Big[  -32 \Lambda ^2 r^5+
   64 \Lambda  r^3-32 r \Big].
\end{align}


\section{Variation with respect to the Riemann tensor}
\label{appb2}

Here, we derive the derivative of the Lagrangian of the theory (\ref{action3}) with respect to
the Riemann tensor. A simple way to do this is to take the derivative  ignoring the symmetries,
that the final expression should possess, and restore them afterwards. For example, if
$A_{abcd}$ is a 4-rank tensor and $A$ is the corresponding scalar quantity, we may write:
\begin{align}
\frac{\partial A}{\partial A_{abcd}}&=\frac{\partial}{\partial A_{abcd}}\left(g^{\mu\rho}g^{\nu\sigma}A_{\mu\nu\rho\sigma}\right)=
g^{\mu\rho}g^{\nu\sigma}\delta^a_\mu\delta^b_\nu\delta^c_\rho\delta^d_\sigma=g^{ac}g^{bd}\,.
\end{align}
Now, if $A_{abcd}=R_{abcd}$, it should satisfy the following relations:
\begin{equation}\label{101}
A_{abcd}=A_{cdab}=-A_{abdc}\; \text{and}\; A_{abcd}+A_{acdb}+A_{adbc}=0.
\end{equation}
Restoring the symmetries, we arrive at:
\begin{equation}
\frac{\partial R}{\partial R_{abcd}}=\frac{1}{2}\left(g^{ac}g^{bd}-g^{bc}g^{ad}\right).
\end{equation}
Alternatively, we could have explicitly written:
\begin{align}
\frac{\partial R}{\partial R_{abcd}}&=\frac{\partial}{\partial R_{abcd}}\left(g^{\mu\rho}g^{\nu\sigma}R_{\mu\nu\rho\sigma}\right)=\frac{1}{2}g^{\mu\rho}g^{\nu\sigma}\frac{\partial}{\partial R_{abcd}}\left(R_{\mu\nu\rho\sigma}-R_{\nu\mu\rho\sigma}\right)\nonumber\\[1mm]
&=\frac{1}{2}g^{\mu\rho}g^{\nu\sigma}\left(\delta^a_\mu\delta^b_\nu\delta^c_\rho\delta^d_\sigma-\delta^a_\nu\delta^b_\mu\delta^c_\rho\delta^d_\sigma\right)=\frac{1}{2}\left(g^{ac}g^{bd}-g^{bc}g^{ad}\right),
\end{align}
which clearly furnishes the same result. 

We now proceed to the higher derivative terms. Let us start with the Kretchmann scalar for which we find
\begin{equation}
\frac{\partial R_{\mu\nu\rho\sigma} R^{\mu\nu\rho\sigma}}{\partial R_{abcd}}=2R^{\mu\nu\rho\sigma}\frac{\partial R_{\mu\nu\rho\sigma}}{\partial R_{abcd}}=2R^{abcd},
\end{equation}
The above result does not need any correction as it is already proportional to $R_{abcd}$,
and satisfies all the desired identities. We now move to the $R_{\mu\nu} R^{\mu\nu}$
term, and employ again the simple method used above. Then: 
\begin{equation}
\frac{\partial A_{\mu\nu}A^{\mu\nu}}{\partial A_{abcd}}=2A^{\mu\nu}\frac{\partial A_{\mu\nu}}{\partial A_{abcd}}=2A^{\mu\nu}g^{\kappa\lambda}\frac{\partial A_{\kappa\mu\lambda\nu}}{\partial A_{abcd}}=g^{ac}A^{bd}-g^{bc}A^{ad}.
\end{equation}
If $A_{abcd}=R_{abcd}$ and $A_{\mu\nu}=R_{\mu\nu}$, the above result will have all the right
properties if it is rewritten as
\begin{equation}
\frac{\partial R_{\mu\nu}R^{\mu\nu}}{\partial R_{abcd}}=\frac{1}{2}\left(g^{ac}R^{bd}-g^{bc}R^{ad}-g^{ad}R^{bc}+g^{bd}R^{ac}\right),
\end{equation}
which is indeed the correct result. Finally, we easily derive that
\begin{equation}
\frac{\partial R^2}{\partial R_{abcd}}=R\left(g^{ac}g^{bd}-g^{bc}g^{ad}\right).
\end{equation}

In order to compute the integral appearing in Eq. (\ref{S2}), we use the near-horizon
solution (\ref{A-rh3}-\ref{B-rh3}) for the metric functions and scalar field. Then
recalling that, near the horizon, the relations $A''\approx -A'^2$ and $B'\approx -A'$
also hold, we find the results
\begin{align*}
R^{0101}\big|_{\mathcal{H}} = -\frac{1}{4} e^{-A-2 B} \left(-2 A''+A'
   B'-A'^2\right)\big|_{\mathcal{H}} \rightarrow 0\,, &\\[1mm]
-2\left(g^{00}R^{11}-g^{10}R^{01}-g^{01}R^{10}+g^{11}R^{00}\right)\big|_{\mathcal{H}}
\rightarrow \frac{4}{r_h}e^{-A-2B}A'\big|_\mathcal{H}\approx -\frac{4b_1^2}{a_1r_h}\,,& \\[1mm]
g^{00}g^{11}R\big|_{\mathcal{H}} \rightarrow \frac{e^{-A-2B}}{r^2}\left(4rA'-2e^B\right)\big|_{\mathcal{H}}\approx \frac{4b_1^2}{a_1r_h} -\frac{2b_1}{a_1 r_h^2}\,,& \\[2mm]
g_{00}g_{11}\big|_{\mathcal{H}}=e^{A+B}\big|_{\mathcal{H}}\rightarrow a_1/b_1\,.&
\end{align*}
Substituting these into Eq. (\ref{S2}), we readily obtain the result (\ref{S2-final}).

\clearpage
\thispagestyle{empty}


\chapter{Wormholes in Einstein–scalar-Gauss-Bonnet theory}\label{appC}

\section{Field Equations} \label{App-equations}

Employing the metric (\ref{metric4}) in Eqs. (\ref{Einstein-eqs})-(\ref{scalar-eq}), the $(tt)$, $(\eta\eta)$
and $(\theta\theta)$ components of Einstein's equations take the form

 \scriptsize

\bea
 &\eta _0^6 \left(2 f'_1 \left(\phi ' \left(\dot{F} \left(f'_1{}^2-4
   f''_1\right)-2 f'_1 \ddot{F} \phi '\right)-2 \dot{F} f'_1 \phi ''\right)+e^{f_1}
   \left(f'_1{}^2+4 f''_1+\phi '^2\right)\right)+\eta^4\Big[  
   -4 \dot{F} \eta  f'_1 \phi '' \left(\eta  f'_1+4\right)    \nonumber\\[1mm]
   &+e^{f_1} \eta  \left(\eta  \left(4 f_1''+\phi '^2\right)+\eta 
  f'_1{}^2+8 f'_1\right) + 2 \phi ' \left(\dot{F} \left(f'_1 \left(\eta ^2  f'_1 {}^2-8\right)-4
   \eta  f''_1 \left(\eta  f'_1+2\right)\right)-2 \eta  f'_1 \ddot{F} \phi '
   \left(\eta  f'_1+4\right)\right)\Big]\nonumber\\[1mm]
&+\eta_0^4 \Big[ 2 \phi ' \left(\dot{F} \left(3 f'_1 \left(\eta ^2  f'_1 {}^2-4\right)-4
   \eta  f''_1 \left(3 \eta  f'_1+2\right)\right)-2 \ddot{F} \phi ' \left(\eta 
   f'_1+2\right) \left(3 \eta  f'_1-2\right)\right)-4 \dot{F} \phi '' \left(\eta 
   f'_1+2\right) \left(3 \eta  f'_1-2\right)\nonumber\\[1mm]
&+e^{f_1} \left(\eta  \left(3 \eta  \left(4 f''_1+ \phi '^2\right)+3 \eta
     f'_1 {}^2+8 f'_1\right)+4\right) \Big] +\eta \eta_0^2\Big[-4 \eta  \ddot{F}  \phi '^2 \left(\eta  f'_1 \left(3 \eta 
   f'_1+8\right)-4\right)+e^{f_1}\eta(4 +\eta(16 f_1' \nonumber\\[1mm]
 &+3\eta f_1'^2 +3\eta(\phi'^2+4f_1''))) +   2 \dot{F} \left(\phi ' \left(3 \eta ^3 \left(f'_1\right){}^3-4 \eta ^2 f''_1 
   \left(3 \eta  f_1'+4\right)-20 \eta  f_1'-16\right)-2 \eta  \phi '' \left(\eta 
   f_1' \left(3 \eta  f_1'+8\right)-4\right)\right)=0,\nonumber\\ \label{tteq}
\eea
\bea
  &e^{f_1}\Big[\eta ^3 \left(2 f_0' \left(\eta  f_1'+2\right)+f_1' \left(\eta  f_1'+4\right)-\eta 
   \phi'^2\right)+2 \eta _0^2 \left(\eta  \left(2 f_0' \left(\eta 
   f_1'+1\right)+f_1' \left(\eta  f_1'+2\right)-\eta  \phi'^2\right)-2\right)\nonumber\\
  &+ \eta _0^4 \left(f_1'{}^2+2 f_0' f_1'-\phi'^2\right)\Big]-2 \dot{F} f_0' \phi ' \left(12 \left(\eta ^2+\eta _0^2\right) \eta  f_1'+3
   \left(\eta ^2+\eta _0^2\right){}^2 f_1'{}^2+8 \eta ^2-4 \eta
   _0^2\right)=0,\label{hheq}
\eea
\bea
   &\eta^3\Big[ -4 \dot{F} f_0' \phi '' \left(\eta  f_1'+2\right)-4 f_0' \ddot{F}  \phi'^2 
   \left(\eta  f_1'+2\right)-2 \dot{F} \phi ' \left(-2 f_0' \left(\eta 
    f_1'{}^2-\eta  f_1''+f_1'\right)+2 f_0'' \left(\eta 
   f_1'+2\right)+f_0'{}^2 \left(\eta  f_1'+2\right)\right)\nonumber\\[1mm]
  &+ e^{f_1} \left(\eta  \left(2 \left(f_0''+f_1''\right)+ \phi'^2\right)+\eta   f_0'{}^2+2 f_0'+2 f_1'\right)\Big]
  +2\eta_0^2\Big[ e^{f_1} \left(\eta  \left(\eta  \left(2 \left(f_0''+f_1''\right)+\phi'^2\right)+\eta  f_0'{}^2+f_0'+f_1'\right)+2\right)\nonumber\\[1mm]
  &-2 \dot{F} \phi ' \left(f_0'
   \left(2-\eta  \left(2 \eta  f_1'{}^2-2 \eta 
   f_1''+f_1'\right)\right)+2 \eta  f_0'' \left(\eta  f_1'+1\right)+\eta 
   f_0'{}^2 \left(\eta  f_1'+1\right)\right) -4  \eta  f_0'   \left(\eta  f_1'+1\right)\left(\dot{F}  \phi ''+\ddot{F}
   \phi'^2  \right)\Big]\nonumber\\[1mm]
   &+\eta_0^4\Big[-4 f_0' f_1' \ddot{F}  \phi'^2-4 \dot{F} f_0' f_1' \phi ''-2 \dot{F} \phi ' \left(f_1'
   \left( f_0'{}^2-2 f_1' f_0'+2 f_0''\right)+2 f_0'
   f_1''\right)+e^{f_1} \left( f_0'{}^2+2
   \left(f_0''+f_1''\right)+ \phi'^2\right)\Big]=0, \nonumber\\ \label{ffeq}
\eea

 \normalsize
respectively. The scalar equation in turn yields 

 \scriptsize

\bea
  & \eta _0^6 \left(f_0' \left(4 \dot{F} f_1' f_1''-\dot{F}
   f_1'{}^3+e^{f_1} \phi '\right)+2 \dot{F} f_1'{}^2
   f_0''+\dot{F} f_0'{}^2 f_1'{}^2+2 e^{f_1} \phi
   ''+e^{f_1} f_1' \phi '\right) +\eta^4\Big[ \dot{F} \eta  f_0'{}^2 f_1' \left(\eta  f_1'+4\right)\nonumber\\[1mm]
   &+ f_0' \left(4 \dot{F} f_1' \left(\eta ^2 f_1''+2\right)+\eta  \left( \dot{F} \left(8f_1''  - \eta
    f_1'{}^3   \right)+e^{f_1} \eta  \phi
   '\right)\right)+\eta  \Big(2 \left(\dot{F} \left(\eta  f_1'{}^2 f_0''+4
    f_1' f_0''\right)+e^{f_1} \eta  \phi ''\right)\nonumber\\[1mm]
 &+e^{f_1} \phi ' \left(\eta  f_1'+4\right)\Big] +
 \eta_0^4\Big[ f_0' \left(12 \dot{F} f_1' \left(\eta ^2 f_1''+1\right)-3 \dot{F} \eta ^2
   f_1'{}^3+\eta  \left(8 \dot{F} f_1''+3 e^{f_1} \eta  \phi
   '\right)\right)+6 \dot{F} \eta ^2 f_1'{}^2 f_0''+8 \dot{F} \eta 
   f_1' f_0''\nonumber\\[1mm]
   &-8 \dot{F} f_0''+6 e^{f_1} \eta ^2 \phi '' +\dot{F} f_0'{}^2 \left(3 \eta ^2  f_1'{}^2+4 \eta 
   f_1'-4\right)+e^{f_1} \eta  \phi ' \left(3 \eta  f_1'+4\right)\Big] +\eta\eta_0^2\Big[\dot{F}
   \eta  f_0'{}^2 \left(3 \eta ^2 f_1'{}^2+8 \eta 
   f_1'-4\right) \nonumber\\[1mm]
   &+f_0' \Big(\dot{F} \left(-3 \eta ^3 f_1'{}^3+4
   \eta  f_1' \left(3 \eta ^2 f_1''+5\right)+16 \left(\eta ^2
   f_1''+1\right)\right)+3 e^{f_1} \eta ^3 \phi '\Big)+\eta  \Big(2 \dot{F} f_0'' \left(3 \eta ^2 f_1'{}^2+8 \eta 
   f_1'-4\right)\nonumber\\[1mm]
   &+6 e^{f_1} \eta ^2 \phi ''+e^{f_1} \eta  \phi ' \left(3 \eta 
   f_1'+8\right)\Big)\Big]=0.\label{sceq}
\eea
\normalsize

In the above equations, the prime denotes differentiation with respect to the radial coordinate $\eta$.


\clearpage
 
\thispagestyle{empty}




\addcontentsline{toc}{chapter}{\numberline{}Bibliography}
\bibliographystyle{utphys}
\bibliography{Bibliography}

\end{document}